\documentclass[useAMS,usenatbib]{mnras}

\usepackage{mathptmx}
\usepackage{mathtools}
\usepackage[T1]{fontenc}
\usepackage{txfonts}
\usepackage{ae,aecompl}
\usepackage{soul}
\usepackage{array}
\usepackage{amssymb}
\usepackage{hyperref}
\usepackage{pdflscape}
\usepackage{cancel}
\usepackage{empheq}

\usepackage[T1]{fontenc}
\usepackage{ae,aecompl}

\usepackage{soul}
\usepackage{indentfirst}
\usepackage{array}
\usepackage{times}

\usepackage{float}
\usepackage{graphicx}	% Including figure files

\usepackage{amsmath}	% Advanced maths commands
\usepackage{amssymb}	% Extra maths symbols
\usepackage{threeparttable}
\usepackage{subfigure}
\usepackage[all]{hypcap}

\usepackage{color}
\usepackage{pdflscape}
\usepackage[normalem]{ulem}
\def\mnras{MNRAS}
\def\apjl{ApJL }
\def\aj{AJ }
\def\apj{ApJ }
\def\pasj{PASJ}
\def\pasp{PASP }

\def\apjs{ApJS }
\def\araa{ARAA }
\def\aap{A\&A }

\hypersetup{colorlinks=true,citecolor=blue,linkcolor=blue,filecolor=black,runcolor=black,breaklinks=true}
\definecolor{purple}{RGB}{160,32,240}
\definecolor{Cerulean}{RGB}{0,123,167}
\usepackage{etoolbox}

\newcommand{\mbh}{$M_{\bullet}$}
\newcommand{\mstar}{$M_{*}$}
\newcommand{\Msun}{M_{\odot}}
\newcommand{\mbulge}{$M_\mathrm{bulge}$}
\newcommand{\mpeak}{$M_\mathrm{peak}$}
\newcommand{\errorbars}{The error bars show the 68\% confidence intervals inferred from the model posterior distribution.}
\newcommand{\shadedregions}{The shaded regions show the 68\% confidence intervals inferred from the model posterior distribution.}
\newcommand{\halocurves}{The white solid lines are the average mass growth curves of haloes with $M_{\rm peak}=10^{12},10^{13},10^{14}$, and $10^{15} M_{\odot}$ at $z=0$.}
\newcommand{\sfcurve}{The yellow dashed line shows the halo mass at which the galaxy star-forming fraction $f_\mathrm{SF}$ is $0.5$ as a function of $z$.}
\newcommand{\waitforjwst}{The scaling relations at $z\ge 8$ are shown in dashed lines, which remain to be verified by future observations (by, e.g., \textit{JWST}).}
\newcommand{\nohalos}{The grey area shows where the number densities of dark matter haloes are negligible, and is therefore labeled as ``No Haloes.''}
\newcommand{\bhbm}{$M_\bullet$--$M_\mathrm{bulge}$}
\newcommand{\bhsm}{$M_\bullet$--$M_*$}
\newcommand{\bhhm}{$M_\bullet$--$M_\mathrm{peak}$}
\newcommand{\eff}{$\epsilon_\mathrm{tot}$}
\newcommand{\slx}{s$L_\mathrm{X}$}
\newcommand{\alphactk}{$\alpha_\mathrm{CTK}$}
\newcommand{\rhobh}{$\rho_\mathrm{BH}$}

\title[\textsc{Trinity I}: Halo--Galaxy--SMBH Connection from $z=0-10$]{\textsc{Trinity} I: Self-Consistently Modeling the Dark Matter Halo--Galaxy--Supermassive Black Hole Connection from $z=0-10$}

\author[H. Zhang et al.]{
Haowen Zhang,$^{1}$\thanks{E-mail: hwzhang0595@email.arizona.edu}
Peter Behroozi,$^{1,2}$
Marta Volonteri,$^{3}$
Joseph Silk,$^{3,4,5}$
\newauthor{
Xiaohui Fan,$^{1}$
Philip F. Hopkins,$^{6}$
Jinyi Yang,$^{1,\ast}$
and James Aird$^{7,8}$
}
\\
$^{1}$University of Arizona, 933 N Cherry Ave., Tucson, AZ 85721, USA, \\
$^{2}$Division of Science, National Astronomical Observatory of Japan, 2-21-1 Osawa, Mitaka, Tokyo 181-8588, Japan, \\
$^{3}$Institut d'Astrophysique de Paris (UMR 7095: CNRS \& Sorbonne Universite), 98 bis Bd. Arago, F-75014, Paris, France\\
$^{4}$Department of Physics and Astronomy, Johns Hopkins University, Baltimore, MD 21218, USA\\
$^{5}$BIPAC, Department of Physics, University of Oxford, Keble Road, Oxford OX1 3RH, UK\\
$^{6}$TAPIR, Mailcode 350-17, California Institute of Technology, Pasadena, CA 91125, USA\\
$^{7}$Institute for Astronomy, University of Edinburgh, Royal Observatory, Edinburgh EH9 3HJ, UK\\
$^{8}$Department of Physics and Astronomy, University of Leicester, University Road, Leicester LE1 7RH, UK\\
$^{\ast}$Strittmatter Fellow
}

\date{Accepted XXX. Received YYY; in original form ZZZ}

\pubyear{2022}

\begin{document}
\label{firstpage}
\pagerange{\pageref{firstpage}--\pageref{lastpage}}
\maketitle

\begin{abstract}
We present \textsc{Trinity}, a flexible empirical model that self-consistently infers the statistical connection between dark matter haloes, galaxies, and supermassive black holes (SMBHs). \textsc{Trinity} is constrained by galaxy observables from $0<z<10$ (galaxies' stellar mass functions, specific and cosmic SFRs, quenched fractions, and UV luminosity functions) and SMBH observables from $0<z<6.5$ (quasar luminosity functions, quasar probability distribution functions, active black hole mass functions, local SMBH mass--bulge mass relations, and the observed SMBH mass distributions of high redshift bright quasars). The model includes full treatment of observational systematics (e.g., AGN obscuration and errors in stellar masses).  From these data, \textsc{Trinity} infers the average SMBH mass, SMBH accretion rate, merger rate, and Eddington ratio distribution as functions of halo mass, galaxy stellar mass, and redshift. Key findings include: 1) the normalization and the slope of the SMBH mass--bulge mass relation increases mildly from $z=0$ to $z=10$; 2) The best-fitting AGN radiative$+$kinetic efficiency is $\sim 0.05-0.06$, but can range from $\sim 0.035-0.07$ with alternative input assumptions; 3) AGNs show downsizing, i.e., the Eddington ratios of more massive SMBHs start to decrease earlier than those of lower-mass objects;  4) The average ratio between average SMBH accretion rate and SFR is $\sim 10^{-3}$ for low-mass galaxies, which are primarily star-forming. This ratio increases to $\sim 10^{-1}$ for the most massive haloes below $z\sim 1$, where star formation is quenched but SMBHs continue to accrete.
\end{abstract}


%%%%%%%%%%%%%%%%%%%%%%%%%%%%%%%%%%%%%%%%%%%%%%%%%%

%%%%%%%%%%%%%%%%% BODY OF PAPER %%%%%%%%%%%%%%%%%%

\section{Introduction}
\label{s:introduction}

It is widely accepted that supermassive black holes (SMBHs) exist in the centres of most galaxies \citep{Kormendy1995,Magorrian1998,Ferrarese2000,Gebhardt2000,Tremaine2002,Ho2008,Gultekin2009,Kormendy2013,Heckman2014}. SMBHs are called active galactic nuclei (AGNs) during phases when they are accreting matter and releasing tremendous amounts of energy. With their potential for high energy output, SMBHs are leading candidates to regulate both the star formation of their host galaxies and their own mass accretion \citep{Silk1998,Bower2006,Somerville2008,Sijacki2015}. At the same time, galaxies may also influence SMBH growth via the physics of how gas reaches the central SMBH as well as via galaxy mergers. Hence, it is possible for both SMBHs and their host galaxies to influence each others' growth, also known as ``coevolution.''  As a result, constraining the interaction between SMBHs and their host galaxies is critical to our understanding of both galaxy and SMBH assembly histories \citep[see, e.g.,][]{Hopkins2006,Ho2008,Alexander2012,Kormendy2013,Heckman2014,Brandt2015}. 

The coevolution scenario is consistent with two key observations. First, relatively tight scaling relations ($\sim 0.3$ dex scatter) exist between SMBH masses, \mbh{}, and host galaxy dynamical properties (e.g., velocity dispersion, $\sigma$, or bulge mass, \mbulge{}, at $z\sim0$;  see \citealt{Haring2004,Gultekin2009,Kormendy2013,McConnell2013,Savorgnan2016}). Second, the cosmic SMBH accretion rate (CBHAR) density tracks the cosmic star formation rate (CSFR) density over $0<z<4$, with a roughly constant CBHAR/CSFR ratio between $10^{-4}-10^{-3}$ \citep{Merloni2004b, Silverman2008, Shankar2009, Aird2010, Delvecchio2014, Yang2018}.  At the same time, other predictions of the coevolution model (e.g., tight galaxy--SMBH property relationships at higher redshifts) have remained more difficult to verify. 

In the local Universe, galaxy--SMBH scaling relations (e.g., \bhbm{} or $M_\bullet-\sigma$) have been measured via high spatial resolution spectroscopy and dynamics modeling \citep[e.g.,][]{Magorrian1998,Ferrarese2005,McConnell2013}. Total (i.e., active$+$dormant) SMBH mass functions can be obtained by convolving these scaling relations with the distributions of galaxy properties, such as galaxy bulge mass function, or velocity dispersion functions \citep[e.g.,][]{Salucci1999,Marconi2004}. Beyond the local Universe, lower spatial resolution makes it impractical to measure individual SMBH masses in the same way. Hence,  SMBH mass measurements at $z>0$ rely on indirect methods such as reverberation mapping \citep{Blandford1982,Peterson2004} and empirical relations between SMBH mass, spectral line width, and AGN luminosity (i.e., ``virial'' estimates; \citealt{Vestergaard2006}).  All such indirect methods work only on actively-accreting SMBHs, which: 1) imposes a selection bias on the SMBHs included, and 2) makes it difficult to measure host galaxy masses at the same time. As a result, it has been even harder to obtain unbiased measurements of the galaxy--SMBH mass connection beyond $z=0$. 

There has also been great interest in measuring SMBH luminosity distributions, as these carry information about mass accretion rates.  At $z>0$, surveys have been carried out in X-ray, optical, infrared, and radio bands to identify AGNs and study their collective properties (see \citealt{Hopkins2007}, \citealt{Shen2020}, and references therein). As redshift increases (e.g., at $z\gtrsim2$), the AGN sample is biased towards brighter and rarer objects, due to the evolution of AGN populations and/or limited instrument capability.  Nonetheless, for lower-luminosity AGNs, it is often possible to measure both the SMBH luminosity and the mass of the host galaxy \citep[e.g.,][]{Bongiorno2012,Aird2018}.

Besides observational efforts, the galaxy--SMBH connection is a key ingredient in galaxy formation theory. Supernova feedback becomes inefficient in massive haloes; hence, to reproduce these haloes' low observed star formation rates, AGN feedback is widely implemented in hydrodynamical simulations and semi-analytic models (SAMs) for galaxy evolution \citep[see, e.g.,][]{Croton2006, Somerville2008,Dubois2012, Sijacki2015, Schaye2015, Weinberger2017}. These simulations allow studying the evolution of the galaxy--SMBH connection for individual galaxies. However, numerical simulations must make assumptions about physical mechanisms below their resolution limits, which complicates the interpretation of their results \citep[see, e.g.,][]{Habouzit2020}.

Empirical models are a complementary tool to study SMBHs. Instead of assuming specific physics, these models use observations to self-consistently and empirically characterize the properties of SMBHs and/or their connection with host galaxies. There are broadly two different categories of empirical models involving SMBHs. 

The first group of models solves the continuity equation for the SMBH mass function, linking the mass growth histories of SMBHs to their energy outputs. By comparing the local cosmic BH mass density with the total AGN energy output, these models provide estimates of the average radiative efficiency, duty cycles, and Eddington ratio distributions of AGNs \citep[see, e.g.,][]{Soltan1982, Small1992, Cavaliere2000, Yu2002, Steed2003, Marconi2004, Yu2004, Merloni2008, Shankar2009, Shankar2013, Aversa2015, Tucci2017}.

The second group of models focuses on the galaxy--SMBH or (galaxy--AGN) connection  \citep[e.g.,][]{Conroy2013,Caplar2015,Caplar2018,Yang2018,Comparat2019, Georgakakis2019,Carraro2020,Shankar2020Nat,Shankar2020MNRAS,Allevato2021}. Some of these models jointly infer the galaxy--SMBH mass scaling relation and SMBH accretion rate distributions. Previous models differ in terms of the flexibility in connecting the accretion rate distribution and the galaxy properties, as well as the datasets they try to fit. For example, \citet{Veale2014} used quasar luminosity functions (QLFs) to constrain several halo--galaxy--SMBH models, e.g., assigning AGN luminosities based on SMBH masses or accretion rates, and assuming log-normal or truncated power-law Eddington ratio distributions. They found that all these models could fit QLFs nearly equally well over $1<z<6$. This model degeneracy implies the need for data constraints beyond QLFs to fully characterize the galaxy--SMBH connection. 

In this paper, we present \textsc{Trinity}, an empirical model connecting dark matter haloes, galaxies, and SMBHs from $z=0-10$; \textsc{Trinity} extends the empirical DM halo--galaxy model from \citet{Behroozi2013}. Compared to previous empirical models, \textsc{Trinity} is constrained by a larger compilation of galaxy and AGN data, including not only quasar luminosity functions (QLFs), but also quasar probability distribution functions (QPDFs), active black hole mass functions (ABHMFs), the local bulge mass--SMBH mass relations, the observed SMBH mass distribution of high redshift bright quasars, galaxy stellar mass functions (SMFs), galaxy UV luminosity functions (UVLFs), galaxy quenched fractions (QFs), galaxy specific star formation rates (SSFRs), and cosmic star formation rates (CSFRs).  The enormous joint constraining power of this dataset allows \textsc{Trinity} to have both a more flexible parameterization as well as better constraints on the model parameters. In addition, \textsc{Trinity} features more realistic modeling of AGN observables by including, e.g., SMBH mergers and kinetic AGN luminosities in the model.

Similar to the model in \citet{Behroozi2013}, \textsc{Trinity} is built upon population statistics from a dark matter N-body simulation. Specifically, the model makes a guess for how haloes, galaxies and SMBHs evolve over time. This guess is then applied to the haloes in the simulation, resulting in a mock universe. This mock universe is compared with the real Universe in terms of the observables above, quantified by a Bayesian likelihood. With this likelihood, a Markov Chain Monte Carlo (MCMC) algorithm is used to explore model parameter space until convergence. The resultant parameter posterior distribution tells us the optimal way to connect galaxies and SMBHs to their host haloes, as well as the uncertainties therein.

This work is the first in a series of \textsc{Trinity} papers, and it covers the \textsc{Trinity} methodology. The second paper (Paper II) discusses quasar luminosity functions, the radiative vs.\ kinetic energy output from AGNs, and the buildup of SMBHs across cosmic time; the third paper (Paper III) provides predictions for quasars and other SMBHs at $z>6$; the fourth paper (Paper IV) discusses the SFR--BHAR correlation as a function of halo mass, galaxy mass, and redshift; and the fifth paper (Paper V) covers SMBH merger rates and \textsc{Trinity}'s predictions for gravitational wave experiments. The sixth (Paper VI) and seventh (Paper VII) papers present the AGN autocorrelation functions and AGN--galaxy cross-correlation functions from \textsc{Trinity}, respectively. They also discuss whether/how well AGN clustering signals can be used to constrain models like \textsc{Trinity}. Mock catalogues containing full information about haloes, galaxies, and SMBHs will be introduced in the sixth paper.

The paper is organized as follows. In \S \ref{s:method}, we describe the methodology. \S \ref{s:sims_and_data} covers the simulation and observations used in \textsc{Trinity}. \S \ref{s:results} presents the results of our model, followed by the discussion and comparison with other models in \S \ref{s:discussion}. Finally, we discuss the caveats of and the future directions for \textsc{Trinity} in \S\ref{s:caveats}, and present conclusions in \S \ref{s:conclusions}.  In this work, we adopt a flat $\Lambda$CDM cosmology with parameters ($\Omega_m=0.307$, $\Omega_{\mathrm{\Lambda}}=0.693$, $h=0.678$, $\sigma_8=0.823$, $n_s=0.96$) consistent with \textit{Planck} results \citep{Planck2016}. We use datasets that adopt the Chabrier stellar initial mass function \citep[IMF, ][]{Chabrier2003}, the \citet{Bruzual2003} stellar population synthesis (SPS) model, and the Calzetti dust attenuation law \citep{Calzetti2000}. Halo masses are calculated following the virial overdensity definition from \citet{Bryan1998}.

\section{Methodology}
\label{s:method}

\subsection{Overview}
\label{ss:overview}

\begin{figure*}
\includegraphics[width=1.5\columnwidth]{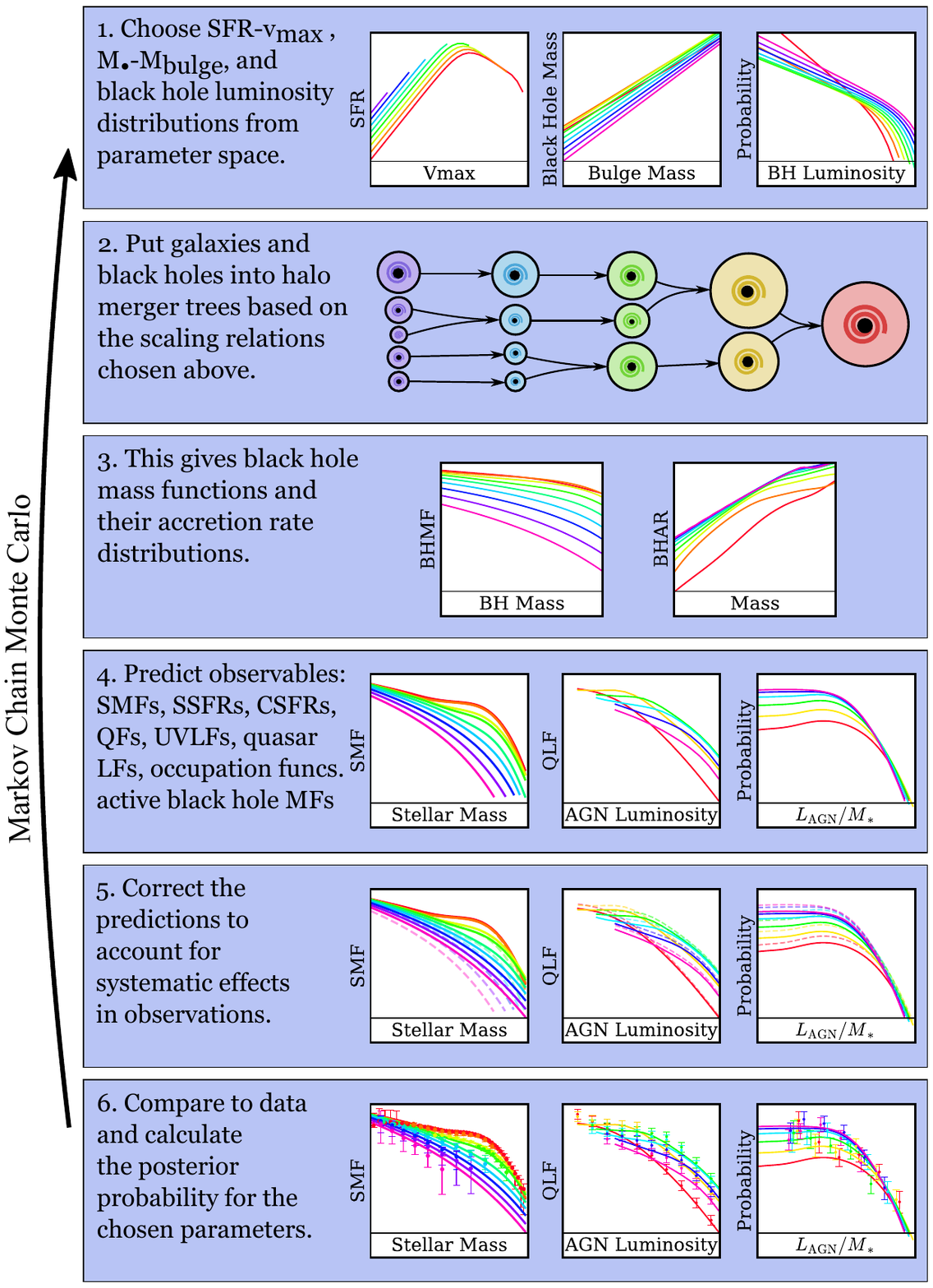}
\vspace{-3ex}
\caption{Visual summary of the methodology used to constrain the halo--galaxy--SMBH connection. See \S\ref{s:method} for details.}
\label{f:flow_chart}
\end{figure*}

\textsc{Trinity} is an empirical model that self-consistently infers the halo--galaxy--SMBH connection from $z=0-10$. In \textsc{Trinity}, we make this statistical connection in several steps (Fig.\ \ref{f:flow_chart}). We first parametrize the star formation rate (SFR) as a function of halo mass and redshift. For a given choice in this parameter space, we can integrate the resulting star formation rates (SFRs) over average halo assembly histories to get the stellar mass--halo mass (SMHM) relation (\S \ref{ss:halo_galaxy_connection}). We then convert total galaxy mass to \emph{bulge} mass with a scaling relation from observations. Next, we connect SMBHs with galaxies by parameterizing the redshift evolution of the SMBH mass--bulge mass (\bhbm{}) relation (\S \ref{ss:galaxy_smbh_connection}).  A given choice of this relation will determine average SMBH accretion rates, because average galaxy growth histories are set by the SFR--halo relationship. Lastly, we parameterize the Eddington ratio distributions and mass-to-energy conversion efficiency, which determines how SMBH growth translates to the observed distribution of SMBH luminosities. In brief, this modeling process gives the distribution of galaxy and SMBH properties. After modeling AGN radiative and kinetic luminosities (\S\ref{ss:bher_kin_rad}) as well as correcting for systematic effects, these properties are used to predict the galaxy and AGN observables (\S \ref{ss:agn_observables}). We compare these predictions to observations to compute a likelihood function, and use a Markov Chain Monte Carlo (MCMC) algorithm to obtain the posterior distribution of model parameters that are consistent with observations. Each choice of model parameters fully specifies the halo--galaxy--SMBH connection, and the posterior distribution provides the plausible range of uncertainties in this connection given observational constraints.

Of note, \textsc{Trinity} models ensemble populations of haloes, galaxies, and SMBHs by following different halo mass bins along average halo growth tracks (as in \citealt{Behroozi2013}), instead of tracking \emph{individual} halo and galaxy histories (as in the \textsc{UniverseMachine}; \citealt{Behroozi2019}). Given this statistical nature, \textsc{Trinity} is not yet able to provide object-specific growth histories. For calculation of average star formation histories in different halo mass bins, we refer readers to Appendix B of \citet{Behroozi2013}. In Appendix \ref{a:mbh_old_unmerged}, we lay out the procedure to calculate SMBH masses that: 1) were inherited from the most-massive progenitor (MMP) haloes; 2) came in with infalling satellite haloes. While a future version of \textsc{Trinity} will be integrated into the \textsc{UniverseMachine}, the present version requires only halo population statistics (i.e., halo mass functions and merger rates) from dark matter simulations (like \citealt[][]{Grylls2019}), as opposed to individual halo merger trees. As a result, \textsc{Trinity} allows extremely efficient computation of observables, and hence, rapid model exploration. 

\subsection{Connecting galaxies to haloes}
\label{ss:halo_galaxy_connection}

We adopt a very similar parameterization for the halo--galaxy connection in \textsc{Trinity} as was shown to work successfully in the \textsc{UniverseMachine} \citep{Behroozi2019}. Although simpler parameterizations exist, this choice makes future integration with the \textsc{UniverseMachine} easier.  The \textsc{UniverseMachine} modeled star-forming and quiescent haloes individually, but \textsc{Trinity} models halo population averages, and we maintain this parameterization in \textsc{Trinity}.  In practice, however, \textsc{Trinity} only depends on the total star formation rate of all haloes in a given mass bin, which depends almost exclusively on the star formation rate for star-forming galaxies and the quiescent fraction as a function of halo mass and redshift.

Our model assumes that the median star formation rates (SFRs) of star-forming galaxies are a function of both the host halo mass and redshift. In this work, we adopt the maximum circular velocity of the halo ($v_{\rm max}= \mathrm{max}\left(\sqrt{GM\left( < R\right)/R}\right)$) at the time when it reaches its peak mass, $v_{\mathrm{Mpeak}}$, as a proxy for the peak halo mass \mpeak{}. This choice reduces the sensitivity to pseudo-evolution in halo mass definitions and to spikes in $v_\mathrm{max}$ during mergers \citep{Behroozi2019}.  Our parameterization is:
\begin{eqnarray}
\mathrm{SFR}_{\mathrm{SF}} & = &\frac{\epsilon}{v^{\alpha} + v^{\beta}} \label{e:sfr_sf}\\
v & = &\frac{v_{\mathrm{Mpeak}}}{V\cdot\mathrm{km\ s}^{-1}} \label{e:v_1_normed}\\
a & = &\frac{1}{1+z} \label{e:aexp}\\ 
\log_{10}\left(V\right) & = &V_0 + V_{a}\left(a - 1)\right) + V_{z1}\ln \left(1 + z\right) + V_{z2}z\label{e:v_1} \\
\log_{10}\left(\epsilon\right) & = &\epsilon_0 + \epsilon_1 \left(a - 1\right) + \epsilon_{z1}\ln \left(1 + z\right) + \epsilon_{z2} z \label{e:epsilon}\\
\alpha & = &\alpha_0 + \alpha_a \left(a - 1\right) + \alpha_{z1} \ln \left(1 + z \right) + \alpha_{z2} z \label{e:alpha}\\
\beta & = &\beta_0 + \beta_a \left(a - 1\right) + \beta_z z \label{e:beta}\ .
\end{eqnarray}
The \emph{median} SFRs of star-forming galaxies ($\mathrm{SFR}_{\mathrm{SF}}$) are a power-law with slope $-\alpha$ for $v_{\rm Mpeak} \ll V$, and another power-law with slope $-\beta$ for $v_{\rm Mpeak} \gg V$. The parameter $\epsilon$ is the characteristic SFR when $v_{\rm Mpeak} \sim V$.  We remove the Gaussian boost in SFR at $v_{\rm Mpeak} \sim V$ in the \textsc{Universemachine}, because the \textsc{UniverseMachine}'s posterior distribution of model parameters suggested no need for such a boost.

We adopt the following parametrization for the fraction of quiescent galaxies, $f_{\mathrm{Q}}$, as a function of redshift and $v_\mathrm{Mpeak}$:
\begin{eqnarray}
\label{e:q_frac}
f_{\mathrm{Q}} & = & 1 - \frac{1}{1 + \exp\left(x\right)}\label{e:f_quenched}\\
x & = & \frac{\log _{10} (v_{\mathrm{Mpeak})} - v_{\rm Q}}{w_{\rm Q}}\\
v_{\rm Q} & = & v_{\mathrm{Q},0} + v_{\mathrm{Q},a}\left(a - 1\right) + v_{\mathrm{Q},z} z\label{e:m_quenched}\\
w_{\rm Q} & = & w_{\mathrm{Q},0} + w_{\mathrm{Q},a}\left(a - 1\right) + w_{\mathrm{Q},z} z\label{e:w_quenched}\ .
\end{eqnarray}
For quiescent galaxies, we assign a median SSFR of $10^{-11.8}$ yr$^{-1}$ to match SDSS values \citep{Behroozi2015}. We also set the log-normal scatter of the SFRs in star-forming and quiescent galaxies to be $\sigma_{\rm SFR, SF} = 0.30$ dex and $\sigma_{\rm SFR, Q} = 0.42$ dex, respectively \citep{Speagle2014}. Thus, the \emph{average total} SFR in each given $M_{\mathrm{peak}}$ (or $v_{\mathrm{Mpeak}}$) bin is simply:
\begin{equation}
\label{e:sfr_tot}
\begin{aligned}
\mathrm{SFR}_{\mathrm{tot}} &=\ \mathrm{SFR}_{\mathrm{SF}} \times (1 - f_{\mathrm{Q}}) \times \exp(0.5(\sigma_{\rm SFR,SF}\ln{10})^2) \\
&+ \mathrm{SSFR}_{\rm Q} \times M_* \times f_{\mathrm{Q}} \times \exp(0.5(\sigma_{\rm SFR,Q}\ln{10})^2)\ ,
\end{aligned}
\end{equation}
where the exponentials reflect the difference between the \emph{average} and \emph{median} values of log-normal distributions.

Aside from star formation, galaxies also gain stellar mass via mergers, where stars from incoming galaxies are transferred to central galaxies. In this work, we assume that a certain fraction, $f_{\rm merge}$, of the stars from incoming galaxies are merged into the central galaxies. As in \citet{Behroozi2019}, we assume $f_{\rm merge}$ to be independent of halo mass due to the approximately self-similar nature of haloes. We also assume $f_{\rm merge}$ to be redshift-independent.  The average stellar mass in a given halo mass bin at a given redshift $z$ is correspondingly:

\begin{eqnarray}
\label{e:mstar_sfr}
& & M_{*}\left(t\right)\ = \int_0^{t}
\left(1 - f_{\mathrm{loss}}\left(t - t'\right)\right)
\mathrm{SFR}_{\mathrm{tot}}\left(t'\right)dt'\nonumber\\
& & +\ f_{\rm merge}\int_0^{t}\int_0^{t'}\left(1 - f_{\mathrm{loss}}\left(t - t''\right)\right)\dot{M}_{*,\rm inc}\left(t',\ t''\right)dt''dt'\label{e:sfh_integral}\\
& & f_{\mathrm{loss}}\left(T\right)\ = 0.05\ \mathrm{ln}\left(1 + \frac{T}{1.4\ \mathrm{Myr}}\right)\label{e:sm_loss}\ ,
\end{eqnarray}
where $f_{\mathrm{loss}}\left(T\right)$ is the stellar mass loss fraction as a function of stellar age $T$ from \citet{Behroozi2013}, $\mathrm{SFR}_{\mathrm{tot}}$ is the total average SFR from Eq.\ \ref{e:sfr_tot}, and $\dot{M}_{*,\rm inc}\left(t',\ t''\right)$ is the rate at which the incoming satellite galaxies merge into central galaxies., as a function of the time of disruption $t'$ and the time that the stellar population formed, $t''$. For a given halo mass bin around the descendant halo mass, $M_\mathrm{desc}$, $\dot{M}_{*,\rm inc}\left(t',\ t''\right)$ can be calculated by convolving the halo merger rates from the \textsc{UniverseMachine} ($d^2 N(M_\mathrm{desc},\ \theta,\ z(t'))/(d\log \theta\ dz)$, see Appendix \ref{a:halo_merger_rates}) with the star formation histories of merged satellite haloes:

\begin{eqnarray}
\label{e:galaxy_merger_rate}
\dot{M}_{*,\rm inc}\left(t',\ t''\right) & = & \int_{0}^{1}\frac{d^2 N(M_\mathrm{desc},\ \theta,\ z(t'))}{d\log \theta\ dz} \mathrm{SFR}(M_\mathrm{sat},\  t'')d\log \theta\nonumber\\
& \times & \frac{dz}{dt'}\ ,
\end{eqnarray}
where $M_\mathrm{sat}$ is the mass of the satellite halo, and $\theta=M_\mathrm{sat}/M_\mathrm{desc}$ is the mass ratio between the satellite halo and the descendant halo.

It is also well-known that there is scatter in stellar mass at fixed halo mass \citep[see, e.g.,][]{Wechsler2018}. We parametrize this scatter as a log-normal distribution with a width $\sigma_*$ that is redshift-independent, with a flat prior on $\sigma_{*,0}$ of 0--0.3 dex. 

\begin{figure}
\vspace{-0.35cm}
\includegraphics[width=0.48\textwidth]{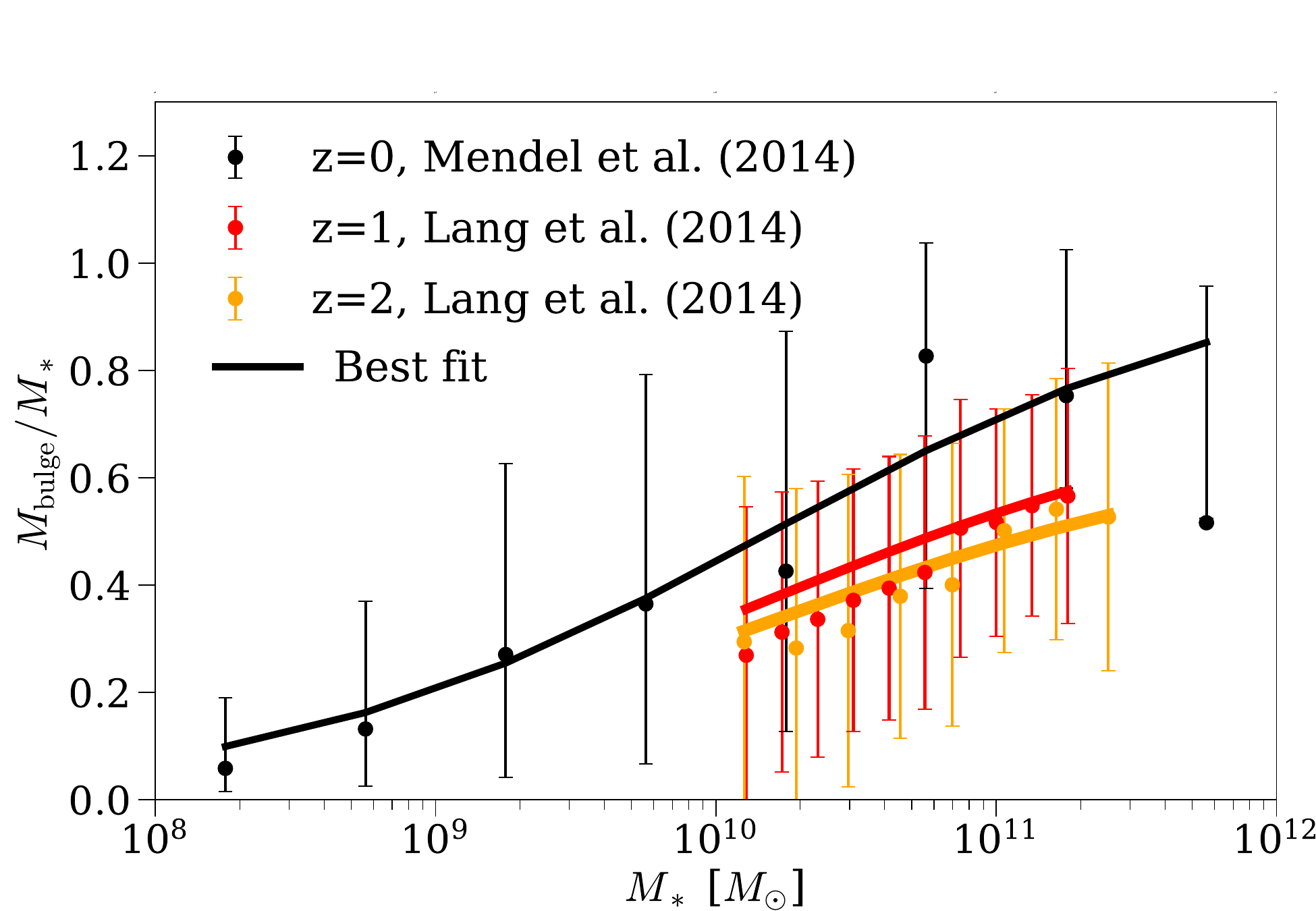}
\caption{The fit to the median galaxy bulge mass--total mass relation for $z=0-2$ (solid lines, Eqs.\ \ref{e:bm_sm}-\ref{e:bm_sm_fz}).  Observed data points are from \citet{Mendel2014} and \citet{Lang2014}. The error bars from \citet{Mendel2014} represent the 16-84$^{\mathrm{th}}$ percentile range of the $M_\mathrm{bulge}/M_*$ ratios from the SDSS catalog, and those from \citet{Lang2014} are based on the 68\% confidence intervals of bulge-to-total ratio ($B/T$) as a function of stellar mass. All the data used to make this plot (including individual data points and our best-fitting model) can be found \href{https://github.com/HaowenZhang/TRINITY/tree/main/plot_data}{here}.}
\label{f:bmsm_fit}
\end{figure}

The galaxy--SMBH connection is made via the SMBH mass--bulge mass (\bhbm{}, \S \ref{ss:galaxy_smbh_connection}) relation. To make the halo--galaxy--SMBH connection, we need to convert \emph{total} galaxy mass $M_*$ to the \emph{bulge} mass $M_{\rm bulge}$. In this work, we fit the median bulge mass--total mass relations from SDSS \citep{Mendel2014} and CANDELS \citep{Lang2014} galaxies with:
\begin{eqnarray}
 M_{\mathrm{bulge}} & = & \frac{f_z(z) M_*}{1 + \exp\{k_\mathrm{SB}\left[\log_{10} (M_* / M_\mathrm{SB})\right]\}}\label{e:bm_sm}\\ 
 f_z(z) & = & \frac{z + 2}{2z + 2}\label{e:bm_sm_fz}\ ,
\end{eqnarray}
where $k_\mathrm{SB}=-1.13$ determines how fast \mbulge{} converges to \mstar{} at the massive end, and $M_\mathrm{SB}=10^{10.2}M_\odot$ is a characteristic stellar mass. This fit is shown in Fig.\ \ref{f:bmsm_fit}. It should be noted that no data points exist beyond $z=2.5$, so Eq.\ \ref{e:bm_sm} is extrapolated at $z>2.5$.  With the functional form chosen here, $M_\mathrm{bulge}/M_*$ asymptotes at high redshifts to half the value of $M_\mathrm{bulge}/M_*$ at $z=0$. We discuss how alternative assumptions for the \mbulge{}--\mstar{} relation would affect our results in Appendix \ref{a:alt_galaxy_smbh}.

Disk--bulge decompositions are sensitive to the fitting method used, and it is also difficult to estimate how much of the scatter in bulge-to-total mass ratios is intrinsic vs.\ observational.  As a result, we subsume the scatter in the bulge-to-total mass relation into the scatter of the \bhbm{} relation, as the two scatters are degenerate in \textsc{Trinity} given current data constraints.

At $0<z<8$, stellar mass functions (SMFs) primarily constrain the halo--galaxy connection. Beyond $z=8$, SMFs are not available, so we constrain the halo--galaxy connection with galaxy UV luminosity functions instead. This requires generating UV luminosities from SFRs as a function of host halo mass and redshift. To do so, we fit the median UV magnitude, $\widetilde{M}_{\rm UV}$, and the log-normal scatter, $\sigma_{M_{\rm UV}}$, as functions of SFR, $M_{\rm peak}$, and redshift from the output of the \textsc{UniverseMachine}:
\begin{eqnarray}
\widetilde{M}_{\rm UV} & = & k_{\rm UV} \times \log_{10} \mathrm{SFR} + b_{\rm UV} \label{e:uv_med}\\
\sigma_{M_{\rm UV}} & = & k_{\sigma_{\rm UV}} \times \log_{10} M_{\rm peak} + b_{\sigma_{\rm UV}}\\
k_{\rm UV} & = & 
\begin{aligned}
    &a_k\left(\log_{10}M_{\rm peak}\right)^2 + b_k\log_{10}M_{\rm peak}\\
    & + c_k\left(a - 1\right) + d_k
\end{aligned}\\
b_{\rm UV} & = & 
\begin{aligned}
    &a_b\left(\log_{10}M_{\rm peak}\right)^2 + b_b\log_{10}M_{\rm peak}\\
    & + c_b\left(a - 1\right) + d_b
\end{aligned}\\
k_{\sigma_{\rm UV}} & = & a_{k_{\sigma}} z + b_{k_{\sigma}}\\
b_{\sigma_{\rm UV}} & = & a_{b_{\sigma}} z + b_{b_{\sigma}}\label{e:b_std_uv}\ .
\end{eqnarray}
Details of the fitting process are shown in Appendix \ref{a:UV_SFR_fit}.  \textsc{UniverseMachine} models UV luminosities using the Flexible Stellar Population Synthesis code \citep[FSPS;][]{Conroy2013}, and Eqs.\ \ref{e:uv_med}-\ref{e:b_std_uv} provide a rapid way to obtain statistically equivalent results. We hence use these scaling relations to assign UV magnitude distributions to haloes given their masses, SFRs and redshifts, allowing us to calculate UVLFs at $z=9$ and $z=10$.

\subsection{Observational systematics for galaxies}
\label{ss:galaxy_sys}

Following \citet{Behroozi2019}, we model several observational systematics when predicting galaxy observables.  We include a mass-independent systematic offset $\mu$ between the observed ($M_{*,\mathrm{obs}}$) and the true stellar mass ($M_{*,\mathrm{true}}$) to model uncertainties from the IMF, SPS model, the dust model, the star formation history (SFH) model, assumed metallicities, and redshift errors:
\begin{equation}
\label{e:mu}
    \log_{10}\left(\frac{M_{*,\mathrm{obs}}}{M_{*,\mathrm{true}}}\right) = \mu\ .
\end{equation}
The offset $\mu$ has the following redshift scaling:
\begin{equation}
\begin{aligned}
\label{e:mu_z}
\mu &= \mu_0 + \mu_a \left(a - 1\right)\ .
\end{aligned}
\end{equation}
Following \citet{Behroozi2013}, we set the prior width on $\mu_0$ and $\mu_a$ to 0.14 and 0.24 dex, respectively (see Table \ref{t:priors}).

As described in Appendix C of \citet{Behroozi2019}, there are systematic offsets between observed and true specific star formation rates that peak near $z\sim 2$, which are most evident when comparing observed specific star formation rates to the evolution of observed SMFs.  As in \citet{Behroozi2019}, we include another redshift-dependent offset $\kappa$ to account for this systematic offset in star formation rates. The total offset between the observed ($\mathrm{SFR}_{*,\mathrm{obs}}$) and true SFRs ($\mathrm{SFR}_{*,\mathrm{true}}$) is:
\begin{equation}
\label{e:mu_kappa}
    \log_{10}\left(\frac{\mathrm{SFR}_{*,\mathrm{obs}}}{\mathrm{SFR}_{*,\mathrm{true}}}\right) = \mu + \kappa\exp{\left(-\frac{(z-2)^2}{2}\right)}\ .
\end{equation}
The prior width on $\kappa$ is set to 0.24 dex (Table \ref{t:priors}), again from \cite{Behroozi2019}.

We also model a redshift-dependent, log-normal scatter in the measured stellar mass relative to the true mass:
\begin{equation}
\label{e:sm_scatter}
    \sigma\left(z\right)=\min (\sigma_0 + \sigma_z z, 0.3) .
\end{equation}
This scatter causes an Eddington bias \citep{Eddington1913} in the SMF, which enhances the number density of massive galaxies because there are more small galaxies that can be scattered up than massive galaxies that can be scattered down. Following \citet{Conroy2013}, we fix $\sigma_0=0.07$ dex. We adopt a Gaussian prior on $\sigma_z$ with centre 0.05 and width 0.015 dex, respectively (see Table \ref{t:priors}), following \citet{Behroozi2019}. 

Finally, the correlation between scatter in the star formation rate and scatter in the stellar mass at fixed halo mass affects the calculation of SSFRs as a function of stellar mass. To account for this correlation $\rho$, we adopt the following formula from \citet{Behroozi2013}:
\begin{equation}
\label{e:rho05}
\rho\left(a\right) = 1 + \left(4\rho_{0.5} - 3.23\right)a + \left(2.46 - 4\rho_{0.5}\right)a^2\ ,
\end{equation}
where $\rho_{0.5}$ is a free parameter that represents the correlation between the SSFR and stellar mass at $z=1$ (i.e., $a=0.5$). The details of this correction are in Appendix C.2 of \citet{Behroozi2013}. Following \citet{Behroozi2013}, we set the prior on $\rho_{0.5}$ to be a uniform distribution between 0.23 and 1.0 (Table \ref{t:priors}).

\subsection{Connecting SMBHs to galaxies}
\label{ss:galaxy_smbh_connection}

\subsubsection{SMBH occupation fractions}
\label{sss:smbh_focc}

In the real Universe, not every halo and galaxy host central SMBHs. That is, the occupation fraction of SMBHs, $f_\mathrm{occ}$, is likely below unity. At $z=0$, we find that most massive galaxies host central SMBHs, but it is still debated how many smaller and/or earlier galaxies are SMBH-occupied (see \citealt{Greene2020} and refereneces therein). Theoretical studies suggest that $f_\mathrm{occ}$ could be a sigmoid function of halo mass with potential redshift evolution (e.g., \citealt[][]{Volonteri2010,Bellovary2011,Dunn2018}). Therefore, we adopt this functional form in \textsc{Trinity}, and allow the following redshift dependence of 1) a minimum SMBH occupation fraction, $f_\mathrm{occ,min}$, 2) the characteristic halo mass, $M_\mathrm{h,c}$; and 3) the (log-)halo mass range, $w_\mathrm{h,c}$, over which $f_\mathrm{occ}$ changes significantly:

\begin{eqnarray}
f_\mathrm{occ}  & = & \frac{\exp{(x)}}{1+\exp{(x)} }\times (1 - f_\mathrm{occ,min}) + f_\mathrm{occ,min}\label{e:focc}\\
x & = & \frac{\log_{10}(M_\mathrm{peak}) - \log_{10}(M_\mathrm{h,c})}{w_\mathrm{h,c}}\label{e:focc_x}\\
\log_{10}(f_\mathrm{occ,min}) & = & f_\mathrm{occ,min,0} + f_\mathrm{occ,min,a}(a - 1)\label{e:focc_min}\\
\log_{10} (M_\mathrm{h,c}) & = & M_{\mathrm{h,c},0} + M_{\mathrm{h,c},a} \left(a - 1\right)\label{e:mh_c}\\
w_\mathrm{h,c} & = & w_{\mathrm{h,c},0} + w_{\mathrm{h,c},a} \left(a - 1\right)\label{e:wh_c}\ .
\end{eqnarray}
$f_\mathrm{occ,min}$ is motivated by the calculation of characteristic \mbh{} for \emph{host} galaxies, where $f_\mathrm{occ}$ is used as a denominator (see Eq. \ref{e:mbh_host}).

However, all the posterior parameter distributions of \textsc{Trinity} models--the fiducial models and the variants covered in the Appendix--predict $f_\mathrm{occ}\sim 1$ in the halo/galaxy mass ranges covered by \textsc{Trinity}. The physical reason is that, without new SMBH seeds at lower redshifts, $f_\mathrm{occ}$ at a fixed halo mass can only decrease as less massive, unseeded haloes grow in mass. On the other hand, a uniformly high $f_\mathrm{occ}$ down to $M_\mathrm{peak}\sim 10^{11} M_\odot$ in the local universe is required to explain AGN observations such as ABHMFs. As a result, $f_\mathrm{occ}$ can only be higher at $z>0$ for $M_\mathrm{peak} > 10^{11} M_\odot$, which leads to $f_\mathrm{occ}\sim 1$. 
This result is also consistent with earlier simulations of SMBH formation (e.g., \citealt{Tremmel2017PhDT,Habouzit2017}), which found $f_\mathrm{occ}\sim 1$ in haloes with $M_\mathrm{peak} > 10^{11} M_\odot$.

\subsubsection{Redshift-dependent \bhbm{} relation}
\label{sss:bhbm_relation}

There are multiple known empirical scaling relations between $M_\bullet$ and galaxy properties, with strong debate over which is most fundamental \citep{Ferrarese2000,Ferrarese2002,Novak2006,Aller2007,Hu2008,Beifiori2012,Shankar2016,vdB2016}.  Here, we parameterize the relation between SMBHs and galaxy bulge mass. Specifically, the \emph{median} \mbh{}--\mbulge{} relation is a redshift-dependent power-law:

\begin{eqnarray}
\log_{10} \widetilde{M}_{\bullet}  & = & \beta_{\mathrm{BH}} + \gamma_{\mathrm{BH}} \log_{10} \left(\frac{M_{\mathrm{bulge}}}{10^{11}M_{\odot}}\right) \label{e:bhbm}\\
\beta_{\mathrm{BH}} & = & \beta_{\mathrm{BH},0} + \beta_{\mathrm{BH},a} \left(a - 1\right) + \beta_{\mathrm{BH},z} z \label{e:beta_bh}\\
\gamma_{\mathrm{BH}} & = & \gamma_{\mathrm{BH},0} + \gamma_{\mathrm{BH},a} \left(a - 1\right) + \gamma_{\mathrm{BH},z} z \label{e:gamma_bh}\ .
\end{eqnarray}
We set Gaussian priors on $\beta_{\rm BH,0}$ and $\gamma_{\rm BH,0}$ from constraints on the local \bhbm{} relation, which will be discussed in \S\ref{sss:smbh_data}.  With Eqs.\ \ref{e:bhbm}-\ref{e:gamma_bh}, some parameter values could result in unphysical (i.e., negative) growth of SMBHs; we hence exclude such parts of parameter space from MCMC exploration.

There is also log-normal scatter in SMBH mass at fixed bulge mass ($\sigma_{\mathrm{BH}}$). We assume $\sigma_{\mathrm{BH}}$ to be redshift-independent. This is because a redshift dependent $\sigma_{\mathrm{BH}}$ will be unphysically small in the early Universe, if the Poisson prior probability of detecting low-mass bright quasars at $z\sim 6$ is applied. See \S\ref{sss:smbh_data} for more details. 

 Since the scatter in bulge mass at fixed stellar mass is subsumed in $\sigma_{\mathrm{BH}}$, this is in effect the scatter in SMBH mass at fixed \emph{total} stellar mass. We also note that this scatter is effectively the combined scatter that accounts for both the variance in the intrinsic \bhbm{} relation, as well as random error in direct  SMBH mass measurements (e.g., dynamical modelling or reverberation mapping, but not virial estimates). Combining the scatter in SMBH mass at fixed stellar mass with the scatter in stellar mass at fixed halo mass, the scatter in SMBH mass at fixed halo mass is:
\begin{equation}
\label{e:scatter_bhhm}
\sigma_{\mathrm{tot}} = \sqrt{\left(\sigma_* \times \gamma_{\mathrm{BH}}\right)^2 + \sigma_{\mathrm{BH}}^2}\ .
\end{equation}
Such a calculation effectively assumes that the bulge mass fraction of galaxies is fixed at fixed halo mass. This log-normal scatter results in a difference between the \emph{mean} ($\overline{M_\bullet}$) and \emph{median} SMBH masses  ($\widetilde{M}_\bullet$) at fixed halo mass:

\begin{equation}
\label{e:mbh_avg_med}
\overline{M_\bullet} = \widetilde{M}_\bullet \times \exp(0.5(\sigma_{\rm tot} \ln 10)^2)\ .
\end{equation}

We note that the median and average \mbh{}'s calculated above are for \emph{all} the galaxies, whether they host SMBHs or not. Generally, these masses are different from those for SMBH \emph{host galaxies}. With an SMBH occupation fraction $f_\mathrm{occ}<1$,  the median and average \mbh{}'s for SMBH host galaxies, $\overline{M_{\bullet}}_{,\mathrm{host}}$ and $\widetilde{M_{\bullet}}_{,\mathrm{host}}$, would be: 

\begin{eqnarray}
\overline{M_{\bullet}}_{,\mathrm{host}} & = & \frac{\overline{M_{\bullet}}}{f_\mathrm{occ}}\label{e:mbh_host}\\
\widetilde{M_{\bullet}}_{,\mathrm{host}} & = & \frac{\widetilde{M_{\bullet}}}{f_\mathrm{occ}}\ .
\end{eqnarray}

However, as we noted in \S\ref{sss:smbh_focc}, all the posterior parameter distributions of \textsc{Trinity} models predict $f_\mathrm{occ}\sim 1$ for $M_\mathrm{peak} > 10^{11} M_\odot$. Therefore, Eq.\ \ref{e:mbh_host} results in effectively identical SMBH properties for \emph{all} vs.\ \emph{host} haloes/galaxies, so we do not provide separate results for all vs.\ host haloes/galaxies in the rest of this work.

\subsection{SMBH mergers and accretion}
\label{ss:bh_mergers}

Similar to their host galaxies, SMBHs grow in mass via accretion and mergers. We parameterize the fraction of SMBH growth due to mergers as $f_{\mathrm{merge,BH}}$, the formula for which is provided later in this section. The average black hole merger rate ($\overline{\mathrm{BHMR}}$) for a certain halo mass bin is by definition:
\begin{equation}
\label{e:bhmr_def}
\centering
\begin{aligned}
\overline{\mathrm{BHMR}}\cdot\Delta t &= (\mathrm{Average\ BH\ Mass\ Now}\\
&- \mathrm{Average\ BH\ Mass\ Inherited\ from}\\ &\mathrm{Most\ Massive\ Progenitors} )\\
&\times f_{\mathrm{merge,BH}}\ ,
\end{aligned}
\end{equation}
where $\Delta t$ is the time interval between two consecutive snapshots, and the inherited and new BH masses are calculated using the halo--galaxy--SMBH connection (see Appendix \ref{a:mbh_old_unmerged} for full details). Similarly, the average black hole accretion rate ($\overline{\mathrm{BHAR}}$) for a certain halo mass bin is:
\begin{equation}
\label{e:bhar_def}
\centering
\begin{aligned}
\overline{\mathrm{BHAR}}\cdot\Delta t &= (\mathrm{Average\ BH\ Mass\ Now}\\
&- \mathrm{Average\ BH\ Mass\ Inherited\ from}\\ &\mathrm{Most\ Massive\ Progenitors} )\\
&\times \left(1 - f_{\mathrm{merge,BH}}\right)\ .
\end{aligned}
\end{equation}

In this work, we assume that the fractional merger contribution to the total SMBH growth ($f_{\mathrm{merge,BH}}$) is proportional to the fraction of galaxy growth due to mergers:
\begin{equation}
\label{e:f_merge_bh}
    f_{\rm merge,BH} = f_\mathrm{scale}\times \frac{f_{\rm merge}\dot{M}_{*,\rm inc}}{\mathrm{SFR} + f_{\rm merge}\dot{M}_{*,\rm inc}}\ ,
\end{equation}
where $f_{\rm merge}$ is the fraction of the incoming satellite galaxies' mass that is merged into central galaxies, and $\dot{M}_{*,\rm inc}$ is the mass rate at which satellite galaxies are disrupted in mergers (see Eq.\ \ref{e:sfh_integral}). The proportionality factor, $f_{\rm scale}$, has the following redshift dependency:
\begin{equation}
\label{e:f_merge_bh_z}
    \log_{10} (f_{\rm scale}) = f_\mathrm{scale,0} + f_\mathrm{scale, 1}  (a - 1)\ .
\end{equation}
While we do not exclude $f_\mathrm{scale}>1$ when exploring parameter space, we find $f_{\rm scale}$ to be consistently smaller than unity in the posterior distribution (see Appendix \ref{a:param_values} for model extremes where $f_\mathrm{scale} = 0$ or $f_\mathrm{scale}=1$).

In \textsc{Trinity}, not all infalling SMBH mass merges with the central SMBH immediately. Physically, this could be due to several reasons: 1) some SMBHs orbit with the disrupted satellite (i.e., in a tidal stream) outside the host galaxy and have very long dynamical friction timescales, 2) some SMBHs experience recoils and are ejected from the central galaxy; 3) some SMBHs may stall in the final parsec before merging with the central SMBH; or 4) some SMBHs may remain in the host galaxy but stay offset from the centre. Given the lack of direct observational evidence, we cannot distinguish between these possible scenarios here. Instead, we label all such objects as ``wandering SMBHs'' for the rest of this work. The average mass in wandering SMBHs ($\overline{M}_{\bullet, \rm wander}$) for each halo mass bin is thus:
\begin{equation}
\label{e:wandering_bh}
\centering
\begin{aligned}
\overline{M}_{\bullet, \rm wander} = &\mathrm{Total\ Infalling\ BH\ Mass}\\
- & \int_0^t \overline{\mathrm{BHMR}}\cdot dt\ .
\end{aligned}
\end{equation}
Although wandering SMBHs do not contribute to the observed \bhbm{} relation, we assume that they do contribute to quasar luminosity functions during their formation. For full details about calculating the average inherited SMBH mass from the previous timestep ($\overline{M}_{\bullet,\mathrm{inherit}}$) and the average infalling SMBH mass ($\overline{M}_{\bullet, \rm infall}$), see Appendix \ref{a:mbh_old_unmerged}.

\subsection{AGN duty cycles, Eddington ratio distributions, and energy efficiencies}
\label{ss:bher_dist}

As noted in \S\ref{ss:overview}, \textsc{Trinity} is not designed to follow the growth histories of individual haloes, galaxies, or SMBHs. Instead, \textsc{Trinity} gives their \emph{average} growth histories. To model AGN accretion rate distributions, it is therefore necessary to parametrize both the AGN duty cycles (i.e., the fraction of galaxies that host active SMBHs, $f_{\rm duty}$) and the shapes of their Eddington ratio distributions. $f_{\rm duty}$ is a function of \mpeak{} and $z$:
\begin{eqnarray}
f_{\mathrm{duty}}\left(M_{\mathrm{peak}},z\right) & = & \min\left\{\left(\frac{M_\mathrm{peak}}{M_\mathrm{duty}}\right)^{\alpha_\mathrm{duty}}, 1\right\}  \label{e:duty_cycle}\\
M_\mathrm{duty} & = & M_\mathrm{duty,0} + M_\mathrm{duty,z}\log (1 + z)\label{e:duty_cycle_mh}\\
\alpha_\mathrm{duty} & = & \alpha_\mathrm{duty,0} + \alpha_\mathrm{duty,z}\log (1 + z) \label{e:duty_cycle_alpha}\ ,
\end{eqnarray}
In this work, we define $f_\mathrm{duty}$ to be the fraction of active SMBH hosts relative to \emph{all} galaxies. But given that the posterior distributions of all \textsc{Trinity} models predict $f_\mathrm{occ}\sim 1$ at  $M_\mathrm{peak} > 10^{11}$ and $0\leq z \leq 10$, $f_\mathrm{duty}$ is effectively the fraction of SMBH \emph{host} galaxies whose SMBHs are active.

At a fixed \emph{halo} mass, the Eddington ratio distribution function (ERDF) is assumed to have a double power-law shape:
\begin{eqnarray}
P\left(\eta|\eta_0, c_1, c_2\right) & = & f_\mathrm{duty}\frac{P_0}{\left(\frac{\eta}{\eta_0}\right)^{c_1}+\left(\frac{\eta}{\eta_0}\right)^{c_2}} + (1 - f_\mathrm{duty}) \delta(\eta) \label{e:bher_dist_shape}\\
c_{1} & = & c_{1,0} + c_{1,a} \left(a - 1\right) \label{e:erdf_c1}\\
c_{2} & = & c_{2,0} + c_{2,a} \left(a - 1\right) \label{e:erdf_c2}\ ,
\end{eqnarray}
where $\eta$ is the Eddington ratio, $P_0$ is the normalization of the ERDF for \emph{active} SMBHs, $c_1$ and $c_2$ are the two power-law indices, $\eta_0$ is the break point of the double power-law, and $\delta(\eta)$ is the ERDF for dormant SMBHs, which is a Dirac delta function centred at $\eta=0$. The constant of proportionality $P_0$ is calculated such that
\begin{equation}
    \int_{0}^{\infty}\frac{P_0}{\left(\frac{\eta}{\eta_0}\right)^{c_1} + \left(\frac{\eta}{\eta_0}\right)^{c_2}}d\log \eta = 1\ .
\end{equation}
This functional form is flexible enough to approximate many past assumptions for the shape of the ERDF (e.g., Gaussian distributions and Schechter functions). 

The characteristic Eddington ratio $\eta_0$ in Eq.\ \ref{e:bher_dist_shape} is \emph{not} a free parameter, but is constrained by the parametrizations in Eqs.\ \ref{e:duty_cycle}--\ref{e:bher_dist_shape}.  Letting $\overline{\eta}$ be the average Eddington ratio, we have from Eq.\ \ref{e:bher_dist_shape} that:
\begin{equation}
\label{e:avg_eta}
\overline{\eta} = f_\mathrm{duty}\int_0^{\infty} \eta P\left(\eta|\eta_0, c_1, c_2\right) d\log \eta\ ,
\end{equation}
and by definition
\begin{equation}
\label{e:eta_calc}
\overline{\eta} = \frac{\epsilon_{\rm tot}\overline{\mathrm{BHAR}}\times 4.5\times10^8\ \mathrm{yrs}}{ (1 - \epsilon_{\rm tot}) \overline{ M_{\bullet}}}\ ,
\end{equation}
where $\overline{ M_{\bullet}}$ and $\overline{\mathrm{BHAR}}$ (Eqs.\ \ref{e:mbh_avg_med} and \ref{e:bhar_def}) are the average SMBH mass and black hole accretion rate, respectively. The parameter \eff{} is the efficiency of releasing energy (both radiative and kinetic) through accretion. We hence solve for $\eta_0$ by combining Eqs.\ \ref{e:avg_eta} and \ref{e:eta_calc}. In this work, $\log_{10} (\epsilon_{\rm tot}$) is assumed to be redshift-independent.

Given the non-zero scatter in SMBH mass at fixed halo mass (Eq.\ \ref{e:scatter_bhhm}), different SMBHs with the same host halo mass may have different Eddington ratio distributions. Without joint observational constraints as a function of SMBH mass and galaxy mass, we assume that SMBHs with the same host halo mass share the same Eddington ratio distribution \emph{shapes} (Eq.\ \ref{e:bher_dist_shape}), but can have different average Eddington ratios. To quantify the systematic change in average Eddington ratio with \mbh{} at fixed halo mass, we parametrize the correlation coefficient between $\overline{\mathrm{BHAR}}$ and \mbh{} as a function of redshift:

\begin{eqnarray}
\hspace{-3ex}\log_{10}\overline{\mathrm{BHAR}}\left(M_\bullet|M_\mathrm{peak}\right) &=& \overline{\mathrm{BHAR}}\left(\overline{M_\bullet}(M_\mathrm{peak})\right)\nonumber\\
&+& \rho_\mathrm{BH} \log_{10}\left(\frac{M_\bullet}{\overline{M_\bullet}(M_\mathrm{peak})}\right)\\
\rho_\mathrm{BH} &=& \rho_\mathrm{BH,0} + \rho_\mathrm{BH,a}(a-1) + \rho_\mathrm{BH,z}z \label{e:rho_bh}\ .
\end{eqnarray}

For example, $\rho_\mathrm{BH}=1$ means that different SMBHs at fixed halo mass share identical \emph{Eddington ratio distribution}, while $\rho_\mathrm{BH}=0$ means that these SMBHs have identical \emph{absolute accretion rate} distributions. Here, we allow \rhobh{} to take a value within $[-1,1]$. Any \rhobh{} above(below) 1($-1$) is capped at 1($-1$).

\subsection{Kinetic and radiative Eddington ratios}
\label{ss:bher_kin_rad}

\begin{figure*}
\subfigure{
\includegraphics[width=0.48\textwidth]{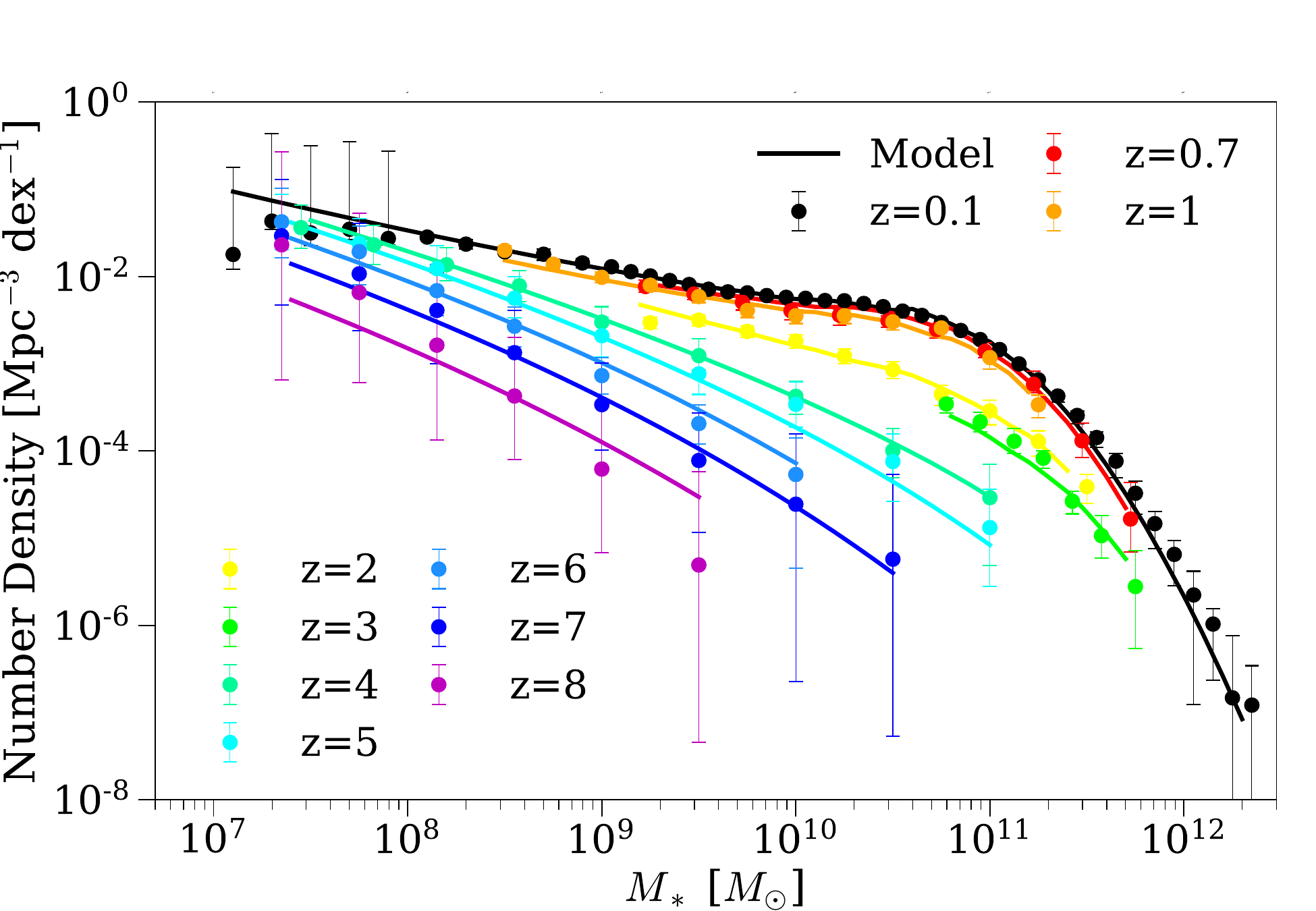}
}
\subfigure{
\includegraphics[width=0.48\textwidth]{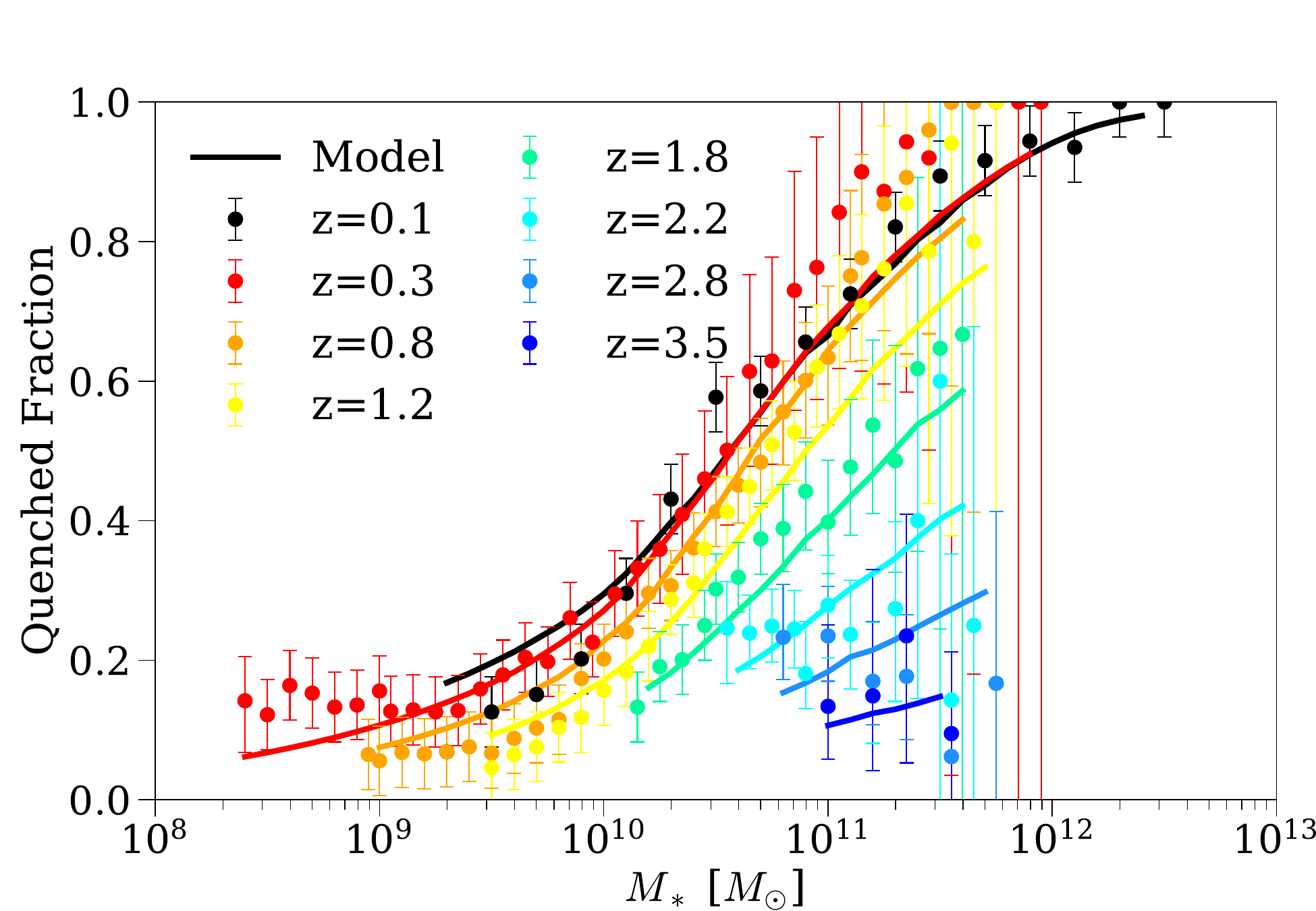}
}
\caption{\textbf{Left Panel:} Comparison between observed galaxy stellar mass functions (SMFs) and our best-fitting model from $z=0-8$. The observed stellar mass functions are listed in Table \ref{t:smf}. \textbf{Right Panel:} Comparison between observed galaxy quenched fractions (QFs) and our best-fitting model from $z=0-4$. The observed quenched fractions are listed in Table \ref{t:qf}. All the data used to make this plot (including individual data points and our best-fitting model) can be found \href{https://github.com/HaowenZhang/TRINITY/tree/main/plot_data}{here}.}
\label{f:smf_qf}
\end{figure*}

\begin{figure*}
\subfigure{
\includegraphics[width=\columnwidth]{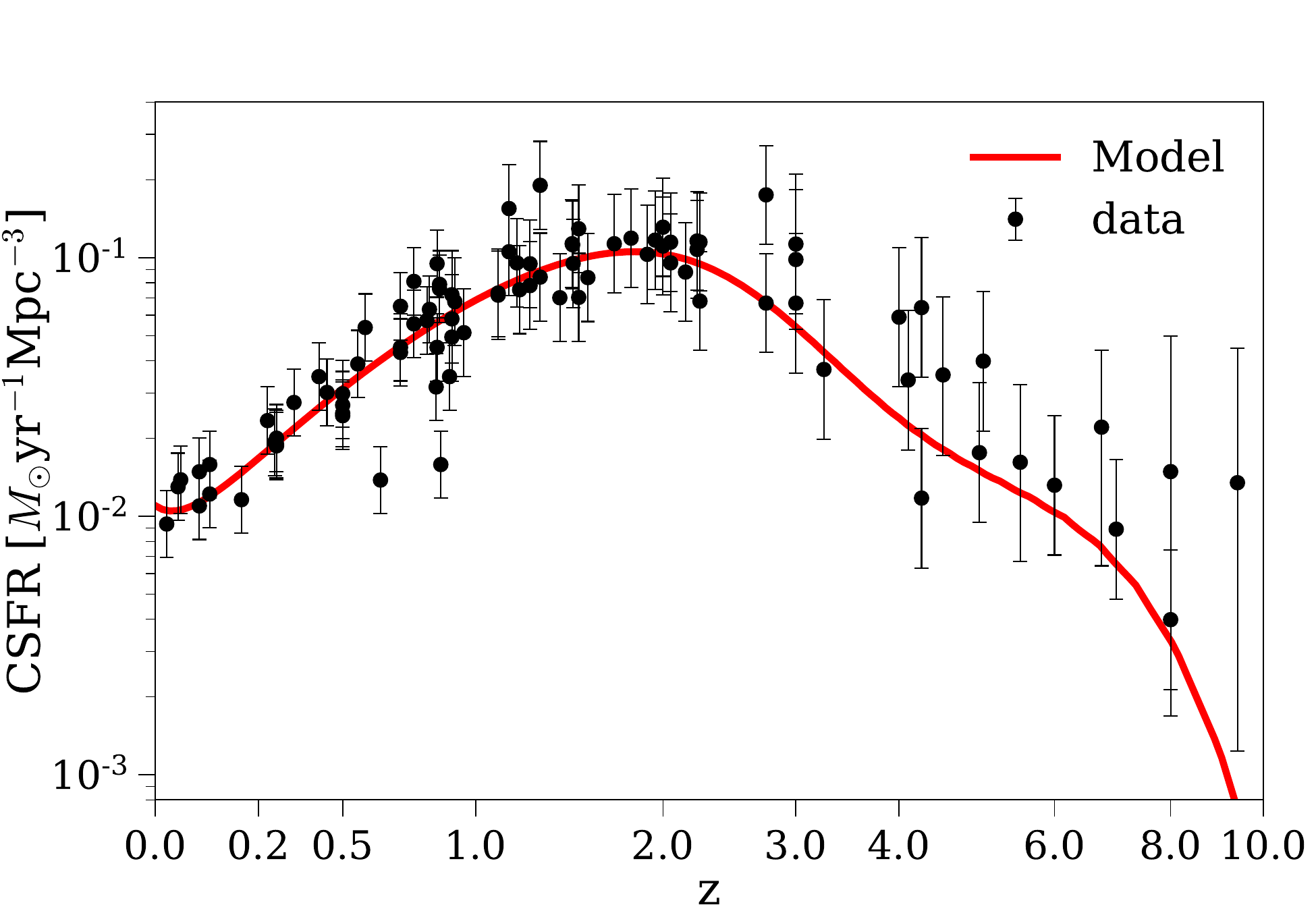}
}
\subfigure{
\includegraphics[width=\columnwidth]{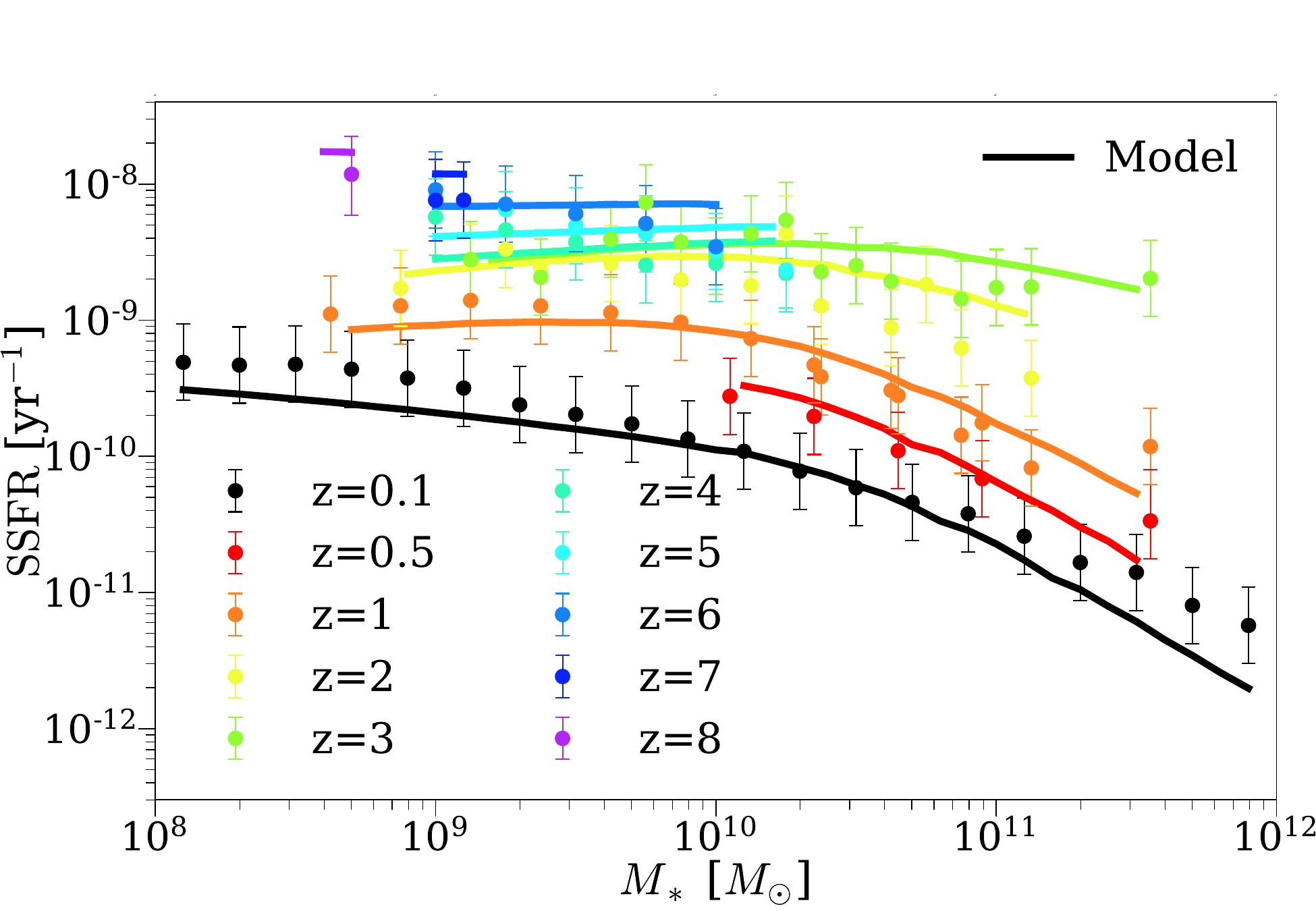}
}
\caption{\textbf{Left Panel:} Comparison between observed cosmic star formation rates (CSFRs) and our best-fitting model from $z=0-10$. The references for observations are listed in Table \ref{t:csfr}. \textbf{Right Panel:} Comparison between observed galaxy specific star formation rates (SSFRs) as a function of stellar mass and our best-fitting model from $z=0-8$. The references for observations are listed in Table \ref{t:ssfr}. All the data used to make this plot (including individual data points and our best-fitting model) can be found \href{https://github.com/HaowenZhang/TRINITY/tree/main/plot_data}{here}.}
\label{f:csfr_ssfr}
\end{figure*}

\begin{figure}
\includegraphics[width=\columnwidth]{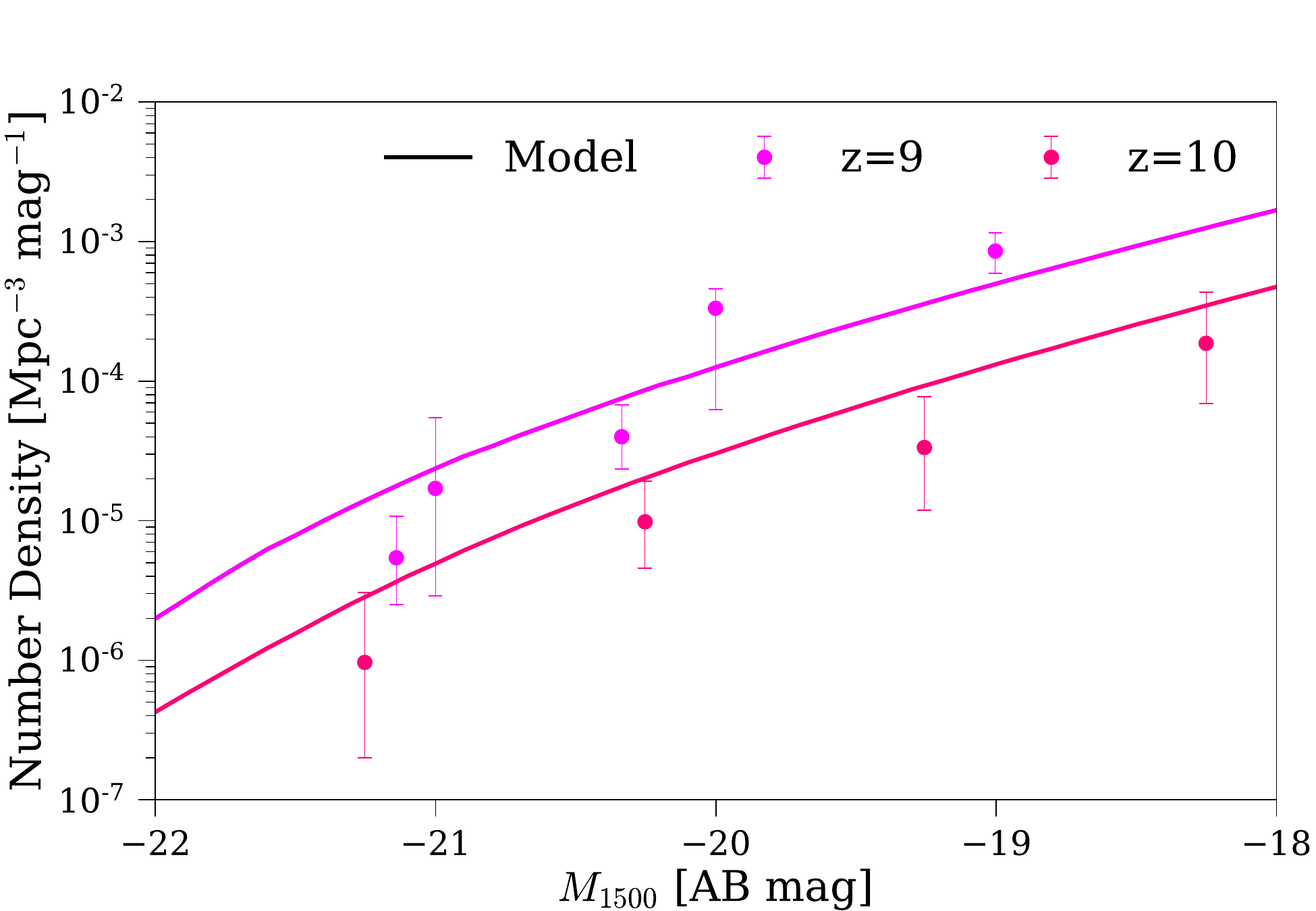}
\caption{Comparison between observed galaxy UV luminosity functions (UVLFs) and our best-fitting model from $z=9-10$. The references for observations are listed in Table \ref{t:uvlf}. All the data used to make this plot (including individual data points and our best-fitting model) can be found \href{https://github.com/HaowenZhang/TRINITY/tree/main/plot_data}{here}.}
\label{f:uvlf}
\end{figure}

\begin{figure}
\includegraphics[width=\columnwidth]{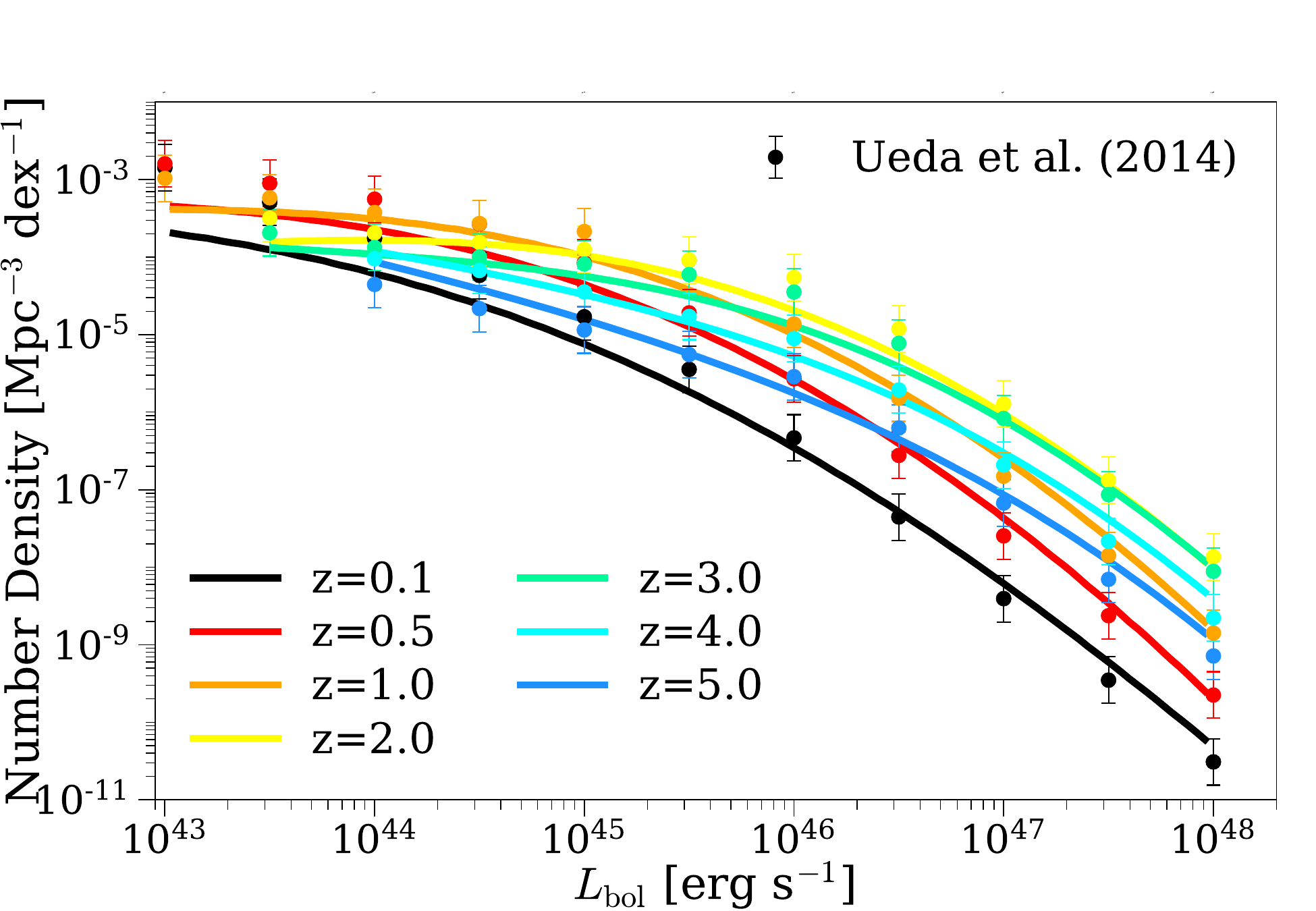}
\caption{Comparison between the observed quasar luminosity functions (QLFs) from \citet{Ueda2014} and our best-fitting model from $z=0-5$. All the data used to make this plot (including individual data points and our best-fitting model) can be found \href{https://github.com/HaowenZhang/TRINITY/tree/main/plot_data}{here}.}
\label{f:qlf}
\end{figure}

\begin{figure}
\includegraphics[width=\columnwidth]{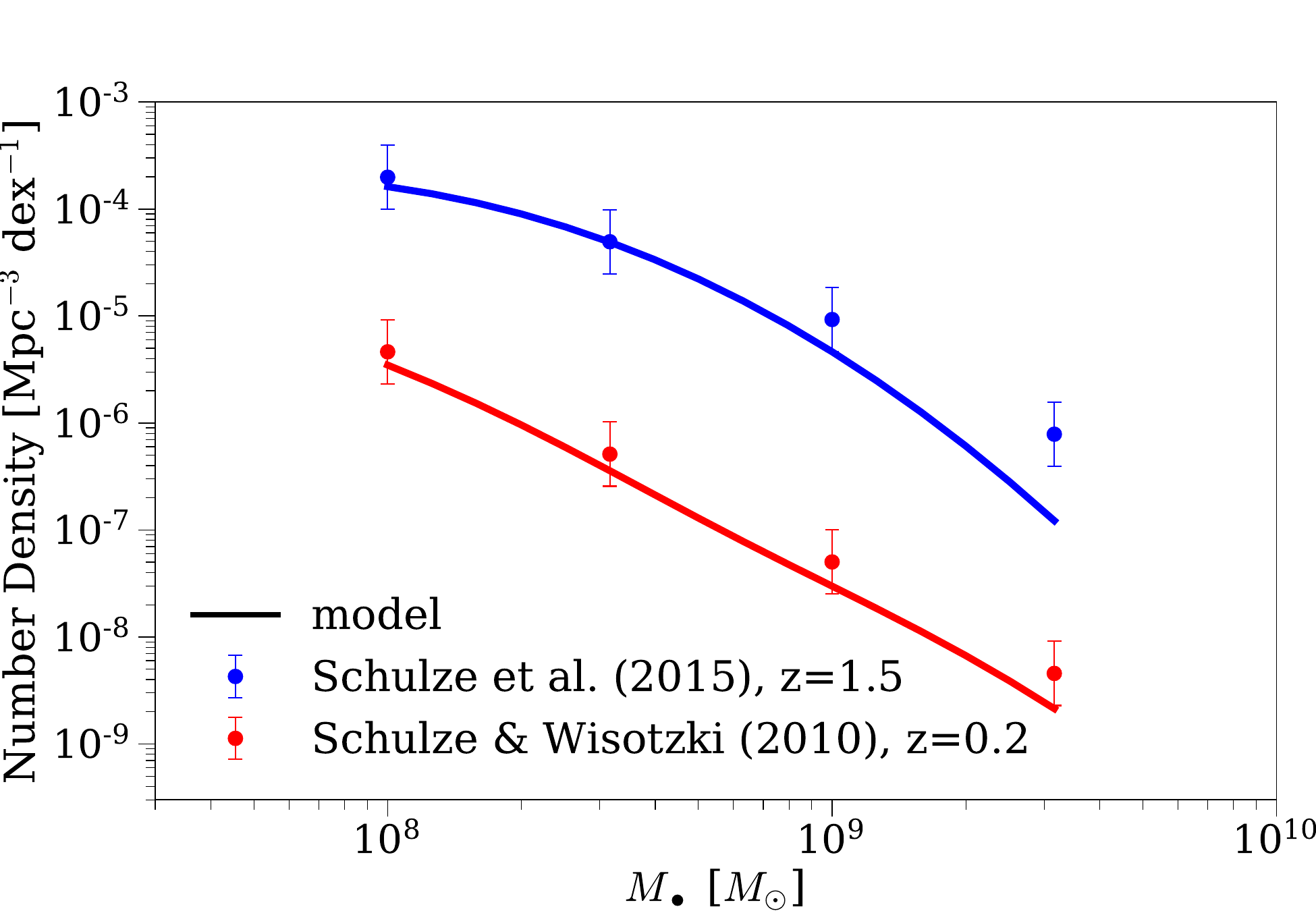}
\caption{Comparison between the observed active black hole mass functions (ABHMFs) from \citet{Schulze2010}, \citet{Schulze2015}, and our best-fitting model at $z=0.2$ and $z=1.5$. All the data used to make this plot (including individual data points and our best-fitting model) can be found \href{https://github.com/HaowenZhang/TRINITY/tree/main/plot_data}{here}.}
\label{f:abhmf}
\end{figure}

\begin{figure}
\includegraphics[width=\columnwidth]{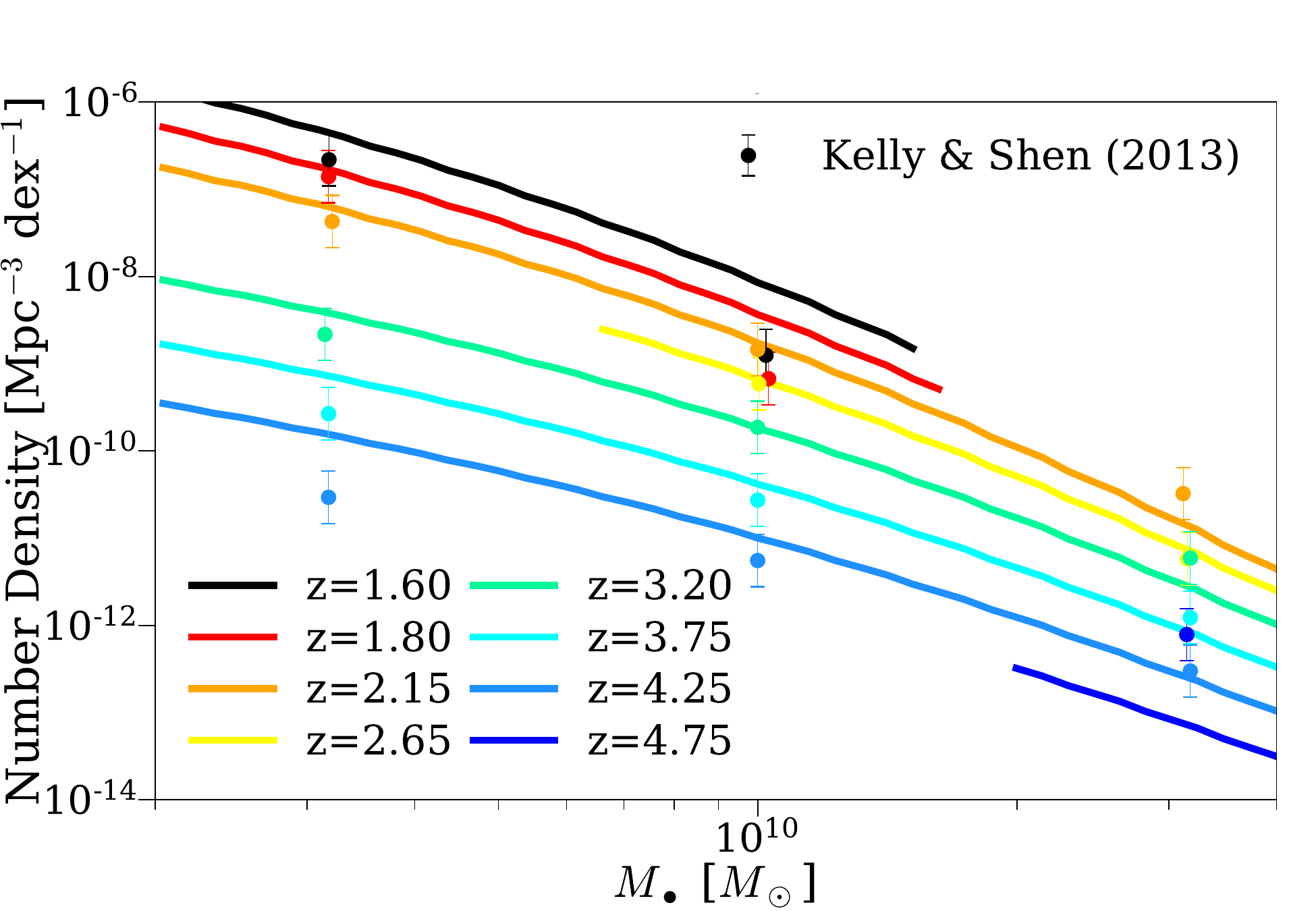}
\caption{Comparison between the observed active black hole mass functions (ABHMFs) from \citet{Kelly2013} and our best-fitting model from $z=1.5-5$. The data points and the best fitting models in each higher redshift bin are shifted downwards by 0.5 dex incrementally for the sake of clarity. All the data used to make this plot (including individual data points and our best-fitting model) can be found \href{https://github.com/HaowenZhang/TRINITY/tree/main/plot_data}{here}.}
\label{f:abhmf_kelly}
\end{figure}

\begin{figure*}

\subfigure{
\includegraphics[width=0.48\textwidth]{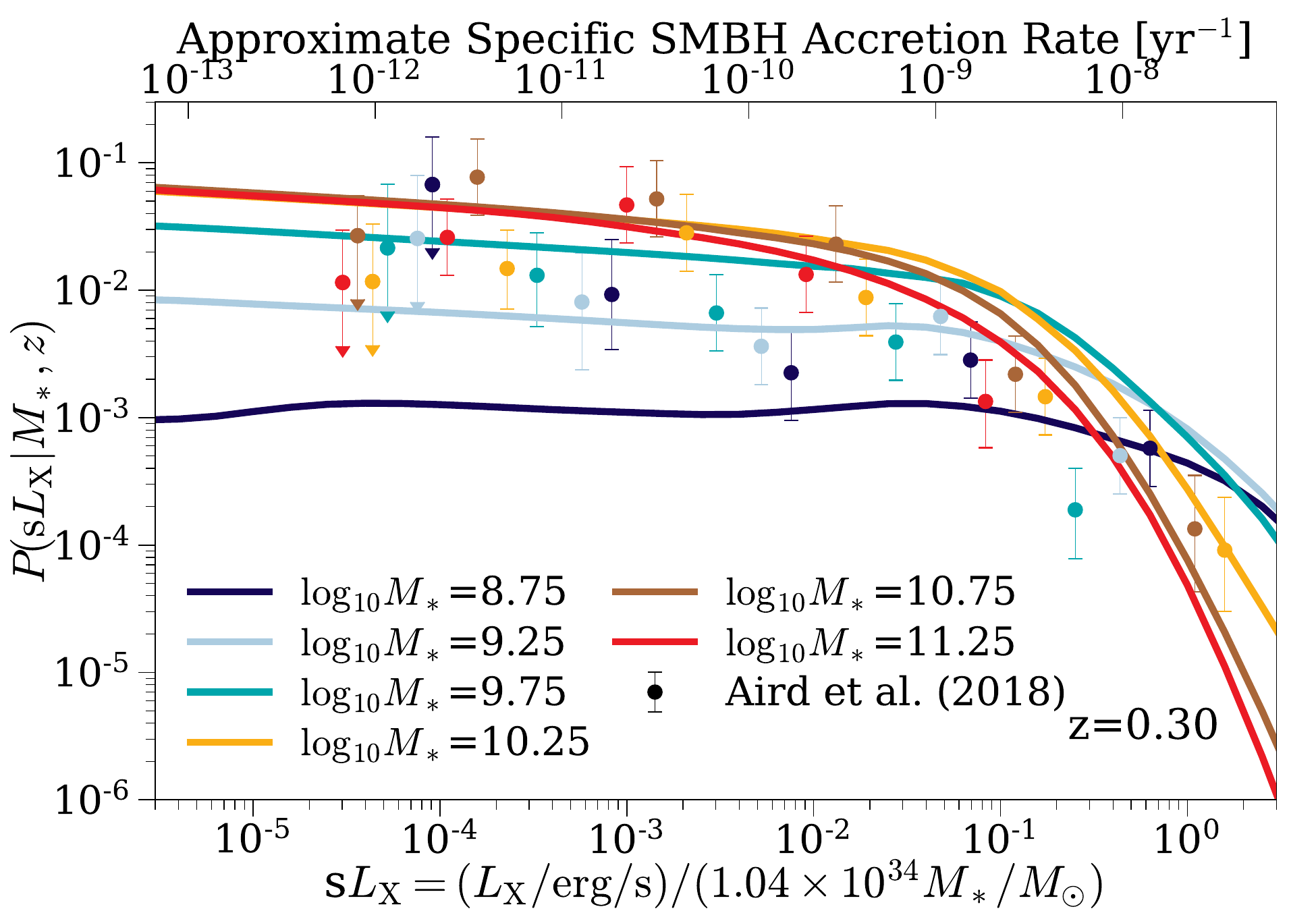}
}
\subfigure{
\includegraphics[width=0.48\textwidth]{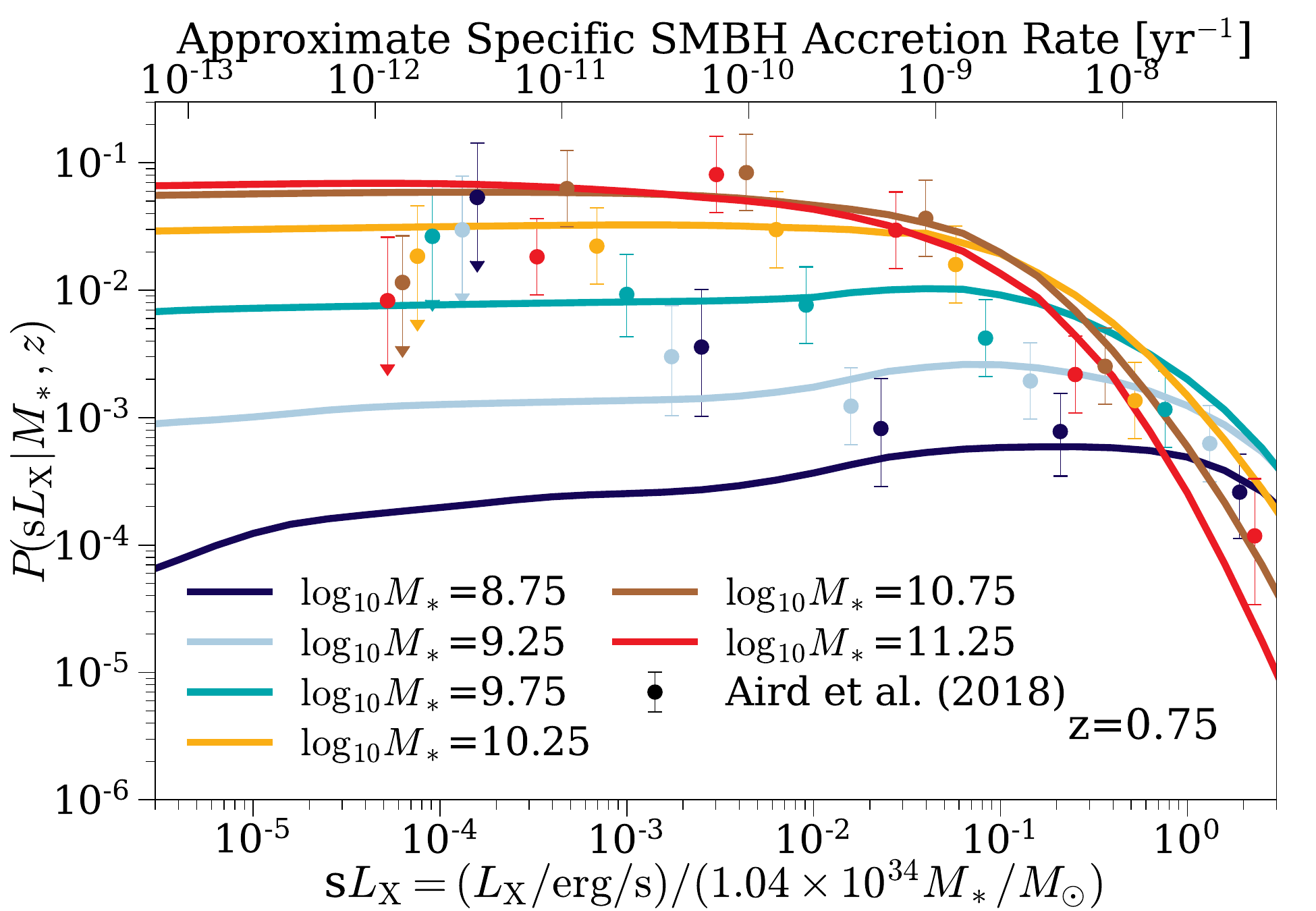}
}
\subfigure{
\includegraphics[width=0.48\textwidth]{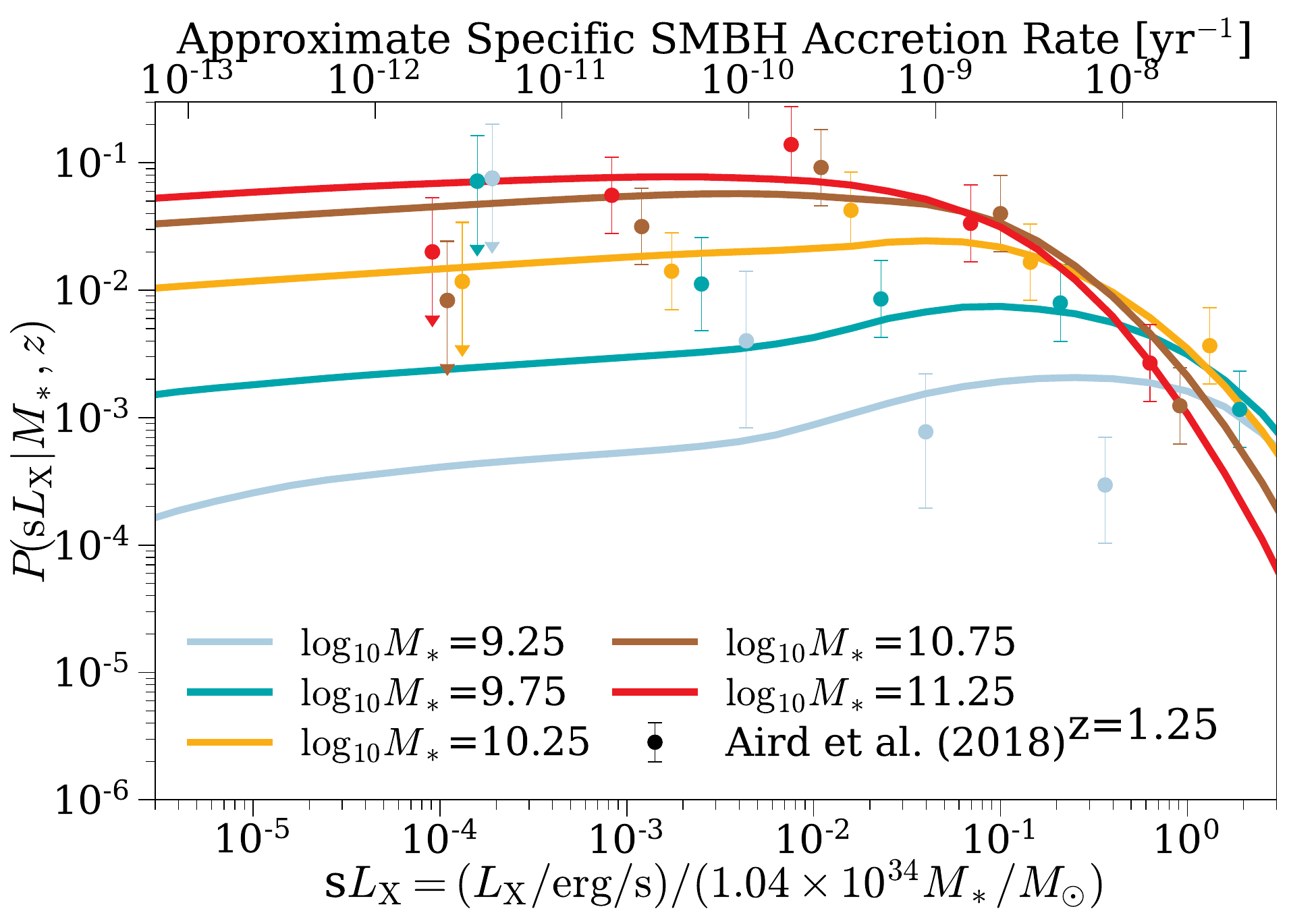}
}
\subfigure{
\includegraphics[width=0.48\textwidth]{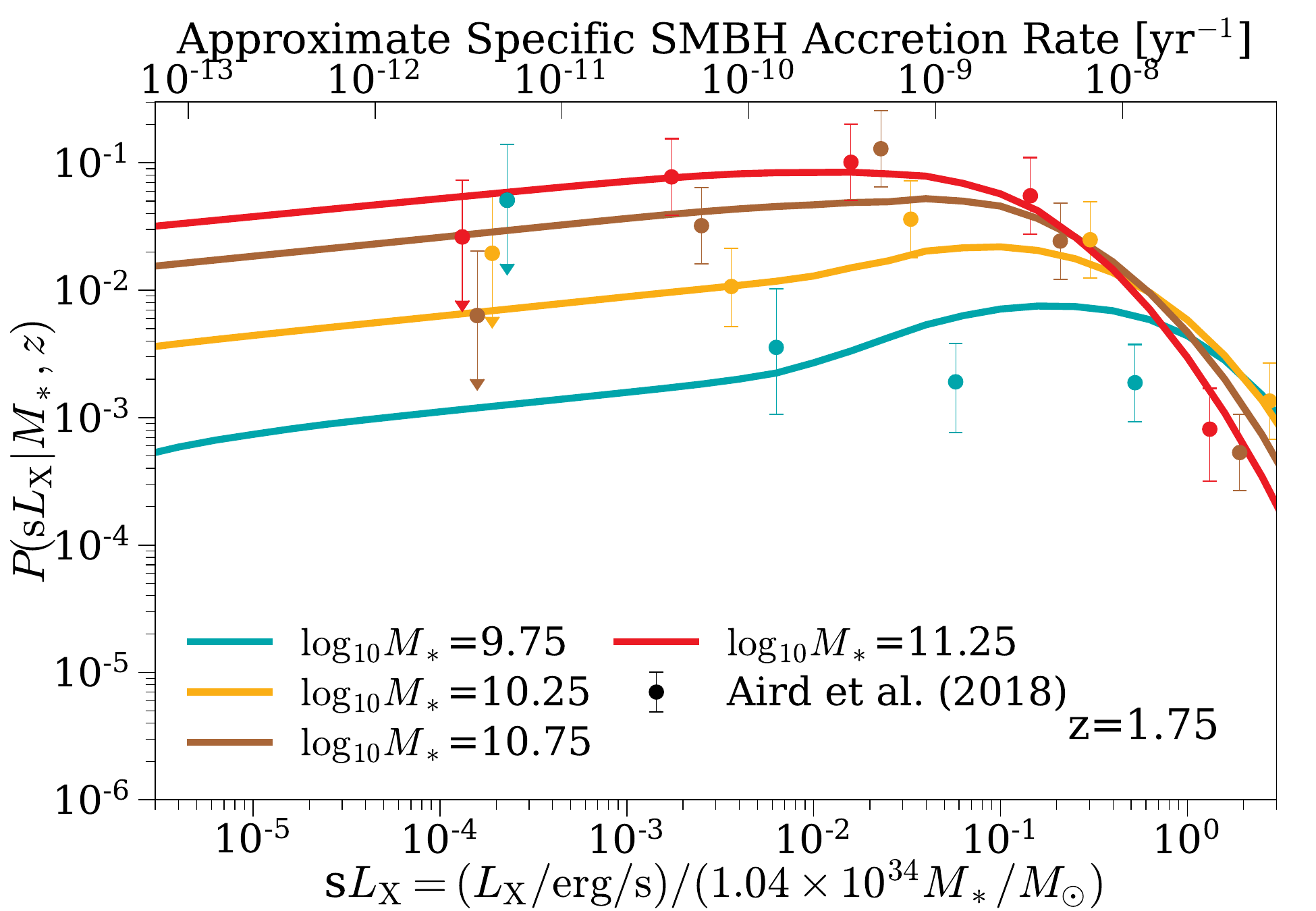}
}
\subfigure{
\includegraphics[width=0.48\textwidth]{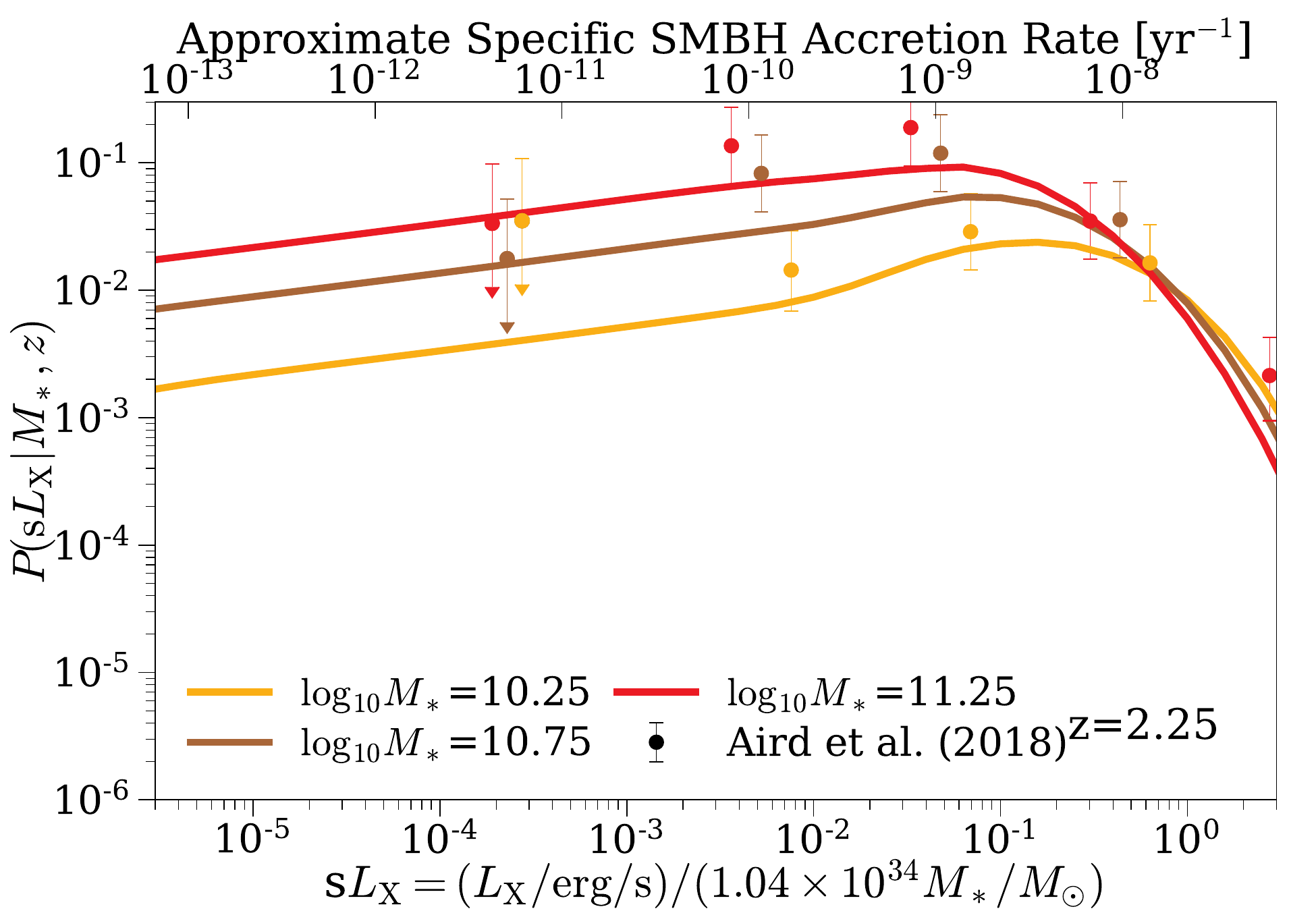}
}

\caption{The comparison between the observed quasar probability distribution functions (QPDFs) from \citet{Aird2018} and our best-fitting model from $z=0-2.5$. The data points include Compton-thin AGNs only, so the model values are corrected for direct comparison. All the data used to make this plot (including individual data points and our best-fitting model) can be found \href{https://github.com/HaowenZhang/TRINITY/tree/main/plot_data}{here}.}
\label{f:qpdf}
\end{figure*}

SMBH accretion produces both radiative and kinetic energy \citep[see, e.g.,][]{Merloni2008}, and the latter dominates the total energy output at low accretion rates. The radiative and kinetic luminosities depend on the efficiency of mass conversion into the two different forms of energies, $\epsilon_{\rm rad}$ and $\epsilon_{\rm kin}$. In analogy with this, we can recast the Eddington ratio in terms of its radiative and kinetic components. To forward model these observables, we adopt the following empirical relation between the total Eddington ratio $\eta$ and its radiative component $\eta_{\rm rad}$:
\begin{equation}
\label{e:eta_tot_rad}
    \eta_{\rm rad} = 
    \begin{cases}
      \eta^2/0.03, & \eta \leq 0.03\\
      \eta, & 0.03 < \eta \leq 2\\
      2\left[1 + \ln(\eta/2)\right], & \eta > 2\\
      
    \end{cases}\ .
\end{equation}
For $\eta \leq 2$, the scaling between $\eta_\mathrm{rad}$ and $\eta$ is similar to the one used by \citet{Merloni2008}. \citet{Merloni2008} adopted a more complex scaling relation between AGN radiative luminosity, X-ray luminosity, and SMBH mass that had substantial scatter. Rather than using the same complex model, we choose to adopt the simpler, more transparent scaling in Eq.\ \ref{e:eta_tot_rad}. For $\eta \geq 2$, we adopt a logarithmic scaling to account for the fact that at such high accretion rates, the accretion disk becomes thick, trapping part of the outgoing radiation \citep{Mineshige2000}. The kinetic component $\eta_{\rm kin}$ is, by definition:
\begin{equation}
\label{e:erdf_kin_rad}
    \eta_{\rm kin} = \eta - \eta_{\rm rad}\ ,\ \eta < 0.03\ .
\end{equation}
At a given $\eta<0.03$, Eq.\ \ref{e:erdf_kin_rad} produces $\sim 0.3-0.5$ dex more kinetic energy than \citet{Merloni2008}. We also ignore the kinetic energy output from active SMBHs with $\eta > 0.03$, due to a lack of observational constraints. Thus, the AGN radiative and kinetic efficiencies are:

\begin{equation}
\label{e:eff_tot_rad}
    \epsilon_{\rm rad} = \epsilon_\mathrm{tot}\times
    \begin{cases}
      \eta/0.03, & \eta \leq 0.03\\
      1, & 0.03 < \eta \leq 2\\
      2/\eta \left[1 + \ln(\eta/2)\right], & \eta > 2\\
      
    \end{cases}\ ,
\end{equation}
and:
\begin{equation}
\label{e:eff_kin_rad}
    \epsilon_{\rm kin} = 
    \begin{cases}
        \epsilon_\mathrm{tot}(1 - \eta/0.03), & \eta < 0.03\\
        0, & \eta > 0.03\\
    \end{cases}\ ,
\end{equation}
respectively. The radiative and kinetic luminosities and Eddington ratio distributions are:
\begin{eqnarray}
    \frac{L_\mathrm{(\cdot)}}{\mathrm{erg/s}} & = & 10^{38.1} \times \frac{M_\bullet}{M_\odot} \times \eta_\mathrm{(\cdot)}\\
    P(\eta_\mathrm{(\cdot)}) & = & P\left(\eta\right)\frac{d\log\eta}{d\log\eta_\mathrm{(\cdot)}}\ ,
\end{eqnarray}
where $(\cdot)$ is either ``rad'' or ``kin'' and $d\log\eta/d\log\eta_\mathrm{(\cdot)}$ is calculated using Eqs.\ \ref{e:eta_tot_rad}-\ref{e:erdf_kin_rad}.

\begin{figure}
\vspace{-0.35cm}
\includegraphics[width=\columnwidth]{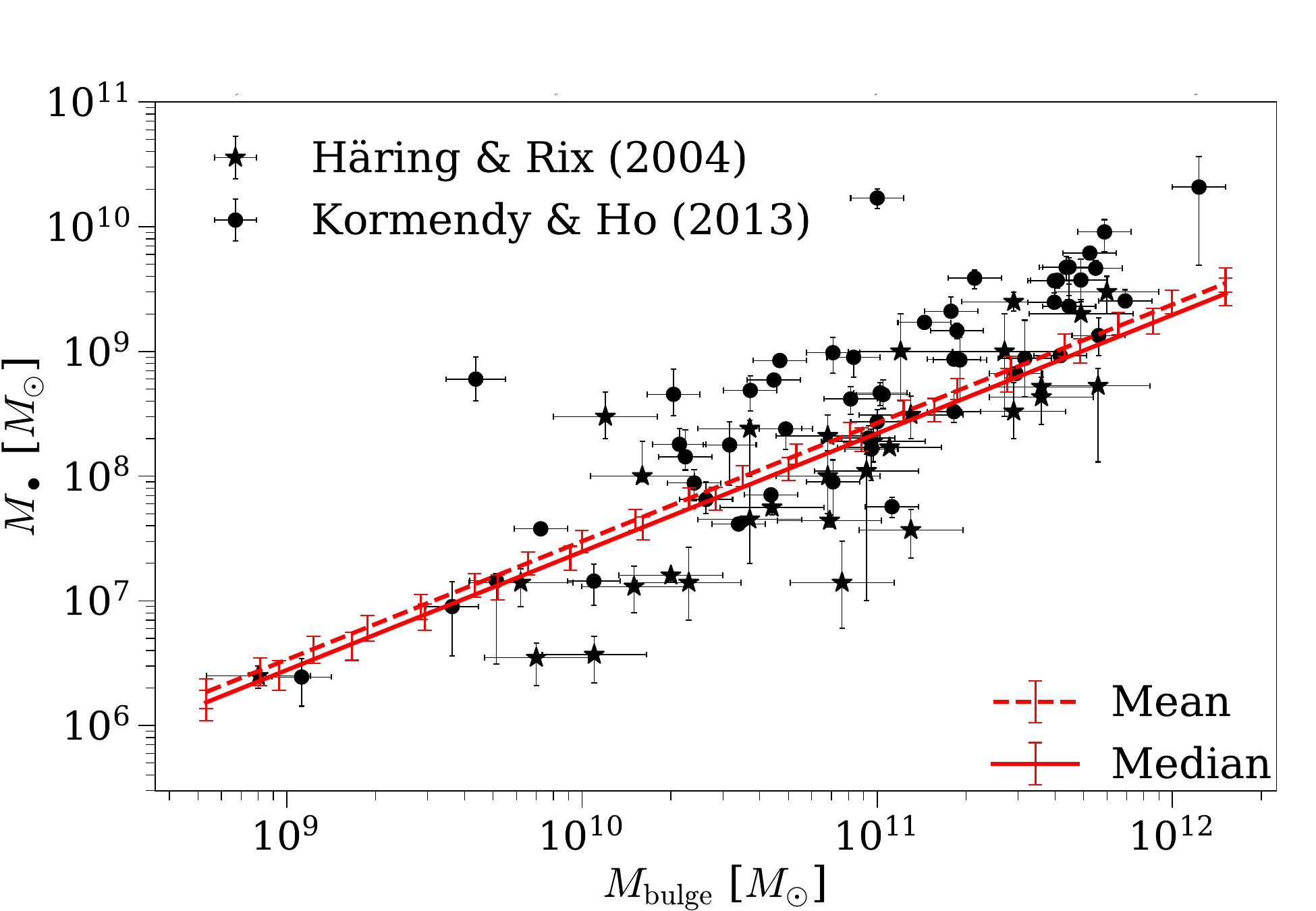}
\caption{The local \bhbm{} relation. The filled circles are the data compiled by \citet{Kormendy2013}, and the stars are those compiled by \citet{Haring2004}. The red solid line is the median SMBH--bulge mass relation, and the red dashed line is the mean relation. These lines are offset because log-normal distributions are positively skewed, with the mean being greater than the median. All the data used to make this plot (including individual data points and our best-fitting model) can be found \href{https://github.com/HaowenZhang/TRINITY/tree/main/plot_data}{here}.}
\label{f:bhbm_fit}
\end{figure}

\begin{table*}
\caption{Summary of Parameters}
\label{t:params}
\begin{tabular}{lcccc}
\hline
Symbol & Description & Equation & Parameters & Section\\
\hline
$V(z)$ & Characteristic $v_{\mathrm{Mpeak}}$ in $\rm SFR$--$v_{\mathrm{Mpeak}}$ relation & \ref{e:v_1} & 4 & \ref{ss:halo_galaxy_connection} \\
$\epsilon(z)$ & Characteristic SFR in $\rm SFR$--$v_{\mathrm{Mpeak}}$ relation & \ref{e:epsilon} & 4 & \ref{ss:halo_galaxy_connection}\\
$\alpha(z)$ & Low-mass slope of the $\rm SFR$--$v_{\mathrm{Mpeak}}$ relation & \ref{e:alpha} & 4 & \ref{ss:halo_galaxy_connection} \\
$\beta(z)$ &  Massive-end slope of the $\rm SFR$--$v_{\mathrm{Mpeak}}$ relation & \ref{e:beta} & 3 & \ref{ss:halo_galaxy_connection} \\
$v_Q(z)$ & Typical $v_{\mathrm{Mpeak}}$ for star formation quenching, in dex & \ref{e:m_quenched} & 3 & \ref{ss:halo_galaxy_connection}\\
$w_Q(z)$ & Typical width in $v_{\mathrm{Mpeak}}$ for star formation quenching, in dex & \ref{e:w_quenched} & 3 & \ref{ss:halo_galaxy_connection}\\
$f_{\rm merge}$ & Fraction of incoming satellite galaxy mass that is merged into central galaxies & - & 1 & \ref{ss:halo_galaxy_connection}\\
$\sigma_*$ & Scatter in true stellar mass at fixed halo mass, in dex & - & 1 & \ref{ss:halo_galaxy_connection}\\
\hline
$\mu(z)$ & Systematic offset between true and observed stellar masses, in dex & \ref{e:mu_z} & 2 & \ref{ss:galaxy_sys} \\
$\kappa(z)$ & Additional systematic offset in observed vs. true SFRs, in dex & \ref{e:mu_kappa} & 1 & \ref{ss:galaxy_sys} \\
$\sigma(z)$ & Scatter between measured and true stellar masses, in dex & \ref{e:sm_scatter} & 1 & \ref{ss:galaxy_sys}\\
$\rho_{0.5}$ & Correlation between SFR and stellar mass at fixed halo mass at $z=1$ ($a=0.5$) & \ref{e:rho05} & 1 & \ref{ss:galaxy_sys}\\
\hline
$f_\mathrm{occ,min}(z)$ & Minimum SMBH occupation fraction & \ref{e:focc_min} & 2 & \ref{sss:smbh_focc}\\
$M_\mathrm{h,c}(z)$ & Characteristic halo mass where SMBH occupation fraction changes significantly & \ref{e:mh_c} & 2 & \ref{sss:smbh_focc}\\
$w_\mathrm{h,c}(z)$ & Log-halo mass range over which SMBH occupation fraction changes significantly & \ref{e:wh_c} & 2 & \ref{sss:smbh_focc}\\
$\beta_{\mathrm{BH}}(z)$ & Median SMBH mass for galaxies with $M_{\mathrm{bulge}}=10^{11} M_{\odot}$, in dex & \ref{e:beta_bh} & 3 & \ref{sss:bhbm_relation}\\
$\gamma_{\mathrm{BH}}(z)$ & Slope of the SMBH mass--bulge mass (\bhbm{}) relation & \ref{e:gamma_bh} & 3 & \ref{sss:bhbm_relation}\\
$\sigma_{\mathrm{BH}}$ & Scatter in SMBH mass at fixed bulge mass, in dex & - & 1 & \ref{sss:bhbm_relation}\\
$f_\mathrm{scale}(z)$ & Ratio between the fractions of SMBH and galaxy growth coming from mergers & \ref{e:f_merge_bh_z} & 2 & \ref{ss:bh_mergers}\\
$f_{\mathrm{duty}}(M_{\mathrm{peak}}, z)$ & AGN duty cycle & \ref{e:duty_cycle} & 4 & \ref{ss:bher_dist}\\
$c_1(z)$, $c_2(z)$ & Faint- and bright-end slopes of the AGN Eddington ratio distribution functions & \ref{e:erdf_c1},\ref{e:erdf_c2} & 4 & \ref{ss:bher_dist}\\
$\epsilon_{\mathrm{tot}}$ & Total energy efficiency (radiative and kinetic) of mass accretion onto SMBHs & - & 1 & \ref{ss:bher_dist}\\
$\rho_\mathrm{BH}(z)$ & Correlation coefficient between SMBH accretion rate and mass at fixed halo mass & \ref{e:rho_bh} & 3 & \ref{ss:bher_dist}\\
\hline
$\xi$ & Systematic offset in Eddington ratio when calculating AGN probability distribution functions, in dex & \ref{e:xi} & 1 & \ref{ss:agn_observables} \\
\hline
\multicolumn{2}{l}{Total Number of Galaxy Parameters} & & 28  & \\
\hline
\multicolumn{2}{l}{Total Number of SMBH Parameters} & & 28  & \\
\hline
\multicolumn{2}{l}{Total Number of Parameters} & & 56  & \\
\hline
\end{tabular}
\parbox{14.5cm}{\textbf{Notes.} $v_\mathrm{Mpeak}$: the maximum circular velocity at the time when the halo reaches its peak mass (see \S\ref{ss:halo_galaxy_connection}).}
\end{table*}

\begin{table*}
\caption{Summary of Priors}
\label{t:priors}
\begin{tabular}{lccc}
\hline
Symbol & Description & Equation & Prior\\
\hline
$\sigma_{*,0}$ & Value of $\sigma_*$ at $z=0$, in dex & - & $U(0, 0.3)$ \\
$\mu_0$ & Value of $\mu$ at $z=0$, in dex & \ref{e:mu_z} & $G(0, 0.14)$ \\
$\mu_a$ & Redshift scaling of $\mu$, in dex & \ref{e:mu_z} & $G(0, 0.24)$ \\
$\kappa$ & Additional systematic offset in observed vs. true SFRs, in dex & \ref{e:mu_kappa} & $G(0, 0.24)$ \\
$\sigma_z$ & Redshift scaling of $\sigma$, in dex & \ref{e:sm_scatter} & $G(0.05, 0.015)$\\
$\rho_{0.5}$ & Correlation between SFR and stellar mass at fixed halo mass at $z=1$ ($a=0.5$) & \ref{e:rho05} & $U(0.23, 1)$\\
\hline
$\beta_{\mathrm{BH},0}$ & SMBH mass at $M_{\mathrm{bulge}}=10^{11} M_{\odot}$ and $z=0$ & \ref{e:beta_bh} & $G\left(8.46, 0.20\right)$\\
$\gamma_{\mathrm{BH},0}$ & Slope of the \bhbm{} relation at $z=0$ & \ref{e:gamma_bh} & $G\left(1.05, 0.14\right)$\\
\hline
\end{tabular}
\parbox{14.5cm}{\textbf{Notes.} $G\left(\mu, \sigma\right)$ denotes a Gaussian with median $\mu$ and width $\sigma$, and $U\left(x_1, x_2\right)$ denotes a uniform distribution between $x_1$ and $x_2$.}
\end{table*}

\subsection{Calculating AGN observables}
\label{ss:agn_observables}

Having specified SMBH growth histories and ERDFs, we can now predict AGN observables. Although there are different observables in our data compilation, all of them involve counting the number densities of the host haloes/galaxies of SMBHs with certain properties.

The SMBH mass function at each redshift is the number density of haloes that host SMBHs of a given mass:
\begin{equation}
\label{e:bhmf}
\begin{aligned}
    &\phi_{\mathrm{BH}}\left(M_{\bullet},z\right) = \int_{0}^{\infty}\phi_{\mathrm{h}}\left(M_{\mathrm{peak}},z\right) P\left(M_{\bullet}|M_{\mathrm{peak}},z\right)d M_{\mathrm{peak}}\ ,
\end{aligned}
\end{equation}
where $\phi_{\mathrm{h}}\left(M_{\mathrm{peak}},z\right)$ is the halo mass function at redshift $z$, and $P\left(M_{\bullet}|M_{\mathrm{peak}},z\right)$ is specified by the halo--galaxy--SMBH connection (see \S \ref{ss:halo_galaxy_connection} and \S \ref{ss:galaxy_smbh_connection}).

To model active black hole mass functions from \citet{Schulze2010} and \citet{Schulze2015}, we apply the same selection criteria and remove SMBHs with radiative Eddington ratios below 0.01. Thus, the active black hole mass function is:
\begin{equation}
\label{e:abhmf}
\begin{aligned}
\phi_{\mathrm{ABH}}\left(M_{\bullet},z\right) = \int_{0}^{\infty}\int_{\eta_{\rm rad,min}=0.01}^{\infty}&\phi_{\mathrm{h}}\left(M_{\mathrm{peak}},z\right) P\left(M_{\bullet}|M_{\mathrm{peak}},z\right) \times\\
&P\left(\eta_{\rm rad}|M_{\bullet},M_{\mathrm{peak}},z\right)d \eta_{\rm rad}\ d M_{\mathrm{peak}}\ .
\end{aligned}
\end{equation}
For the type I quasar SMBH mass functions from \citet{Kelly2013}, we include all SMBHs with $\eta > 0$. This is because modeling of the underlying $M_\mathrm{BH}-L_\mathrm{bol}$ distributions showed little incompleteness induced by the SDSS luminosity cut at $\log_{10} M_\bullet \gtrsim 9.5$, and we only use data above this mass. To account for obscured type II quasars, we use an empirical formula for the obscured fraction $F_\mathrm{obs}$ as a function of X-ray luminosity from \citet{Merloni2014}:
\begin{equation}
    F_\mathrm{obs}(L_\mathrm{X}) = 0.56 + \frac{1}{\pi}\arctan \left(\frac{43.89 - \log L_\mathrm{X}}{0.46}\right)\ .
\end{equation}
Thus, the type I quasar BHMF is:
\begin{equation}
\label{e:abhmf_kelly}
\begin{aligned}
\phi_{\mathrm{ABH}}\prime \left(M_{\bullet},z\right) = \int_0^\infty \int_0^\infty &\phi_{\mathrm{h}}\left(M_{\mathrm{peak}},z\right) P\left(M_{\bullet}|M_{\mathrm{peak}},z\right) \times\\
&P\left(\eta_{\rm rad}|M_{\bullet},M_{\mathrm{peak}},z\right) \times \\
&(1 - F_\mathrm{obs}(L_\mathrm{X})) d\eta_{\rm rad}\ d M_{\mathrm{peak}}\ ,
\end{aligned}
\end{equation}
where $L_X$ is the X-ray luminosity that is calculated using the bolometric correction from \citet{Ueda2014}:
\begin{eqnarray}
\label{e:bol_correction}
L_\mathrm{X} &=& \frac{L_\mathrm{bol}}{k_\mathrm{bol}(L_\mathrm{bol})}\\
L_\mathrm{bol}/\mathrm{erg\cdot s}^{-1} &=& 10^{38.1} \cdot M_\bullet \cdot \eta_\mathrm{rad}\\
k_{\mathrm{bol}} \left(L_{\mathrm{bol}}\right) &=& 10.83\left(\frac{L_{\mathrm{bol}}}{10^{10}L_{\odot}}\right)^{0.28} + 6.08\left(\frac{L_{\mathrm{bol}}}{10^{10}L_{\odot}}\right)^{-0.020}\ .
\end{eqnarray}

Similarly, QLFs are given by the number density of haloes hosting SMBHs with a given luminosity:
\begin{equation}
\label{e:qlf_1}
\begin{aligned}
\phi_{\mathrm{L}}\left(L_{\mathrm{bol}}, z\right) &= \int_0^\infty \phi_{\mathrm{h}}\left(M_{\mathrm{peak}}\right)P\left(L_{\rm bol}|M_{\rm peak}, z\right)d M_{\mathrm{peak}}\ ,
\end{aligned}
\end{equation}
where $P\left(L_{\rm bol}|M_{\rm peak}, z\right)$ is calculated by counting the number density of SMBHs with the corresponding Eddington ratio:
\begin{equation}
\begin{aligned}
P\left(L_{\rm bol}|M_{\rm peak}, z\right) = \int_0^\infty &P\left(\eta_{\rm rad}\left(L_{\rm bol}, M_{\bullet}\right)|M_{\bullet},M_{\rm peak},z\right)\times\\
&P\left(M_{\bullet}|M_{\rm peak},z\right)d M_{\bullet}\label{e:qlf_2}\ .
\end{aligned}
\end{equation}
Finally, for quasar probability distribution functions, \citet{Aird2018} expressed \emph{Compton-thin} QPDFs in terms of the specific $L_\mathrm{X}$ (\slx{}):
\begin{equation}
    \mathrm{s}L_\mathrm{X} =\frac{L_{\mathrm{X}} / \mathrm{erg\cdot s}^{-1}}{1.04\times10^{34}\ \times M_* / M_{\odot}}\ .
\end{equation}
The distribution of \slx{} at fixed stellar mass and redshift is:
\begin{eqnarray}
P\left(\mathrm{s}L_\mathrm{X}|M_*,z\right) & = & \left(1 - f_{\mathrm{CTK}}\left(L_\mathrm{X}, z\right)\right)\times P\left(L'_{\rm bol}|M_{*}, z\right)\\
L'_{\rm bol} & = & \frac{L_{\rm bol}}{\xi}\label{e:xi}\\
L_{\mathrm{bol}}/\mathrm{erg\cdot s}^{-1} & = & 1.04\times 10^{34} \times M_*/M_\odot\ \times \mathrm{s}L_\mathrm{X} \times k_\mathrm{bol}(L_{\rm bol})\\
P(L_\mathrm{bol}|M_*,z) & = & \int_0^{\infty}dM_\mathrm{peak}\int_0^{\infty}dM_\bullet P(\eta_\mathrm{rad}(L_\mathrm{bol}, M_\bullet)|M_\mathrm{peak},z)\nonumber\\
& & P(M_\bullet|M_*,z) P(M_*|M_\mathrm{peak}z),
\end{eqnarray}
where the Compton-thick fraction $f_{\mathrm{CTK}}\left(L_X, z\right)$ and the bolometric correction $k_{\mathrm{bol}}\left(L_{\mathrm{bol}}\right)$ are both given by \citet{Ueda2014} (see Appendix \ref{aa:ctk_corr} for full details about $f_{\mathrm{CTK}}$), and $\xi$ is the systematic offset in bolometric luminosity when calculating the AGN probability distribution functions in terms of \slx{}. This free parameter accounts for a residual inconsistency between the QPDFs from \citet{Aird2018} and the QLFs from \citet{Ueda2014} after the data point downsampling and exclusion as described in Appendix \ref{aa:aird_qpdf}.

\subsection{Methodology summary}
\label{ss:method_summary}

Here, we summarize the major steps to constrain the halo--galaxy--SMBH connection as shown in Fig.\ \ref{f:flow_chart}:
\begin{itemize}
    \item[1.] Choose a point in parameter space (Table \ref{t:params}), which fully specifies the halo--galaxy--SMBH connection (\S\ref{ss:halo_galaxy_connection}, \S\ref{ss:galaxy_smbh_connection}), SMBH merger contributions (\S\ref{ss:bh_mergers}), and the BHAR--AGN luminosity conversion (\S\ref{ss:bher_dist}, \ref{ss:bher_kin_rad}).
    \item[2.] Put galaxies and SMBHs into haloes accordingly, which determines galaxy and SMBH growth histories.
    \item[3.] Calculate SMBH mass functions and Eddington ratio distributions (\S\ref{ss:bher_dist}).
    \item[4.] Predict galaxy and AGN observables (\S\ref{ss:agn_observables} and Table \ref{t:summary_of_obs}).
    \item[5.] Correct these predictions for systematic effects in real observations, e.g., systematic offsets in measured vs.\ true stellar masses (\S\ref{ss:galaxy_sys}) as well as Compton-thick obscuration (\S\ref{ss:agn_observables} and Appendix \ref{a:agn_data}).
    \item[6.] Compare these predictions with real data to calculate the posterior probability $P(\boldsymbol{\theta}|\mathbf{d})=\pi (\boldsymbol{\theta}) \times \mathcal{L}(\boldsymbol{\theta}|\mathbf{d})$ of the parameters $\boldsymbol{\theta}$ given the observational constraints $\mathbf{d}$. The likelihood $\mathcal{L}(\boldsymbol{\theta}|\mathbf{d})$ is calculated with the $\chi^2(\boldsymbol{\theta}|\mathbf{d})$ from the comparison between our predictions with real data: $\mathcal{L}(\boldsymbol{\theta}|\mathbf{d}) \propto \exp[-\chi^2(\boldsymbol{\theta}|\mathbf{d})]$.
    \item[7.] Repeat steps 1--6, using an MCMC algorithm to determine the posterior distribution of the model parameters.
\end{itemize}

In this work, we use a custom implementation of the adaptive Metropolis MCMC method \citep{Haario2001}. A chain length of $2\times10^6$ steps was chosen to ensure the convergence of the posterior distribution. We have verified that this choice of chain length is at least $\sim$50 times longer than the autocorrelation length for every model parameter.

\section{Simulations and Data Constraints}
\label{s:sims_and_data}

\subsection{Dark Matter Halo Statistics}
\label{ss:dm_sims}

As noted in \S \ref{ss:overview}, \textsc{Trinity} requires only halo population statistics from dark matter simulations, as opposed to individual halo merger trees.  We use the peak historical mass (\mpeak{}) halo mass functions from \citet{Behroozi2013} for the cosmology specified in the introduction. These mass functions are based on central halo mass functions from \citet{Tinker2008}, with adjustments to include satellite halo number densities as well as to use \mpeak{} instead of the present day mass. These adjustments were based on the Bolshoi \& Consuelo simulations \citep{Klypin2011}. We refer readers to Appendix G of \citet{Behroozi2013} for full details. With these calibrations, the halo statistics used in this work are suitable for studying the evolution of halos from $10^{10} M_\odot$ to $10^{15} M_\odot$. For average halo mass accretion histories, we use the fitting formulae in Appendix H of \citet{Behroozi2013}. For halo mergers, we fit merger rates from the \textsc{UniverseMachine} \citep{Behroozi2019}, with full details and formulae in Appendix \ref{a:halo_merger_rates}.

\subsection{Observational Data Constraints}
\label{ss:obs_data}

We have compiled galaxy and AGN observables from $z=0-10$, which are summarized in Table \ref{t:summary_of_obs}. The following sections provide brief descriptions of these data.

\begin{table*}
\caption{Summary of Observational Constraints}
\begin{tabular}{lccc}
\hline
Type & Redshifts & Primarily Constrains & References\\
\hline
Stellar mass functions & 0-8 & SFR--$v_{\mathrm{Mpeak}}$ relation & Table \ref{t:smf} \\
Galaxy quenched fractions & 0-4 & Quenching--$v_{\mathrm{Mpeak}}$ relation & Table \ref{t:qf}\\
Cosmic star formation rates & 0-10 & SFR--$v_{\mathrm{Mpeak}}$ relation & Table \ref{t:csfr}\\
Specific star formation rates & 0-9 & SFR--$v_{\mathrm{Mpeak}}$ relation & Table \ref{t:ssfr}\\
Galaxy UV luminosity functions & 9-10 & SFR--$v_{\mathrm{Mpeak}}$ relation & Table \ref{t:uvlf}\\
Quasar luminosity functions & 0-5 & Total SMBH accretion & \citet{Ueda2014} \\
Quasar probability distribution functions & 0-2.5 & AGN duty cycle, BHAR distributions  & \citet{Aird2018}\\
Active SMBH mass functions & 0-5 & AGN energy efficiency & Table \ref{t:agn_data}\\
SMBH mass -- bulge mass relation & 0 & Galaxy--SMBH connection & Table \ref{t:bhbm_data}\\
Observed SMBH mass distribution of bright quasars & 5.8-6.5 & Galaxy-SMBH connection & \citet{Shen2019}\\
\hline
\end{tabular}
\label{t:summary_of_obs}

\parbox{17cm}{\textbf{Notes.} $v_{\mathrm{Mpeak}}$ is the maximum circular velocity of the halo at the time when it reaches its peak mass, \mpeak{}. This is used as a proxy for the halo mass in \textsc{Trinity}.  BHAR is the SMBH accretion rate.}

\end{table*}

\subsubsection{Galaxy data}
\label{sss:galaxy_data}
Five different observables are used to constrain the halo--galaxy connection in \textsc{Trinity}: stellar mass functions (SMFs, Table \ref{t:smf}), quenched fractions (QFs, Table \ref{t:qf}), cosmic star formation rates (CSFRs, Table \ref{t:csfr}), specific star formation rates (SSFRs, Table \ref{t:ssfr}), and UV luminosity functions (UVLFs, Table \ref{t:uvlf}). In this work, we adopt the compilation of these observables from \citet{Behroozi2019}. Here, we briefly introduce the data sources and the conversions made to ensure consistent physical assumptions across different datasets. For full details, we refer readers to Appendix C of \citet{Behroozi2019}.

\begin{table}
\caption{Observational Constraints on Galaxy Stellar Mass Functions} \label{t:smf}
\begin{tabular}{lccc}
\hline
Publication & Redshifts & Wavebands & Area (deg$^2$)\\
\hline
\citet{Baldry2012} & 0.002-0.06 & $ugriz$ & 143 \\
\citet{Moustakas2013} & 0.05-1 & UV-MIR & 9 \\
\citet{Tomczak2014} & 0.2-3 & UV-K$_{\mathrm{S}}$ & 0.08 \\
\citet{Ilbert2013} & 0.2-4 & UV-K$_{\mathrm{S}}$ & 1.5 \\
\citet{Muzzin2013} & 0.2-4 & UV-K$_{\mathrm{S}}$ & 1.5 \\
\citet{Song2016} & 4-8 & UV-MIR & 0.08 \\
\hline
\end{tabular}
\end{table}

Stellar mass functions at $z=0-8$ come from the following surveys: the Sloan Digital Sky Survey (SDSS, \citealt{York2000}), the PRIsm MUlti-object Survey (PRIMUS, \citealt{Coil2011,Cool2013}), UltraVISTA \citep{McCracken2012}, the Cosmic Assembly Near-infrared Deep Extragalactic Legacy Survey (CANDELS, \citealt{Grogin2011,Koekemoer2011}), and the FourStar Galaxy Evolution Survey (ZFOURGE, \citealt{Straatman2016}). Data points were converted to be consistent with the \citet{Chabrier2003} IMF, the \citet{Bruzual2003} SPS model, and the \citet{Calzetti2000} dust model. Additional corrections were made to homogenize photometry for massive galaxies \citep[see Appendix C of][]{Behroozi2019}. 

Constraints on galaxy quenched fractions as a function of stellar mass are taken from \citet{Bauer2013}, \citet{Moustakas2013} and \citet{Muzzin2013}. Each group calculated quenched fractions in a different way, but we assume that they all refer to galaxies with negligible global star formation rates (see \S \ref{ss:halo_galaxy_connection}). Although this results in some uncertainty in the interpretation of galaxy quenched fractions, it does not affect the main analysis, which only depends on the average star formation rate as a function of halo mass.

\begin{table}
\caption{Observational Constraints on Galaxy Quenched Fractions}
\label{t:qf}
\begin{tabular}{rcc}
\hline
Publication & Redshifts & Definition of Quenching\\
\hline
\citet{Bauer2013} & 0-0.3 & Observed SSFR\\
\citet{Moustakas2013} & 0.2-1 & Observed SSFR\\
\citet{Muzzin2013} & 0.2-4 & UVJ diagram\\
\hline
\end{tabular}
\end{table}

SSFRs and CSFRs at $0 < z < 10.5$ are obtained from multiple surveys (including SDSS, GAMA, UltraVISTA, CANDELS, and ZFOURGE) and techniques (UV, IR, radio, H$\alpha$, SED fitting, and gamma-ray bursts). These data points were only corrected to ensure the same initial mass function (the \citealt{Chabrier2003} IMF), because aligning other physical assumptions does not improve the self-consistency between SFRs and the growth of SMFs \citep{Madau2014,Leja2015,Tomczak2016}. 

\begin{table}
\caption{Observational Constraints on the Cosmic Star Formation Rate} \label{t:csfr}
\begin{tabular}{rccc}
\hline
Publication & Redshifts & Waveband & Area (deg$^2$)\\
\hline
\citet{Robotham2011} & 0-0.1 & UV & 833 \\
\citet{Salim2007} & 0-0.2 & UV & 741 \\
\citet{Gunawardhana2013} & 0-0.35 & H$\alpha$ & 144 \\
\citet{Ly2011a} & 0.8 & H$\alpha$ & 0.8 \\
\citet{Zheng2007} & 0.2-1 & UV/IR & 0.46 \\
\citet{Rujopakarn2010} & 0-1.2 & FIR & 0.4-9 \\
\citet{Drake2015} & 0.6-1.5 & [OII] & 0.63 \\
\citet{Shim2009} & 0.7-1.9 & H$\alpha$ & 0.03 \\
\citet{Sobral2014} & 0.4-2.3 & H$\alpha$ & 0.02-1.7 \\
\citet{Magnelli2011} & 1.3-2.3 & IR & 0.08 \\
\citet{Karim2011} & 0.2-3 & Radio & 2 \\
\citet{Santini2009} & 0.3-2.5 & IR & 0.04 \\
\citet{Ly2011b} & 1-3 & UV & 0.24 \\
\citet{Kajisawa2010} & 0.5-3.5 & UV/IR & 0.03 \\
\citet{Schreiber2015} & 0-4 & FIR & 1.75 \\
\citet{Planck2014} & 0-4 & FIR & 2240 \\
\citet{Dunne2009} & 0-4 & Radio & 0.8 \\
\citet{Cucciati2012} & 0-5 & UV & 0.6 \\
\citet{LeBorgne2009} & 0-5 & IR-mm & varies \\
\citet{vanderBurg2010} & 3-5 & UV & 4 \\
\citet{Yoshida2006} & 4-5 & UV & 0.24 \\
\citet{Finkelstein2015} & 3.5-8.5 & UV & 0.084 \\
\citet{Kistler2013} & 4-10.5 & GRB & varies\\
\hline
\end{tabular}
\parbox{1.0\columnwidth}{\textbf{Notes.} The technique of \citet{LeBorgne2009} (parametric derivation of the cosmic SFH from counts of IR-sub mm sources) uses multiple surveys with different areas. \citet{Kistler2013} used GRB detections from the \textit{Swift} satellite, which has fields of view of $\sim$3000 deg$^2$ (fully coded) and $\sim$10000 deg$^2$ (partially coded).}
\end{table}

\begin{table}
\caption{Observational Constraints on Galaxy Average Specific Star Formation Rates}
\label{t:ssfr}
\begin{tabular}{rccc}
\hline
Publication & Redshifts & Type & Area (deg$^2$)\\
\hline
\citet{Salim2007} & 0-0.2 & UV & 741 \\
\citet{Bauer2013} & 0-0.35 & H$\alpha$ & 144 \\
\citet{Whitaker2014} & 0-2.5 & UV/IR & 0.25 \\
\citet{Zwart2014} & 0-3 & Radio & 1 \\
\citet{Karim2011} & 0.2-3 & Radio & 2 \\
\citet{Kajisawa2010} & 0.5-3.5 & UV/IR & 0.03 \\
\citet{Schreiber2015} & 0-4 & FIR & 1.75 \\
\citet{Tomczak2016} & 0.5-4 & UV/IR & 0.08 \\
\citet{Salmon2015} & 3.5-6.5 & SED & 0.05 \\
\citet{Smit2014} & 6.6-7 & SED & 0.02 \\
\citet{Labbe2013} & 7.5-8.5 & UV/IR & 0.04 \\
\citet{McLure2011} & 6-8.7 & UV & 0.0125\\
\hline
\end{tabular}
\end{table}

In this work, we also use UV luminosity functions from \citet{Ishigaki2018}, \citet{Oesch2018}, and \citet{Bouwens2019} at $z=9-10$ to constrain the halo--galaxy connection beyond the redshift coverage of SMFs. 

\begin{table}
\caption{Observational Constraints on Galaxy UV Luminosity Functions}
\label{t:uvlf}
\begin{tabular}{rcc}
\hline
Publication & Redshifts & Area (deg$^2$)\\
\hline
\citet{Bouwens2019} & 8-9 & 0.24\\
\citet{Ishigaki2018} & 8-9 & 0.016\\
\citet{Oesch2018} & 10 & 0.23\\
\hline
\end{tabular}
\end{table}

In this paper, we have assumed a non-evolving IMF from \citet{Chabrier2003}. With IMFs from \citet{Kroupa2001} and \citet{Salpeter1955}, the inferred stellar masses would be factors of 1.07 and 1.7 higher than using the \citet{Chabrier2003} IMF, respectively.  For SFRs, these factors are 1.06 and 1.58, respectively \citep{Salim2007}.  More generally, a top-heavy IMF would produce a higher fraction of massive stars, decreasing the mass-to-UV light ratios of galaxies, and ultimately the inferred stellar masses and SFRs from stellar population synthesis. There is some observational evidence that the IMF becomes more top-heavy with increasing SFR (e.g., \citealt[][]{Gunawardhana2011}), but it remains an open issue whether IMF varies with environment or redshift \citep[][]{Conroy2009,Bastian2010,vanDokkum2012,Krumholz2014,Lacey2016}. Therefore, we opt to use a universal IMF in this paper; for discussion on the potential effects of non-universal IMFs, we refer readers to Appendix G of \citet{Behroozi2019}.

\subsubsection{Supermassive black hole data}
\label{sss:smbh_data}

 There are five different kinds of SMBH observables in our compiled dataset: quasar luminosity functions (QLFs), quasar probability distribution functions (QPDFs), active black hole mass functions (ABHMFs), the local SMBH mass--bulge mass (\mbh{}--\mbulge{}) relation, and the observed SMBH mass distribution of high redshift bright quasars. These SMBH data are summarized in Table \ref{t:agn_data} (QLFs, QPDFs, and ABHMFs) and Table \ref{t:bhbm_data} (\mbh{}--\mbulge{}).

\begin{table*}
\caption{Observational Constraints on AGNs}
\label{t:smbh_data}
\begin{tabular}{rcccc}
\hline
Publication & Type & Redshifts & Waveband & Area (deg$^2$)\\
\hline
\citet{Ueda2014} & Luminosity functions & 0-5 & X-ray & 0.12-34000 \\
\citet{Aird2018} & AGN probability distribution functions& 0.1-2.5 & X-ray & 0.22-1.6 \\
\citet{Schulze2010} & Active black hole mass functions & 0-0.3 & Optical & 9500 \\
\citet{Schulze2015} & Active black hole mass functions & 1-2 & Optical & 0.62-6250 \\
\citet{Kelly2013} & Active black hole mass functions & 1.5-5 & Optical & 6250\\
\citet{Shen2019} & Observed SMBH mass distribution of bright quasars & 5.8-6.5 & Optical & 14000\\
\hline
\end{tabular}
\label{t:agn_data}

\parbox{14.5cm}{\textbf{Notes.} ``Waveband'' indicates the waveband used to measure SMBH properties.  \citet{Aird2018} additionally used UV, optical, and IR data to constrain host galaxy properties.}
\end{table*}

\begin{table}
\caption{Observational Constraints on the SMBH mass--bulge mass (\bhbm{}) relation at $z=0$}
\begin{tabular}{rcc}
\hline
Publication & $\beta_{\rm BH}$ & $\gamma_{\rm BH}$\\
\hline
\citet{Haring2004} & 8.20 & 1.12\\
\citet{Beifiori2012} & 8.25 & 0.79\\
\citet{Kormendy2013} & 8.69 & 1.15\\
\citet{McConnell2013} & 8.46 & 1.05\\
\citet{Savorgnan2016} & 8.55 & 1.05\\
\hline
Median & 8.46 & 1.05\\
\hline
Standard deviation & 0.20 & 0.14\\
\hline
\end{tabular}
\label{t:bhbm_data}

\textbf{Notes.} The median \bhbm{} relation is assumed to be a power-law: $\log_{10} (M_\bullet/M_\odot) = \beta_{\rm BH} + \gamma_{\rm BH} \log_{10} (M_{\rm bulge} / 10^{11} M_{\odot})$.
\end{table}

We have used bolometric quasar luminosity functions (QLFs) at $z=0-5$ from \citet{Ueda2014}, which are based on a series of X-ray surveys. There are also QLFs based on observations in other wavebands \citep[e.g., UV luminosity functions from][]{Kulkarni2019}, but we use those from X-ray surveys due to their uniformity in AGN selection and robustness against (moderate) obscuration. We adopted the empirical correction scheme from \citet{Ueda2014} to account for Compton-thick AGN populations (see Appendix \ref{aa:ctk_corr} for full details). We also tested using bolometric QLFs from multiple wavebands from \citet{Shen2020}, and found no qualitative changes in our results. The posterior distribution of model parameters does change significantly if assuming quasar luminosity functions and Compton-thick corrections from \citet{Ananna2019}. However, there is strong inconsistency between these luminosity functions and the QPDFs from \citet{Aird2018}. In light of this, we do not adopt Ananna et al.\ QLFs in the main text. For further details, we refer readers to Appendix \ref{aa:ctk_corr}.

QLFs constrain the total radiative energy output of active SMBHs \citep[][]{Conroy2013,Caplar2015}. To constrain the mass-dependence of AGN luminosity distributions, we included quasar probability distribution functions (QPDFs) from \citet{Aird2018}. These functions are expressed as the conditional probability distributions of \slx{}$\equiv L_{\mathrm{X}} / (1.04\times10^{34}\mathrm{erg\ s}^{-1}\times M_*/M_\odot)$. These distributions are given as functions of stellar mass ($M_*$) and redshift, and are obtained by modeling the X-ray luminosities of galaxies in the CANDELS and UltraVISTA surveys. \citet{Aird2018} did not correct for the presence of Compton-thick AGNs in their modeling, so we adopted the empirical scheme given by \citet{Ueda2014} to correct our predicted QPDFs for this selection bias (see Appendix \ref{aa:ctk_corr} for more details).

In modeling how AGN luminosity connects to SMBH growth, there is a degeneracy between the SMBH accretion rate and the radiative efficiency. To break this degeneracy, we include 1) active black hole mass functions (ABHMFs) from $z=0.2-5$ from \citet{Schulze2010,Kelly2013,Schulze2015}; and 2) the local \bhbm{} relation to constrain the total amount of SMBH mass accreted over cosmic time. Given the different sample selection criteria and data reduction schemes used by different groups, we decided not to use individual data points for the \bhbm{} relation. Instead, we picked five commonly-used local \bhbm{} relations and calculated the medians and standard deviations of their slopes and intercepts (see Table \ref{t:bhbm_data}). We then apply Gaussian priors on both the slope and the intercept at $z=0$ in \textsc{Trinity}, with the centres and widths set to these medians and standard deviations.

Given the capability of contemporary telescopes, the sample of $z\gtrsim5$ AGNs is likely biased against faint objects. However, the observed SMBH mass distribution of these high redshift quasars still provides useful constraints on \textsc{Trinity}. Specifically, we know from observations that few quasars with $L_\mathrm{bol} > 10^{47}$ erg/s at $5.8<z<6.5$ have \emph{observed} $M_\bullet < 10^8 M_\odot$ \citep{Shen2019}. Therefore, the expected number of these quasars in \textsc{Trinity}, $N_\mathrm{exp}$, should also be small. Assuming Poisson statisics, the prior probability that we detect no low-mass bright quasars with a survey like SDSS is:
\begin{eqnarray}
P(N_\mathrm{obs}=0|N_\mathrm{exp})
& = & \exp{(-N_\mathrm{exp})} \\
N_\mathrm{exp} & = &  \int_{0}^{10^8} P(M_{\bullet,\rm obs} | M_{\bullet,\rm int}) d M_{\bullet, \rm obs}\\ \nonumber 
& \times & \int_{0}^{\infty} d M_{\bullet, \rm int} \int_{10^{47}}^{\infty} d L_\mathrm{bol} P(L_\mathrm{bol} | M_{\bullet, \rm int}) \phi_\mathrm{BH}(M_{\bullet, \rm int}) \\ \nonumber
& \times & \mathcal{S}_\mathbf{SDSS} \times \Delta z\\
P(M_{\bullet,\rm obs} | M_{\bullet,\rm int}) & = & \frac{1}{\sqrt{2\pi} \sigma_{\rm BH,obs}}\\ \nonumber
& \times & \exp{\left[-\frac{(\log M_{\bullet,\rm obs} - \log M_{\bullet,\rm int})^2}{2\sigma_{\rm BH,obs}^2}\right]}\ ,
\end{eqnarray}
where $M_{\bullet,\rm int}$ and $M_{\bullet,\rm obs}$ are the intrinsic and observed SMBH masses, respectively, and $\sigma_{\rm BH,obs}=0.4$ dex is the random scatter in SMBH mass as induced by virial estimates \citep{Park2012}. $\mathcal{S}_\mathrm{SDSS}=14000$ deg$^2$ is the survey area of SDSS. Here, we take  $\Delta z = 6.5- 5.8 = 0.7$ to keep consistency with \citet{Shen2019}. In the MCMC process, we included this prior to prevent \textsc{Trinity} from producing too many low-mass and super-Eddington quasars, which are not supported by observations \citep[e.g.,][]{Mazzucchelli2017,Trakhtenbrot2017}.

In the process of compiling these data, we found systematic discrepancies between some observational datasets, which are addressed in Appendices \ref{aa:aird_qpdf} (quasar X-ray luminosities) and \ref{aa:abhmf} (active black hole mass functions).
 
\begin{figure}
\subfigure{
\includegraphics[width=0.48\textwidth]{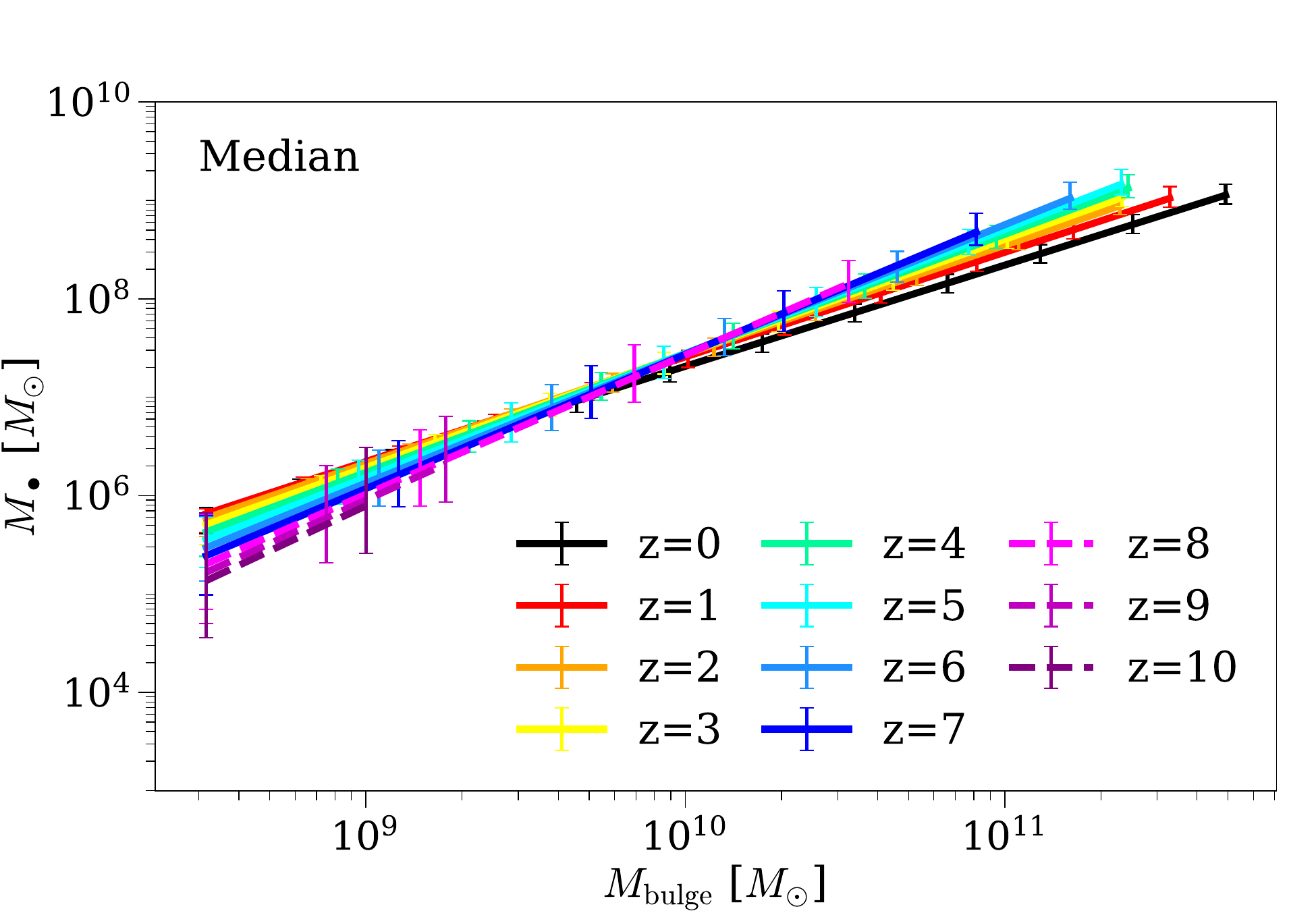}
}
\subfigure{
\includegraphics[width=0.48\textwidth]{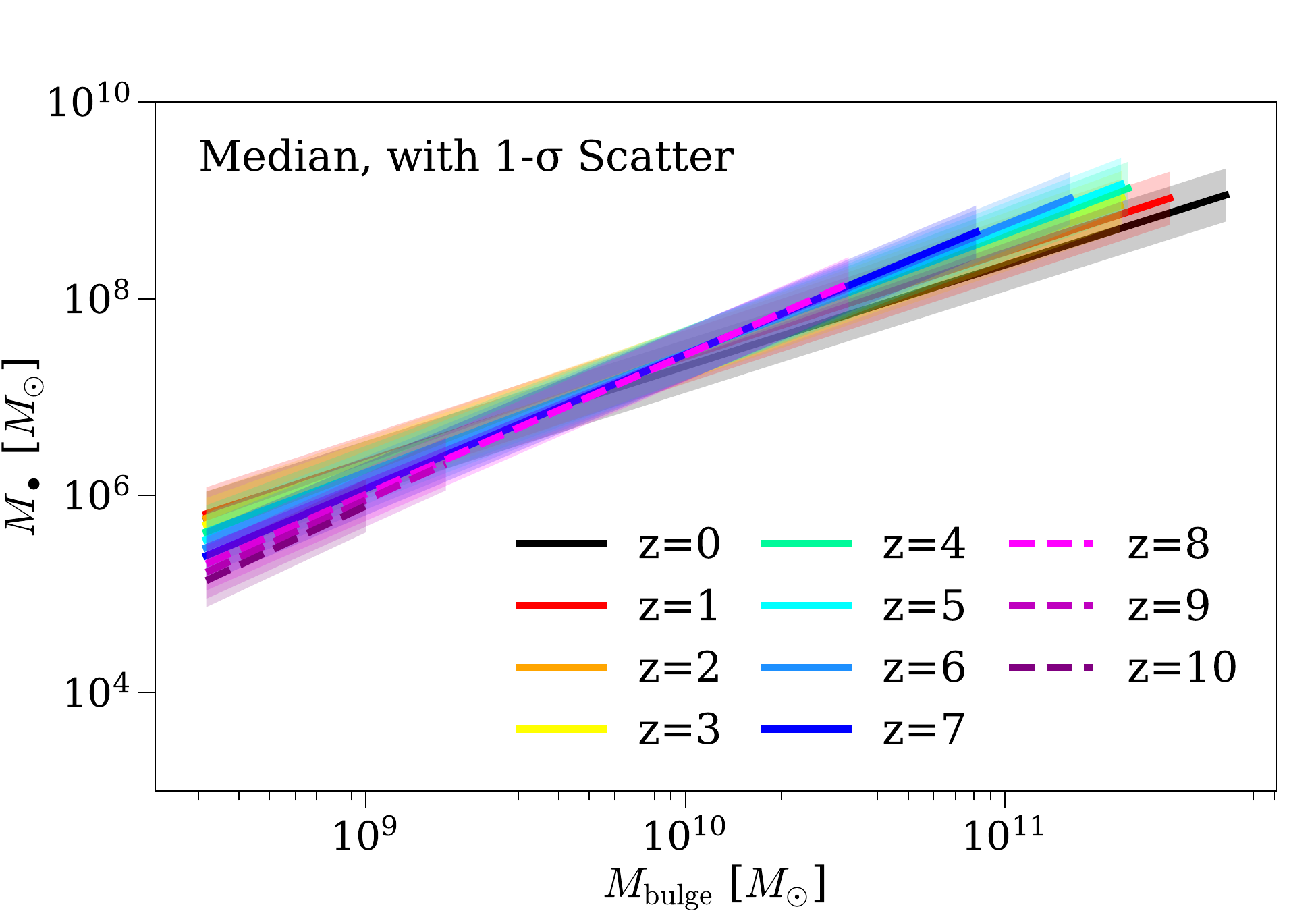}
}
\caption{The evolution of the median \mbh{}--\mbulge{} relation and the corresponding log-normal scatter from $z=0-10$. \textbf{Top Panel:} the median relations (see \S\ref{ss:results_bhbm}). \errorbars{} \textbf{Bottom Panel:} The same median relations, except that the shaded regions show the \emph{log-normal scatter} around the median relations. \waitforjwst{} All the data used to make this plot can be found \href{https://github.com/HaowenZhang/TRINITY/tree/main/plot_data}{here}.}
\label{f:bhbm_median}
\end{figure}

\section{Results}
\label{s:results}

\begin{figure}
\includegraphics[width=0.48\textwidth]{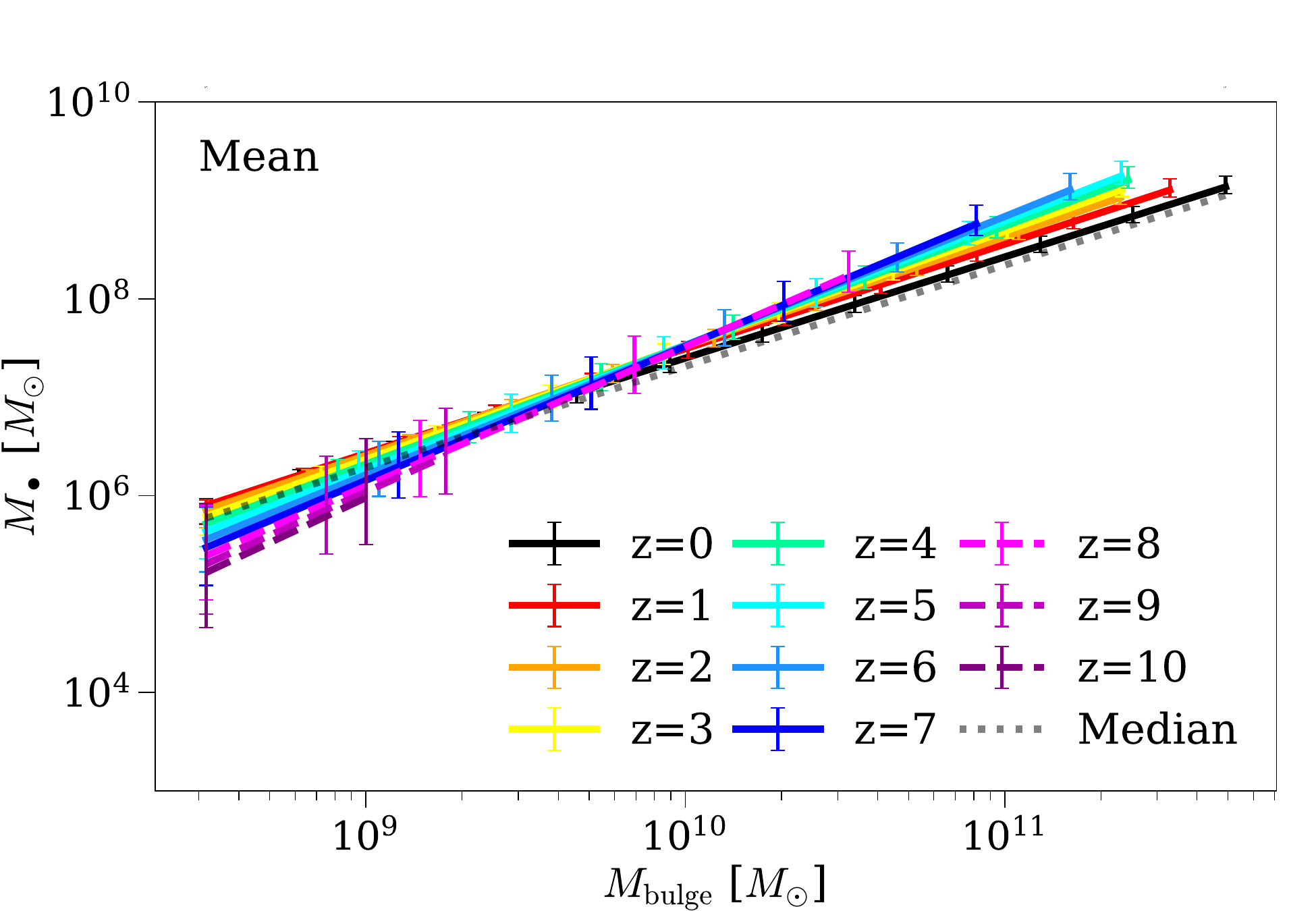}
\caption{The evolution of the mean \mbh{}--\mbulge{} relation from $z=0-10$ (see \S\ref{ss:results_bhbm}). The grey dotted line shows  the median relation at $z=0$ for comparison. \errorbars{} \waitforjwst{} All the data used to make this plot can be found \href{https://github.com/HaowenZhang/TRINITY/tree/main/plot_data}{here}.}
\label{f:bhbm_mean}
\end{figure}

We present the best fitting parameters and the comparisons to observations in \S\ref{ss:results_best_fit}, as well as results for the evolution of the \bhbm{} relation in \S\ref{ss:results_bhbm}, black hole accretion rates and Eddington ratio distributions in \S\ref{ss:results_bhar_bher}, the SMBH mass function in \S\ref{ss:results_bhmf}, SMBH mergers in \S\ref{ss:results_mergers}, AGN energy efficiency as well as systematic uncertainties in \S\ref{ss:results_systematics}, and the correlation coefficient between average SMBH accretion rate and \mbh{} at fixed halo mass in \S\ref{ss:rho_bh}.

\subsection{Best fitting parameters and comparison to observables}
\label{ss:results_best_fit}

\begin{figure}
\subfigure{
\includegraphics[width=0.48\textwidth]{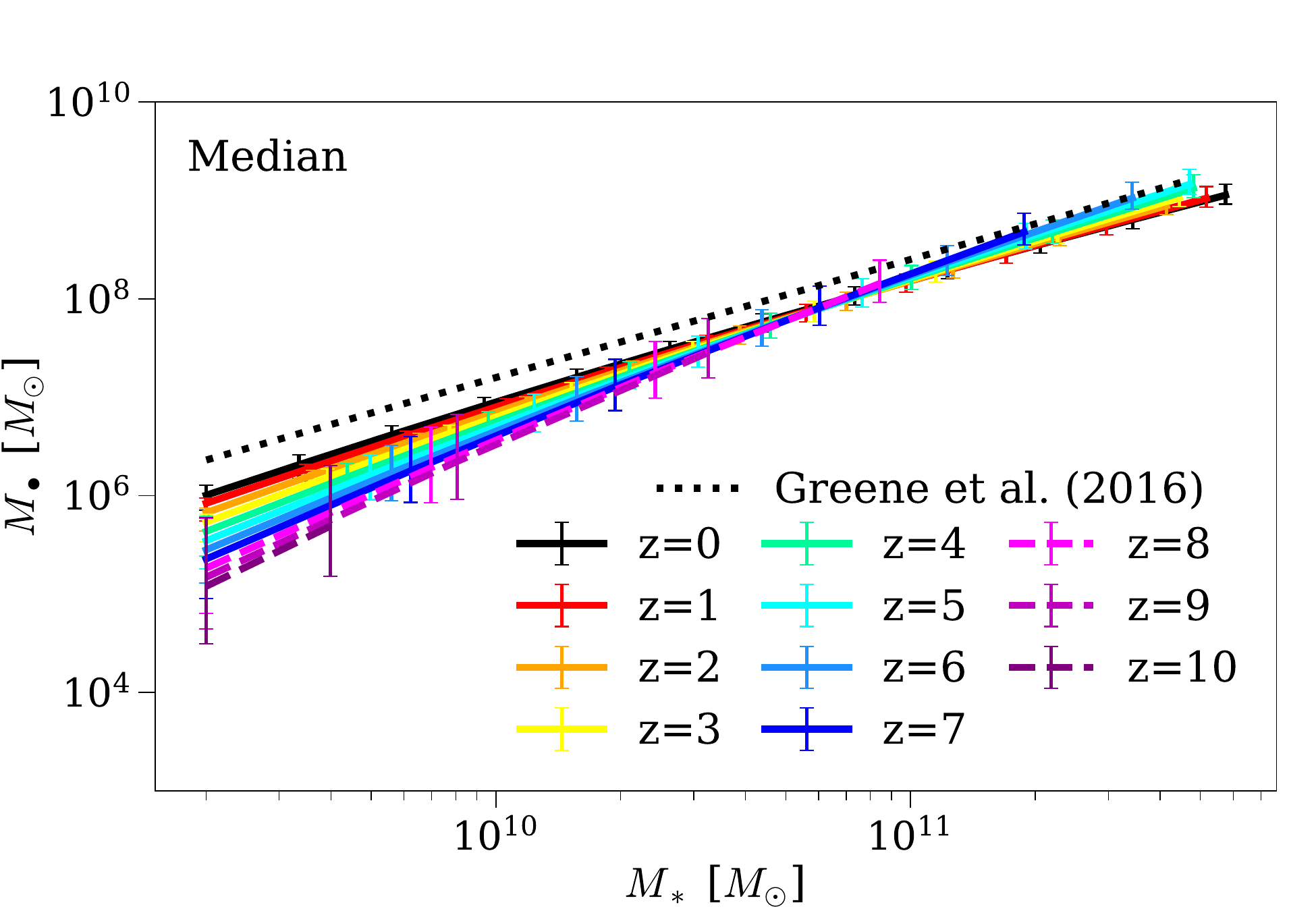}
}
\subfigure{
\includegraphics[width=0.48\textwidth]{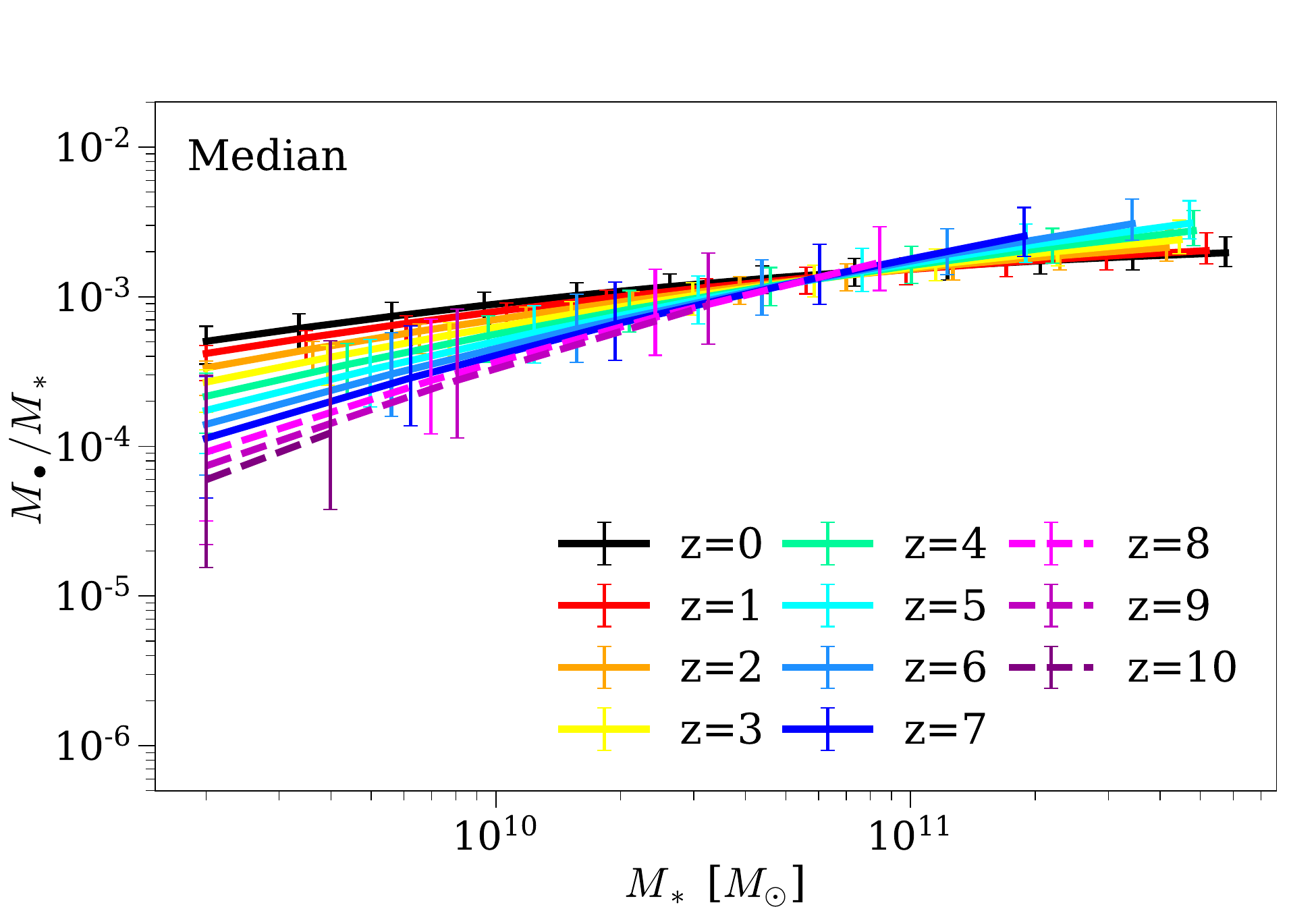}
}
\caption{\textbf{Top Panel:} the best-fitting median \mbh{}--\mstar{} relation from $z=0-10$ (solid lines, see \S\ref{ss:results_bhbm}), and the observed $z=0$ \bhsm{} relation from \citet{Greene2016} (dotted line). \textbf{Bottom Panel:} the best-fitting median \mbh{}/\mstar{} ratios as a function of \mstar{} and $z$. \errorbars{} \waitforjwst{} All the data used to make this plot can be found \href{https://github.com/HaowenZhang/TRINITY/tree/main/plot_data}{here}.}
\label{f:bhsm_median}
\end{figure}

We obtained the posterior distribution of model parameters with an MCMC algorithm (\S \ref{ss:method_summary}). The best fitting model was found by the following two-step procedure: (1) calculate the weighted average of the 2000 highest-probability points in the MCMC chain; (2) starting from this weighted average, run a gradient descent optimization over each dimension of the parameter space, until the model $\chi^2$ stops changing.

Our best fitting model is able to fit all the data in our compilation (\S\ref{s:sims_and_data}), including stellar mass functions (SMFs, Fig.\ \ref{f:smf_qf}, left panel), quenched fractions (QFs, Fig.\ \ref{f:smf_qf}, right panel), cosmic star formation rates (CSFRs, Fig.\ \ref{f:csfr_ssfr}, left panel), specific star formation rates (SSFRs, Fig.\ \ref{f:csfr_ssfr}, right panel), galaxy UV luminosity functions (UVLFs, Fig.\ \ref{f:uvlf}), quasar luminosity functions (QLFs, Fig.\ \ref{f:qlf}), active black hole mass functions (ABHMFs, Figs.\ \ref{f:abhmf} and \ref{f:abhmf_kelly}), quasar probability distribution functions (QPDFs, Fig.\ \ref{f:qpdf}) and the local \mbh{}--\mbulge{} relation (Fig.\ \ref{f:bhbm_fit}). For 1189 data points and 56 parameters, the naive reduced $\chi^2$ is 0.66, which suggests a reasonable fit. The best fitting model and $68\%$ confidence intervals for parameters are presented in Appendix \ref{a:param_values}.

As shown in Fig.\ \ref{f:qpdf}, \textsc{Trinity} largely reproduces the mass-dependence of the QPDFs from \citet{Aird2018}, but it does not fully recover the QPDF shape for galaxies with $M_* < 10^{10} M_\odot$. Specifically, \textsc{Trinity} tends to overpredict active AGNs in these low mass galaxies at $z>1$. Given the complexity of the models adopted by \citet{Aird2018} to calculate these QPDFs, we did not add additional free parameters to fully reproduce their shapes, which reduces the risk of over-fitting.

\subsection{The \bhbm{} relation for $z=0$ to $z=10$}
\label{ss:results_bhbm}

\begin{figure}
\subfigure{
\includegraphics[width=0.48\textwidth]{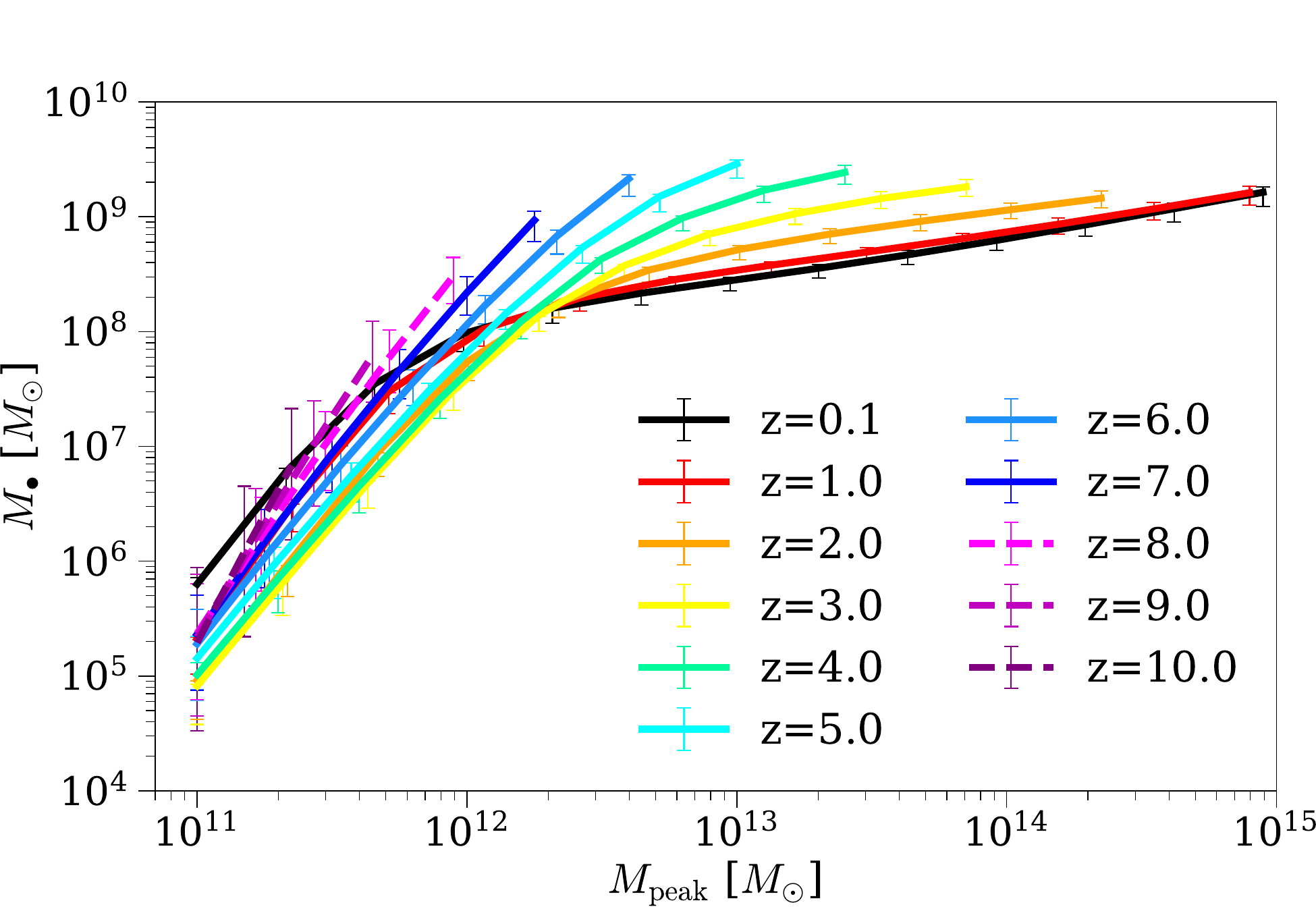}
}
\subfigure{
\includegraphics[width=0.48\textwidth]{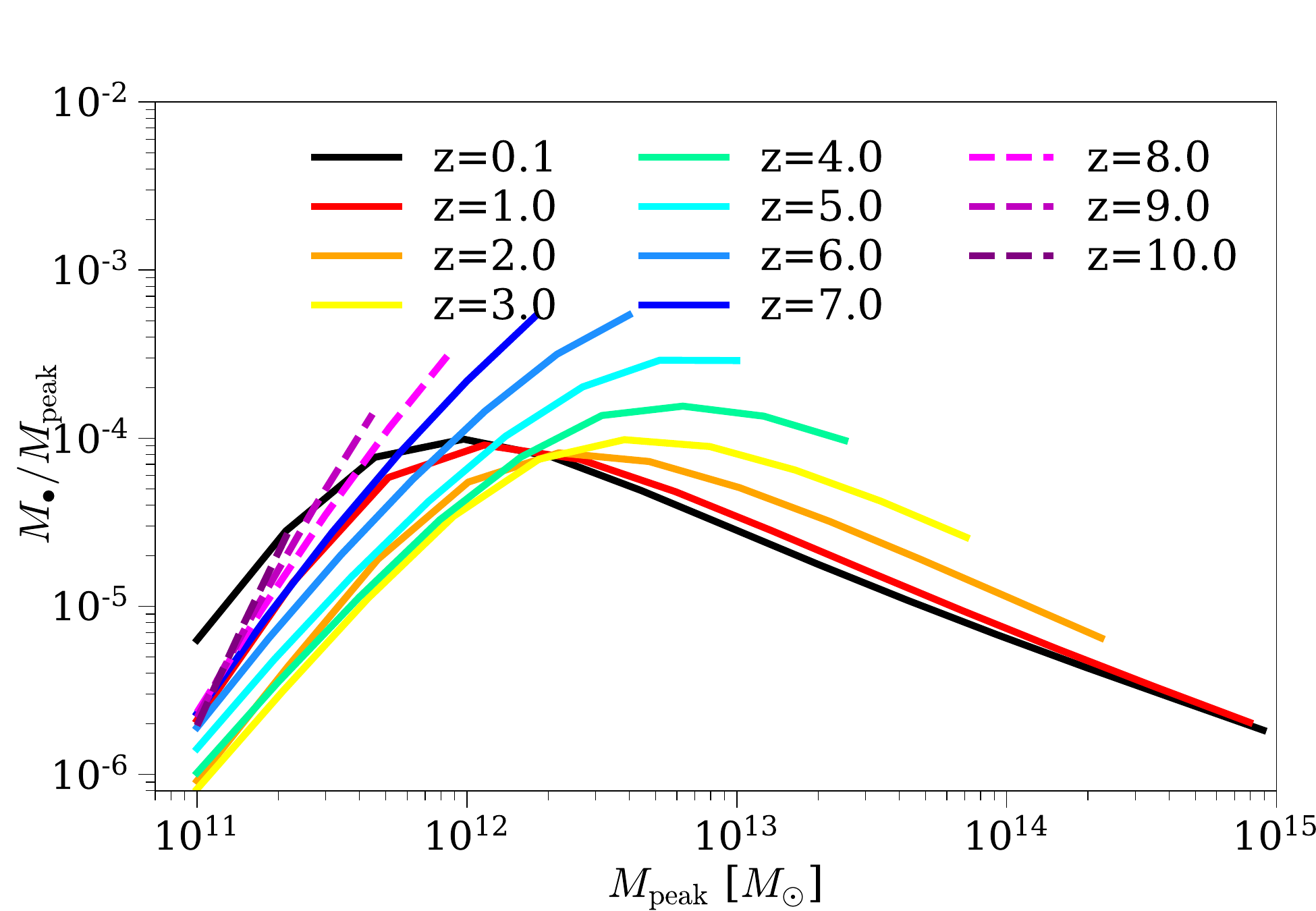}
}
\caption{\textbf{Top Panel:} the best-fitting median \mbh{}--\mpeak{} (peak halo mass) relation from $z=0-10$ (see \S\ref{ss:results_bhbm}). \textbf{Bottom Panel:} the best-fitting $M_\bullet/M_\mathrm{peak}$ ratios  as a function of \mpeak{} and $z$. \errorbars{} \waitforjwst{} All the data used to make this plot can be found \href{https://github.com/HaowenZhang/TRINITY/tree/main/plot_data}{here}.}
\label{f:bhhm_median}
\end{figure}

In Fig.\ \ref{f:bhbm_median}, we show the redshift evolution of the \emph{median} SMBH mass--bulge mass (\bhbm{}) relation (top panel) along with the log-normal scatter (bottom panel) from $z=0-10$. We find that both the slope and the normalization of the median \bhbm{} relation increase mildly from $z=0-10$. From $z=0-3$, the evolution in the median \mbh{} at fixed \mbulge{} is at most $\sim 0.3$ dex, which is within the typical SMBH mass uncertainties. The median \bhbm{} relation beyond $z=0$ is jointly constrained by the quasar luminosity functions (QLFs), quasar probability distribution functions (QPDFs), active black hole mass functions (ABHMFs), and the galaxy stellar mass functions (SMFs). Specifically, QLFs and QPDFs jointly constrain the Eddington ratio distributions and duty cycles of SMBHs. On the other hand, ABHMFs specify the abundances of \emph{active} SMBHs as a function of their masses. Combined with the Eddington ratio distributions and duty cycles, this information helps \textsc{Trinity} infer the number density of \emph{active+dormant} SMBHs at different masses, i.e., the \emph{total} SMBH mass functions. Reproducing these SMBH mass functions given the observed number density of galaxies (i.e., their SMFs) places strong constraints on the \bhbm{} relation. At $z\ge 8$ (shown in Fig.\ \ref{f:bhbm_median} as dashed lines), the median \mbh{} at fixed bulge mass is lower compared to the $z=0$ values, but consistent within the statistical uncertainties from MCMC. Without existing SMBH data at this cosmic era, we expect that future observations (by, e.g., the \emph{James Webb Space Telescope} [\textit{JWST}]) will test our predictions. It is likely that many future observations can only probe the most massive SMBHs at such high redshifts, but they will still provide useful tests as to whether their number densities are consistent with the median \bhbm{} relation and the scatter around it.

The scatter around the median \bhbm{} relation is $\sigma_\mathrm{BH}\approx 0.27$ dex. As described in \S\ref{ss:bh_mergers}, a log-normal scatter of $\sigma_\mathrm{BH}$ causes an offset between the \emph{median} and \emph{mean} SMBH masses (\S\ref{ss:bh_mergers}) at fixed stellar mass. Mean SMBH masses directly influence average BHARs, which are constrained by observed QLFs and QPDFs. Consequently, $\sigma_\mathrm{BH}$ is primarily constrained by (a) the evolution of the median \bhbm{} relation; and (b) the average BHARs inferred from QLFs and QPDFs. Another constraint comes from the shape of ABHMFs, since bigger scatter would produce more over-massive SMBHs in low-mass galaxies than under-massive SMBHs in high-mass galaxies. Therefore, flatter ABHMFs implies a bigger scatter around the \bhbm{} relation.

In Fig.\ \ref{f:bhbm_mean}, we show the evolution of the \emph{mean} \bhbm{} relation from $z=0-10$. With $\sigma_{\rm BH}\approx 0.27$ dex, the mean relation is offset from the median relation by a constant factor of $0.5\sigma_{\rm BH}^2 \ln 10 \approx 0.08$ dex.

\begin{figure}
\subfigure{
\includegraphics[width=0.48\textwidth]{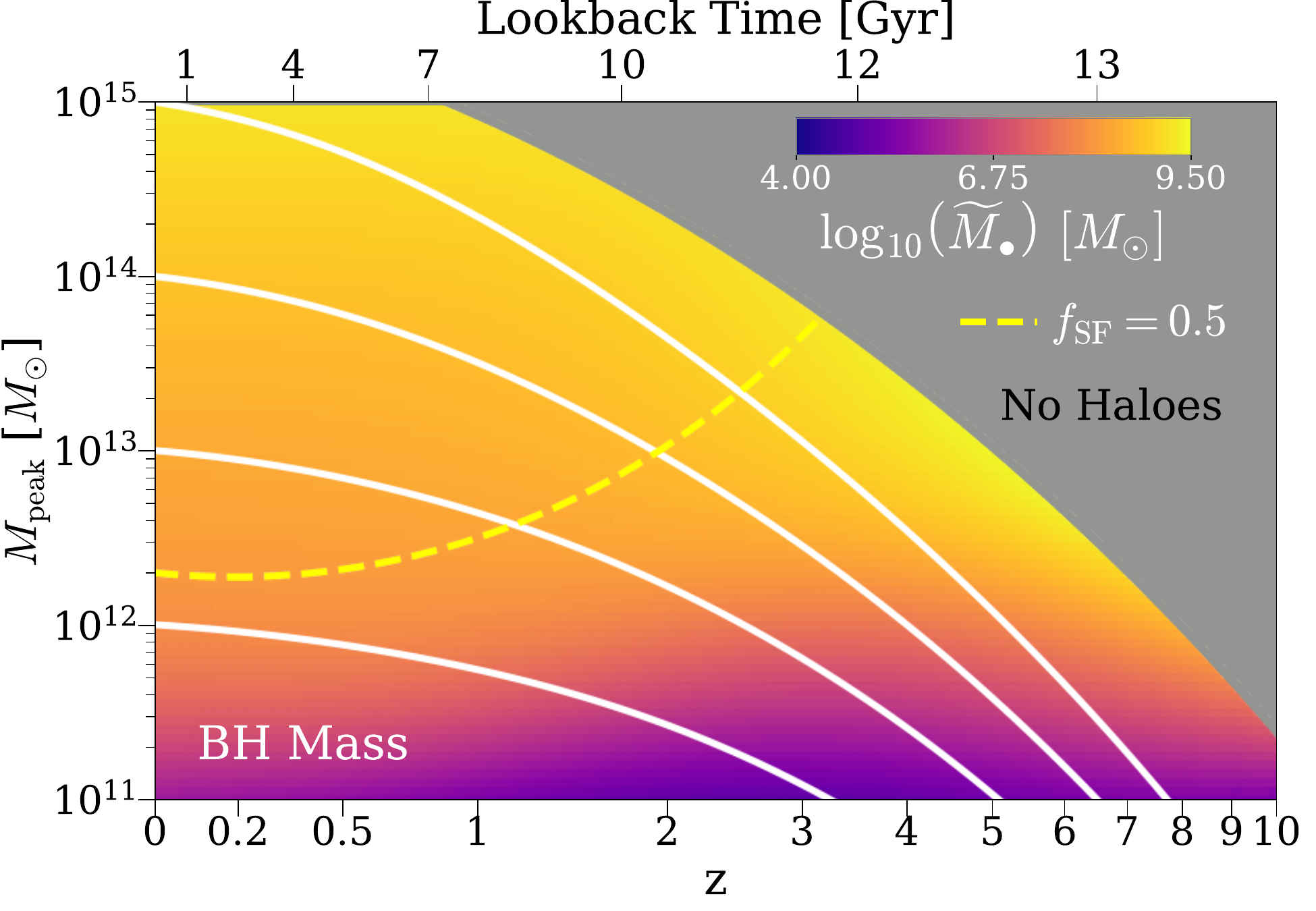}
}
\subfigure{
\includegraphics[width=0.48\textwidth]{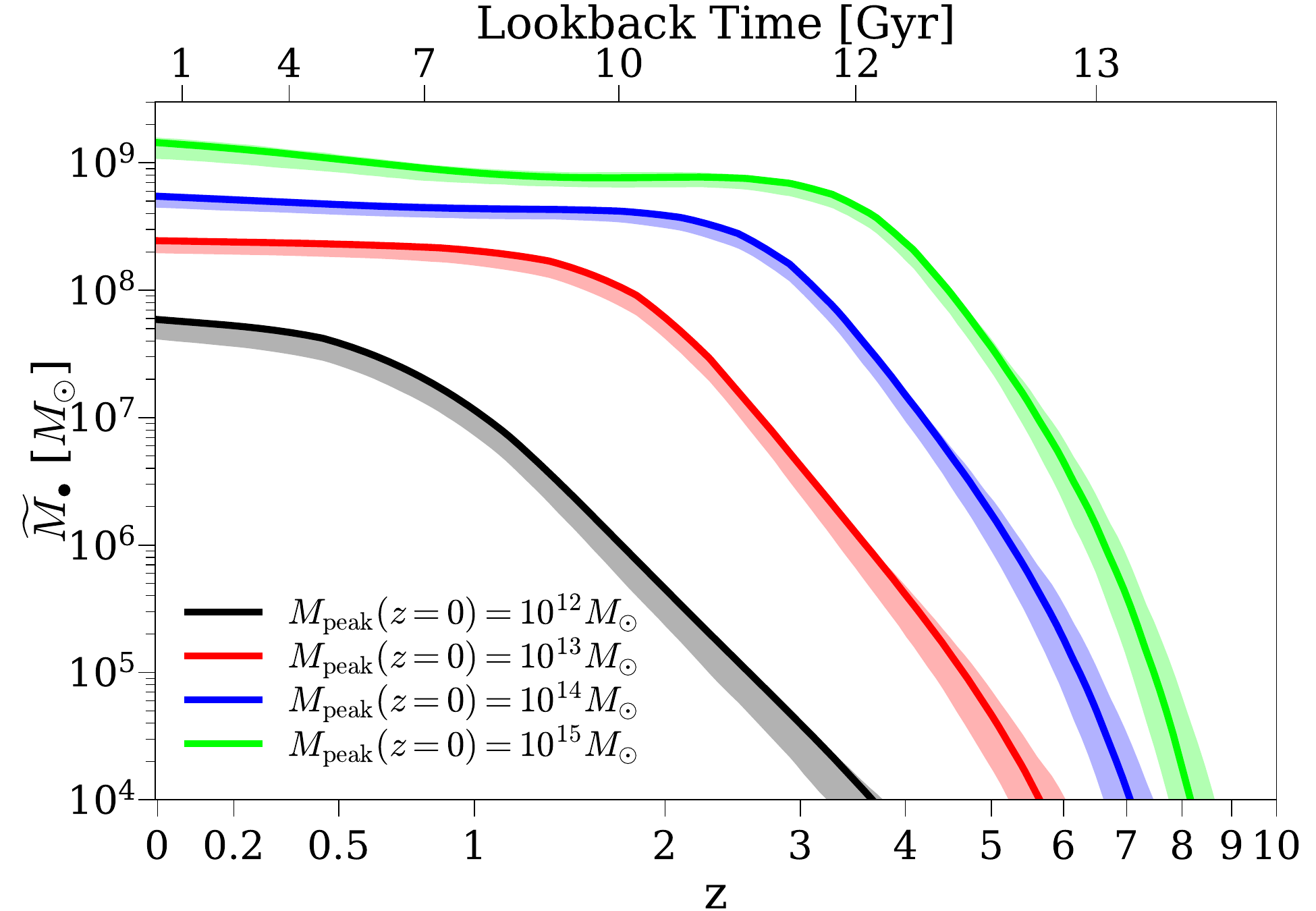}
}

\caption{\textbf{Top Panel:} the median SMBH mass ($\widetilde{M}_\bullet$) as a function of $M_{\rm peak}$ and $z$ (see \S\ref{ss:results_bhbm}). \sfcurve{} \halocurves{} \nohalos{} \textbf{Bottom Panel:} The $\widetilde{M}_\bullet$ histories as a function of halo mass at $z=0$. \shadedregions{} All the data used to make this plot can be found \href{https://github.com/HaowenZhang/TRINITY/tree/main/plot_data}{here}.}
\label{f:mbh_map}
\end{figure}

Fig.\ \ref{f:bhsm_median} shows the best-fitting median SMBH mass--galaxy total stellar mass (\bhsm{}) relation. Our $z=0$ \bhsm{} relation is consistent with measurements by \citet{Greene2016} using water megamaser disk observations. This relation is qualitatively similar to the \bhbm{} relation mainly because of the approximate proportionality between \mbulge{} and \mstar{} (Eq.\ \ref{e:bm_sm}). Quantitatively, the evolution of the \bhsm{} relation between $0<z<2$ is less significant than that of the \bhbm{} relation, due to lower $M_\mathrm{bulge} / M_*$ ratios at higher redshifts, which is also consistent with observational studies like \citet{Ding2020}.  The evolution of the \bhsm{} relation causes the median \mbh{}/\mstar{} ratio (Fig.\ \ref{f:bhsm_median}, bottom panel) to decrease with redshift. Overall, the mild evolution is consistent with observational studies that found no significant redshift 
dependence in the \bhbm{} and \bhsm{} relations between $0<z<2$ (e.g., \citealt[][]{Schramm2013,Sun2015,Suh2020}).

Fig.\ \ref{f:bhhm_median} shows the best-fitting median SMBH mass--halo peak mass (\mbh{}--\mpeak{}) relation. At $z\lesssim 5$, the \mbh{}--\mpeak{} relation can be approximated as a double power-law, connected by a knee at $M_{\rm peak} \sim 10^{12} M_{\odot}$. Above $z=5$, it is roughly a single power-law due to the lack of massive haloes. This halo mass dependence is inherited from the well-known stellar mass--halo mass (\mstar{}--\mpeak{}) relation, because of the approximate single power-law shapes of the \bhsm{} connection (Fig.\ \ref{f:bhsm_median}; see also \citealt{Kormendy2013}).

The top panel of Fig.\ \ref{f:mbh_map} shows the median SMBH mass ($\widetilde{M}_\bullet$) as a function of \mpeak{} and $z$. From $z=0-10$, SMBH masses in haloes with $M_{\rm peak} \sim 10^{11} M_{\odot}$ remain consistently low. But SMBHs do grow in mass along with their host haloes/galaxies, as indicated by the halo growth curves (white solid lines).

The bottom panel of Fig.\ \ref{f:mbh_map} shows the $\widetilde{M}_\bullet$ histories along the growth histories of different haloes. At all halo masses, SMBH growth is very fast in the early Universe, and slows down towards lower redshifts. However, the fast-growth phase ends earlier for more massive black holes. This is consistent with the phenomenon called ``AGN downsizing'' \citep[e.g.,][]{Merloni2004a,Barger2005}, and we discuss this further in \S\ref{ss:results_bhar_bher} and \S\ref{ss:discussions_physical_implications}.

\subsection{Average black hole accretion rates and Eddington ratio distributions}
\label{ss:results_bhar_bher}

\begin{figure}
\subfigure{
\includegraphics[width=0.48\textwidth]{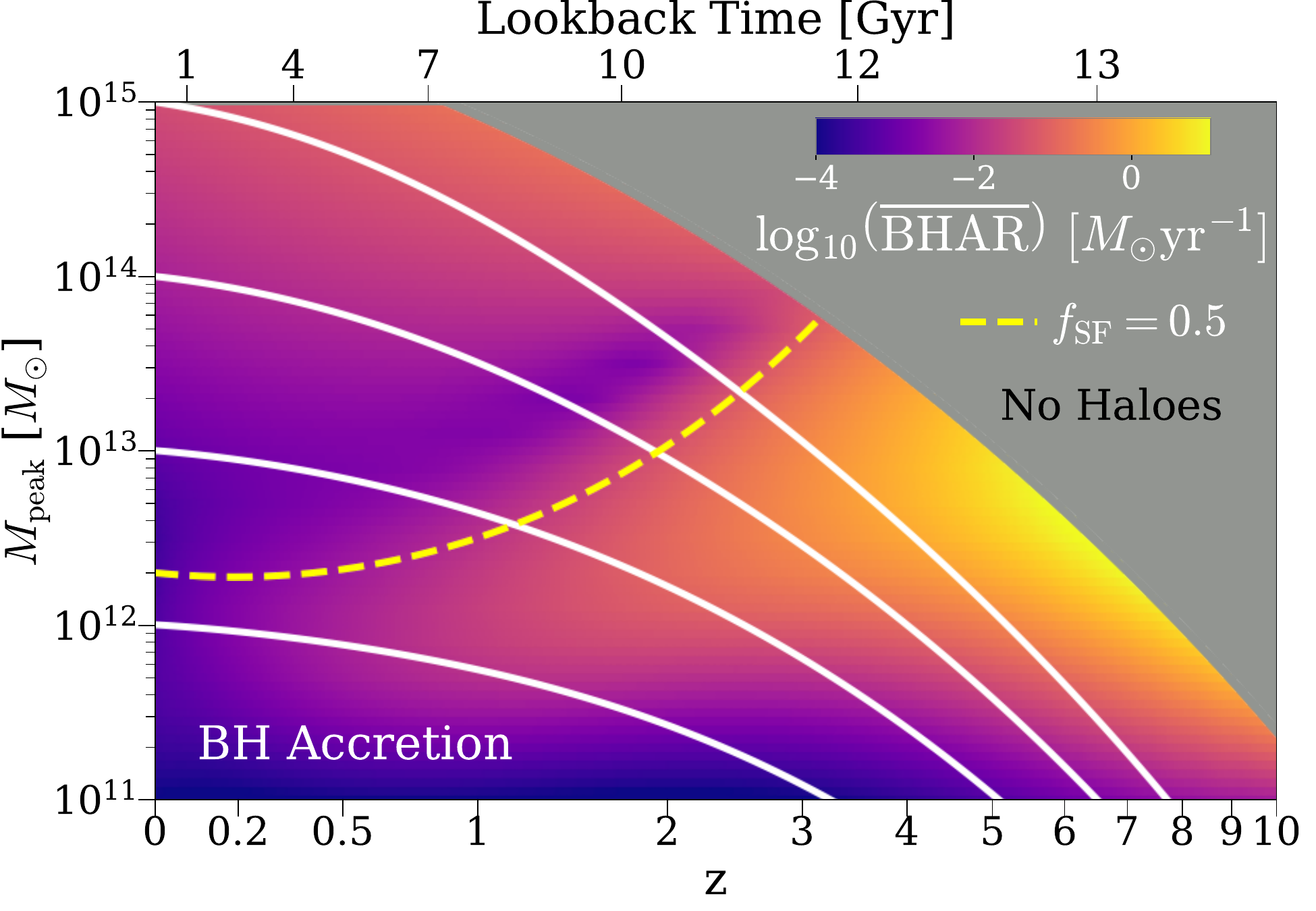}
}
\subfigure{
\includegraphics[width=0.48\textwidth]{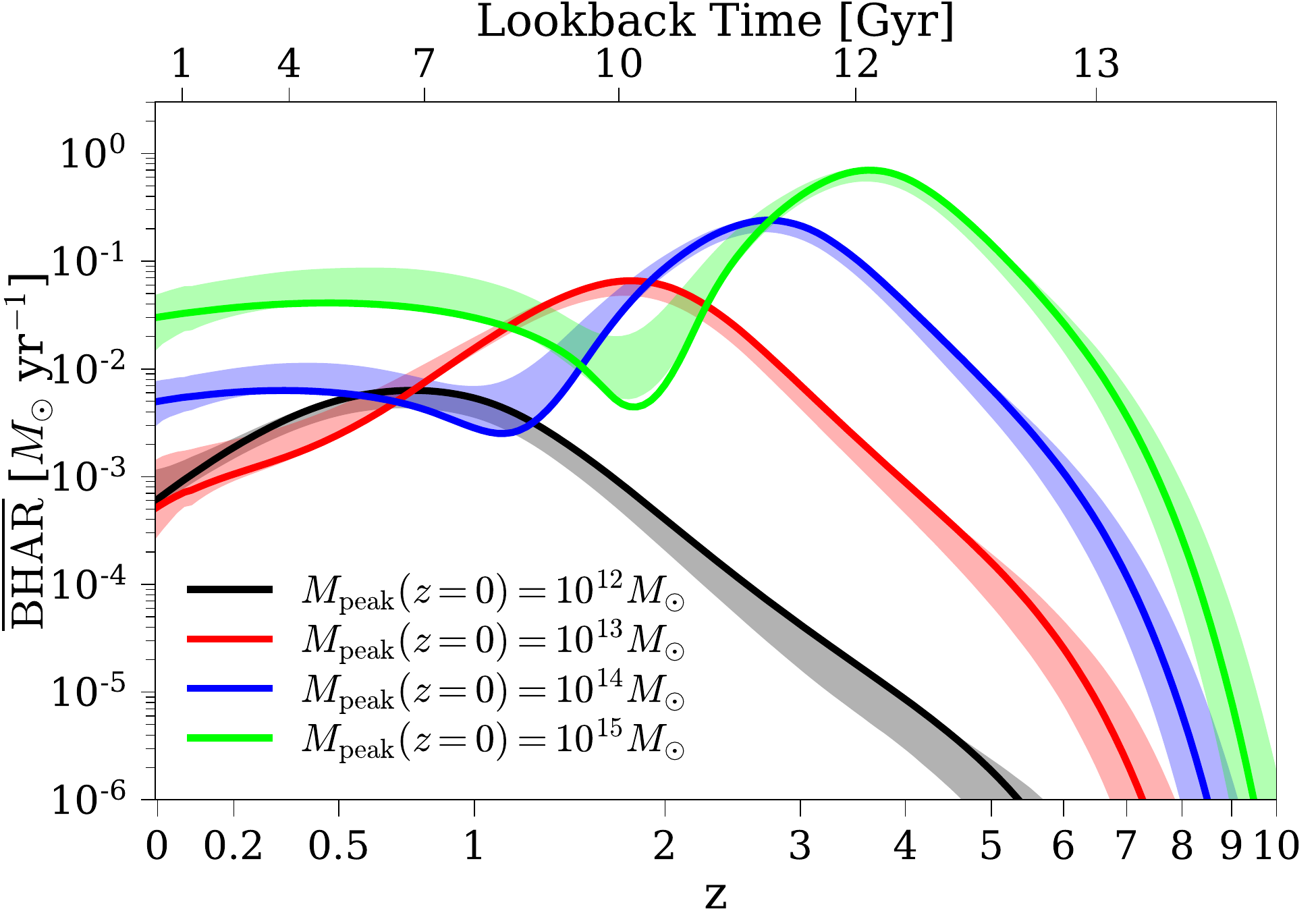}
}
\caption{\textbf{Top Panel:} average black hole accretion rate ($\overline{\mathrm{BHAR}}$) as a function of $M_{\rm peak}$ and $z$ (see \S\ref{ss:results_bhar_bher}). \sfcurve{} \halocurves{} \nohalos{} \textbf{Bottom Panel:} $\overline{\mathrm{BHAR}}$ histories as a function of halo mass at $z=0$. \shadedregions{} All the data used to make this plot can be found \href{https://github.com/HaowenZhang/TRINITY/tree/main/plot_data}{here}.}
\label{f:bhar_map}
\end{figure}

The top panel of Fig.\ \ref{f:bhar_map} shows the average black hole accretion rate ($\overline{\mathrm{BHAR}}$) as a function of $M_{\rm peak}$ and $z$. In general, BHARs peak at $M_{\rm peak} \sim 10^{12} M_{\odot}$, and decrease towards lower and higher masses. Below $z\sim 2$ and $M_\mathrm{peak}\sim 10^{13.5} M_\odot$, BHARs decrease with time at fixed mass. At $z\sim 2$, there is also a slight increase in BHAR towards higher halo mass. The yellow dashed line shows the halo mass at which the galaxy star-forming fraction $f_\mathrm{SF}$ is $0.5$ as a function of redshift. Below (above) this dashed line, the mass growth of SMBHs occurs primarily in star-forming (quenched) galaxies, respectively.  In \textsc{Trinity}, average BHARs are constrained by the total energy output from AGNs, which is mainly inferred from the QPDFs and ABHMFs.

\begin{figure}
\includegraphics[width=0.48\textwidth]{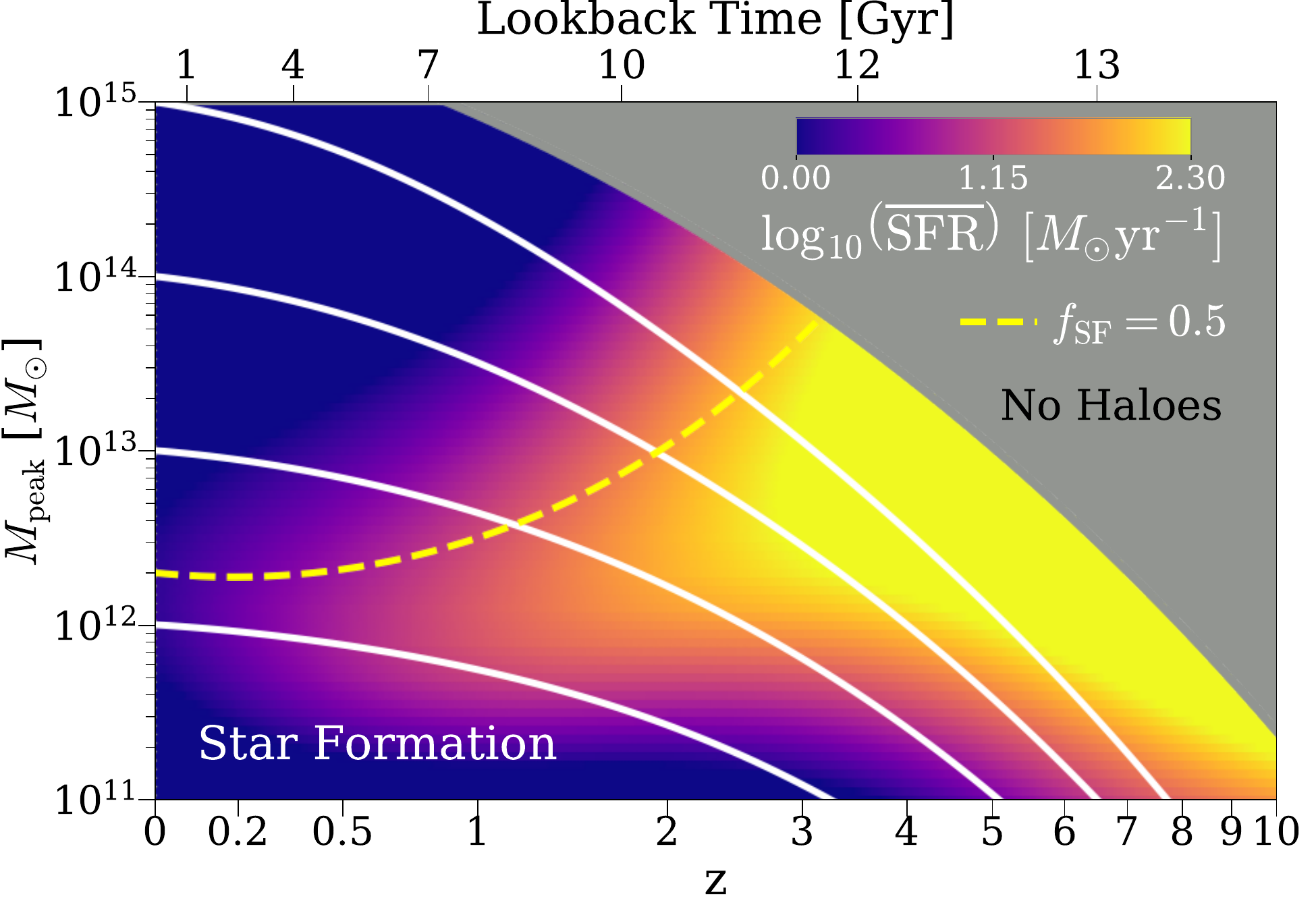}
\caption{The average star formation rates ($\overline{\mathrm{SFR}}$) as a function of $M_{\rm peak}$ and $z$ (see \S\ref{ss:results_bhar_bher}). \sfcurve{} \halocurves{} \nohalos{} All the data used to make this plot can be found \href{https://github.com/HaowenZhang/TRINITY/tree/main/plot_data}{here}.}
\label{f:sfr_map}
\end{figure}

The bottom panel of Fig.\ \ref{f:bhar_map} shows the average BHAR histories of haloes with different masses at $z=0$. At all halo masses, average BHARs keep rising in the early Universe, and then peak and decrease towards lower redshifts. The BHARs of more massive haloes peak at higher redshifts. There is also an increase in BHAR with time below $z\sim2$ among the most massive haloes. This is mainly constrained by the increase in AGN luminosities with stellar mass, as indicated by the low redshift QPDFs from Fig.\ 5 of \citet{Aird2018}.

Fig.\ \ref{f:sfr_map} shows the average galaxy star formation rates (SFRs) as a function of $M_{\rm peak}$ and $z$. The $M_\mathrm{peak}$ and $z$ dependencies of SFR are similar to those of BHAR below $M_\mathrm{peak}\sim 10^{14} M_\odot$. Above $M_\mathrm{peak}\sim 10^{14} M_\odot$, however, SFR decreases monotonically with halo mass at all redshifts, whereas the massive black holes still have detectable accretion rates. In other words, BHARs follow SFRs mainly among less-massive haloes, where star-forming galaxies dominate the population. For massive galaxies at lower redshifts, they are much more likely to be quiescent in their SFRs, but still have significant SMBH activity. This difference between small and large galaxy populations is hidden when we compare the \emph{cosmic} BHARs and SFRs, where less massive objects ($M_\mathrm{peak}\sim 10^{12}\Msun$) dominate the demographics.

\begin{figure}
\subfigure{
\includegraphics[width=0.48\textwidth]{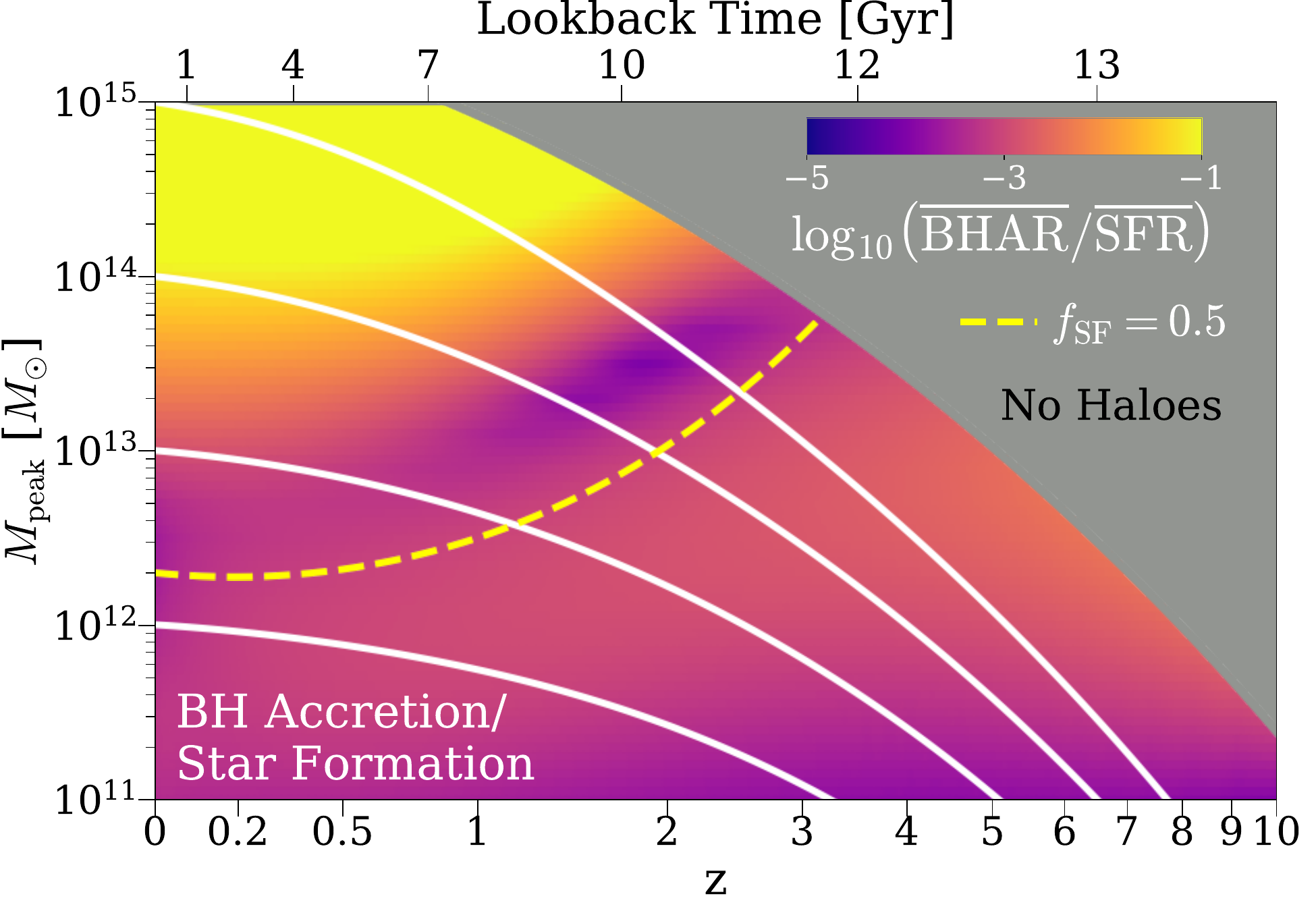}
}
\subfigure{
\includegraphics[width=0.48\textwidth]{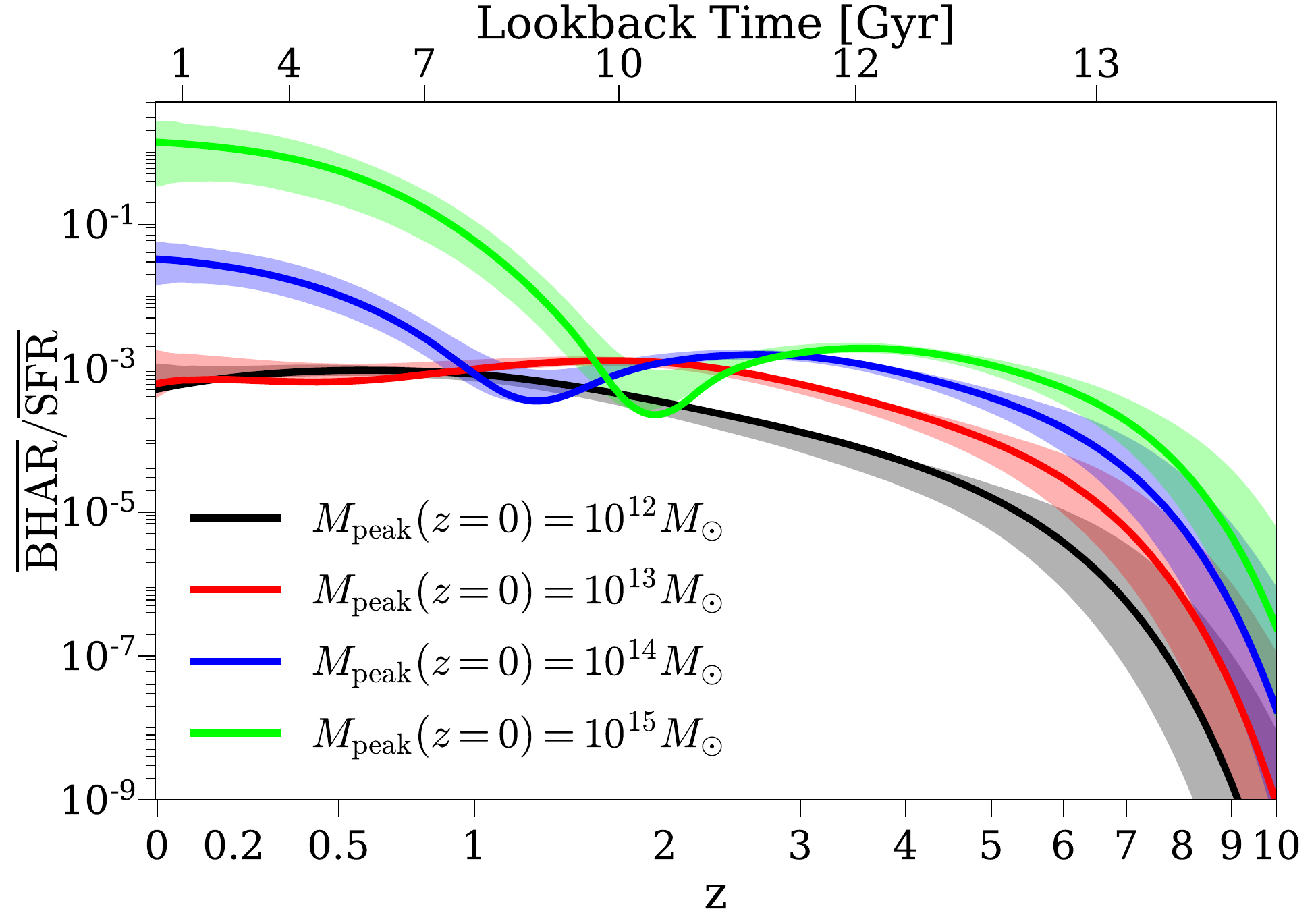}
}
\caption{\textbf{Top} panel: the $\overline{\mathrm{BHAR}}/\overline{\mathrm{SFR}}$ ratio as a function of redshift and \mpeak{} for our best fitting model (see \S\ref{ss:results_bhar_bher}). \sfcurve{} \halocurves{} \textbf{Bottom} panel: the $\overline{\mathrm{BHAR}}/\overline{\mathrm{SFR}}$ ratio histories as a function of \mpeak{} at $z=0$. \shadedregions{} \nohalos{} All the data used to make this plot can be found \href{https://github.com/HaowenZhang/TRINITY/tree/main/plot_data}{here}.}
\label{f:bhar_sfr_ratio}
\end{figure}

The top panel of Fig.\ \ref{f:bhar_sfr_ratio} shows the ratios between the average BHAR and SFR, $\overline{\rm BHAR}/\overline{\rm SFR}$, as a function of \mpeak{} and $z$. At $z\gtrsim 6$, $\overline{\rm BHAR}/\overline{\rm SFR}$ increases with increasing \mpeak{}. Towards lower redshifts, $\overline{\rm BHAR}/\overline{\rm SFR}$ grows more slowly for all haloes, and shows a plateau at $\overline{\rm BHAR}/\overline{\rm SFR}\sim 10^{-3}$. More massive haloes reach this plateau at higher redshifts, which is consistent with the downsizing of SMBH growth. Below $z\sim 2$, however, the mass dependency gets stronger again, in the sense that more massive haloes have higher $\overline{\rm BHAR}/\overline{\rm SFR}$. Physically, this is because massive galaxies are strongly quenched towards lower redshifts, but the mass accretion of massive black holes is not suppressed as much. The bottom panel of Fig.\ \ref{f:bhar_sfr_ratio} shows the $\overline{\rm BHAR}/\overline{\rm SFR}$ histories of different halo populations. At $z\lesssim 2$, $\overline{\rm BHAR}/\overline{\rm SFR}$ either stays at a similar level as $z\gtrsim 2$, or increases with time for essentially all halo populations, indicating that SMBHs are catching up with galaxies in their growth. 

\begin{figure}
\subfigure{
\includegraphics[width=0.48\textwidth]{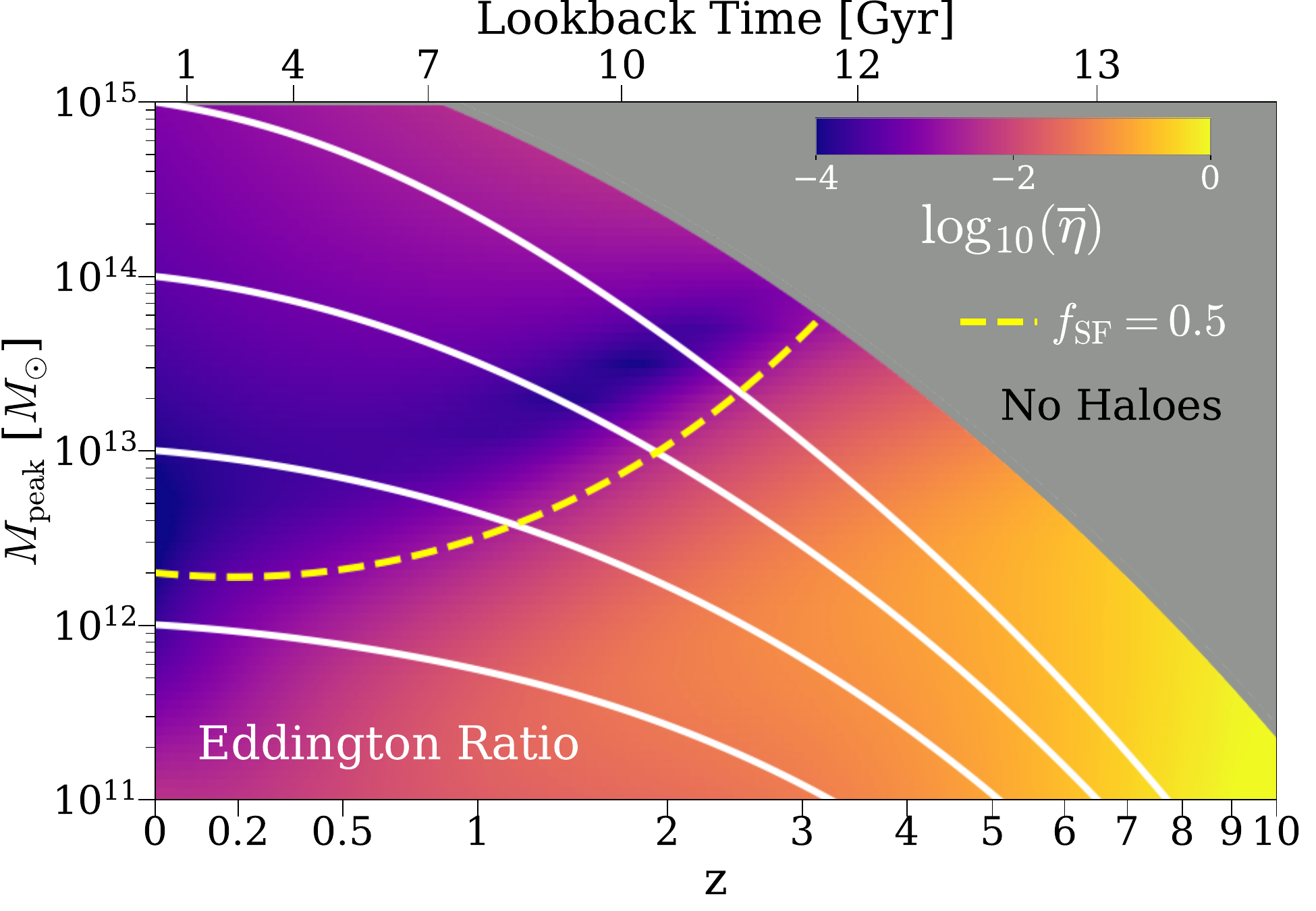}
}
\subfigure{
\includegraphics[width=0.48\textwidth]{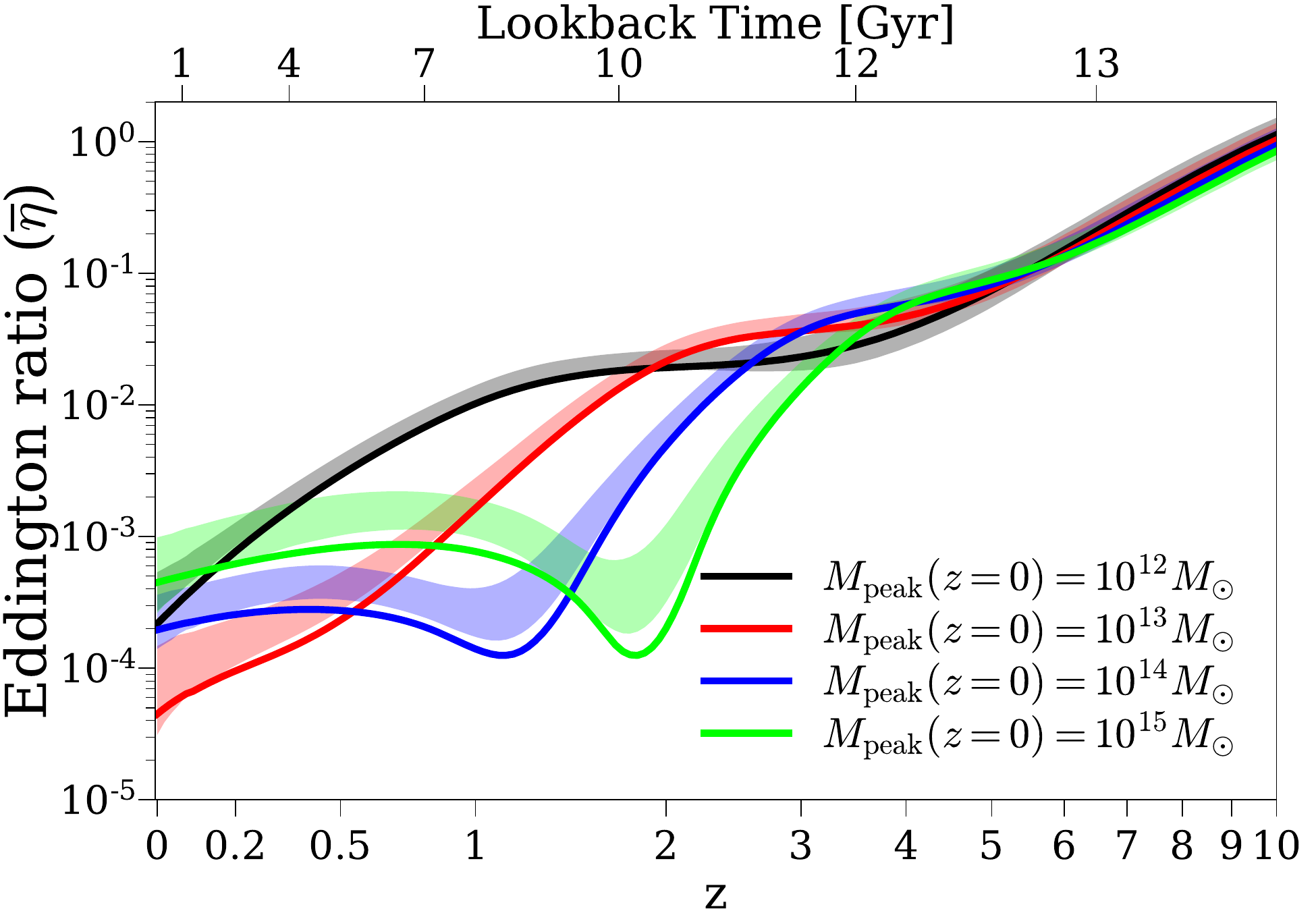}
}
\caption{\textbf{Top Panel:} average SMBH \emph{total} (i.e., radiative$+$kinetic) Eddington ratio ($\overline{\eta}$) as a function of $M_{\rm peak}$ and $z$ (see \S\ref{ss:results_bhar_bher}). \sfcurve{} \halocurves{} \nohalos{} \textbf{Bottom Panel:} $\overline{\eta}$ histories as a function of halo mass at $z=0$. \errorbars{} All the data used to make this plot can be found \href{https://github.com/HaowenZhang/TRINITY/tree/main/plot_data}{here}.}
\label{f:bher_map}
\end{figure}

The top panel of Fig.\ \ref{f:bher_map} shows the average SMBH total Eddington ratio ($\overline{\eta}$) as a function of $M_{\rm peak}$ and $z$. At $z\gtrsim 7$, all SMBHs have $0.1 < \overline{\eta} < 1$ regardless of host halo mass.  At lower redshifts, the average Eddington ratio decreases, with stronger trends for higher halo masses. In other words, SMBHs are less active in massive haloes and/or at later cosmic times. A similar trend can be seen when we follow the growth of different haloes, as shown by the white solid curves. In the bottom panel, we see all SMBHs accreting rapidly at high redshifts, with average Eddington ratios of unity at $z\sim10$. Below $z=10$, Eddington ratios drop with time for all SMBHs, but the exact patterns differ among halo populations. For more massive haloes with $M_{\rm peak} > 10^{13} M_{\odot}$, the average Eddington ratios experience a
two-phase decline before the final slight rejuvenation: an initial, slower decrease, and a later, faster drop. Haloes with $M_{\rm peak} = 10^{12}-10^{13} M_{\odot}$ at $z=0$ do not experience the final flattening phase in Eddington ratio. Below $z\sim 4$, more massive haloes experience the final and faster decline in Eddington ratios earlier compared to less massive ones. As the bottom panel of Fig.\ \ref{f:mbh_map} shows, this also reflects the same ``AGN downsizing'' phenomenon: SMBH activity starts to decline earlier in more massive haloes/galaxies.

\begin{figure}

\includegraphics[width=0.48\textwidth]{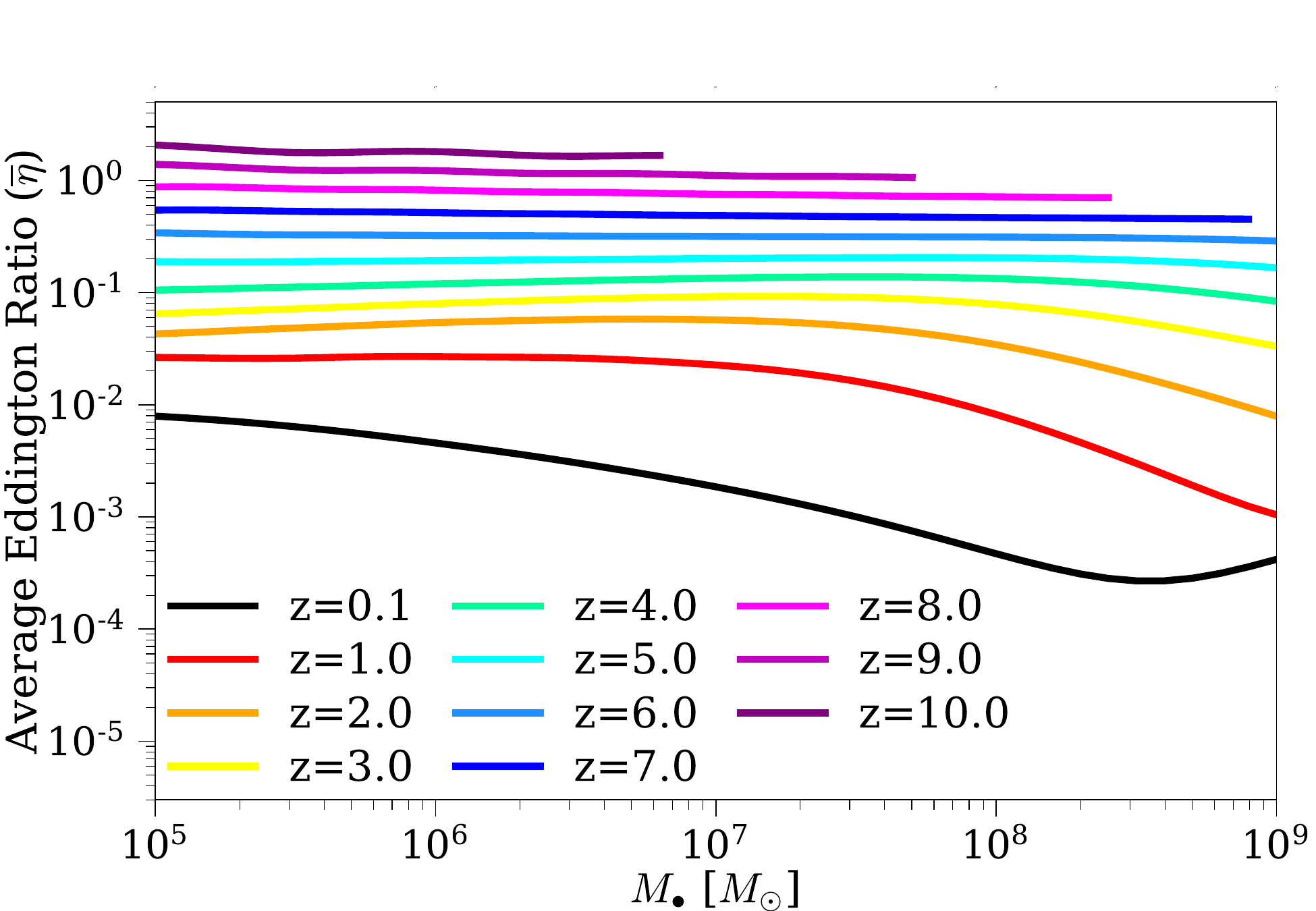}

\caption{Average SMBH \emph{total} (i.e., radiative$+$kinetic) Eddington ratio ($\overline{\eta}$) as a function of $M_{\bullet}$ and $z$. See \S\ref{ss:results_bhar_bher}. All the data used to make this plot can be found \href{https://github.com/HaowenZhang/TRINITY/tree/main/plot_data}{here}.}
\label{f:bher_mbh}
\end{figure}

It should be pointed out that the ``AGN downsizing'' effect exists not only when we look at different halo populations, but also when we look at SMBHs with different masses. Fig.\ \ref{f:bher_mbh} shows the average SMBH \emph{total} (i.e., radiative$+$kinetic) Eddington ratio, $\overline{\eta}$, as a function of \mbh{} and $z$. Again, we see that at high redshifts, SMBHs of different masses accrete at similar Eddington ratios. Below $z\sim 3$, the activity level among more massive black holes starts to decline earlier. Consequently, we see that  $\overline{\eta}$ decreases towards higher \mbh{}.

\subsection{SMBH mass functions}
\label{ss:results_bhmf}

\begin{figure}
\includegraphics[width=0.48\textwidth]{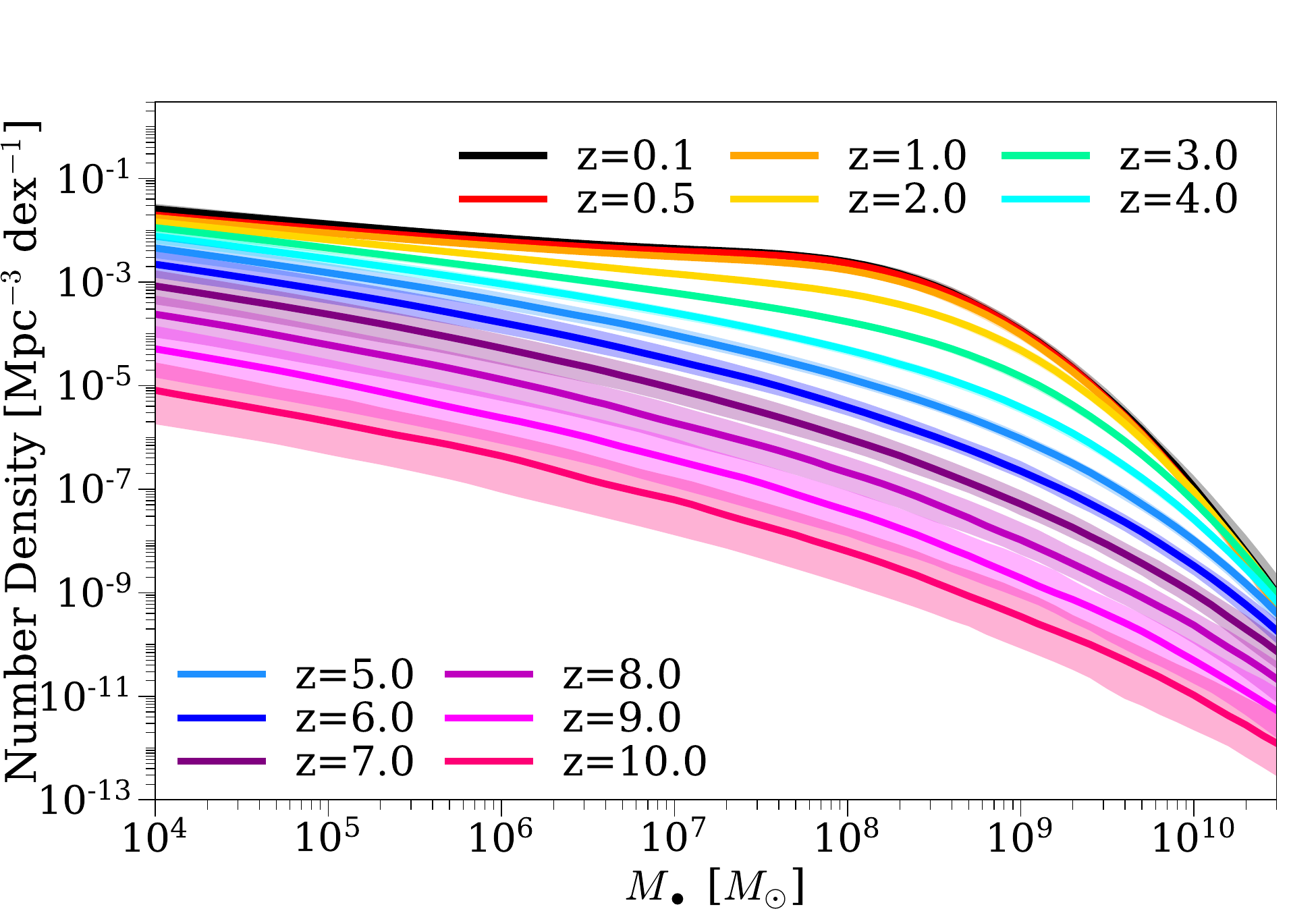}
\caption{The total black hole mass function between $0\leq z\leq 10$ (see \S\ref{ss:results_bhmf}). \shadedregions{} All the data used to make this plot can be found \href{https://github.com/HaowenZhang/TRINITY/tree/main/plot_data}{here}.}
\label{f:bhmf_z0_z10}
\end{figure}

Fig.\ \ref{f:bhmf_z0_z10} shows the total black hole mass functions (BHMFs) for $0\leq z\leq 10$. Similar to the galaxy stellar mass functions, the ``knee'' in the black hole mass function becomes less and less significant towards higher redshifts. This is because, in the early Universe, the \mstar{}--\mpeak{} relation, and therefore the \mbh{}--\mpeak{} relation, can be approximated as a single power-law. We also see strong evolution in the black hole mass function above $z\gtrsim 5$ \emph{regardless of SMBH mass}. This directly results from the universally high Eddington ratios at high redshifts. (see also \S\ref{ss:results_bhar_bher}). At $z<3$, the AGN downsizing effect slows down the evolution of the total BHMF at the massive end. In the meantime, moderately massive SMBHs with $10^8 < M_\bullet < 10^9 M_\odot$ grow significantly. This continued growth builds up the ``knee'' in the BHMF in the low-redshift Universe.

\subsubsection{The host haloes of $M_\bullet > 10^{9.5} M_\odot$ SMBHs}
\label{sss:bhmf_mh}

\begin{figure}
\subfigure{
\includegraphics[width=0.48\textwidth]{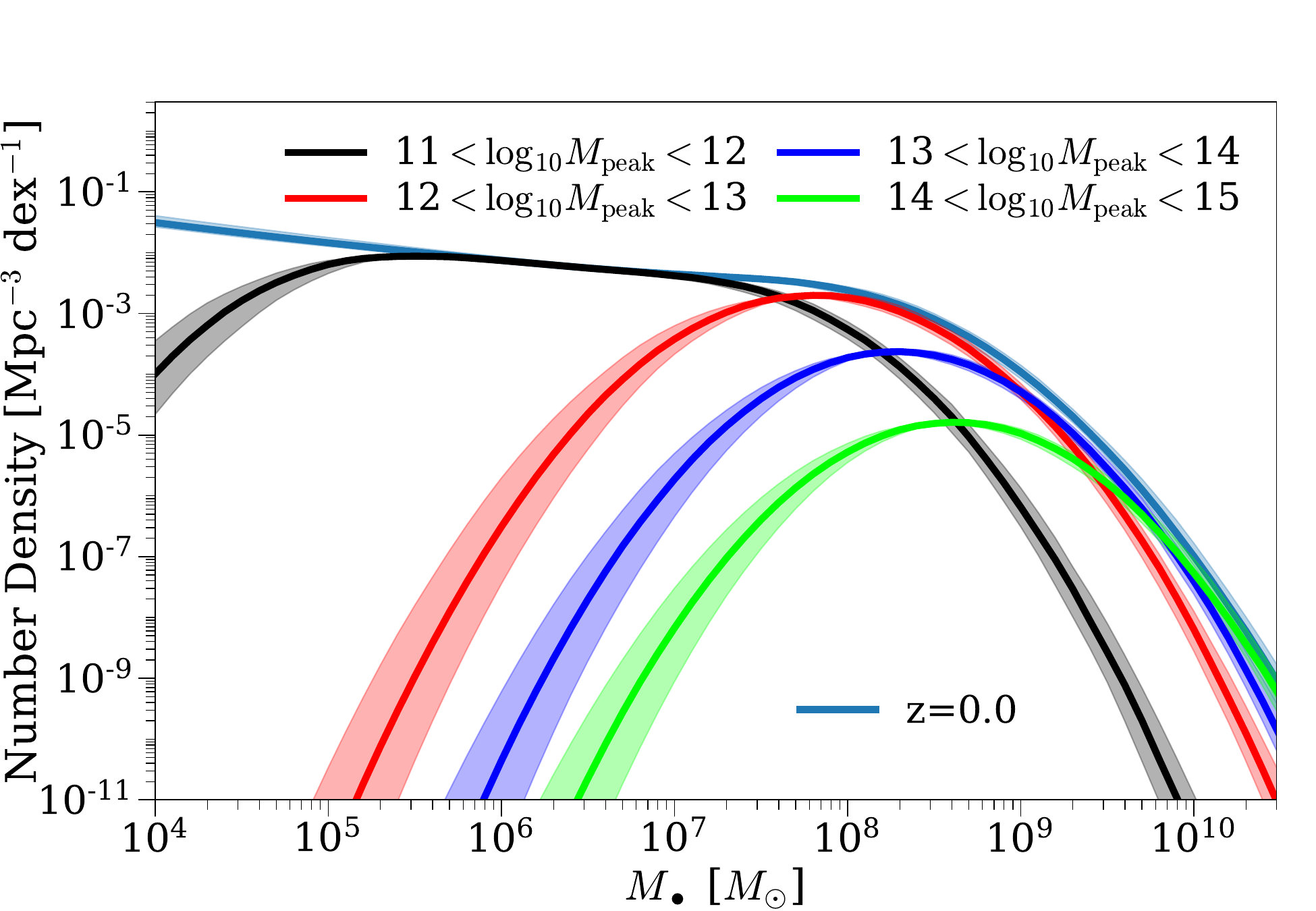}
}
\subfigure{
\includegraphics[width=0.48\textwidth]{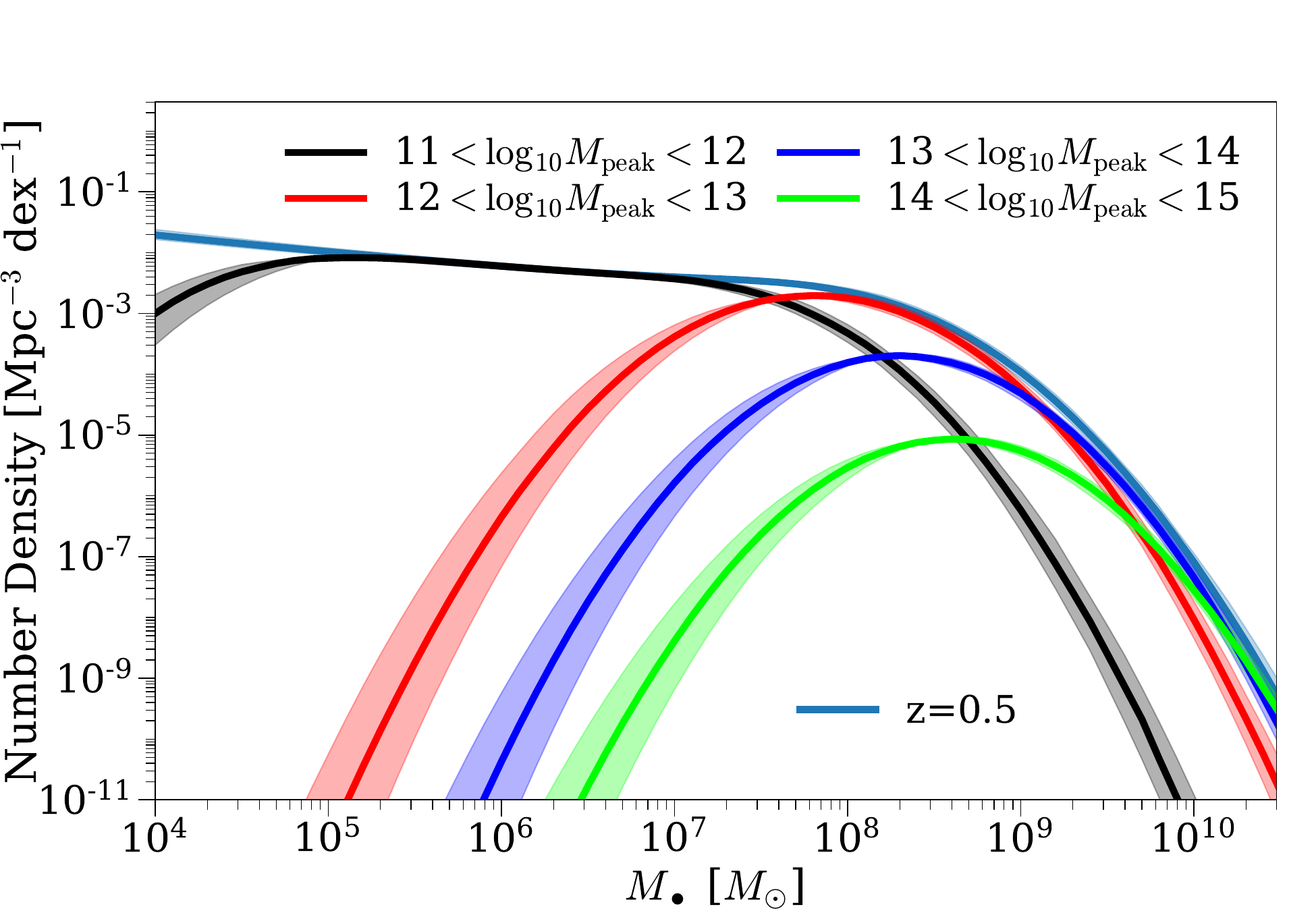}
}
\subfigure{
\includegraphics[width=0.48\textwidth]{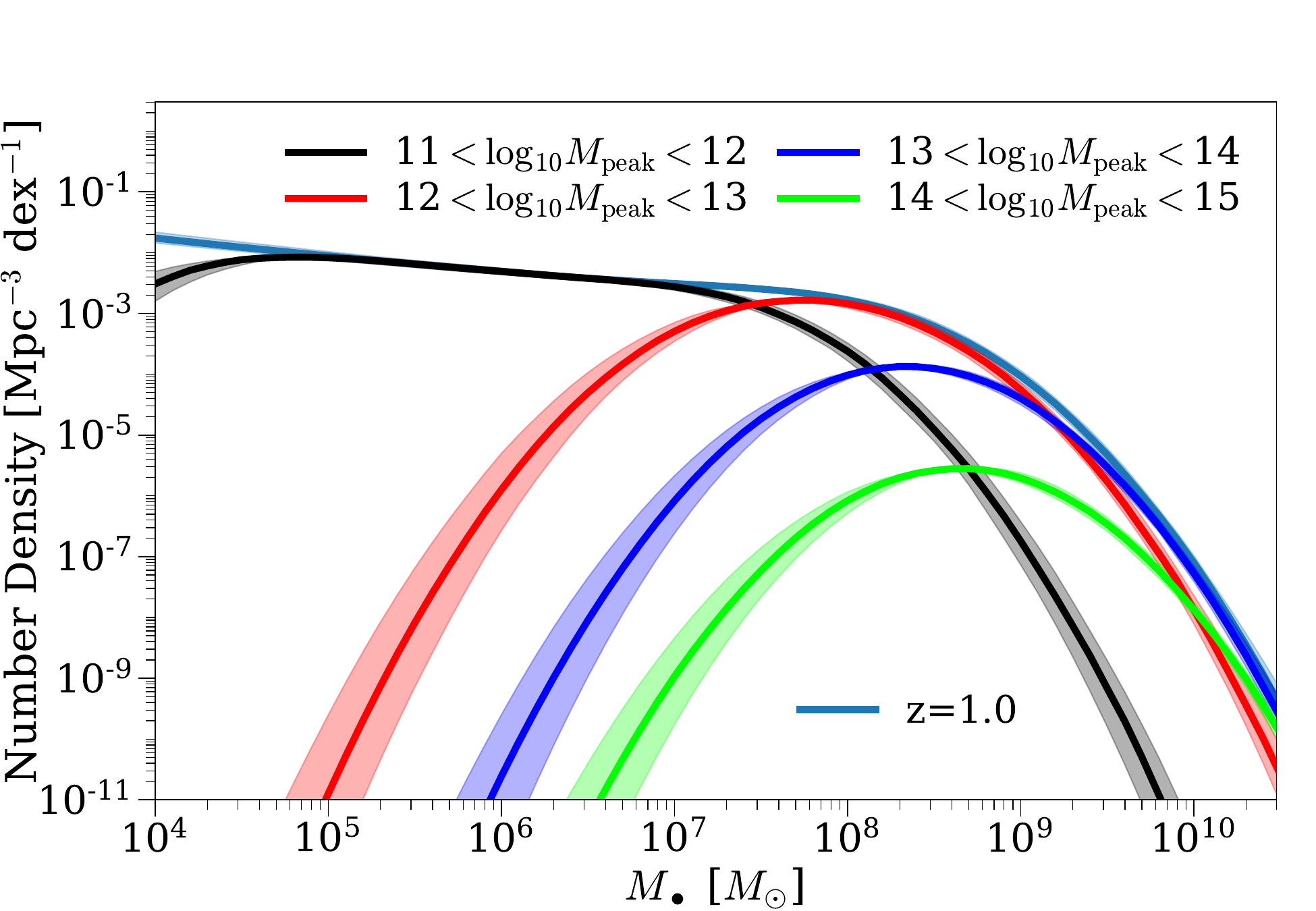}
}
\caption{Total black hole mass functions at $z=0.0,0.5$, and $1.0$ (the top, middle, and bottom panels), split into the contributions from different host dark matter halo mass bins (see \S\ref{sss:bhmf_mh}). \shadedregions{} All the data used to make this plot can be found \href{https://github.com/HaowenZhang/TRINITY/tree/main/plot_data}{here}.}
\label{f:bhmf_mh}
\end{figure}

In Fig.\ \ref{f:bhmf_mh}, we show the total BHMFs at $z=0.0,0.5$ and $1.0$, decomposed into contributions from different host halo masses. Similar to Eq.\ \ref{e:bhmf}, the BHMF contributed by haloes in the mass range $(M_\mathrm{peak,min}, M_\mathrm{peak,max})$ is:

\begin{eqnarray}
    && \phi(M_\bullet, M_\mathrm{peak,min}, M_\mathrm{peak,max}, z) = \nonumber \\
    && \qquad \int^{M_\mathrm{peak,max}}_{M_\mathrm{peak,min}} \phi(M_\mathrm{peak}, z)P(M_\bullet|M_\mathrm{peak}, z)dM_\mathrm{peak}, \
\end{eqnarray}
where $\phi(M_\mathrm{peak}, z)$ is the halo mass function and $P(M_\bullet|M_\mathrm{peak}, z)$ is the probability distribution of \mbh{}, given the host halo mass \mpeak{} at redshift $z$. In \textsc{Trinity}, $P(M_\bullet|M_\mathrm{peak}, z)$ is a log-normal distribution with the median and scatter determined from the halo--galaxy--SMBH connection (\S\ref{ss:halo_galaxy_connection} and \S\ref{ss:galaxy_smbh_connection}). Given the flat \bhhm{} relation at the massive end (see Fig.\ \ref{f:bhhm_median}), $P(M_\bullet|M_\mathrm{peak}, z)$ only changes slightly with increasing halo mass. On the other hand, there are many fewer haloes with $M_\mathrm{peak}>10^{14} M_\odot$ than $M_\mathrm{peak}<10^{14} M_\odot$, due to the exponential decrease in halo number density. Hence, the haloes with $10^{13} M_\odot < M_\mathrm{peak} < 10^{14} M_\odot$, rather than those with $10^{14} M_\odot <  M_\mathrm{peak} < 10^{15} M_\odot$, dominate the BHMF for $M_\bullet > 10^{9.5} M_\odot$ at $z=1.0$. In other words, when looking at a \mbh{}--selected sample with large \mbh{}, we are more likely to observe less massive haloes than indicated by the median \bhhm{} relation. This bias is also discussed in \citet{Lauer2007}. Towards lower redshifts, more and more massive haloes emerge with time. As a result, the high-mass BHMF in the local Universe is composed almost equally of haloes with $13 < \log_{10} M_\mathrm{peak} < 14$ and $14 < \log_{10} M_\mathrm{peak} < 15$. In short, cluster-scale haloes ($\log_{10} M_\mathrm{peak}>14$) are too rare to dominate the massive end of low-redshift BHMFs, mainly due to their own rarity and the flat \bhhm{} at these redshifts.

\subsection{SMBH mergers}
\label{ss:results_mergers}

\begin{figure}
\subfigure{
\includegraphics[width=0.48\textwidth]{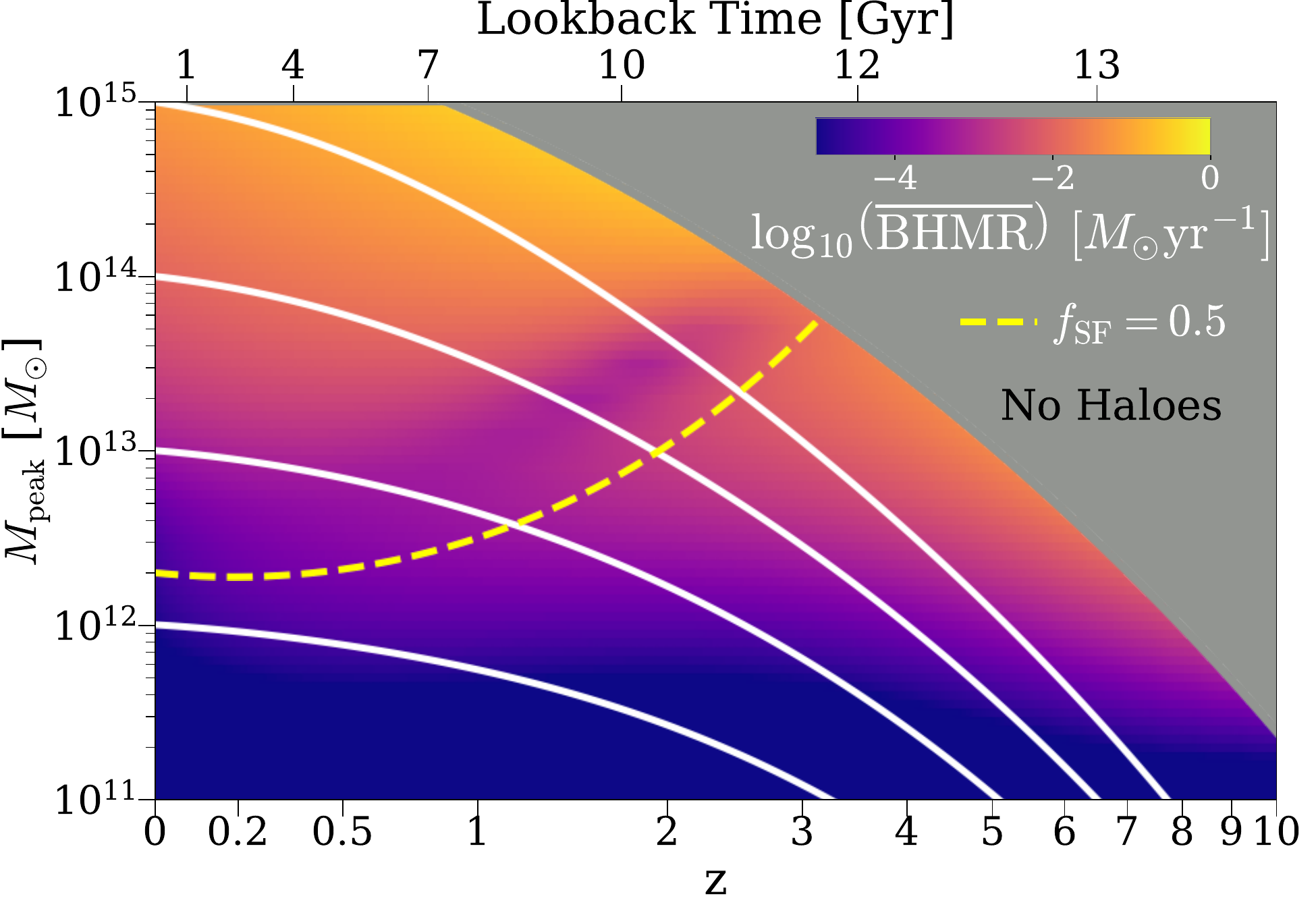}
}
\subfigure{
\includegraphics[width=0.48\textwidth]{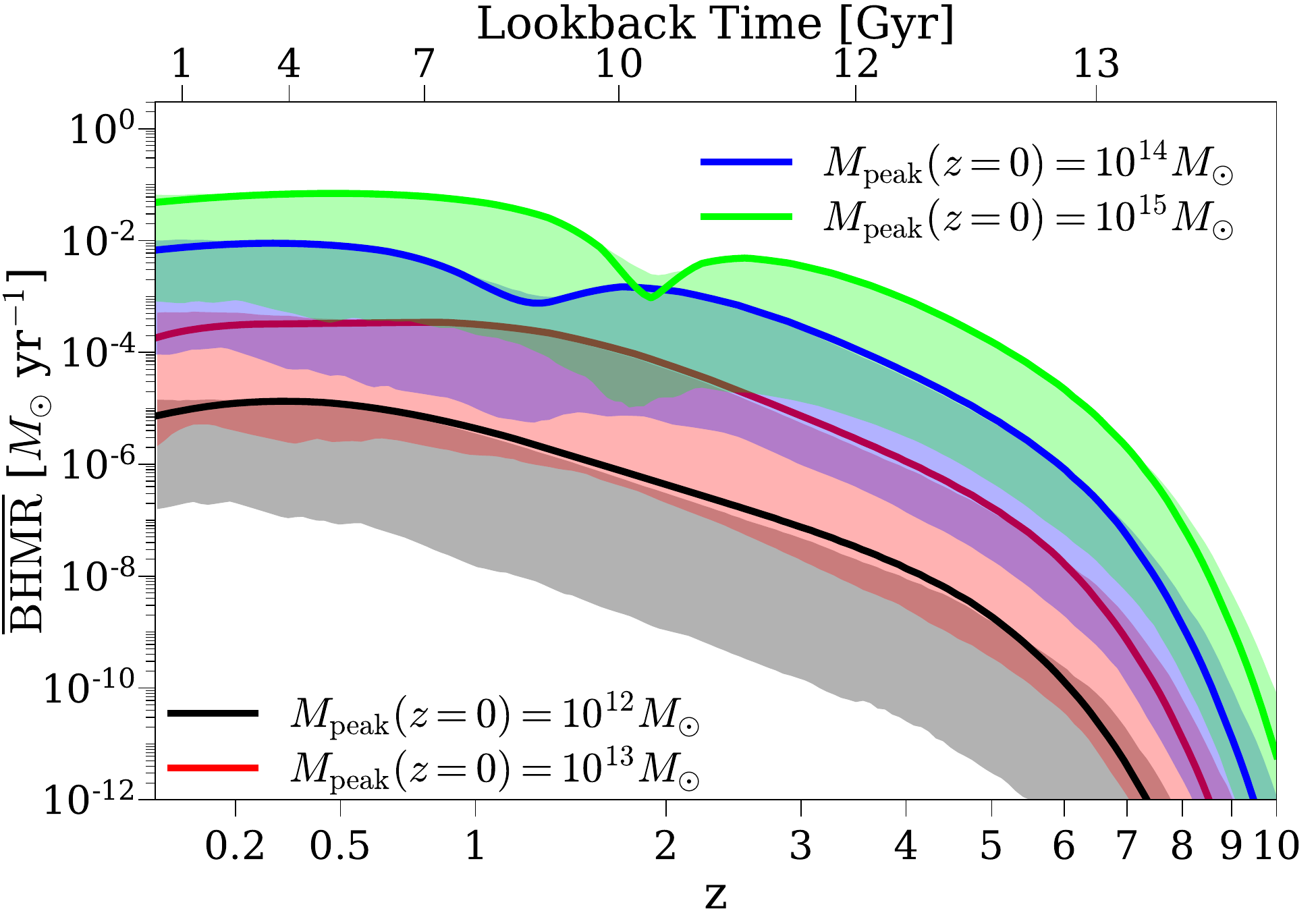}
}
\caption{\textbf{Top Panel:} the average black hole merger rates ($\overline{\mathrm{BHMR}}$) as a function of $M_{\rm peak}$ and $z$ (see \S\ref{ss:results_mergers}). \halocurves{} \nohalos{} \textbf{Bottom Panel:} $\overline{\mathrm{BHMR}}$ histories as a function of halo mass at $z=0$. \shadedregions{} All the data used to make this plot can be found \href{https://github.com/HaowenZhang/TRINITY/tree/main/plot_data}{here}.}
\label{f:bhmr_map}
\end{figure}

 The top panel of Fig.\ \ref{f:bhmr_map} shows the average black hole merger rates (BHMRs) as a function of \mpeak{} and $z$. Note that in this paper, we define BHMR as the \emph{SMBH growth rate due to mergers}, instead of the number of SMBH mergers per unit SMBH, per unit redshift, and per unit (log-) SMBH mass ratio (as presented in Paper V). In general, BHMRs increase monotonically with \mpeak{} and $z$. The same conclusion holds when we look at the average BHMR histories as a function of \mpeak{} at $z=0$, which is shown in the bottom panel of Fig.\ \ref{f:bhmr_map}. The best-fitting model lies on the upper edges of the 68\% confidence intervals. Although the best fitting model uses a significant amount of mergers to fit the data, the dominance of SMBH growth via smooth accretion (see Paper V) means that parameter sets with lower merger rates also fit the data well. As mentioned in \S\ref{ss:bh_mergers}, BHMRs are calculated by allowing a fraction of galaxy mergers (the free parameter $f_\mathrm{scale}$) to result in mergers of their SMBHs. This is done due to continuing uncertainty about SMBH merger time scales \citep[e.g.,][]{Tremmel2018}. Therefore, these BHMRs are constrained by the combination of: a) SMBH total growth rates, which are given by the evolution of active and total black hole mass functions; and b) average black hole accretion rates, which are constrained by the quasar luminosity functions and probability distribution functions. The best-fitting \textsc{Trinity} model predicts $f_\mathrm{scale}$ to be $\log_{10} (f_{\rm scale}) = -0.192^{+0.126}_{-2.285} + (-0.000^{+1.970}_{-0.523}) (a - 1)$. This means that, for example, when the fractional merger contribution to instantaneous galaxy growth is 10\%, the merger contribution to SMBH growth would be $10\%\times 10^{-0.192}\approx 6.4\%$. In Appendix \ref{a:other_merger_models}, we also show the results of models with alternate assumptions about SMBH mergers. Further discussion about SMBH mergers in \textsc{Trinity} and predictions for gravitational wave experiments are presented in Paper V.

\subsection{AGN energy efficiency and systematic uncertainties}
\label{ss:results_systematics}

As described in \S\ref{ss:galaxy_sys} and \S\ref{ss:agn_observables}, we modeled systematic uncertainties in stellar mass, star formation rates, and SMBH Eddington ratios. These uncertainties are propagated into our model predictions, and their values quantify the degree of tension between different datasets. In \textsc{Trinity}, the best fitting values (see Appendix \ref{a:param_values}) of the galaxy systematics are all consistent with those given by \citet{Behroozi2019}. The systematic offset in SMBH Eddington ratios is motivated by the discrepancy between the quasar luminosity functions (QLFs) from \citet{Ueda2014} and the quasar probability distribution functions (QPDFs) from \citet{Aird2018} (see Appendix \ref{aa:aird_qpdf}). This discrepancy can be caused by different assumptions for: 1) differences in \mstar{} estimates used by \citet{Aird2018} and those in our galaxy data compilation (\S\ref{sss:galaxy_data}); 2) the ways in which X-ray photons are counted, including how galaxy contributions are subtracted; 3) the functional forms used to fit the observational data. The net effect is $\eta' - \eta \sim 0.5$ dex, where $\eta$ is the intrinsic Eddington ratio, and $\eta'$ is the Eddington ratio used to calculate the observed QPDFs in \citet{Aird2018}.

The total AGN energy efficiency from \textsc{Trinity} is $\log_{10}\epsilon_\mathrm{tot}= -1.318^{+0.115}_{-0.009}$. In other words, the best-fitting model is consistent with a redshift-independent $\sim 5\%$ mass-to-energy conversion efficiency. However, the exact value of $\epsilon_\mathrm{tot}$ is affected by various input assumptions, such as AGN bolometric corrections, Compton-thin/Compton-thick obscured fractions, and/or the assumed local \bhbm{} scaling relation (if ever assumed). These assumptions alter the amount of radiation to be produced by SMBH accretion, which systematically changes the best-fitting $\epsilon_\mathrm{tot}$. In Appendices \ref{aa:kbol}, \ref{aa:ctk_corr}, \ref{aa:f_obs}, and \ref{a:alt_galaxy_smbh}, we carry out experiments with different bolometric corrections, Compton-thick/Compton-thin obscuration fractions, fixed local \bhbm{} scaling relations and compare with the fiducial \textsc{Trinity} model. When varying these input assumptions, the best-fitting AGN energy efficiency can change from $\sim 0.035-0.07$, i.e., a factor of 2 (or 0.3 dex). In this work, we opt not to allow a systematic offset in the normalization of the \bhbm{} relation, $\beta_\mathrm{BH}$, due to its complete degeneracy with the AGN energy efficiency. Thus, the best-fitting value of the energy efficiency \eff{} should be viewed as \emph{a combination of} the intrinsic average efficiency and any potential systematic offset in $\beta_\mathrm{BH}$. We emphasize that this energy efficiency quantifies how effectively gravitational energy is converted into \emph{radiation and kinetic energy}. Thus, there is no unique link between our efficiency and the average SMBH spin value.

\subsection{Correlation coefficient (\rhobh{}) between average SMBH accretion rate and \mbh{} at fixed halo mass}
\label{ss:rho_bh}

\begin{figure}
\includegraphics[width=0.48\textwidth]{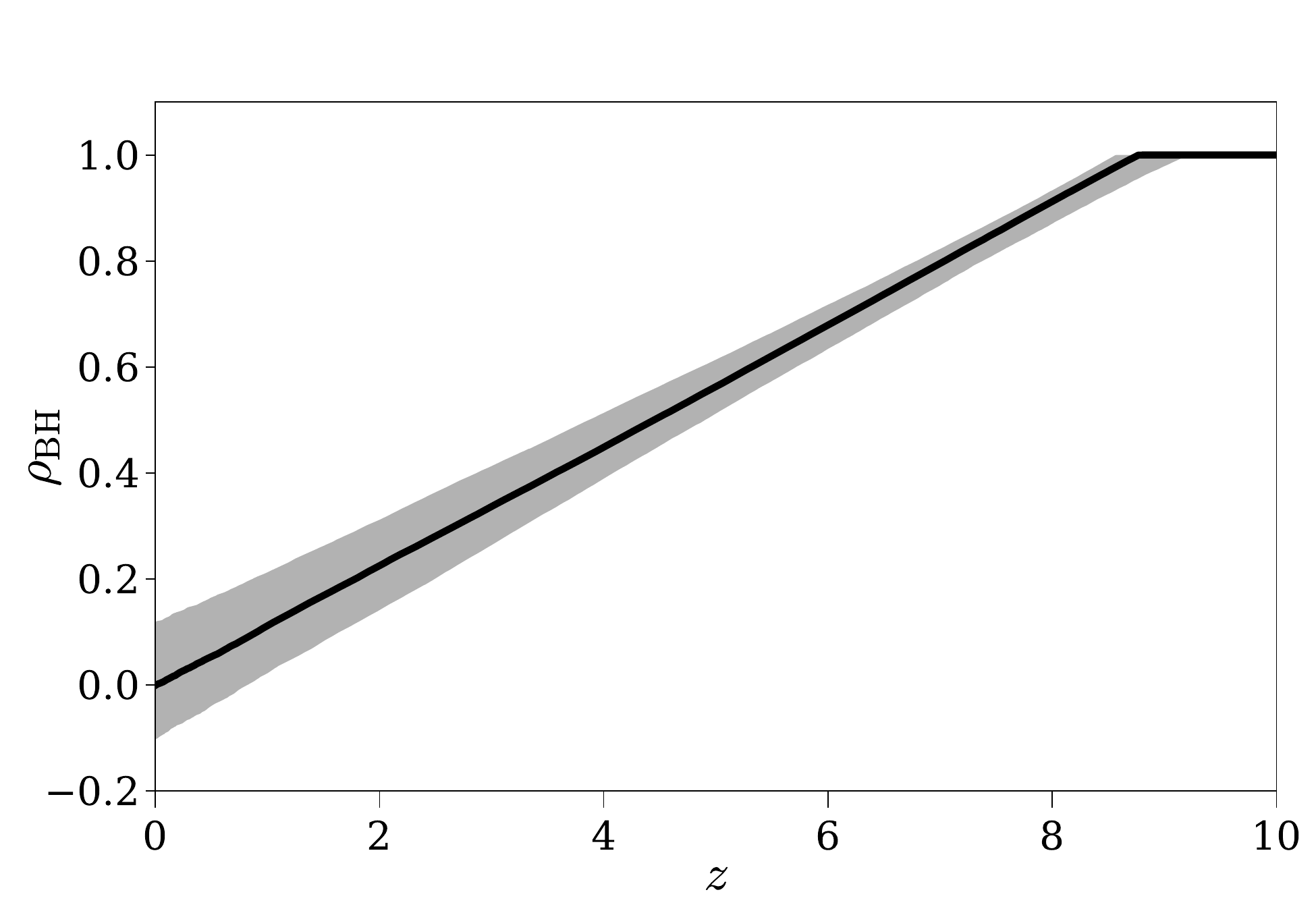}
\caption{The correlation coefficient, \rhobh{}, between average SMBH accretion rate and \mbh{} at fixed halo mass. See \S\ref{ss:rho_bh}. The shaded region shows the 68\% confidence intervals inferred from the model posterior distribution.  The data used to make this plot can be found \href{https://github.com/HaowenZhang/TRINITY/tree/main/plot_data}{here}.}
\label{f:rho_bh}
\end{figure}

Fig. \ref{f:rho_bh} shows the redshift evolution of \rhobh{} from the best-fitting model. At $z\gtrsim8$, the average SMBH accretion rate and \mbh{} are highly correlated at fixed host halo mass. In other words, high-redshift SMBHs share the same \emph{Eddington ratio} distributions, if they are hosted by haloes with similar masses. This correlation fades towards lower redshifts. By $z=0$, there is essentially no correlation between average SMBH accretion rate and \mbh{}, i.e., different SMBHs have the same \emph{absolute} accretion rate distributions, if hosted by similar haloes. Overall, this evolution makes large SMBHs less and less active compared to their smaller counterparts (measured by difference in average Eddington ratio) in the same halo mass bin. Consequently, AGN downsizing effects apply not only to SMBHs in \emph{different host haloes} (as shown in \S\ref{ss:results_bhar_bher}), but also to those hosted by \emph{similar haloes and galaxies.} Although this conclusion holds qualitatively in all the model variants covered in the Appendix, the exact \rhobh{} value at $z=0$ does change significantly in some of these models (see Appendices \ref{aa:kbol} and \ref{aa:f_obs}).

\section{Comparison with previous studies and discussion}
\label{s:discussion}

In this section, we compare \textsc{Trinity} with hydrodynamical simulations as well as discuss the potential physical mechanisms that could reproduce the redshift evolution of the \bhbm{} relation (\S\ref{ss:discussions_bhbm_evolution}); present the cosmic SMBH mass density as a function of redshift (\S\ref{ss:discussions_cosmic_bh_density}); and discuss the physical implications of the best-fitting \textsc{Trinity} model (\S\ref{ss:discussions_physical_implications}).

\subsection{Evolution of the galaxy--SMBH scaling relation}
\label{ss:discussions_bhbm_evolution}

\begin{figure}
\includegraphics[width=0.48\textwidth]{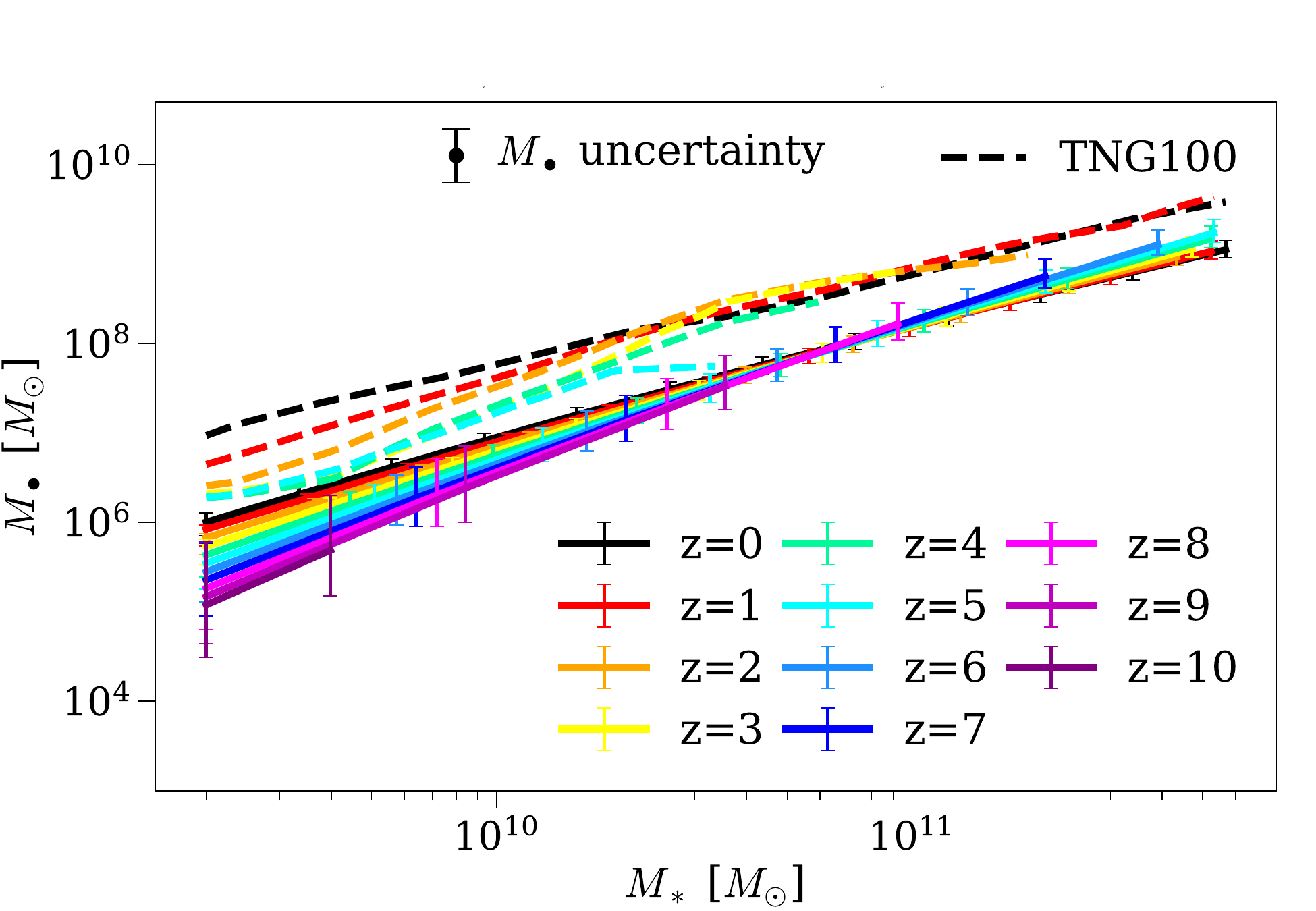}
\caption{The median \bhsm{} relations as functions of $z$ for \textsc{Trinity} (\textit{solid lines}) and the IllustrisTNG100 simulation \citep[\textit{dashed lines};][]{Pillepich2018,Habouzit2020}. See \S\ref{ss:discussions_bhbm_evolution}. The typical uncertainty in the measurement of $M_\bullet$, $0.3$ dex, is shown by the black solid dot. At $z\geq 3$, the dynamical ranges of \mbh{} and \mstar{} in TNG100 are smaller than in \textsc{Trinity}, due to the smaller simulation box size.} All the data used to make this plot (including those from IllustrisTNG and our best-fitting model) can be found \href{https://github.com/HaowenZhang/TRINITY/tree/main/plot_data}{here}.
\label{f:bhbm_evolution}
\end{figure}

The growth of SMBHs and their feedback on host galaxies are important physical mechanisms to capture in hydrodynamical simulations. Although different simulations find similar local \mbh{}--\mbulge{} (or \mstar{}) relations, they differ in the relation's redshift evolution. For example, the IllustrisTNG \citep{Pillepich2018} and SIMBA \citep{Dave2019} simulations predicted increasing normalizations of the scaling with time, whereas the Illustris \citep{Vogelsberger2014}, Horizon-AGN \citep{Dubois2014,Dubois2016}, and EAGLE simulations \citep{Schaye2015} predicted the opposite \citep{Habouzit2020}. This diversity in the redshift evolution results from different sub-grid physics adopted by each simulation. 

\textsc{Trinity} infers the redshift evolution of this scaling relation by extracting information directly from observational data, without any assumptions about the underlying physics. This can help determine which sub-grid physics models give results that are more consistent with observations. We show the the \mbh{}--\mstar{} relations at different redshifts from \textsc{Trinity} and IllustrisTNG100 \citep{Pillepich2018,Habouzit2020} in Fig.\ \ref{f:bhbm_evolution}. Despite the offset, both mass scalings show increasing normalizations with time at $M_* \leq 10^{11}M_\odot$. This implies that SMBH growth becomes increasingly efficient compared to galaxy growth at lower redshifts. For the hydrodynamical simulations listed in \citet{Habouzit2020}, the following sub-grid physics models succeeded in reproducing this trend: a) the strong supernova feedback in low-mass galaxies at high redshifts that reduces early SMBH growth in IllustrisTNG \citep{Dubois2015,Bower2017,Pillepich2018}; and b) the low accretion AGN feedback mode that quenches galaxies but favors further SMBH growth in SIMBA \citep{Dave2019}. That said, SMBH masses depend on many different aspects of sub-grid physics, including cooling, star formation, supernova feedback, magnetic fields, etc.\ beyond those directly related to the growth of the SMBH.  Hence, the success of a given sub-grid recipe at matching properties of SMBHs cannot be taken as evidence in support of its correctness without the context of the recipe's successes and failures at matching other non-SMBH observations.

\subsection{Cosmic SMBH mass density}
\label{ss:discussions_cosmic_bh_density}

\begin{figure}
\includegraphics[width=0.48\textwidth]{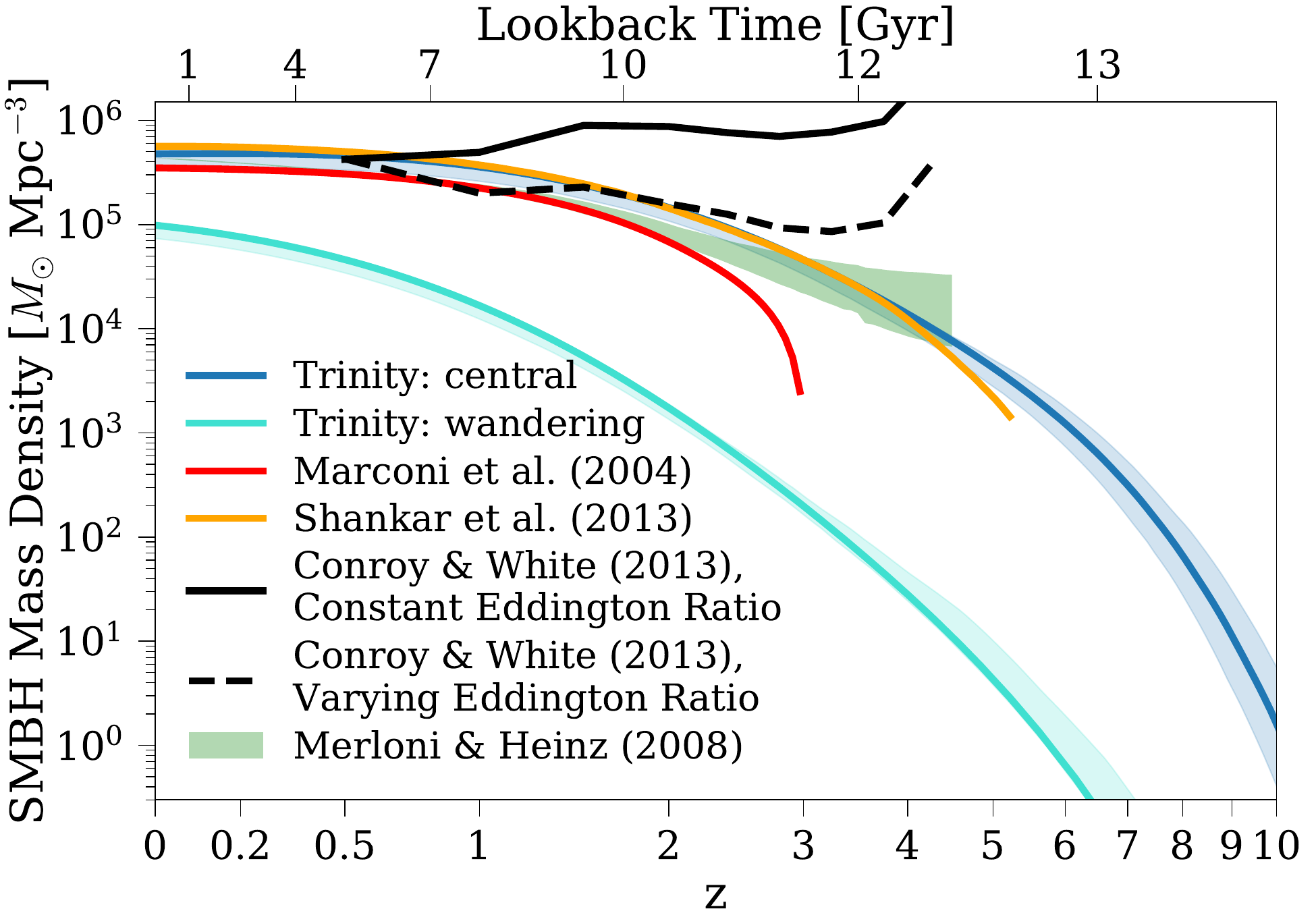}
\caption{Cosmic SMBH mass density as a function of $z$ (see \S\ref{ss:discussions_cosmic_bh_density}). \shadedregions{} All the data used to make this plot (including those from previous studies and our best-fitting model) can be found \href{https://github.com/HaowenZhang/TRINITY/tree/main/plot_data}{here}.}
\label{f:bh_density_z}
\end{figure}

Fig.\ \ref{f:bh_density_z} shows the cosmic SMBH mass density as a function of redshift from \textsc{Trinity} compared to previous studies. Unlike previous studies that tried to solve the continuity equation, in \textsc{Trinity}, we assume that wandering SMBHs also contribute to quasar luminosity functions during their growth. Thus, we include the cosmic wandering SMBH mass density in Fig.\ \ref{f:bh_density_z} for a fair comparison. We also show the cosmic wandering SMBH density separately in cyan, which accounts for $\sim15\%$ of the total SMBH mass density at $z=0$. This is broadly consistent with the results from \citet{Volonteri2003} based on a semi-analytical model and \citet{Ricarte2021} based on the \textsc{Romulus} simulations.

Below $z\sim2$, the offsets in the mass density between different studies are mostly driven by the different AGN energy efficiencies. Above $z\sim2$, the systematic difference with \citet{Marconi2004} increases with redshift. The reason is that \citet{Marconi2004} forward modeled AGN evolution assuming that all SMBH growth occurred at $z<3$. These initial conditions did not consider SMBH assembly histories at higher redshifts, and hence give different SMBH mass functions at $z\sim 3$ from \textsc{Trinity}, in which SMBHs are modeled to start growing from $z=15$.

Compared to other studies, \citet{Conroy2013} inferred quite different SMBH mass density histories. They assumed a mass-independent Eddington ratio distribution and a linear $M_\bullet-M_*$ relation, and tried to fit the quasar luminosity functions \emph{at each individual redshift} with two free parameters: 1) the normalization of the \mbh{}--\mstar{} relation, and 2) the AGN duty cycle. The SMBH mass density at each redshift was then obtained by convolving the galaxy stellar mass function with the \mbh{}--\mstar{} relation. This method does not enforce any continuity equation for SMBH mass. As a result, it cannot guarantee the consistency between the inferred cosmic SMBH mass growth rates and the quasar luminosity functions. This is shown in Fig.\ \ref{f:bh_density_z}, where the SMBH mass density from \citet{Conroy2013} decreases with time at some points in cosmic history for all variations considered. In light of this, we do not make further comparison with \citet{Conroy2013} here.

\begin{figure}
\includegraphics[width=0.48\textwidth]{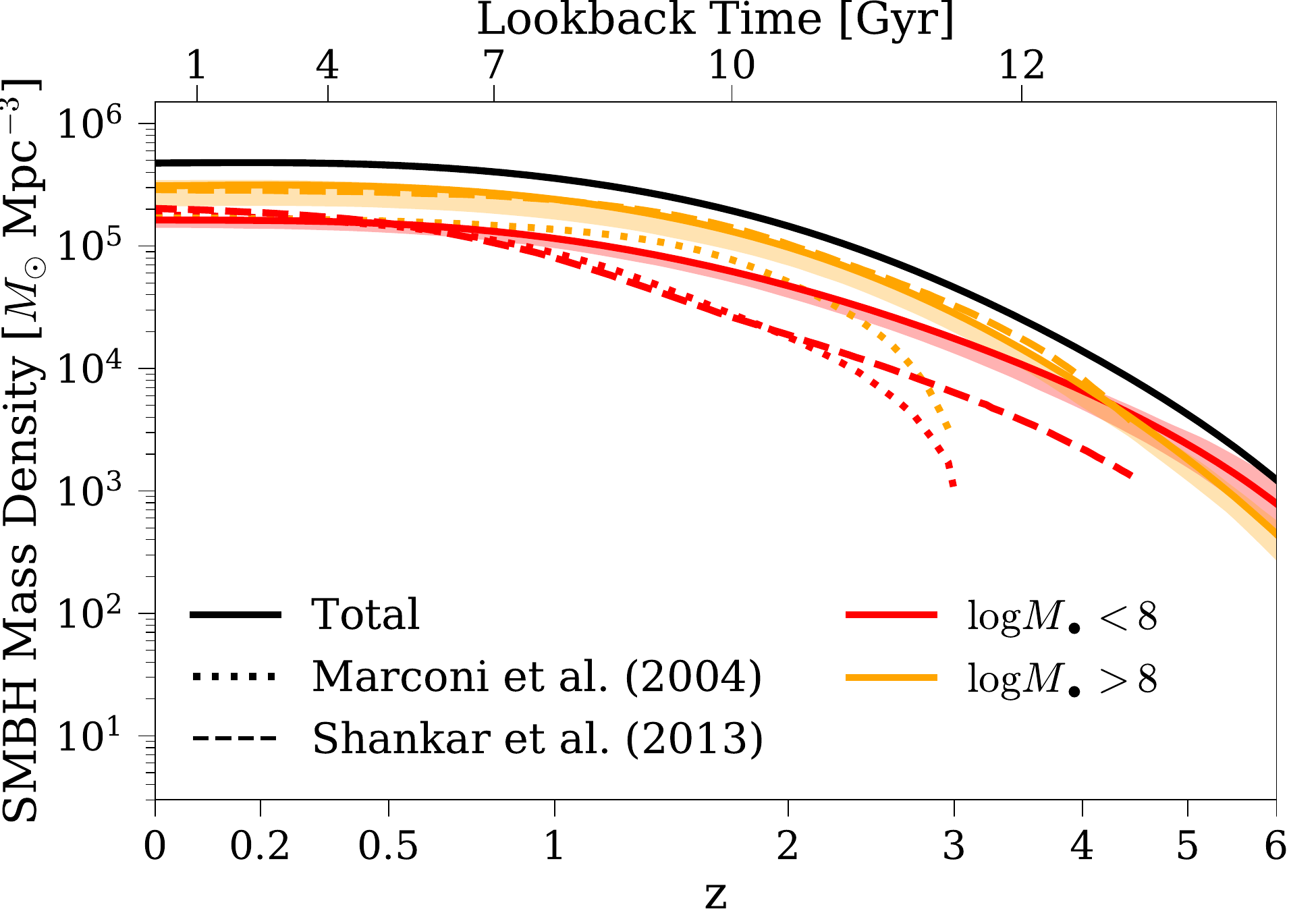}
\caption{Cosmic SMBH mass densities split in different SMBH mass bins as functions of $z$, from \textsc{Trinity} (solid lines), \citet{Marconi2004} (dotted lines), and \citet{Shankar2013} (dashed lines). See \S\ref{ss:discussions_cosmic_bh_density}. All the data used to make this plot (including those from previous studies and our best-fitting model) can be found \href{https://github.com/HaowenZhang/TRINITY/tree/main/plot_data}{here}.}
\label{f:bh_density_z_split}
\end{figure}

Fig.\ \ref{f:bh_density_z_split} shows the cosmic SMBH mass density histories of different SMBH populations from \textsc{Trinity} (solid lines), \citet{Marconi2004} (dotted lines), and \citet{Shankar2013} (dashed lines). The main difference between the results from \textsc{Trinity} and these two studies is the cosmic times when low mass SMBHs ($M_\bullet \leq 10^8 M_\bullet$) experience major growth. Specifically, SMBHs below $10^8 M_\bullet$ nearly stop growing below $z\sim1$ in \textsc{Trinity}, but grow siginificantly from $z=1$ to $z=0$ in the Marconi et al. and Shankar et al. models. One possible reason for this is that \textsc{Trinity} is required to fit the QPDFs for low-mass galaxies at lower redshifts from \cite{Aird2018}, which limit the growth of low-mass black holes. However, neither \citet{Marconi2004} nor \citet{Shankar2013} had access to these QPDFs, so their predictions are not necessarily consistent with these data. Another difference exists at $z>1$: at a fixed redshift, these low mass SMBHs also make up a larger share of the cosmic SMBH mass density in \textsc{Trinity}. This is likely due to \textsc{Trinity}'s self-consistent inference of SMBH growth history from $z=15$, which 
results in non-negligible cosmic SMBH mass densities at the starting redshifts in the Marconi et al. and Shankar et al. models (i.e., $z\sim3$ and $z\sim5$, respectively).

\subsection{Physical implications: AGN downsizing and AGN feedback on galaxy populations}
\label{ss:discussions_physical_implications}

In \S\ref{ss:results_bhar_bher} and \S\ref{ss:rho_bh}, we confirmed the ``AGN downsizing'' effect, in the sense that more massive black holes become less active earlier compared to smaller black holes, whether they are in the same host halo mass bin or not. This is true when the SMBH activity is measured by Eddington ratio (see Figs.\ \ref{f:bher_map} and \ref{f:bher_mbh}). If we instead measure SMBH activity with absolute accretion rate, we see a slight increase in BHAR towards higher masses at $z\lesssim 2$ (see Fig.\ \ref{f:bhar_map}). As mentioned earlier, this is required by the quasar probability distribution functions from \citet{Aird2018}. Physically, this is consistent with AGN feedback \citep{Somerville2008,Croton2006}. That is, in massive haloes, SMBHs still show ongoing accretion, but become less active \emph{relative to their masses} and radiatively inefficient. The energy from their mass accretion is mainly released in the form of kinetic jets and/or outflows, which serves to  maintain quenching in their host galaxies. This picture is also supported by Fig.\ \ref{f:bhar_sfr_ratio}, where the $\overline{\mathrm{BHAR}}/\overline{\mathrm{SFR}}$ ratio increases towards higher mass and lower redshifts.   Although cooling flows are known to exist in massive haloes \citep{Fabian1994}, Fig.\ \ref{f:bhar_sfr_ratio} suggests that the ratio of cold gas reaching the SMBH compared to the galaxy increases for more massive haloes. The same amount of gas also causes much more relative mass growth for SMBHs than galaxies, given their contrast in mass. Other possible fueling channels include gas recycling from stellar mass loss. Regardless of the source, SMBHs in massive haloes plausibly have sufficient material to continue growing (and generating feedback) even as the host galaxy itself is not able to grow. 

Fig.\ \ref{f:bhar_sfr_ratio} also shows that below $z\sim 6$, $\overline{\mathrm{BHAR}}$ and $\overline{\mathrm{SFR}}$ have relatively fixed average ratios for the haloes in which most star formation occurs.  This is consistent with a picture in which the SMBH and the galaxy regulate each others' growth, but it is also consistent with a process in which a separate mechanism (e.g., mass accretion onto the halo) jointly feeds both galaxy and SMBH growth.  Regardless of the mechanism, it must qualitatively change in haloes above masses of $10^{12}-10^{13} M_\odot$ to reproduce the clear upturn in $\overline{\mathrm{BHAR}}/\overline{\mathrm{SFR}}$ for massive haloes.

\section{Caveats and Future directions for empirical modeling of the halo--galaxy--SMBH connection}
\label{s:caveats}

In this section, we discuss caveats in the current version of \textsc{Trinity}, which motivates its future incorporation into \textsc{UniverseMachine}.

\subsection{Bright quasars at $5.7<z<6.5$ below $M_\bullet = 10^8 M_\odot$}

As described in \S\ref{sss:smbh_data}, we applied a Poisson prior on the number of high-redshift bright quasars with masses below $M_\bullet = 10^8 M_\odot$. This is motivated by the fact that few such objects are found in real observations. However, our best-fitting model still predicts $\sim3.5$ such objects in the same area as covered by SDSS, in contrast to current observations. By checking the intrinsic and observed BHMFs of bright quasars produced by \textsc{Trinity}, we found that most of these objects have intrinsically high black hole masses but have lower observed masses due to the random scatter in virial estimates (see \S\ref{sss:smbh_data}). Therefore, even if there are no intrinsically low-mass bright quasars at $z\gtrsim 6$, some should still exist in the observed sample.

\subsection{Future directions}
\label{ss:discussions_future_directions}

Currently, \textsc{Trinity} makes only \emph{statistical} halo--galaxy--SMBH connections. In the future, we plan to incorporate \textsc{Trinity} into the \textsc{UniverseMachine} by modeling SMBHs in \emph{individual} haloes and galaxies. This will allow: a) constraining the correlation between individual galaxy growth and SMBH growth, b) more flexibility in terms of the distributions of physical properties; c) direct modeling of AGN duty cycle timescales; d) study of the environmental effects on galaxy--SMBH coevolution; e) use of more data constraints, including separate probability distribution functions for star-forming and quiescent galaxies as well as quasar correlation functions; and f) enable the generation of more realistic halo--galaxy--SMBH mock catalogues for the whole community.

\section{Conclusions}
\label{s:conclusions}

In this work, we introduce \textsc{Trinity}, which is an empirical model that parametrizes the statistical halo--galaxy--SMBH connection. (\S\ref{s:method}). Compared to previous studies that are typically focused on one or two kinds of observables, \textsc{Trinity} self-consistently matches a comprehensive set of observational data for galaxies and SMBHs from $z=0-10$ (\S\ref{s:sims_and_data}, \S\ref{ss:results_best_fit}). These joint constraints enable \textsc{Trinity} to break degeneracies present in past studies. \textbf{Key results are as follows}:
\begin{itemize}
    \item The normalization and the slope of the median \mbh{}--\mbulge{} relation increase slightly from $z=0$ to $z=10$. At all redshifts, the mild evolution of the median \mbh{} at fixed galaxy total/bulge mass is consistent with existing observational measurements (\S\ref{ss:results_bhbm}, Fig.\ \ref{f:bhbm_median}).
    
    \item The AGN mass-to-energy conversion efficiency \eff{} is $\sim 0.05$. However, the exact value of AGN efficiency depends on the adopted AGN bolometric correction, Compton-thin/Compton-thick obscured fractions, and the assumed local \bhbm{} relation. When these input assumptions are changed, \eff{} can vary from $\sim 0.035-0.07$, i.e., a factor of 2, or 0.3 dex. (\S\ref{ss:results_systematics}, Appendices \ref{aa:kbol}, \ref{aa:ctk_corr}, \ref{aa:f_obs}, and \ref{a:alt_galaxy_smbh}).
    
    \item Average SMBH Eddington ratios are between 0.1 and 1 at $z\gtrsim 6$. This is consistent with the scenario that different SMBH populations at high redshifts are growing at close to the Eddington rate. Towards lower redshifts, their Eddington ratios (and thus specific accretion rates) decline. Therefore, total black hole mass functions (BHMFs) show a strong increase in normalization at all masses from $z\sim 10$ to $z\sim 5$, and the evolution slows down towards lower redshifts. (\S\ref{ss:results_bhar_bher}, Fig.\ \ref{f:bher_map}, \S\ref{ss:results_bhmf}, Fig.\ \ref{f:bhmf_z0_z10}).
    
    \item AGNs experience downsizing, in the sense that average Eddington ratios start to decrease earlier for more massive SMBHs. This applies to SMBHs hosted by either similar haloes/galaxies, or in different host mass bins. However, this AGN downsizing \emph{does not} hold for average SMBH accretion rates, which do not decrease towards higher masses at low redshifts (\S\ref{ss:results_bhar_bher}, \S\ref{ss:rho_bh}, Figs.\ \ref{f:bhar_map}, \ref{f:bher_map}, \ref{f:bher_mbh}, and \ref{f:rho_bh}).
    
    \item The ratio between average SMBH accretion rate and galaxy SFR is $\sim10^{-3}$ for low-mass haloes, where star-forming galaxies dominate the population.  This ratio increases in massive haloes (and galaxies) towards lower redshifts, where galaxies are more likely to be quiescent even as their SMBHs are still growing (\S\ref{ss:results_bhar_bher}, Fig.\ \ref{f:bhar_sfr_ratio}).
    
    \item Sub-grid physics recipes that qualitatively reproduce the \mbh{}--\mbulge{} redshift evolution include but are not limited to: a) strong supernova feedback in high-redshift, low-mass galaxies (IllustrisTNG, \citealt{Dubois2015,Bower2017,Pillepich2018}); b) a low accretion feedback mode that keeps SMBH growing but quenches galaxies (SIMBA, \citealt{Dave2019}).  See \S\ref{ss:discussions_bhbm_evolution} and Fig.\ \ref{f:bhbm_evolution}.

    \item Forbidding super-Eddington accretion as well as non-unity occupation fractions prevents SMBHs from growing sufficiently to match the local \bhbm{} relation. In this scenario, an AGN energy efficiency of $\sim 24\%$ is needed to explain observations like QLFs and QPDFs at high redshifts (Appendix \ref{a:no_super_eddington}, Fig.\ \ref{f:bhbm_no_super_eddington}).
    
    \item Forbidding redshift evolution of the \bhbm{} relation results in a best-fitting \bhbm{} relation that is consistent with the fiducial model, (Appendix \ref{a:const_bhbm}, Fig.\ \ref{f:const_bhbm}), but a much higher correlation coefficient between SMBH accretion rate and BH mass at fixed halo mass (\rhobh{}) is required to reproduce AGN data (Fig.\ \ref{f:rho_bh_const_bhbm}).
    
    \item During galaxy mergers, central SMBHs are unlikely to quickly consume all the infalling satellite SMBHs, otherwise black hole accretion rates would experience a precipitous decline towards lower redshift and higher masses (Appendix \ref{a:inst_merger}, Fig.\ \ref{f:bhar_bhmr_inst_merger}).  Hence, a significant number of ``wandering'' black holes are necessary.
    
    \item The following models make qualitatively consistent predictions with the fiducial \textsc{Trinity} model: a) no SMBH mergers take place; b) the fractional growth contribution to SMBH growth is always the same as that for galaxy growth (Appendix \ref{a:no_or_same_merger}, Figs.\ \ref{f:bhbm_no_merger} and \ref{f:bher_no_merger}).
    
\end{itemize}

This work is the first in a series of \textsc{Trinity} papers. Paper II (H.\ Zhang et al., in prep.) discusses quasar luminosity functions and the buildup of SMBHs across cosmic time; Paper III (H.\ Zhang et al., in prep.) presents predictions for quasars and other SMBHs at $z>6$; Paper IV (H.\ Zhang et al., in prep.) discusses the SFR-BHAR correlation as a function of halo mass, galaxy mass, and redshift; and paper V (H.\ Zhang et al., in prep) covers black hole merger rates and \textsc{Trinity}'s predictions for gravitational wave experiments. Paper VI (O.\ Knox et al, in prep) and Paper VII (Huanian Zhang et al., in prep) present the AGN auto-correlation functions and AGN--galaxy cross-correlation functions from \textsc{Trinity}, respectively.

\section*{Data availability}
\label{s:data_availability}

The parallel implementation of \textsc{Trinity}, the compiled datasets (\S\ref{ss:obs_data}), data for all figures, and the posterior distribution of model parameters (\S\ref{ss:results_best_fit}, Appendix \ref{a:param_values}) are available \href{https://github.com/HaowenZhang/TRINITY}{\textbf{online}}.

\section*{Acknowledgements}
\label{s:acknowledgements}

We thank Stacey Alberts, Rachael Amaro, Gurtina Besla, Haley Bowden, Jane Bright, Katie Chamberlain, Alison Coil, Ryan Endsley, Sandy Faber, Hayden Foote, Dan Foreman-Mackey, Nico Garavito-Camargo, Nickolay Gnedin, Richard Green, Jenny Greene, Kate Grier, Melanie Habouzit, Kevin Hainline, Elaheh Hayati, Andrew Hearin, Julie Hlavacek-Larrondo, Luis Ho, Allison Hughes, Yun-Hsin Huang, Raphael Hviding, Victoria Jones, Stephanie Juneau, Ryan Keenan, Oddisey Knox, David Koo, Andrey Kravtsov, Daniel Lawther, Rixin Li, Joseph Long, Jianwei Lyu, Chung-Pei Ma, Garreth Martin, Karen Olsen, Feryal Özel, Vasileios Paschalidis, Ekta Patel, Dimitrios Psaltis, Joel Primack, Yujing Qin, Eliot Quataert, George Rieke, Marcia Rieke, Paolo Salucci, Jan-Torge Schindler, Spencer Scott, Xuejian Shen, Yue Shen, Dongdong Shi, Irene Shivaei, Rachel Somerville, Fengwu Sun, Wei-Leong Tee, Yoshihiro Ueda, Marianne Vestergaard, Feige Wang, Ben Weiner, Christina Williams, Charity Woodrum, Jiachuan Xu, Minghao Yue, Dennis Zaritsky, Huanian Zhang, Xiaoshuai Zhang, and Zhanbo Zhang for very valuable discussions.

Support for this research came partially via program number HST-AR-15631.001-A, provided through a grant from the Space Telescope Science Institute under NASA contract NAS5-26555. PB was partially funded by a Packard Fellowship, Grant \#2019-69646. PB was also partially supported by a Giacconi Fellowship from the Space Telescope Science Institute.  Finally, PB was also partially supported through program number HST-HF2-51353.001-A, provided by NASA through a Hubble Fellowship grant from the Space Telescope Science Institute, under NASA contract NAS5-26555.

Data compilations from many studies used in this paper were made much more accurate and efficient by the online \textsc{WebPlotDigitizer} code.\footnote{\url{https://apps.automeris.io/wpd/}} This research has made extensive use of the arXiv and NASA's Astrophysics Data System.

This research used the Ocelote supercomputer of the University of Arizona. The allocation of computer time from the UA Research
Computing High Performance Computing (HPC) at the University
of Arizona is gratefully acknowledged. The Bolshoi-Planck simulation was performed by Anatoly Klypin within the Bolshoi project of the University of California High-Performance AstroComputing Center (UC-HiPACC; PI Joel Primack).

{\footnotesize
\bibliography{main}
}

\appendix

\section{Halo Merger Rates}
\label{a:halo_merger_rates}

\begin{figure*}
\subfigure{
\includegraphics[width=0.48\textwidth]{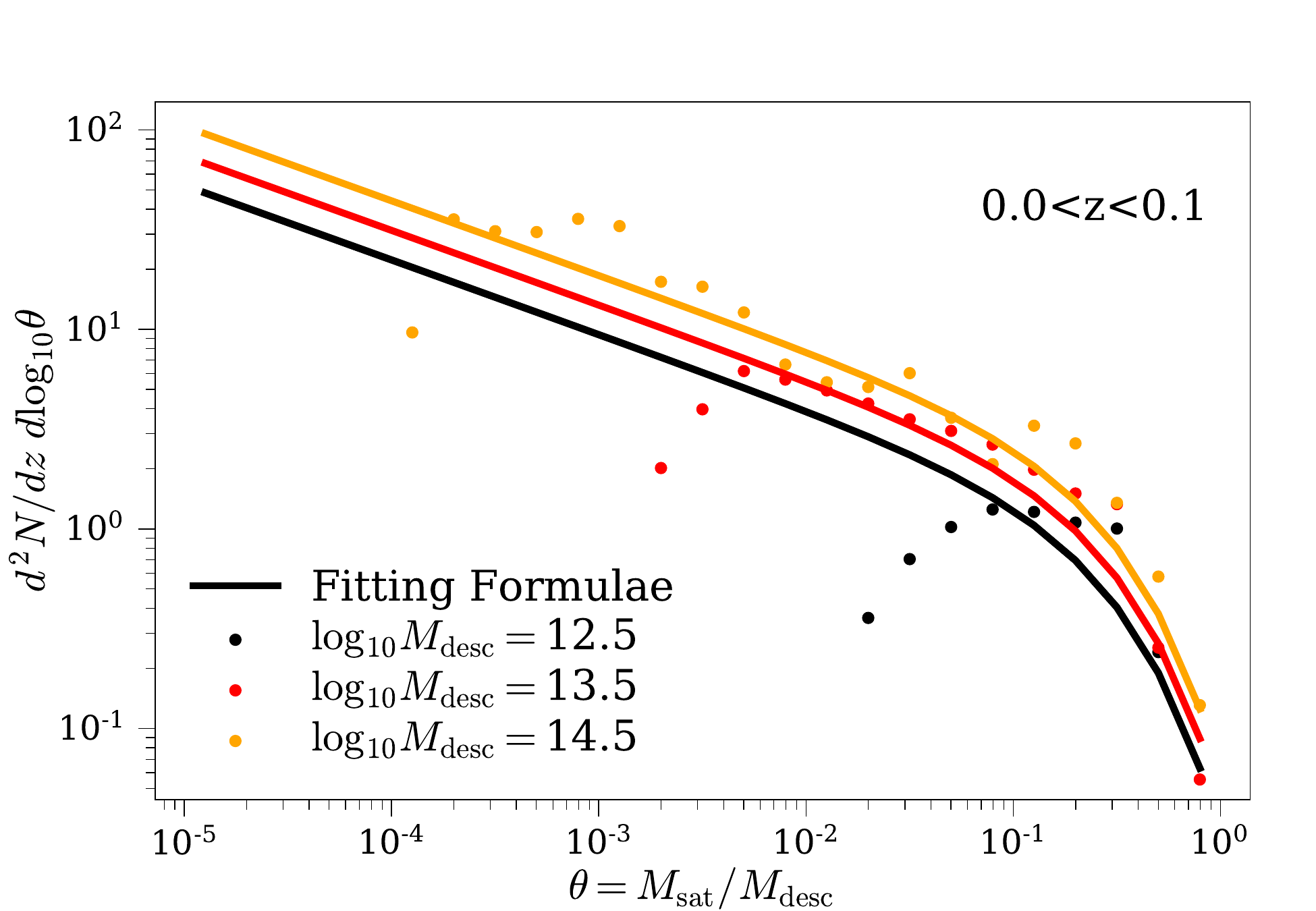}
}
\subfigure{
\includegraphics[width=0.48\textwidth]{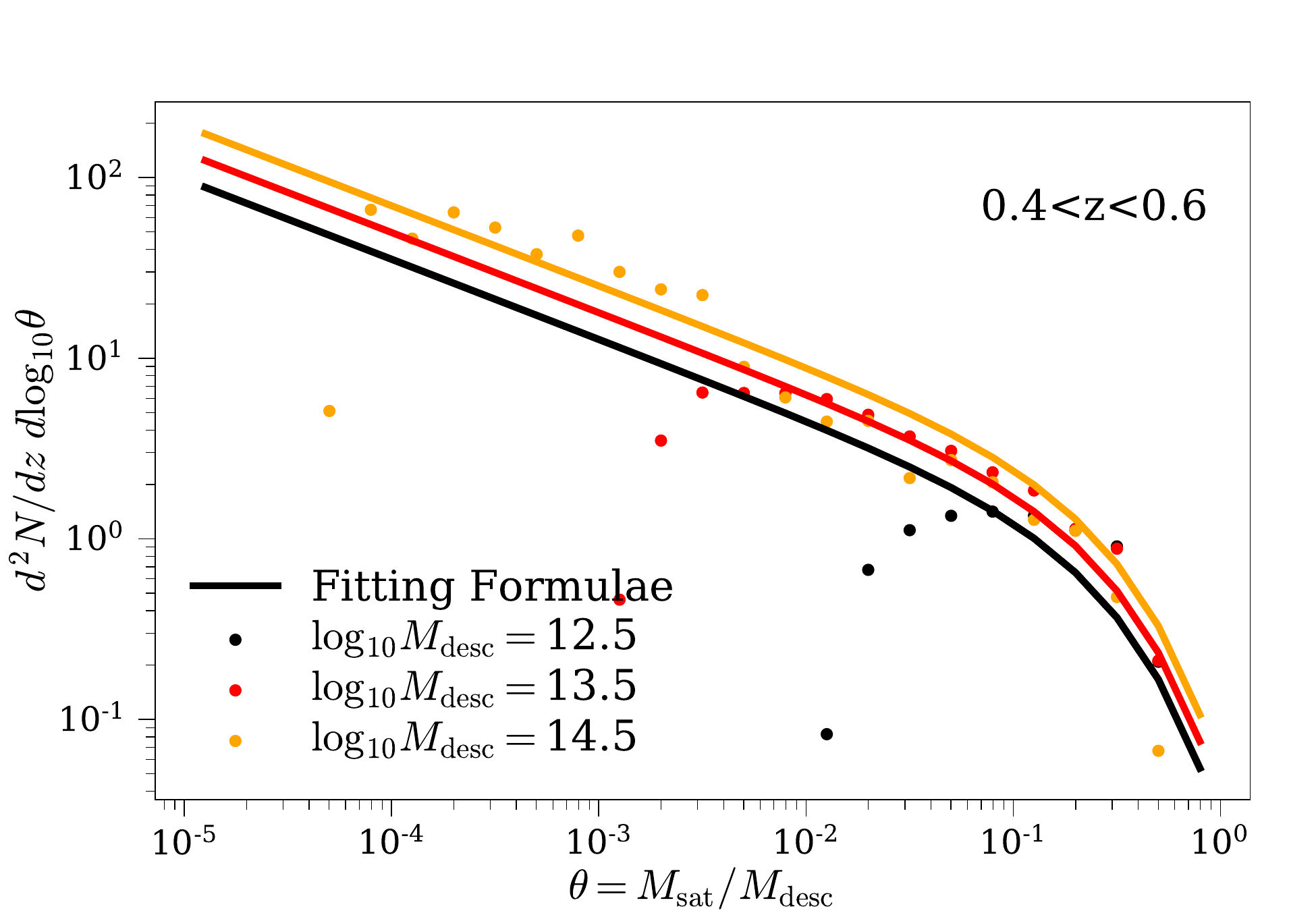}
}
\subfigure{
\includegraphics[width=0.48\textwidth]{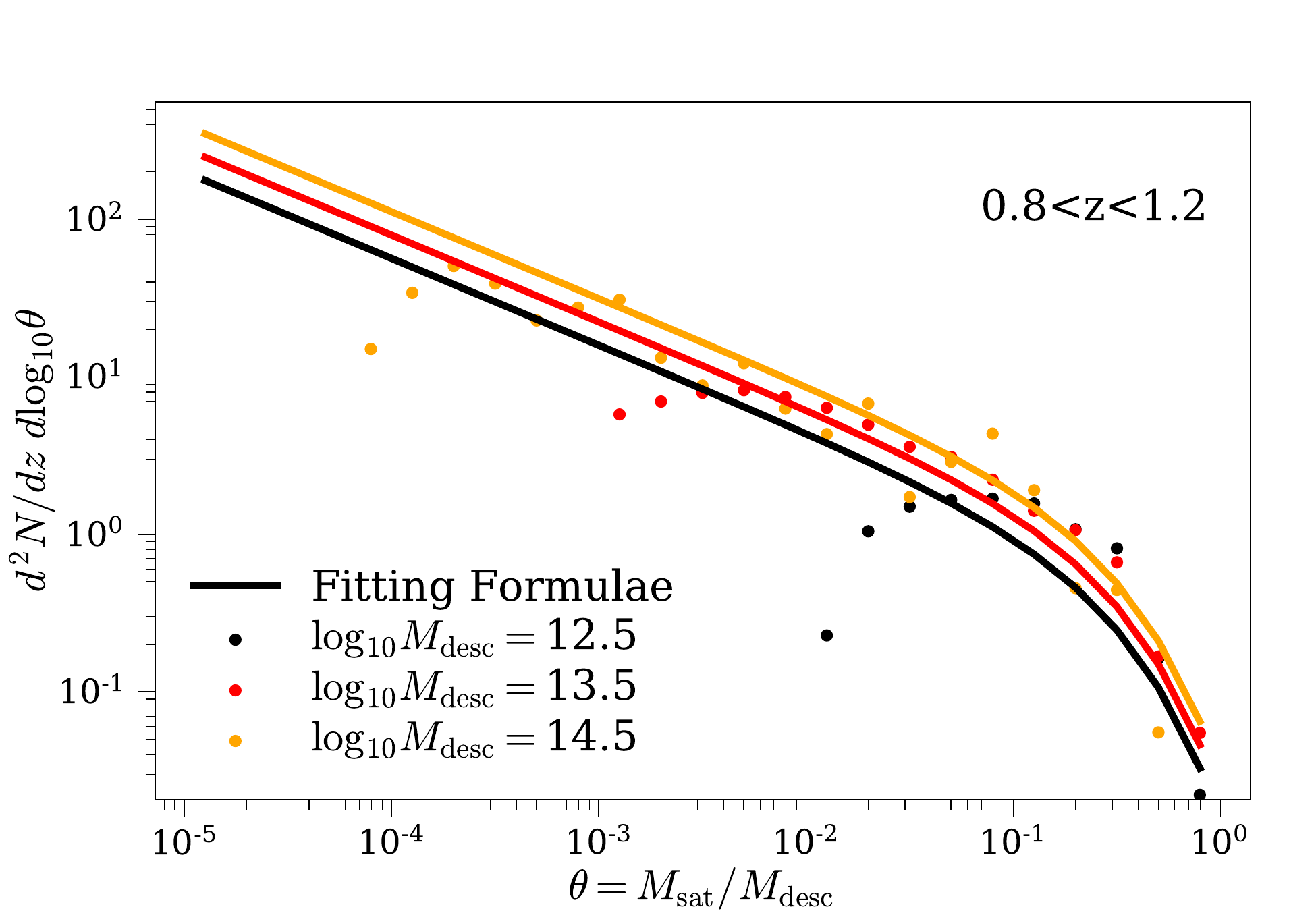}
}
\subfigure{
\includegraphics[width=0.48\textwidth]{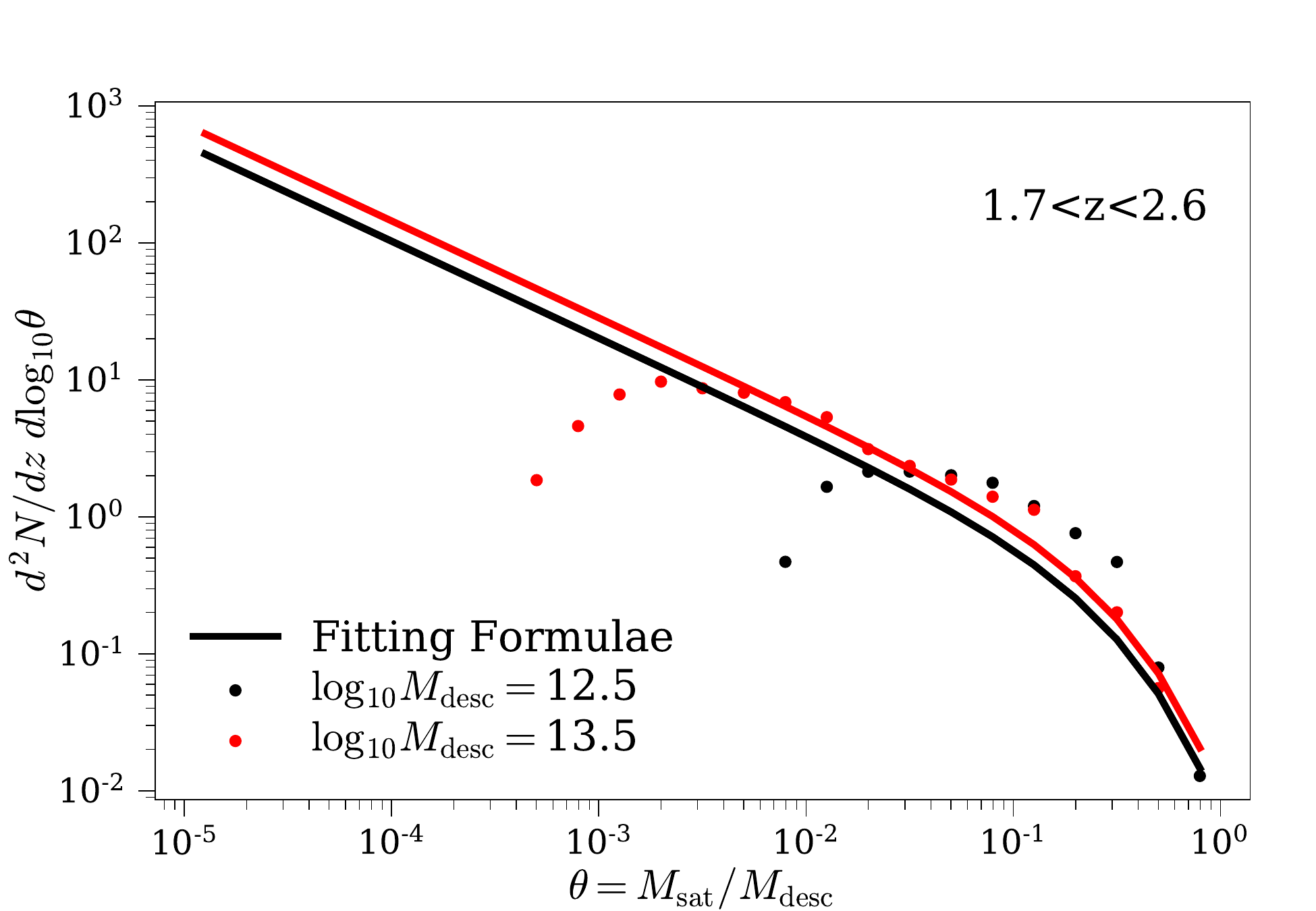}
}
\subfigure{
\includegraphics[width=0.48\textwidth]{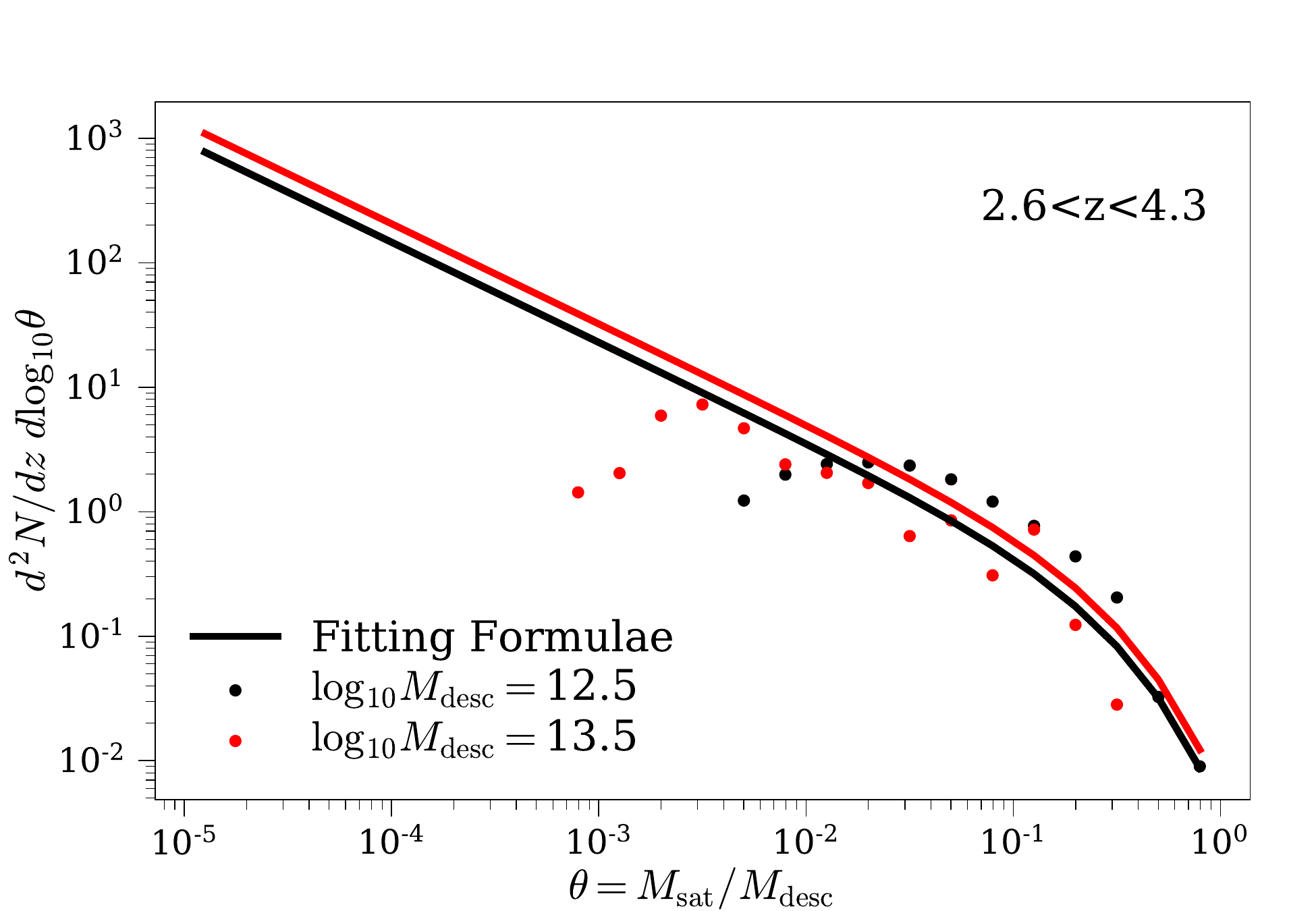}
}
\subfigure{
\includegraphics[width=0.48\textwidth]{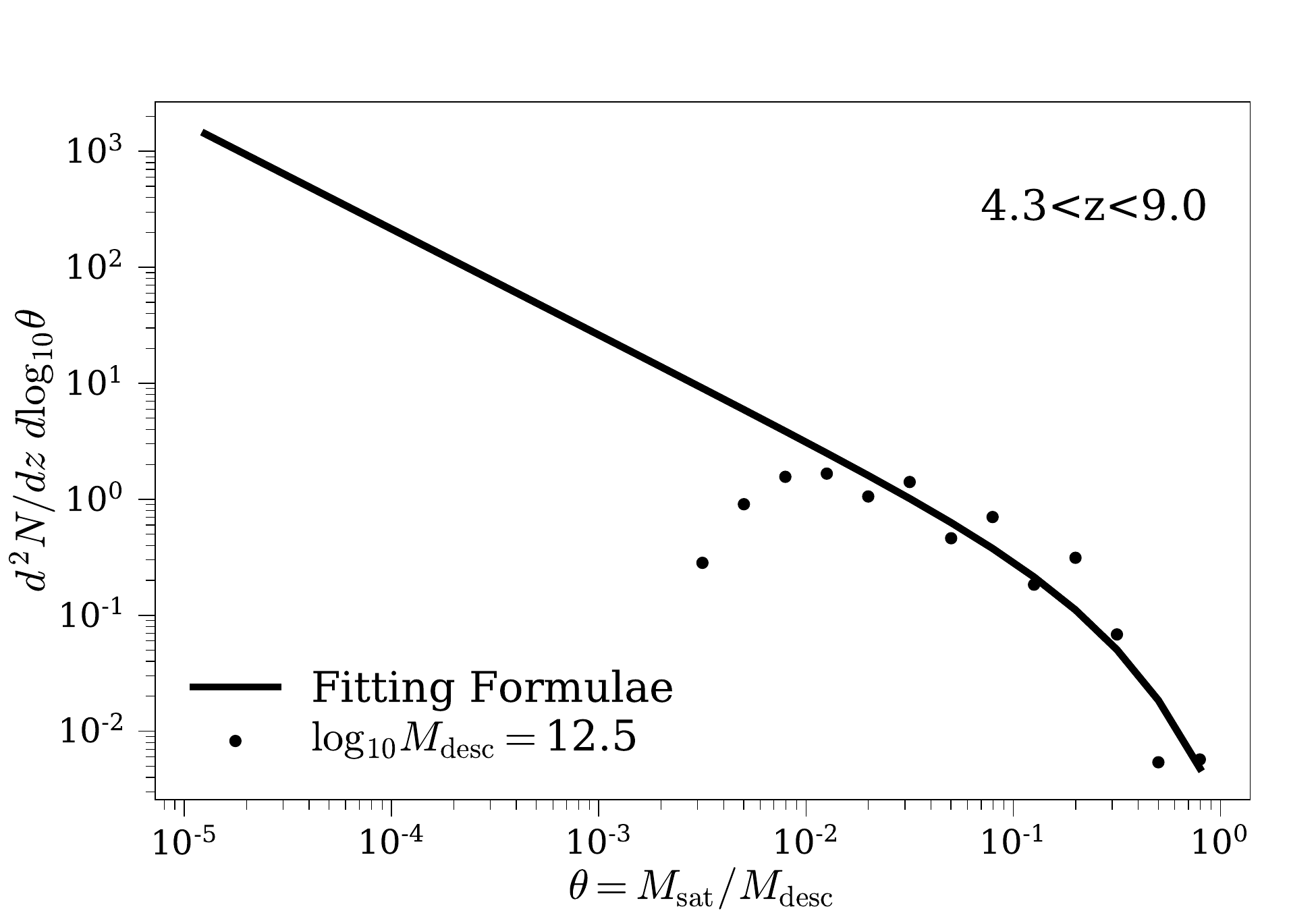}
}
\caption{The rate of satellite galaxy disruption in host haloes in the \textsc{UniverseMachine} as a function of $z$, descendant mass $M_\mathrm{desc}$, and satellite-to-descendant mass ratio $\theta=M_\mathrm{sat}/M_\mathrm{desc}$. The solid symbols are the binned estimates of merger rates, and the solid lines are the fitted results. See Appendix \ref{a:halo_merger_rates}. All the data used to make this plot (including the individual data points and our best-fitting model) can be found \href{https://github.com/HaowenZhang/TRINITY/tree/main/plot_data}{here}.}
\label{f:halo_merger_rates}
\end{figure*}

In \textsc{Trinity}, SMBH mergers are directly linked to galaxy mergers. As shown in Eq. \ref{e:galaxy_merger_rate}, halo merger rates are needed in the calculation of galaxy merger rates. Hence, we use the halo merger rates from the \textsc{UniverseMachine}, where satellite galaxies will disrupt when their $v_\mathrm{max}/v_\mathrm{Mpeak}$ ratios reach a certain threshold (see \S\ref{ss:halo_galaxy_connection} for the definitions of $v_\mathrm{max}$ and $v_\mathrm{Mpeak}$). We refer readers to \S 3.3 and Appendix B of \citet{Behroozi2019} for full details. Here, we fit these merger rates with a set of analytical formulae. Letting $a=1/(1+z)$ be the scale factor, $M_\mathrm{desc}$ the mass of the descendant halo, $M_\mathrm{sat}$ the mass of the satellite halo, and $\theta=M_\mathrm{sat}/M_\mathrm{desc}$ the mass ratio, the merger rate is expressed as the number of mergers per unit descendant halo, per unit redshift per log interval in mass ratio:

\begin{eqnarray}
\label{e:subhalo_merger_rate}
    -\frac{d^2N(M_\mathrm{desc}, \theta, z)}{dzd\log_{10}\theta} & = & 10^{A(M_\mathrm{desc}, a)}\theta^{B(a)} \exp{\left(-3.162\theta\right)}\\
    A(M_\mathrm{desc}, a) & = & A_0(M_\mathrm{desc}) + A_1(a)\\
    A_0(M_\mathrm{desc}) & = & 0.148\log_{10} \left(\frac{M_\mathrm{desc}}{10^{12} M_\odot}\right) - 0.291\\
    A_1(a) & = & -1.609 + 3.816 a + (-2.152)a^2\\
    B(a) & = & -1.114 + 1.498 a + (-0.757)a^2\ .
\end{eqnarray}
We show the quality of these fits in Fig.\ \ref{f:halo_merger_rates}.  Compared to \cite{Behroozi2013}, these merger rates are lower by 15--40\% due to the presence of orphan galaxies in the \textsc{UniverseMachine}.

\section{Median galaxy UV magnitudes and scatter as functions of halo mass and star formation rates}
\label{a:UV_SFR_fit}

To constrain the high-redshift halo--galaxy connection in \textsc{Trinity}, we use the median galaxy UV magnitudes and the corresponding log-normal scatter from the \textsc{UniverseMachine} as functions of redshift, halo mass ($M_{\rm peak}$), and star formation rates to calculate galaxy UV luminosity functions at $z=9$ and $z=10$. Here, we show the best fitting parameters for these scaling relations, as well as the goodness of fitting.

The median galaxy UV magnitudes $\widetilde{M}_{\rm UV}$ have the following dependence on redshift, $M_{\rm peak}$, and SFR:
\begin{eqnarray}
\widetilde{M}_{\rm UV} & = & k_{\rm UV} \times \log_{10} \mathrm{SFR} + b_{\rm UV}\label{e:uv_med_best_fit}\\
k_{\rm UV} & = & 
\begin{aligned}
    &0.154\left(\log_{10}M_{\rm peak}\right)^2 + (-2.876)\log_{10}M_{\rm peak}\\
    & + (-2.378)\left(a - 1\right) + 9.478
\end{aligned}\\
b_{\rm UV} & = & 
\begin{aligned}
    &(-0.347)\left(\log_{10}M_{\rm peak}\right)^2 + 6.853\log_{10}M_{\rm peak}\\
    & + 1.993\left(a - 1\right) + (-50.344)
\end{aligned}\label{e:b_uv_best_fit}\ .
\end{eqnarray}
The log-normal scatter $\sigma_{\rm UV}$ has the following redshift and $M_{\rm peak}$ dependency:
\begin{eqnarray}
\sigma_{\rm UV} & = & k_{\sigma_{\rm UV}} \times \log_{10} M_{\rm peak} + b_{\sigma_{\rm UV}}\\
k_{\sigma_{\rm UV}} & = & -0.031 z + 0.042\\
b_{\sigma_{\rm UV}} & = & 0.319 z + 0.241\label{e:b_std_uv_best_fit}\ .
\end{eqnarray}

\begin{figure*}
\subfigure{
\includegraphics[width=0.48\textwidth]{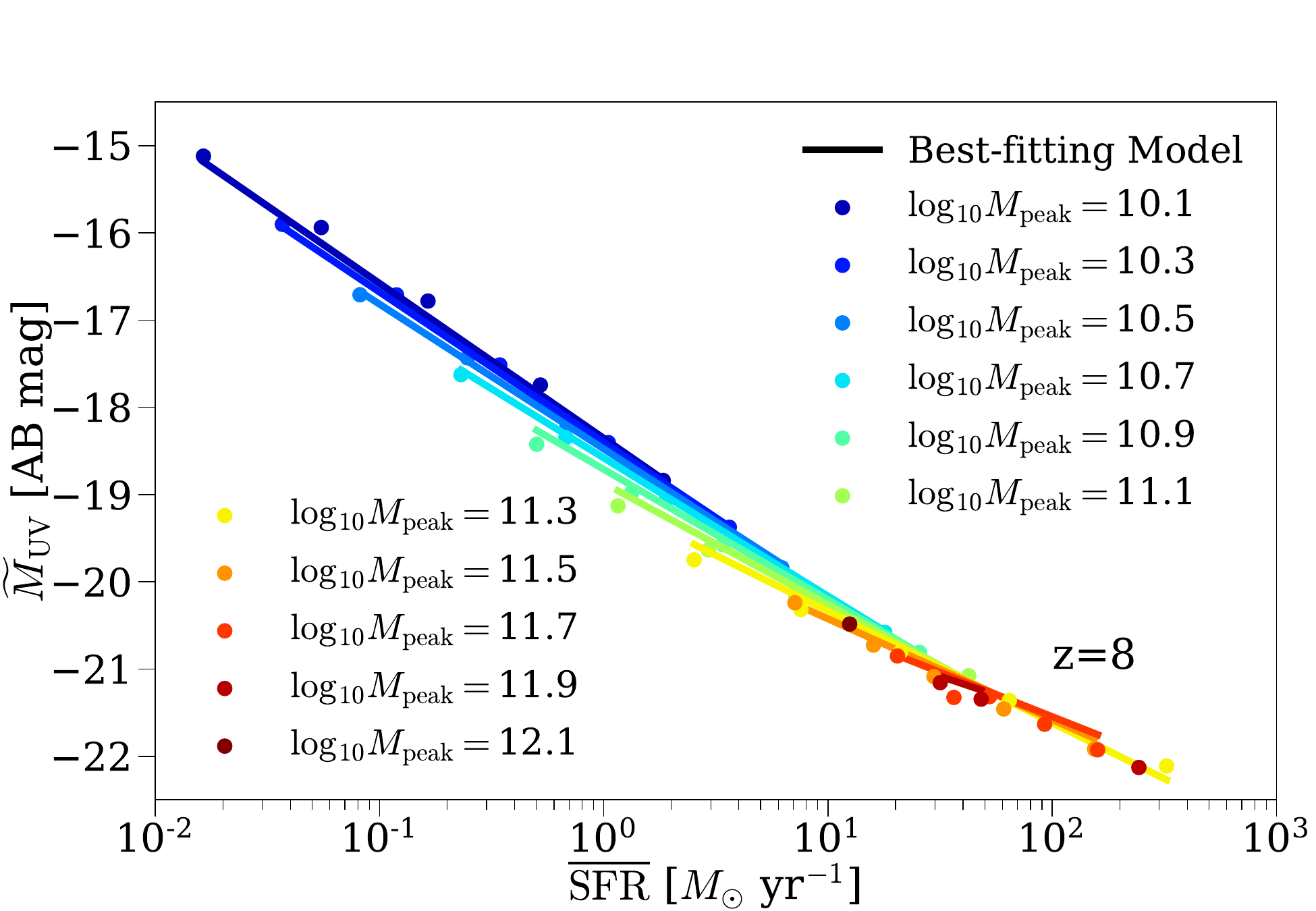}
}
\subfigure{
\includegraphics[width=0.48\textwidth]{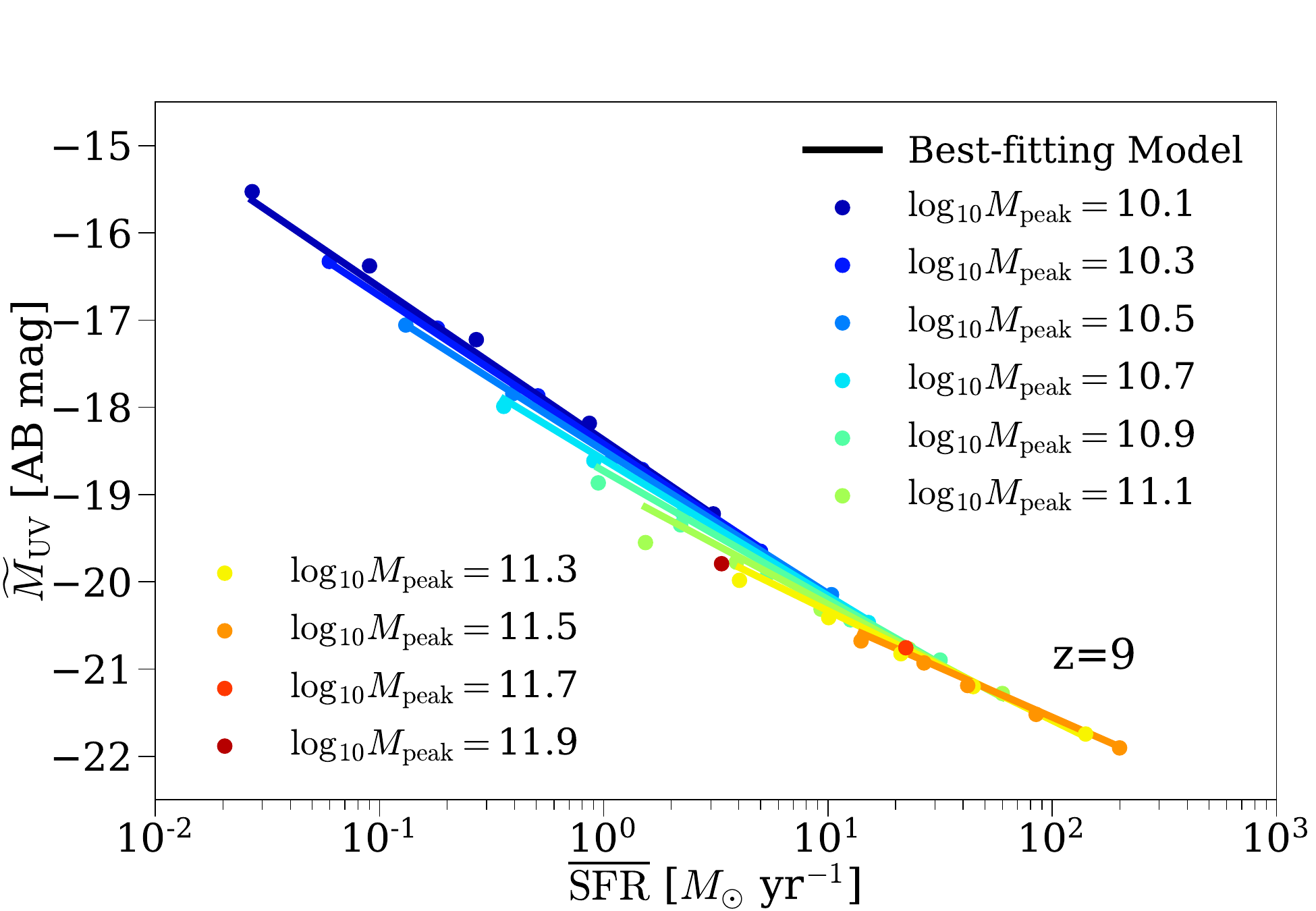}
}
\subfigure{
\includegraphics[width=0.48\textwidth]{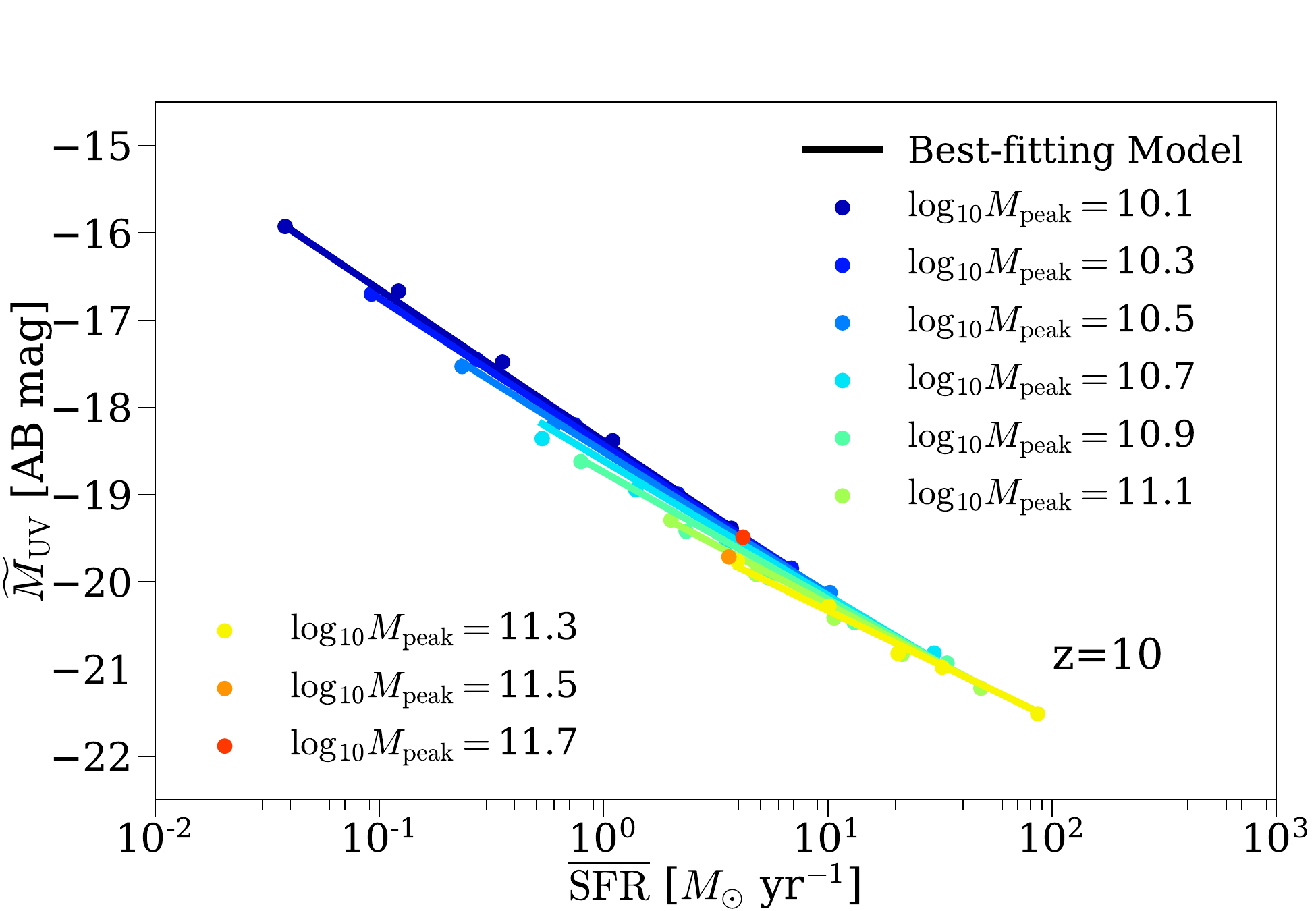}
}
\subfigure{
\includegraphics[width=0.48\textwidth]{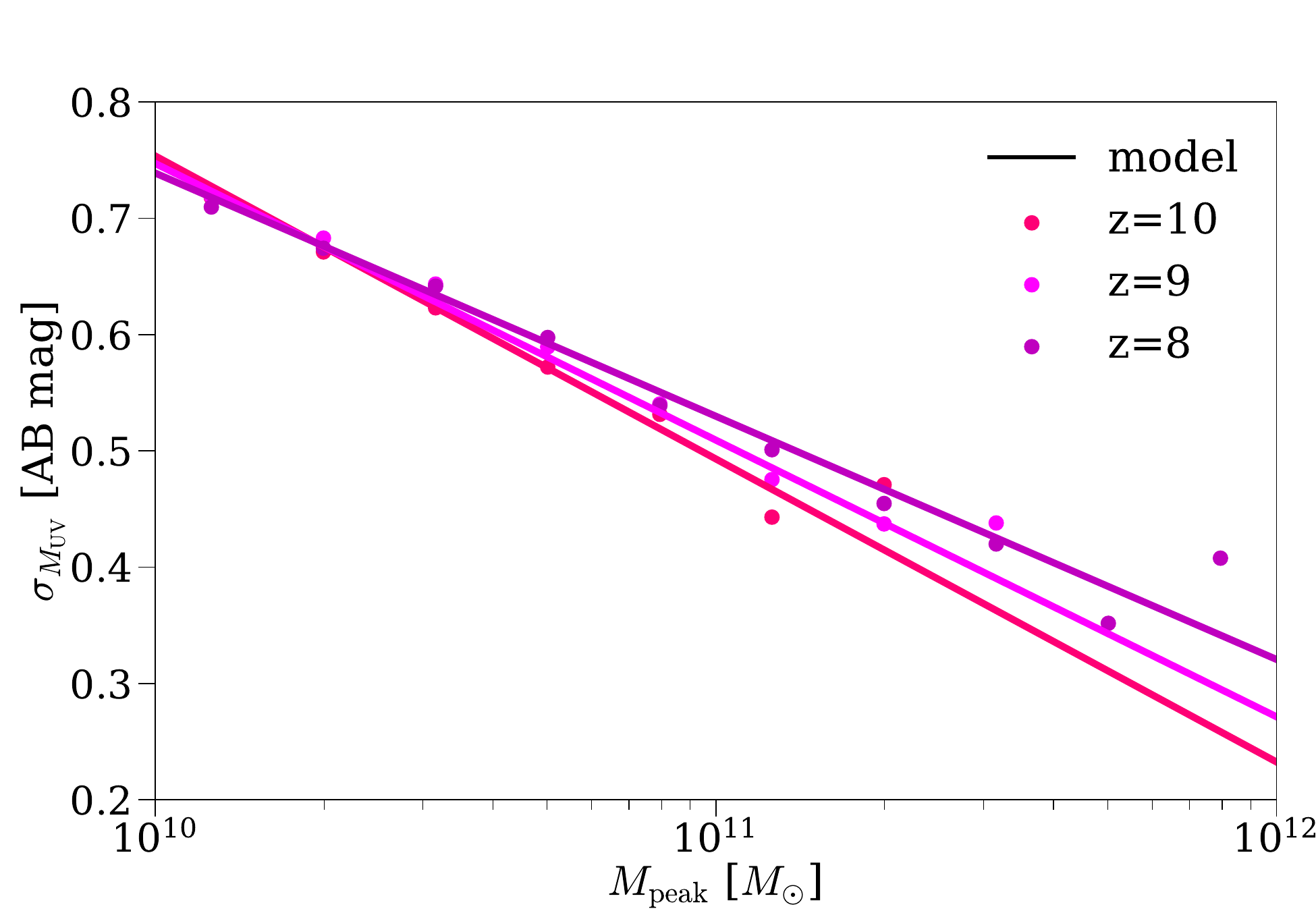}
}
\caption{The fits to median UV magnitude, $\widetilde{M}_{\rm UV}$, as a function of $M_\mathrm{peak}$, $\overline{\mathrm{SFR}}$, and $z$, and the corresponding scatter, $\sigma_{M_{\mathrm{UV}}}$, as a function of $M_\mathrm{peak}$ and $z$, from the  \textsc{UniverseMachine}. The filled circles are the data points from the \textsc{UniverseMachine}, and the solid lines are the best-fitting models in Eqs.\ \ref{e:uv_med_best_fit}-\ref{e:b_std_uv_best_fit}. See Appendix \ref{a:UV_SFR_fit}. All the data used to make this plot (including the individual data points and our best-fitting model) can be found \href{https://github.com/HaowenZhang/TRINITY/tree/main/plot_data}{here}.}
\label{f:uv_sfr_fitting}
\end{figure*}

Fig.\ \ref{f:uv_sfr_fitting} shows the goodness of fit for Eqs.\ (\ref{e:uv_med_best_fit})-(\ref{e:b_std_uv_best_fit}) to both $\widetilde{M}_{\rm UV}$ and $\sigma_{\rm UV}$ from $z=8-10$. Using these fitting functions, \textsc{Trinity} produces SFRs and galaxy UV luminosities that are both consistent with the \textsc{UniverseMachine}.

\section{Calculating inherited and infalling SMBH masses from merger tree statistics}
\label{a:mbh_old_unmerged}

In \textsc{Trinity}, we assign SMBH masses to haloes at all redshifts and then calculate black hole growth rates (BHGRs) by differentiation. This is different from how we model galaxies (where we directly model galaxy growth rates and integrate to obtain stellar masses), because the functional forms for galaxy growth rates in haloes are better known than the functional forms for SMBH growth rates in galaxies. Here, we detail how we calculate the masses of the inherited and infalling (see \S\ref{ss:bh_mergers}) SMBHs.

In \textsc{Trinity}, haloes inherit both central and wandering SMBHs from their most massive progenitors (MMPs). For the $j$th halo mass bin at the $i$th snapshot, the average central SMBH mass inherited from MMPs is:
\begin{eqnarray}
    && \hspace{-4ex} \overline{M}^{j}_{\bullet,\mathrm{inherit},i} = \sum_{k} P^{j,k}_{\mathrm{MMP},i}\overline{M}^{k}_{\bullet, i-1}\ ,
\end{eqnarray}

 where $P^{j,k}_{\mathrm{MMP},i}$ is the probability that haloes in the $j$th halo mass bin at the $i$th snapshot have MMPs in the $k$th mass bin at the $(i-1)$th snapshot. This probability is calculated based on the average halo growth curves from N-body simulations (see \S\ref{ss:dm_sims}). $\overline{M}^{k}_{\bullet, i-1}$ is the average \emph{central} SMBH mass of the haloes in the $k$th mass bin at the $(i-1)$th snapshot, determined by the halo--galaxy--SMBH connection.

As for infalling SMBHs, they come from: 1) wandering SMBHs inherited from MMPs; 2) \emph{all} the SMBHs from infalling satellite haloes. The average mass of infalling SMBHs for the $j$th halo mass bin at the $i$th snapshot is then, by definition:
\begin{eqnarray} \overline{M}^{j}_{\bullet,\mathrm{infall},i} &=& \sum_{k} P^{j,k}_{\mathrm{MMP},i}\overline{M}^{k}_{\bullet, \mathrm{wandering},i-1}\\ \nonumber
    &+& \sum_{k}\mathcal{R}^{j,k}_{\mathrm{merger},i}\overline{M}^{k}_{\bullet, i-1}\ ,
\end{eqnarray}
where $\overline{M}^{k}_{\bullet, \mathrm{wandering},i-1}$ is the average total \emph{wandering} SMBH mass of the haloes in the $k$th mass bin at the $(i-1)$th snapshot, and $\mathcal{R}^{j,k}_{\mathrm{merger},i}$ is the merger rate of satellite haloes in the $k$th mass bin into the descendant haloes in the $j$th mass bin at the $i$th snapshot. This rate is calculated by integrating Eq.\ \ref{e:subhalo_merger_rate} over the redshift dimension:
\begin{equation}
    \mathcal{R}^{j,k}_{\mathrm{merger},i} =  \int_{10^{-0.5\Delta\log_{10} M_\mathrm{peak}}M^{k}_{\mathrm{peak},i}/M^{j}_{\mathrm{peak},i}}^{10^{0.5\Delta\log_{10} M_\mathrm{peak}}M^{k}_{\mathrm{peak},i}/M^{j}_{\mathrm{peak},i}} \frac{d^2 N(M_\mathrm{peak}, \theta, z)}{d\log\theta dz}\bigg\rvert^{M_\mathrm{peak}=M^j_{\mathrm{peak},i}}_{z=z_i}  d\theta\ ,
\end{equation}
where $z_{i}$ is the redshift of the $i$th snapshot, and $M^j_{\mathrm{peak},i}$ is the peak mass of the halo in the $j$th mass bin at the $i$th snapshot.

\section{Corrections, Exclusions, and Uncertainties for AGN Data}
\label{a:agn_data}

\subsection{Bolometric Corrections}
\label{aa:kbol}

\begin{figure}
\includegraphics[width=0.48\textwidth]{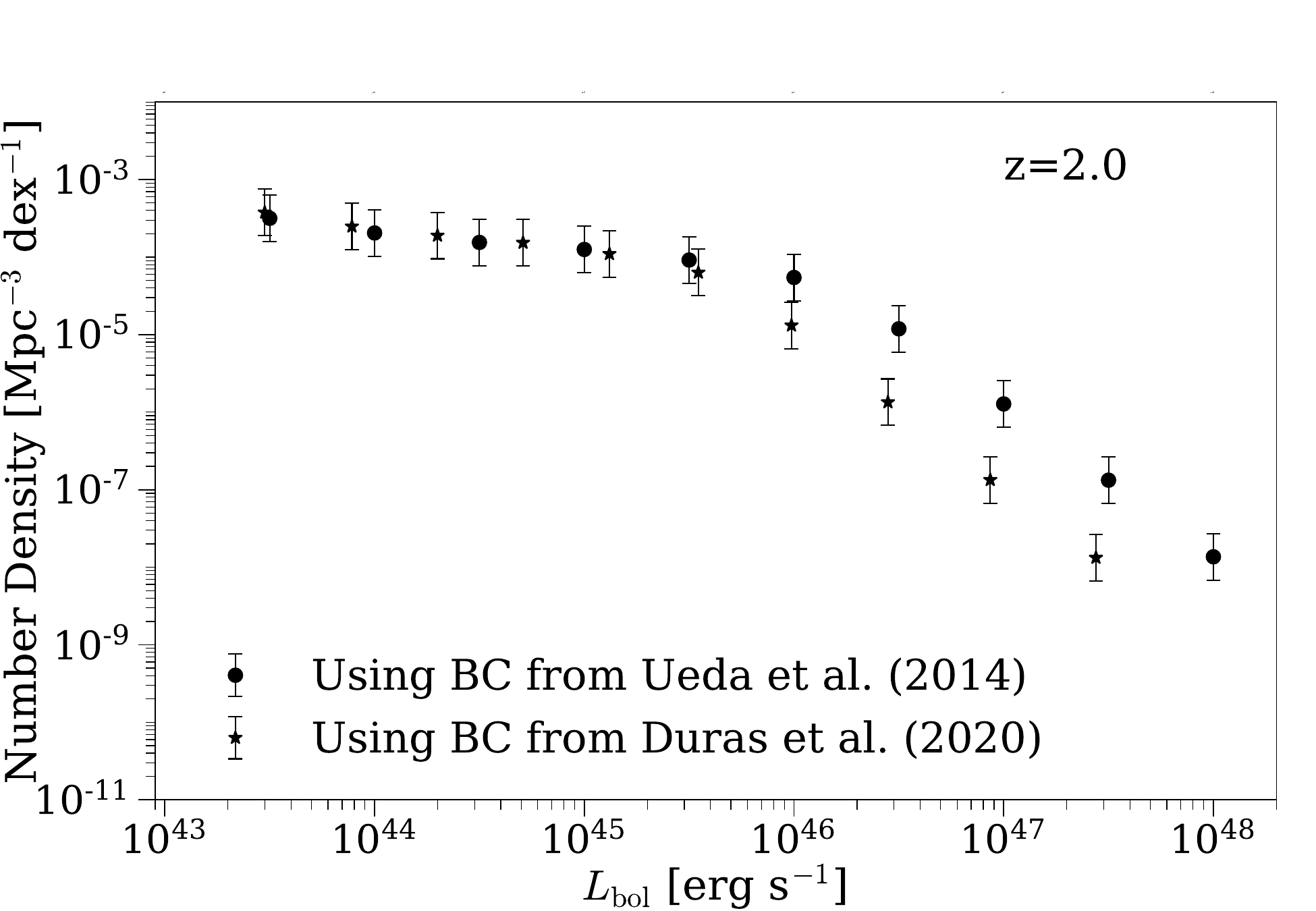}
\caption{The comparison of the QLFs at $z=2$ from \citet{Ueda2014}, when using bolometric corrections (BC) from \citet{Ueda2014} (filled circles) and from \citet{Duras2020} (stars). See Appendix \ref{aa:kbol}. All the data used to make this plot can be found \href{https://github.com/HaowenZhang/TRINITY/tree/main/plot_data}{here}.}
\label{f:qlf_ueda_duras}
\end{figure}

Different bolometric corrections (BC) for the same quasar sample produce different \emph{bolometric} QLFs, which, in principle, could lead to systematic differences in the inferred SMBH properties. Here, we investigate how the systematic difference in bolometric corrections would impact our results in \S\ref{s:results}.

\begin{figure}
\subfigure{
\includegraphics[width=0.48\textwidth]{figs/submit_fiducial/BHBM_median_submit_fiducial.pdf}
}
\subfigure{
\includegraphics[width=0.48\textwidth]{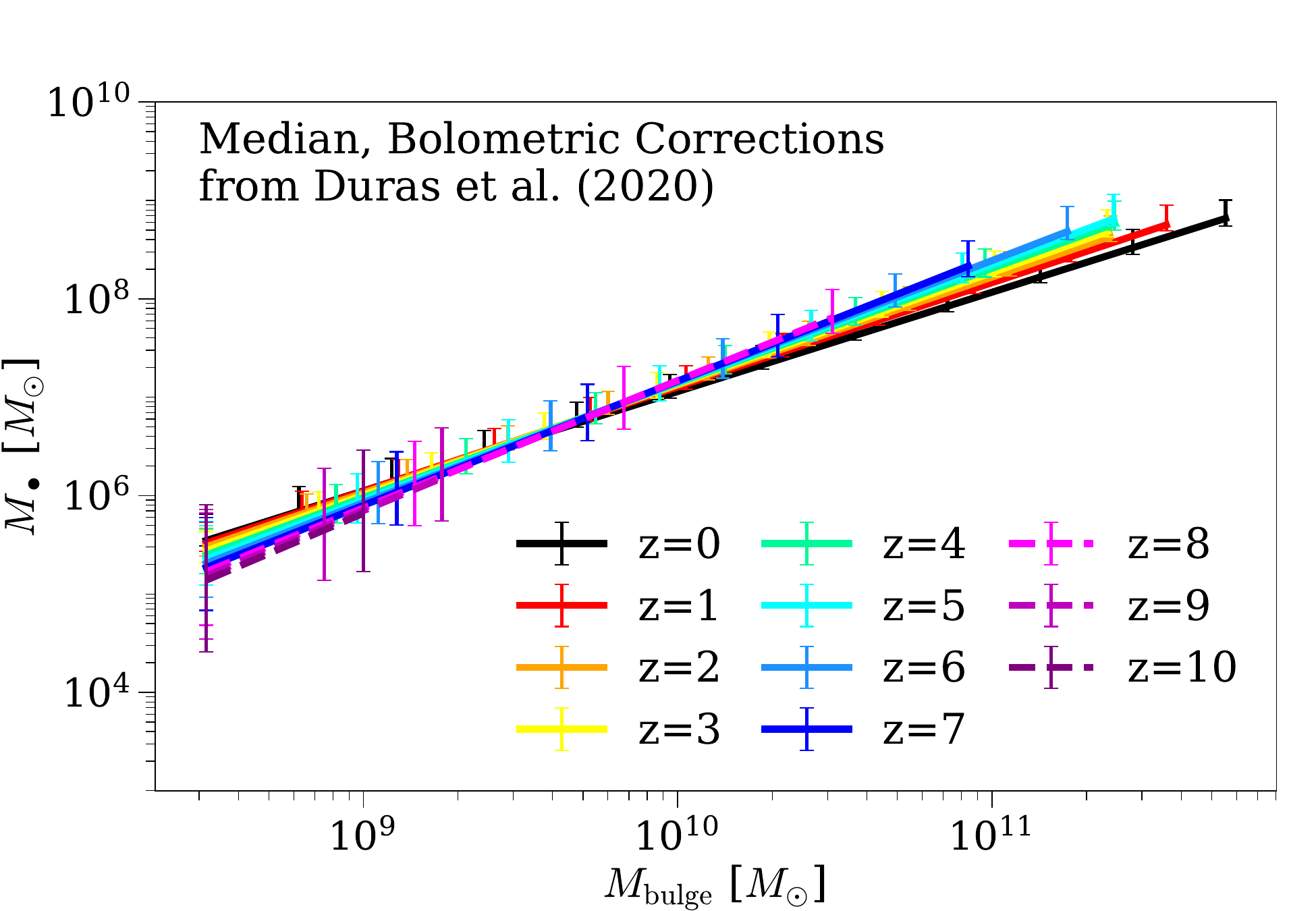}
}
\caption{\textbf{Top Panel:} the best-fitting median \bhbm{} relation from $z=0-10$ assuming the bolometric corrections from \citet{Ueda2014}. \textbf{Bottom Panel:} the best-fitting median \bhbm{} relation from $z=0-10$ assuming the bolometric corrections from \citet{Duras2020}. \errorbars{} \waitforjwst{} All the data used to make this plot can be found \href{https://github.com/HaowenZhang/TRINITY/tree/main/plot_data}{here}.}
\label{f:bhbm_duras_kbol}
\end{figure}

Fig.\ \ref{f:qlf_ueda_duras} shows the different resulting bolometric QLFs at $z=2$ produced by correcting \citet{Ueda2014} QLFs with BCs from \citet{Ueda2014} (filled circles, ``UedaBC'') and \citet{Duras2020} (stars, ``DurasBC''). Due to smaller BC values at high X-ray luminosities, the ``DurasBC'' gives many fewer bright quasars. At the less massive end, the two BCs result in consistent quasar number densities. The low number densities of bright quasars suppress the abundance of more massive SMBHs, because only the latter can produce so much energy with reasonable Eddington ratios. Ultimately, this forces \textsc{Trinity} to choose \bhbm{} relations with lower normalizations ($\beta_\mathrm{BH}$) and slopes ($\gamma_\mathrm{BH}$), as shown in Fig.\ \ref{f:bhbm_duras_kbol}. With the decrease in both the total energy output and the \bhbm{} normalization, the AGN energy efficiency only decreases by $\sim 0.02$ dex if the ``DurasBC'' is adopted.

\begin{figure}
\includegraphics[width=0.48\textwidth]{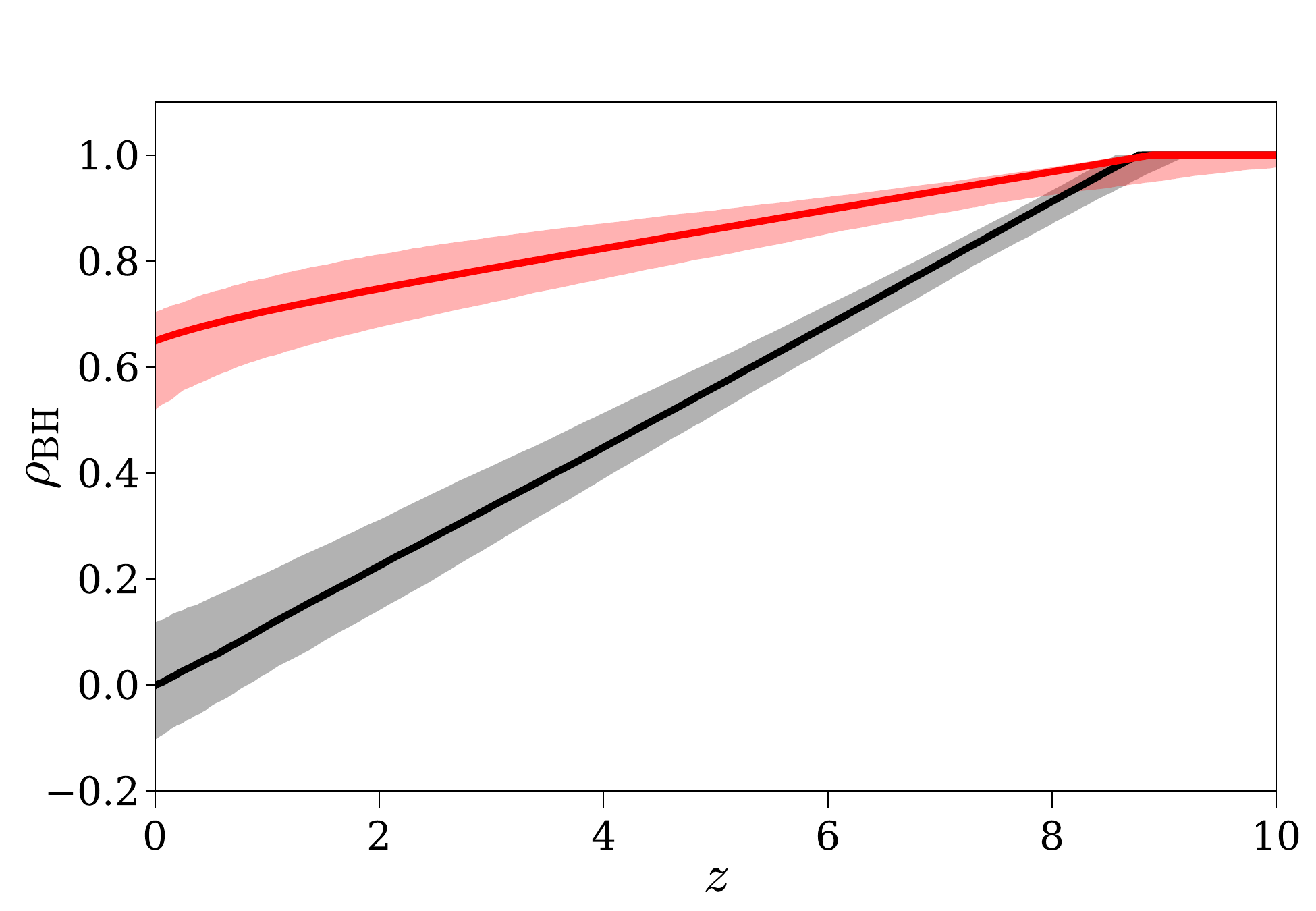}
\caption{The correlation coefficient between average SMBH accretion rate and \mbh{} at fixed host halo mass, \rhobh{}, assuming the bolometric corrections from \citet{Ueda2014} (black solid line) and \citet{Duras2020} (red solid line). All the data used to make this plot can be found \href{https://github.com/HaowenZhang/TRINITY/tree/main/plot_data}{here}.}
\label{f:rho_bh_duras_kbol}
\end{figure}

However, we do find significantly higher values of the correlation coefficient between average SMBH accretion rate and \mbh{} at fixed host halo mass, \rhobh{} (\S\ref{ss:rho_bh}), when adopting the ``DurasBC'' (Fig.\ \ref{f:rho_bh_duras_kbol}). This is because \textsc{Trinity} still has to reproduce similar numbers of quasars with $L_\mathrm{bol} \sim 10^{45}$ erg/s as in the ``UedaBC'' case, but with lower \mbh{}. If \rhobh{} stays as low as in the ``UedaBC'' case, \textsc{Trinity} will inevitably produce more(fewer) low-(high-)mass active black holes with Eddington ratios of $\eta > 0.01$. This would be inconsistent with the ABHMFs from \citet{Schulze2010} and \citet{Schulze2015}.

Other than \rhobh{}, using the bolometric corrections from either \citet{Ueda2014} or \citet{Duras2020} does not make any qualitative differences in our main results.

\subsection{Compton-thick correction}
\label{aa:ctk_corr}

\begin{figure}
\includegraphics[width=0.48\textwidth]{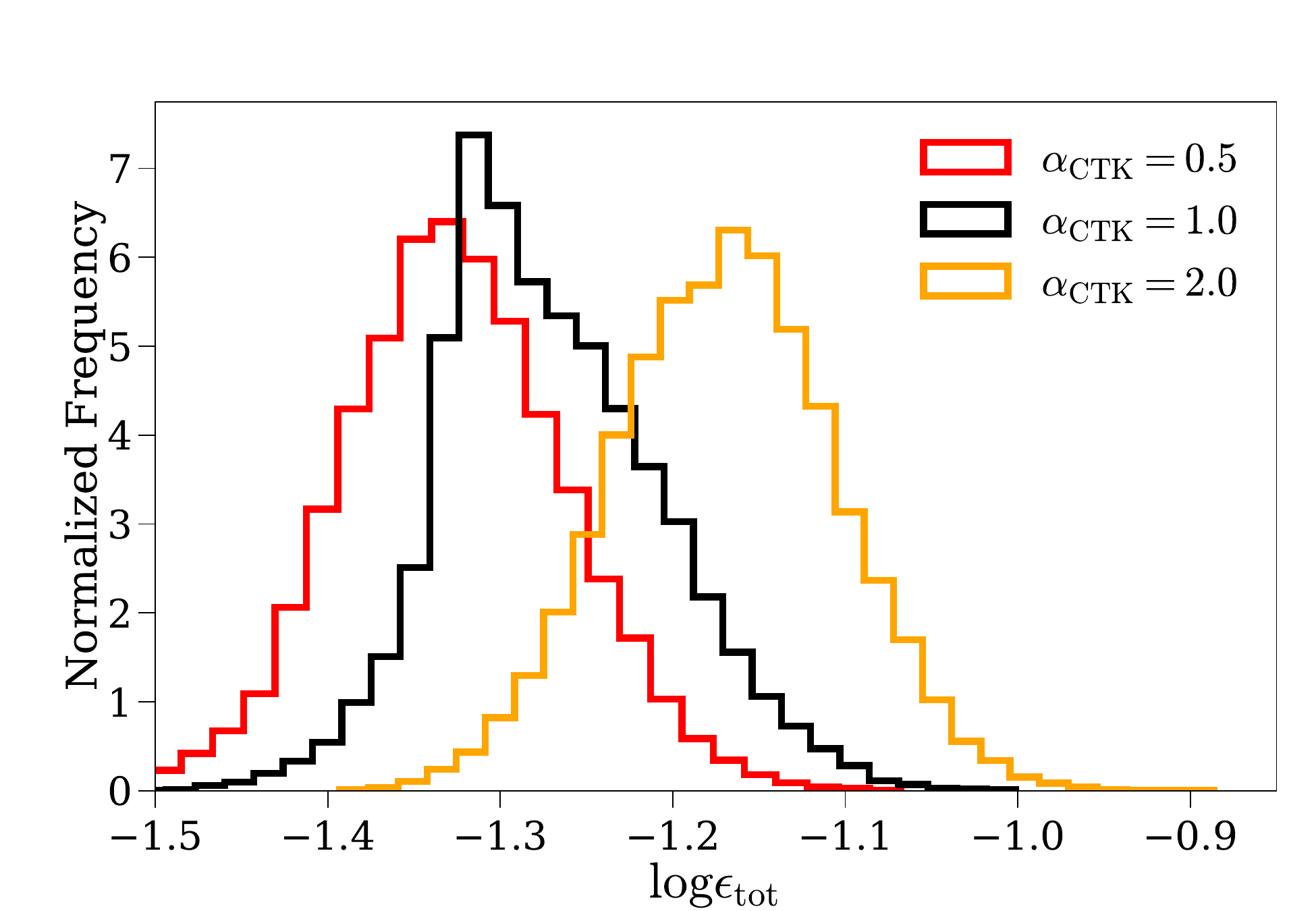}
\caption{The comparison of the posterior distributions of SMBH efficiency $\epsilon_\mathrm{tot}$ between models with $\alpha_\mathrm{CTK}=0.5, 1.0$, and 2.0. See Appendix \ref{aa:ctk_corr}. All the data used to make this plot can be found \href{https://github.com/HaowenZhang/TRINITY/tree/main/plot_data}{here}.}
\label{f:eff_alpha_ctk}
\end{figure}

As mentioned in \S \ref{ss:obs_data}, we have adopted quasar luminosity functions (QLFs) from \citet{Ueda2014} to constrain the total AGN energy budget. However, Ueda et al.\ did not include Compton-thick obscured AGNs in their QLF plots. Hence, we applied the following empirical correction given by \citet{Ueda2014} to convert from Compton-thin-only QLFs to total QLFs:

\begin{equation}
\begin{aligned}
    \Phi_{L,\mathrm{tot}}\left(L_X, z\right) &= \Phi_{L,\mathrm{CTN}}\left(L_X, z\right)\times\left(1 + \alpha_\mathrm{CTK}\psi\left(L_X, z\right)\right)\\
    \psi\left(L_X, z\right) &= \min{\left[0.84,\ \max{\left[\psi_{43.75}\left(z\right) - 0.24L_{43.75},\ \psi_{\mathrm{min}}\right]}\right]}\\
    \psi_{43.75}\left(z\right)&=\left\{
    \begin{aligned}
    &0.43\left(1+z\right)^{0.48}\ \left[z < 2.0\right]\\
    &0.43\left(1+2\right)^{0.48}\ \left[z \geq 2.0\right]
    \end{aligned}
    \right.\\
    L_{43.75} &= \log_{10} \left(L_X/\mathrm{erg\ s}^{-1}\right) - 43.75\ ,
\end{aligned}
\end{equation}
where $\psi\left(L_X, z\right)$ is the fraction of Compton-thin absorbed AGN, and $\alpha_\mathrm{CTK}$ is the number ratio between Compton-thick and Compton-thin AGN. Ueda et al. adopted $\alpha_\mathrm{CTK}=1$ in their main analysis, but their analysis of the cosmic X-ray background radiation shows that there is a $\pm 50\%$ uncertainty in $\alpha_\mathrm{CTK}$. In light of this, we ran \textsc{Trinity} with $\alpha_\mathrm{CTK}=0.5$ and 2.0, aside from the fiducial model where $\alpha_\mathrm{CTK}=1.0$. The \emph{only} model parameter that shows significant differences is the SMBH total efficiency ($\epsilon_\mathrm{tot}$, Fig.\ \ref{f:eff_alpha_ctk}). A higher \alphactk{} implies a larger Compton-thick AGN population, and thus higher QLFs at all redshifts. Consequently, \textsc{Trinity} needs a higher AGN efficiency to account for the larger AGN number densities.

\begin{figure}
\includegraphics[width=0.48\textwidth]{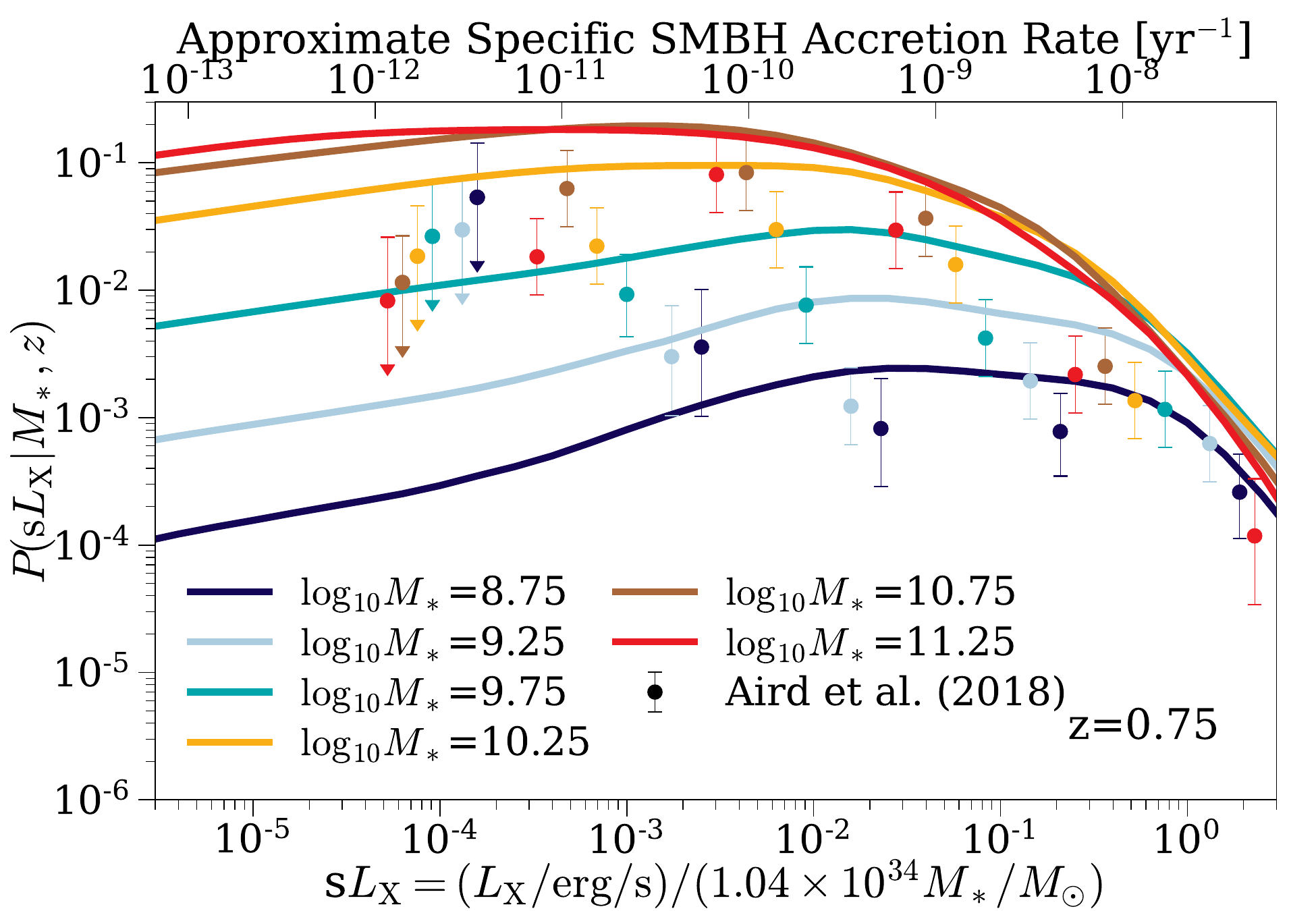}
\caption{The comparison between the observed quasar probability distribution functions (QPDFs) from \citet{Aird2018} and the best-fitting model with quasar luminosity functions and Compton-thick obscuration corrections from \citet{Ananna2019}, at $z=0.75$. All the data used to make this plot (including individual data points and our best-fitting model) can be found \href{https://github.com/HaowenZhang/TRINITY/tree/main/plot_data}{here}.}
\label{f:ananna_ctk}
\end{figure}

Since \citet{Ueda2014}, several studies updated the absorption functions, i.e., the probability distribution of gas column density as a function of X-ray luminosity and redshift, and found much higher Compton-thick obscured fractions, especially for bright AGNs \citep{Buchner2015,Ananna2019}. According to \citet{Ananna2019}, $\gtrsim 80\%$ of the AGNs with $L_\mathrm{X,2-10\ KeV} \gtrsim 10^{45}$ are Compton-thick obscured. This is significantly higher than $\sim 20\%$ as suggested by \citet{Ueda2014}. To explore the potential impact of different Compton-thick corrections on \textsc{Trinity} results, we ran a model with quasar luminosity functions and Compton-thick obscuration corrections from \citet{Ananna2019}. In this experiment, we found significant inconsistency between \citet{Ananna2019} results and other AGN data. Specifically, the high Compton-thick fractions at the bright end 
produces too many bright quasars. In this case, \textsc{Trinity} is unable to reproduce the bright end of the luminosity function with only SMBHs in massive galaxies, given their small number densities. Consequently, \textsc{Trinity} is forced to make SMBHs over-massive in lower-mass galaxies to reproduce the luminosity functions. This ultimately leads to inconsistency with the quasar probability distribution functions for low-mass galaxies from \citet{Aird2018} (see Fig.\ \ref{f:ananna_ctk}). The best fitting model with \citet{Ananna2019} luminosity functions and Compton-thick corrections give a $\chi^2\approx 844.62$, which is significantly worse compared to the fiducial model with data and corrections from \citet{Ueda2014} ($\chi^2\approx 746.70$). We note that such a strong inconsistency is present even when the systematic offset in Eddington ratio, $\xi$, is allowed to vary in the MCMC (see \S\ref{ss:agn_observables}). Given this inconsistency with other AGN data, we choose to keep using the quasar luminosity functions and Compton-thick corrections from \citet{Ueda2014} in the main text. From this experiment, we have shown that \textsc{Trinity} does have the ability to place upper limits on Compton-thick AGN fractions based on inter-dataset consistency. Further discussion into this topic is beyond the scope of this paper, and is thus deferred to a future investigation.

\subsection{Obscured fraction}
\label{aa:f_obs}
In the fiducial \textsc{Trinity} model, we adopted the correction for obscured AGN from \citet{Merloni2014} for ABHMFs. We did not adopt the Compton-thin obscured fraction from \citet{Ueda2014} due to the reported inconsistency between the optical type-I vs.\ type-II and X-ray obscured vs.\ unobscured AGNs \citep[][]{Merloni2014}. Here, we show the quantitative changes in the best-fitting model if the Compton-thin obscured fraction from \citet{Ueda2014} (i.e., $\psi\left(L_X, z\right)$ in Appendix \ref{aa:ctk_corr}) is also applied to ABHMFs.

\begin{figure}
\includegraphics[width=0.48\textwidth]{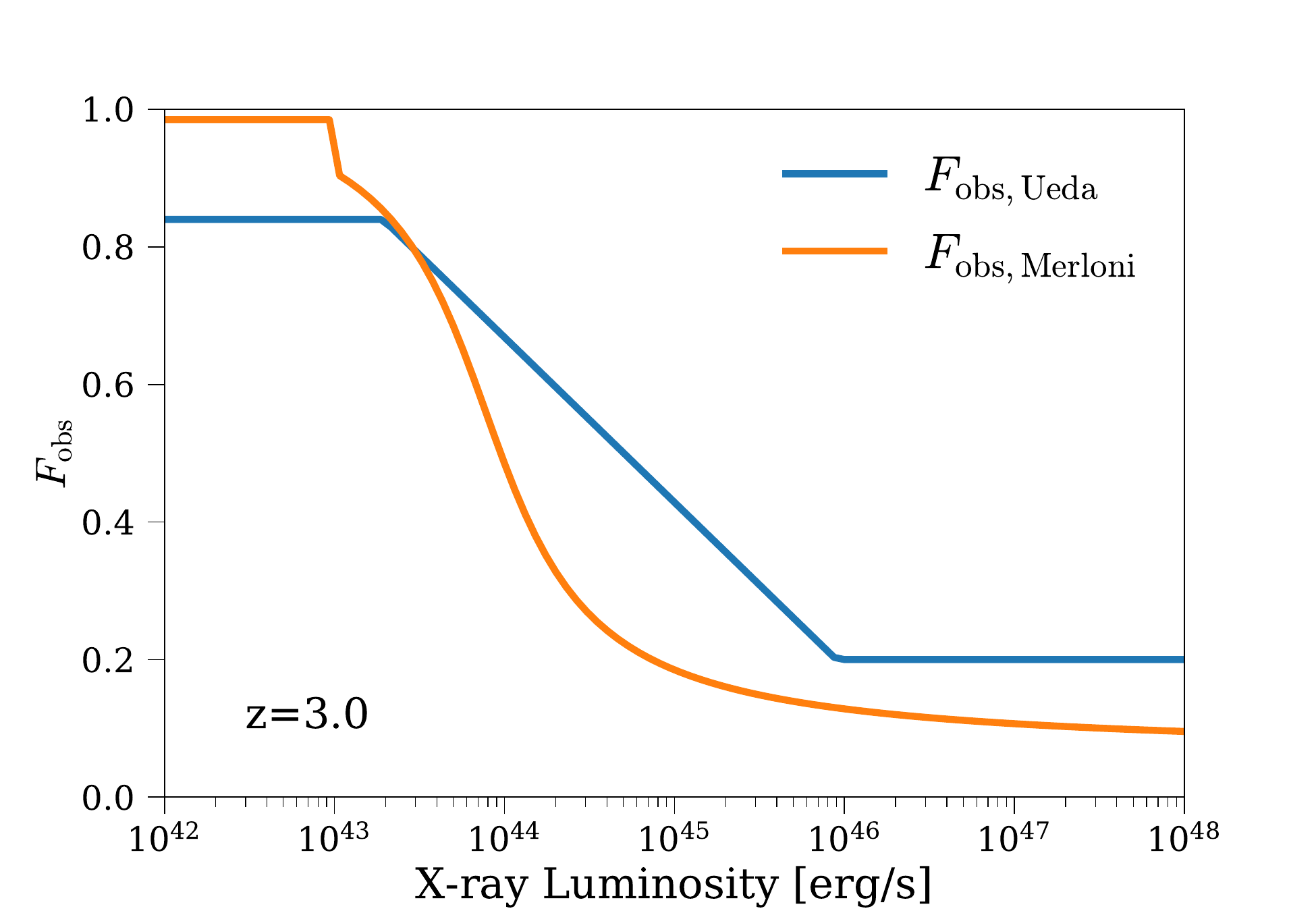}
\caption{The obscured fractions of AGNs as functions of X-ray luminosity from \citet{Ueda2014} (blue solid line) and \citet{Merloni2014} (orange solid line). To save space, we only show the fractions at $z=3$, since there are no qualitative differences across the relevant redshift range. All the data used to make this plot can be found \href{https://github.com/HaowenZhang/TRINITY/tree/main/plot_data}{here}.}
\label{f:f_obs_ueda_kelly}
\end{figure}

\begin{figure}
\subfigure{\includegraphics[width=0.48\textwidth]{figs/submit_fiducial/BHSM_median_submit_fiducial.pdf}}
\subfigure{\includegraphics[width=0.48\textwidth]{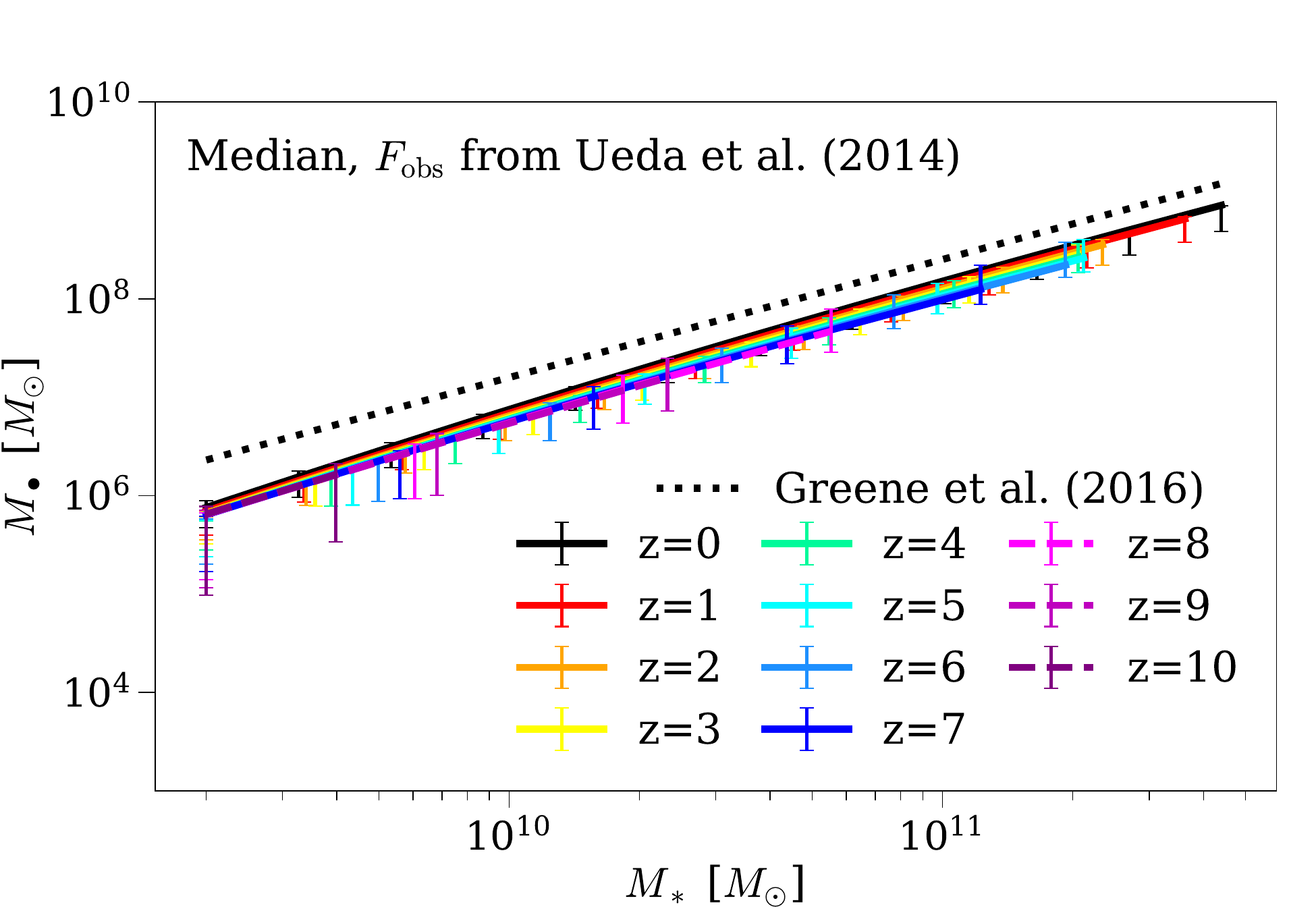}}
\caption{The median \bhsm{} relations from $z=0-10$ from the fiducial model (top panel) and the model where  obscured fractions of AGNs as functions of X-ray luminosity from \citet{Ueda2014} are applied to the ABHMFs (bottom panel). \errorbars{} All the data used to make this plot can be found \href{https://github.com/HaowenZhang/TRINITY/tree/main/plot_data}{here}.}
\label{f:bhbm_f_obs_ueda_kelly}
\end{figure}

Fig.\ \ref{f:f_obs_ueda_kelly} shows the difference in the obscured fraction, $F_\mathrm{obs}$, as a function of X-ray luminosity. We only show the comparison at $z=3$ as an example, and there is no qualitative difference at any other relevant redshift. The obscured fraction from \citet{Ueda2014} is higher than that from \citet{Merloni2014} at any fixed X-ray luminosity above $L_\mathrm{X}\sim 3\times 10^{43}$ erg/s. This leaves fewer unobscured AGN in the type I AGN mass function. To compensate for this deficit, \textsc{Trinity} needs to increase the radiative efficiency from $\sim 5\%$ to $\sim 10\%$ to make more bright AGNs. However, only increasing efficiency will also increase the normalization of QLFs and QPDFs. Thus, \textsc{Trinity} has to simultaneously adjust the redshift evolution of the \bhbm{} relation, as shown in Fig.\ \ref{f:bhbm_f_obs_ueda_kelly}. Compared to the fiducial model, we no longer see significant evolution in the slope of the \bhbm{} relation, whereas its normalization decreases slightly towards higher redshifts. These changes lead to less(more) growth of low-(high-)mass SMBHs, and thus, less(more) contribution to QLFs and QPDFs from low-(high-)mass SMBHs. The ultimate net result is that QLFs and QPDFs are still reproduced while the ABHMFs are corrected by a larger $F_\mathrm{obs}$.

\subsection{AGN probability distribution functions from \citet{Aird2018}}
\label{aa:aird_qpdf}

To use QPDFs from \citet{Aird2018} to constrain our model, we had to account for two factors as below.

Firstly, \citet{Aird2018} modeled the AGN probability distribution functions for each stellar mass and redshift bin as a finite series of gamma distributions. The function values in their public release\footnote{available at \url{https://zenodo.org/record/1009605}.} were evaluated with these model functions over a dense grid of \slx{}. Thus, naively taking all the points in their data release would artificially increase the weight of this dataset. To avoid this, we downsampled their modeled AGN probability distribution functions with 1 dex spacing. This choice is based on the fact that the spacing between two neighboring gamma distributions is 0.2 dex, and that an extra prior was applied to ensure smoothness of the probability distribution functions across neighbouring gamma distributions.

Secondly, in the process of compiling different datasets, we found that there is significant inconsistency between the QLFs from \citet{Ueda2014} and the high-\slx{} and high-z (i.e., $z>2.5$) end of AGN probability distribution functions from \citet{Aird2018}. This may be due to the massive end of the AGN probability distribution functions being affected by the smoothness prior. To ensure consistency between these two datasets, we excluded AGN probability distribution function points with $z > 2.5$ or \slx{}$>1$ from \citet{Aird2018}. After removing the most inconsistent data points, residual inconsistencies on the order of 0.3 dex persist between these two datasets. To address this, we further enlarged the uncertainties in the AGN probability distribution functions to 0.3 dex, and included an extra free parameter $\xi$ to describe the systematic offset in the Eddington ratio in the calculation of probability distribution functions in terms of \slx{} (see Eq.\ \ref{e:xi} in \S \ref{ss:agn_observables}).

\subsection{Active black hole functions}
\label{aa:abhmf}

\subsubsection{Active black hole functions from \citet{Schulze2010} and \citet{Schulze2015}}
\label{aaa:abhmf_schulze}

In \textsc{Trinity}, we use active black hole mass functions (ABHMFs) at $z=0.2$ and $z=1.5$ from \citet{Schulze2010} and \citet{Schulze2015}. However, two issues were addressed before using these ABHMFs as constraints. Firstly, as is shown in Fig.\ 22 of \citet{Schulze2015}, the massive end of the ABHMF varies with different model assumptions due to the different significance of Eddington bias. To avoid this model dependence, we chose to only use the data points in the region where the ABHMF estimate is independent of their model assumptions, i.e., $\log_{10}M_\bullet \lesssim 9.8$. Secondly, \citet{Schulze2010} used virial BH mass estimates that are on average smaller by 0.2 dex than those used in \citet{Schulze2015}. To account for this, we applied a mass shift of $+0.2$ dex for all the ABHMF data points at $z=0.2$ to keep consistency with those at $z=1.5$.

\subsubsection{Systematic Uncertainties in ABHMFs}

Despite the corrections and exclusions for ABHMFs from \citet{Schulze2010} and \citet{Schulze2015}, significant systematic differences remain among ABHMFs from different studies. For example, \citet{Ananna2022} obtained much higher $z\sim 0.2$ ABHMFs compared to \citet{Schulze2010}. The potential causes for such differences include the different wavebands and bolometric corrections that were used (X-ray vs.\ optical), different ways of correcting for obscured AGN, etc.. We note that ABHMFs do provide important constraints on SMBH masses in \textsc{Trinity}. Without any ABHMF data, \textsc{Trinity} would yield a \bhbm{} normalization with $\beta_\mathrm{BH,0}=8.47$, and a too low AGN energy efficiency of $\epsilon_\mathrm{tot}\sim 3\%$. This is because the prior constraint on the local \bhbm{} relation is not stringent enough as the sole constraint on SMBH masses, given the large inter-publication scatter (see Table \ref{t:bhbm_data}). Therefore, we decided to keep ABHMF data in our data constraints.

\begin{figure}
\includegraphics[width=\columnwidth]{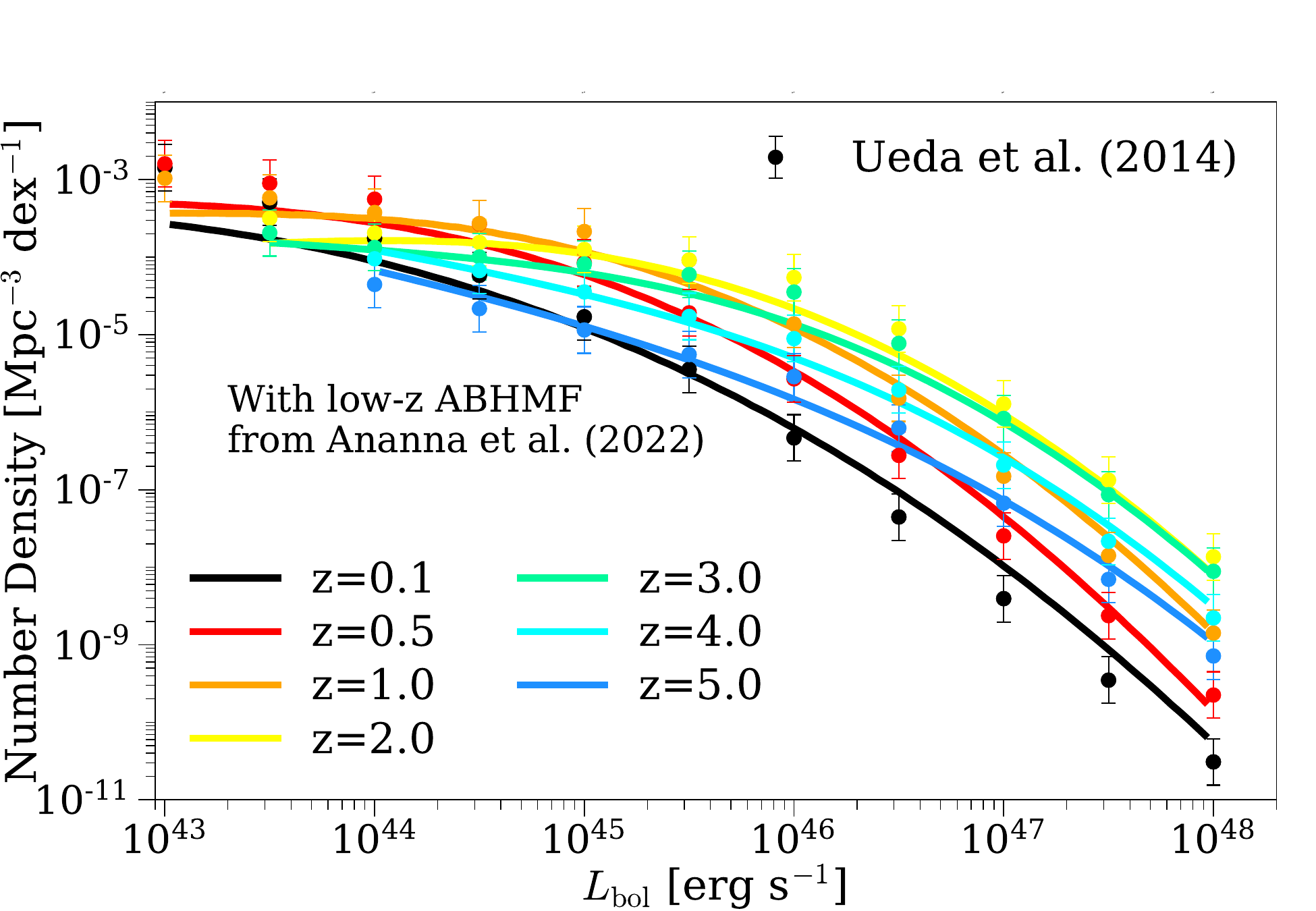}
\caption{Comparison between the observed quasar luminosity functions (QLFs) from \citet{Ueda2014} and our model prediction from $z=0-5$, with $z\sim 0.2$ active black hole mass functions (ABHMFs) from \citet{Schulze2010} replaced by the \citet{Ananna2022} results. Higher-redshift ABHMFs are the same as the fiducial model. All the data used to make this plot (including individual data points and our best-fitting model) can be found \href{https://github.com/HaowenZhang/TRINITY/tree/main/plot_data}{here}.}
\label{f:qlf_ananna_abhmf}
\end{figure}

\begin{figure}
\includegraphics[width=0.48\textwidth]{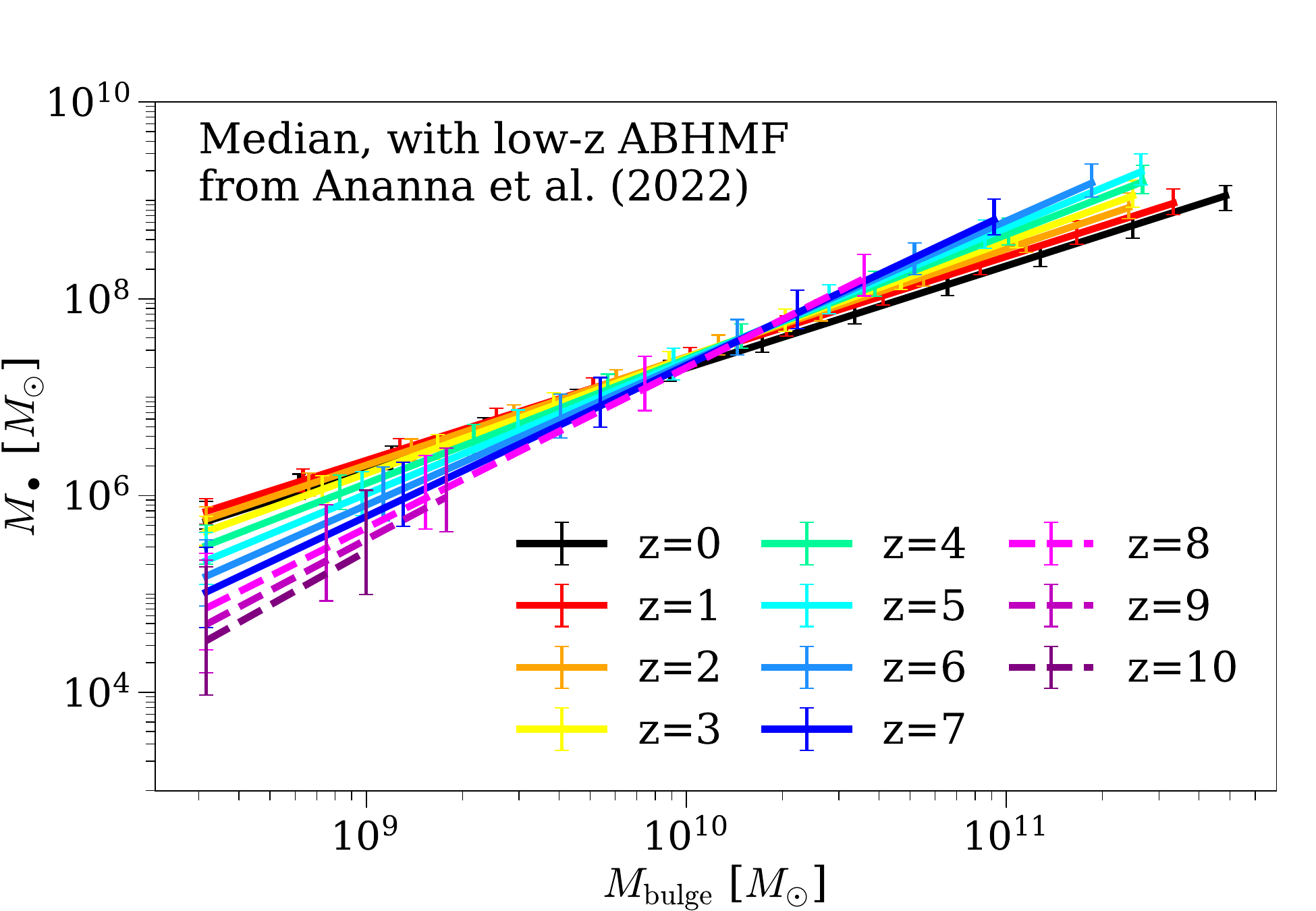}
\caption{The evolution of the median \mbh{}--\mbulge{} relation, with $z\sim 0.2$ active black hole mass functions (ABHMFs) from \citet{Schulze2010} replaced by the \citet{Ananna2022} results. Higher-redshift ABHMFs are the same as the fiducial model. \waitforjwst{} All the data used to make this plot can be found \href{https://github.com/HaowenZhang/TRINITY/tree/main/plot_data}{here}.}
\label{f:bhbm_ananna_abhmf}
\end{figure}

To show the potential effects of adopting different ABHMF measurements, we did an experiment with the fiducial \textsc{Trinity} model, replacing the low-redshift ABHMF from \citet{Schulze2010} with the one from \citet{Ananna2022}. As shown in Fig.\ \ref{f:bhbm_ananna_abhmf}, the resulting redshift evolution of the \bhbm{} relations is still consistent with the fiducial \textsc{Trinity} model, although the difference is more significant at $z=8-10$, where we do not have any AGN data. On the other hand, \textsc{Trinity} needs to produce many more active SMBHs to match much higher number densities as required by \citet{Ananna2022}. Consequently, a higher AGN efficiency of $\epsilon_\mathrm{tot}\sim6.3\%$ is adopted. Such a combination of \bhbm{} relations and AGN efficiency naturally produces higher QLFs at $z\lesssim 2$ compared to the fiducial \textsc{Trinity} results, as shown in Fig.\ \ref{f:qlf_ananna_abhmf}, but the difference is well within the QLF uncertainties. Finally, a higher correlation coefficient between average SMBH accretion rate and \mbh{} at fixed halo mass, \rhobh{}, is also needed to match the higher ABHMF at the massive end. Other than these quantitative changes, all the qualitative results remain invariant.

\section{Alternate Model Parametrizations}

\subsection{Eddington-limited SMBH growth}
\label{a:no_super_eddington}

\begin{figure}
\subfigure{
\includegraphics[width=0.48\textwidth]{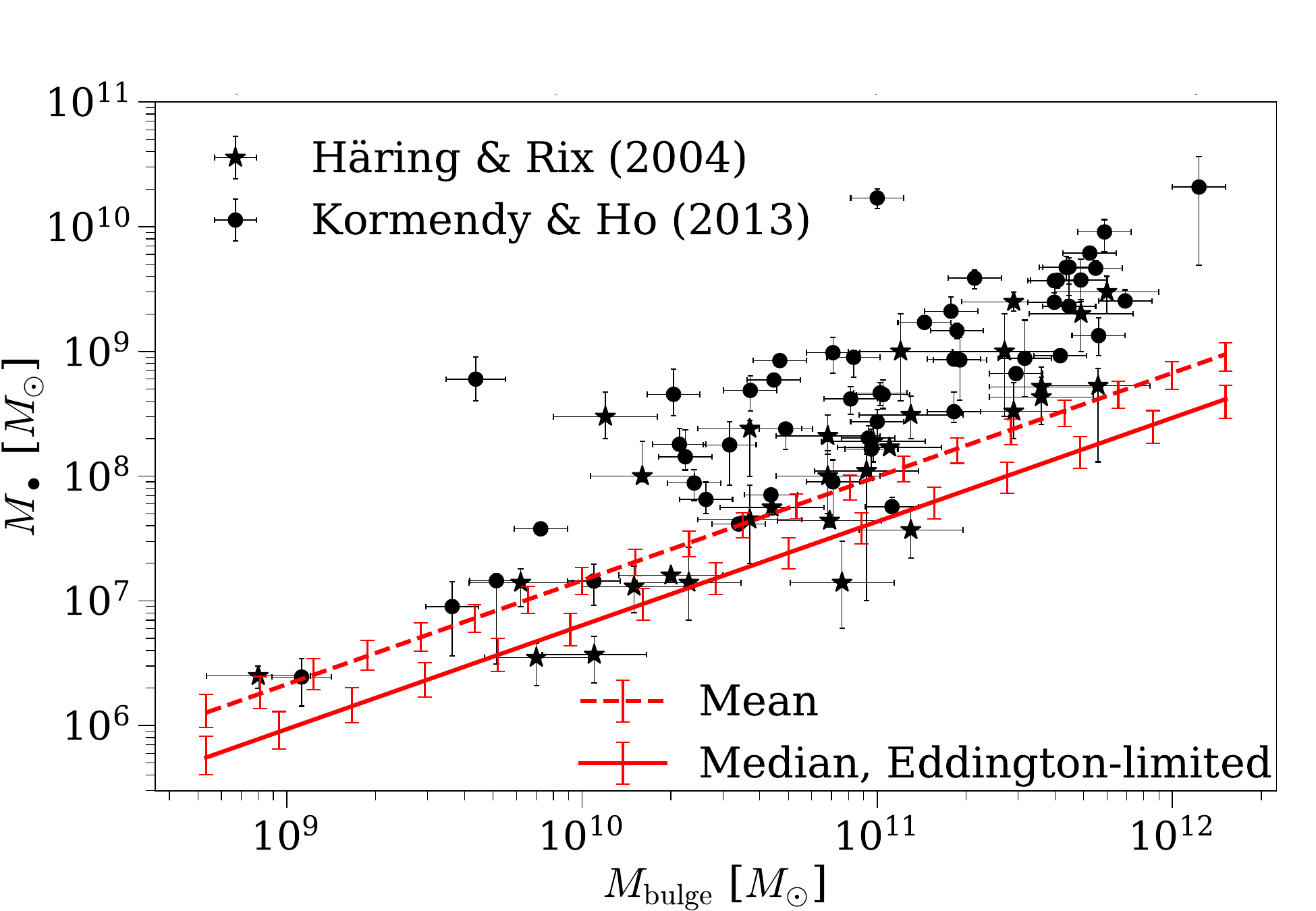}
}
\subfigure{
\includegraphics[width=0.48\textwidth]{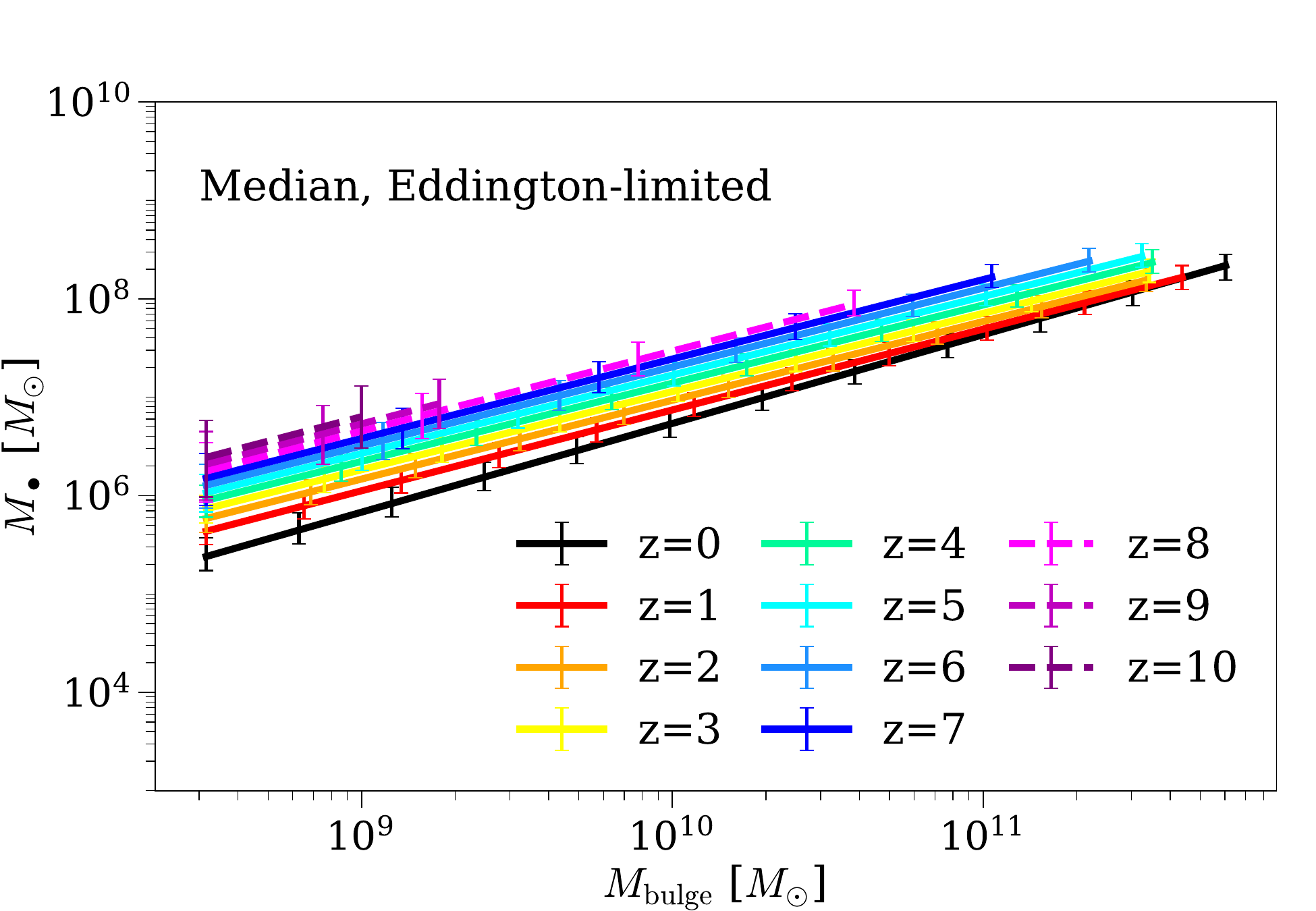}
}
\caption{\textbf{Top Panel:} The comparison between the $z=0$ \bhbm{} relation from the ``Eddington-limited'' model and real data. \textbf{Bottom Panel:} The redshift evolution of the \bhbm{} relation from the ``Eddington-limited'' model, where SMBH accretion is Eddington-limited. See Appendix \ref{a:no_super_eddington}. All the data used to make this plot (including the individual data points and our best-fitting model) can be found \href{https://github.com/HaowenZhang/TRINITY/tree/main/plot_data}{here}.}
\label{f:bhbm_no_super_eddington}
\end{figure}

In the fiducial model, we do not set any upper limit on the specific SMBH accretion rate. We also tested an alternate model where SMBHs cannot accrete at super-Eddington rates (hereafter called the ``Eddington-limited model''). Fig.\ \ref{f:bhbm_no_super_eddington} shows the comparison between the local \bhbm{} relation with observations (top panel), and its redshift evolution (bottom panel). Given the limit in Eddington ratios, SMBHs cannot grow as fast as in the fiducial model. This results in a local \bhbm{} relation that lies significantly below the observed values, and an increase in the normalization with increasing redshift. With limited accretion rates, \textsc{Trinity} is also forced to recruit much higher AGN energy efficiencies--as high as 24\%--to get as many close-to-Eddington objects and reproduce the observations expressed in luminosities. Given the inconsistency with the observations, we do not adopt this model in the main text.

\subsection{Alternative galaxy--SMBH connections}
\label{a:alt_galaxy_smbh}

\begin{figure*}
\subfigure{
\includegraphics[width=0.48\textwidth]{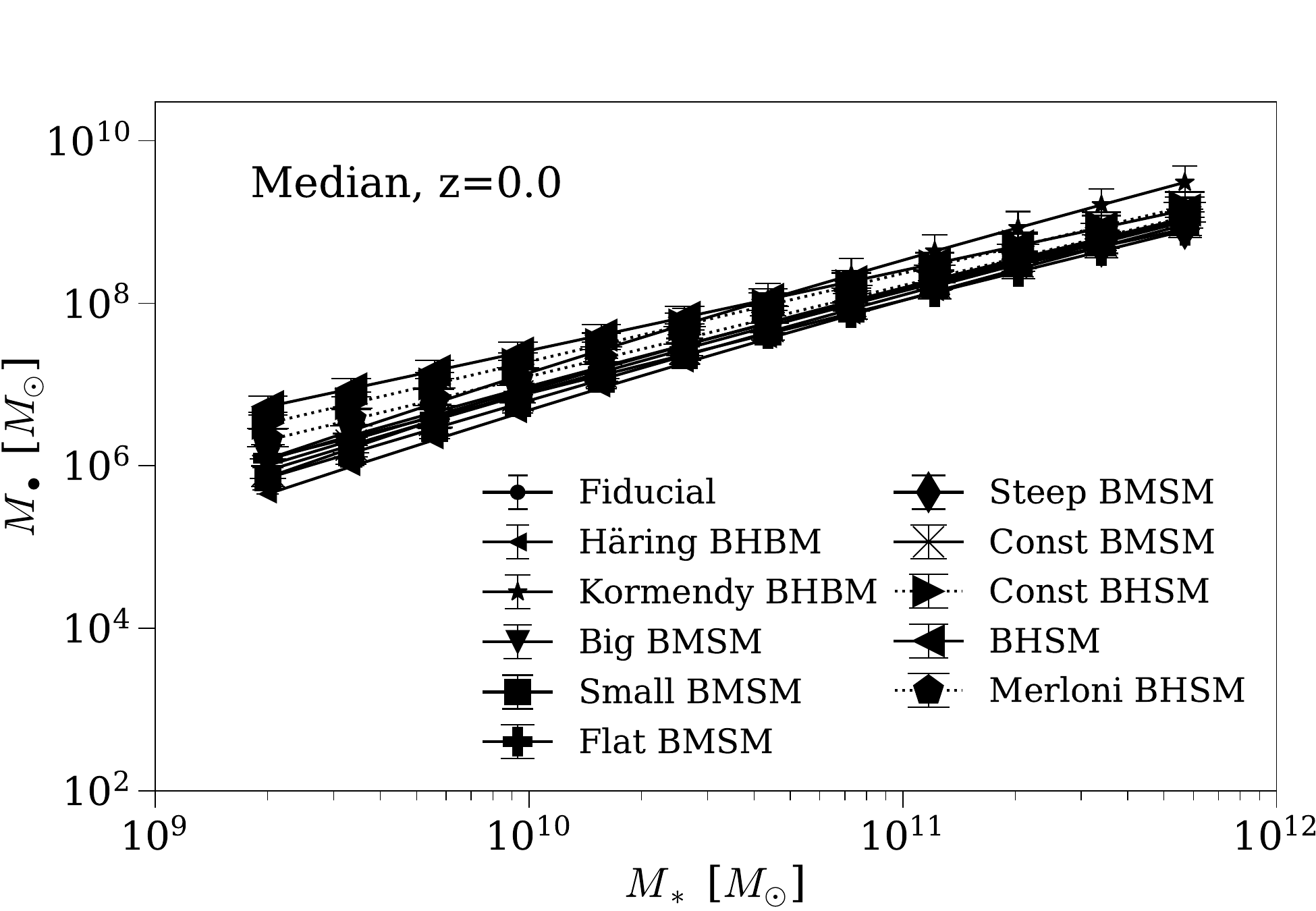}
}
\subfigure{
\includegraphics[width=0.48\textwidth]{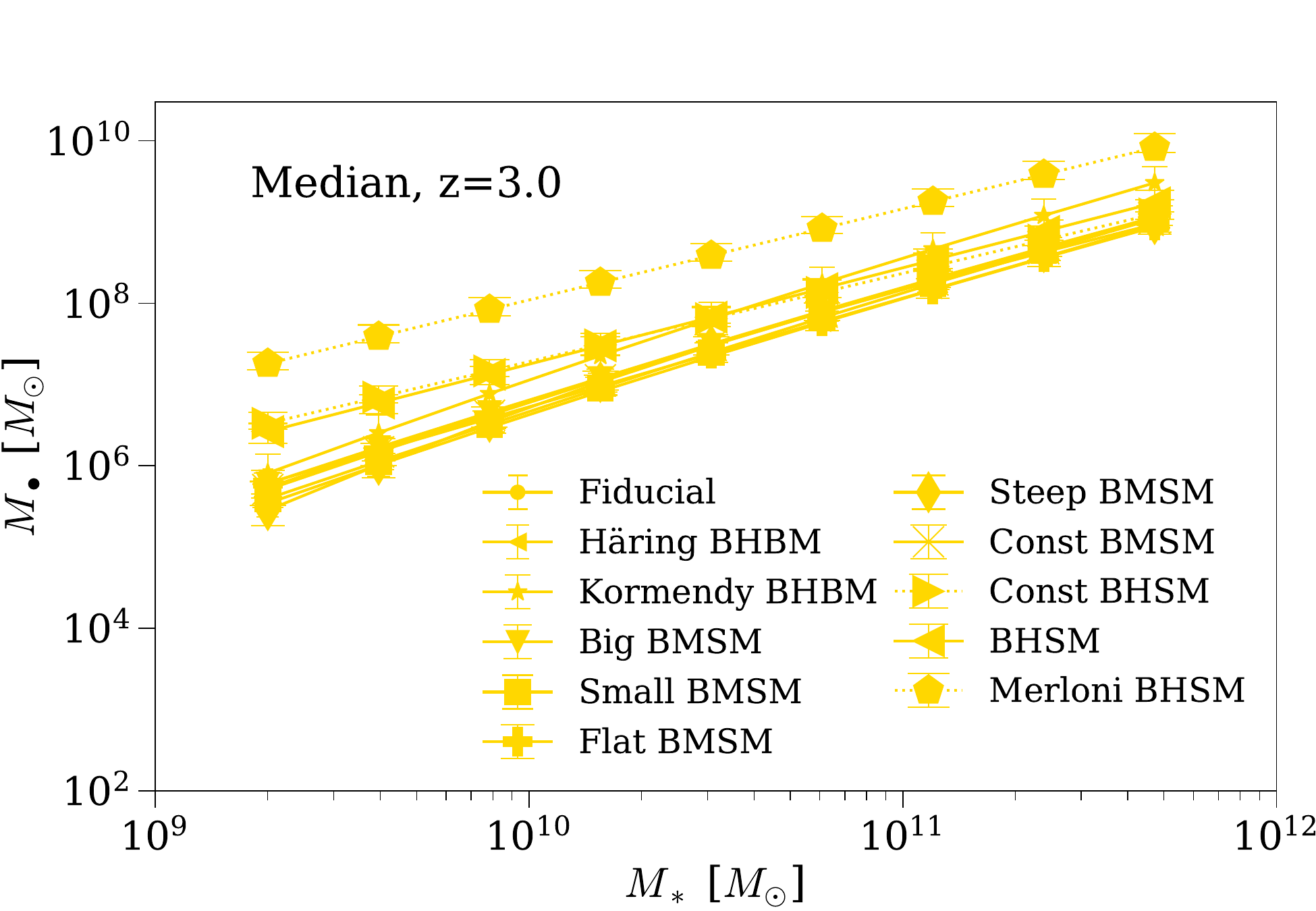}
}
\subfigure{
\includegraphics[width=0.48\textwidth]{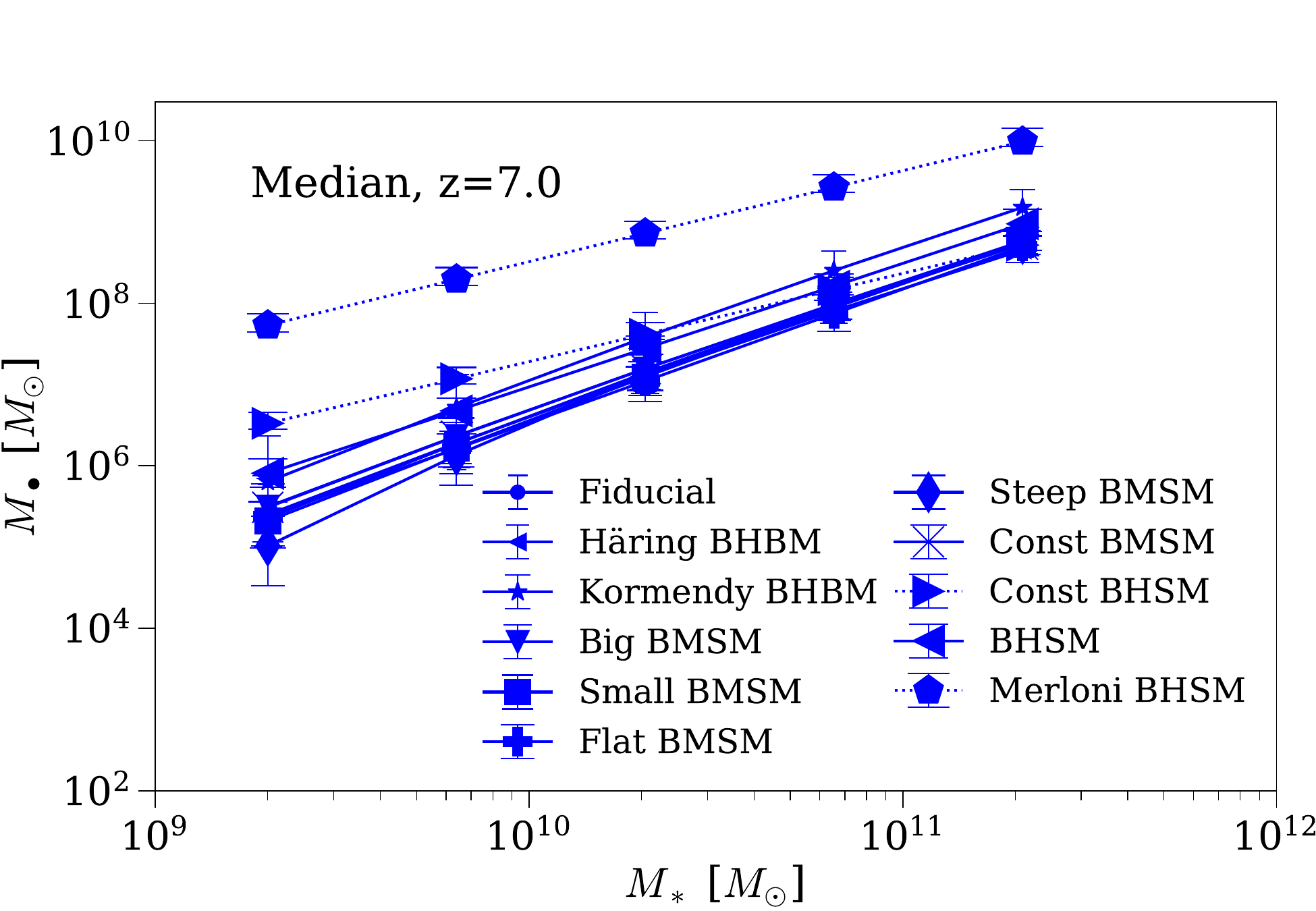}
}
\subfigure{
\includegraphics[width=0.48\textwidth]{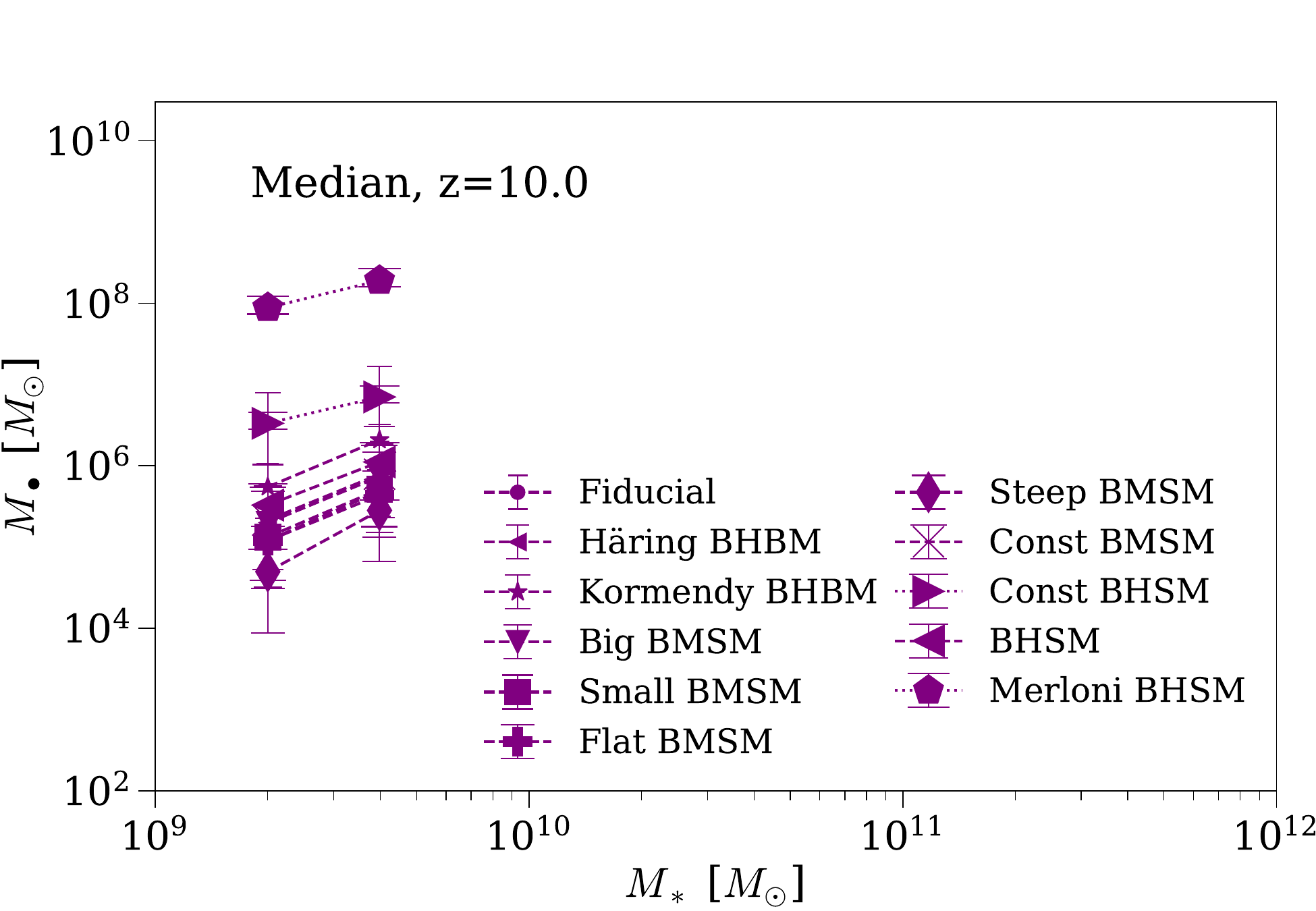}
}

\caption{The median \bhsm{} relations from different variant models at $z=0,3,7,$ and $10$. See Appendix \ref{a:alt_galaxy_smbh}. The ``Const BHSM'' and ``Merloni BHSM'' models (dotted lines) have pre-determined redshift evolution in the median \mbh{}-\mstar{} relation, and are thus shown only for completeness. For all the other redshifts, see \href{https://github.com/HaowenZhang/TRINITY/tree/main/plot_data}{here}.}
\label{f:bhsm_all_models}

\end{figure*}

In the fiducial \textsc{Trinity} model, we make the galaxy--SMBH connection with redshift dependent \mbulge{}--\mstar{} and \mbh{}--\mbulge{} relations. Given the observational uncertainties in these scaling relations, it is necessary to verify the robustness of our main results against these uncertainties. Therefore, we have run \textsc{Trinity} with the following alternative assumptions: a) the \mbulge{}--\mstar{} relation is redshift-independent and set to the observed one at $z=0$ (``Const BMSM''); b) the normalization of the \mbulge{}--\mstar{} relation is lower(higher) by setting $M_\mathrm{SB}=11.5(9.0)$ (see Eq. \ref{e:bm_sm}, ``Small BMSM'' and ``Big BMSM''); c) the \mbulge{}--\mstar{} relation is steeper(flatter) by setting $k_\mathrm{SB}=2.0(0.2)$ (also see Eq. \ref{e:bm_sm}, ``Steep BMSM'' and ``Flat BMSM''\footnote{These alternative $M_\mathrm{SB}$ and $k_\mathrm{SB}$ values are chosen to cover the full range of 1$\sigma$ uncertainties of the observed \mbulge{}--\mstar{} relation. See Fig.\ \ref{f:bmsm_fit}.}); d) The $z=0$ \mbulge{}--\mstar{} relation is fixed to the ones from either \citet{Haring2004} or \citet{Kormendy2013} (``H{\"a}ring BHBM'' and ``Kormendy BHBM''); e) The galaxy--SMBH connection is built by a redshift-dependent power-law \mbh{}--\mstar{} relation, i.e., replacing \mbulge{} with \mstar{} in Eq. \ref{e:bhbm} (``BHSM''); f) The galaxy--SMBH connection is built by a \emph{redshift-independent} power-law \mbh{}--\mstar{} relation (``Const BHSM''); g) The normalization of the \mbh{}--\mstar{} relation has a redshift evolution as given by \citet{Merloni2014}, and its slope is redshift-independent (``Merloni BHSM''). As shown in Fig.\ \ref{f:bhsm_all_models}, most of these alternative models yield mutually consistent \mbh{}--\mstar{} relations even before taking the inter-publication scatter of 0.2 dex (Table \ref{t:bhbm_data}) into account\footnote{The ``Const BHSM'' and ``Merloni BHSM'' models (dotted lines) have pre-determined redshift evolution, and thus are included only for completeness.} The only exceptions are the ``Kormendy BHBM'' and the ``BHSM'' models. The ``Kormendy BHBM'' model is consistent with the rest of the models when the inter-publication spread is included. We do note that the \bhbm{} relation from \citet{Kormendy2013} implies extremely massive black holes at fixed stellar mass. When constrained by galaxy stellar mass functions and QPDFs, \textsc{Trinity} would overproduce ABHMFs. In this sense, the \bhbm{} from \citet{Kormendy2013} is inconsistent with the galaxy data and ABHMFs in our data compilation. But to see the effect of an overall \mbh{} offset on \textsc{Trinity} results, we tried adding an offset in SMBH mass of $8.7 - 8.343 = 0.357$ dex (where 8.343 is the normalization of the local \bhbm{} relation given by the best-fitting fiducial model, also see Appendix \ref{a:param_values}) to all the ABHMF data points, which effectively assumes that the Kormendy \& Ho \bhbm{} relation had been used to calibrate SMBH masses in the ABHMFs. With this offset, the ``Kormendy BHBM'' model gives an AGN energy efficiency of $\epsilon_\mathrm{tot}\sim 3.5\%$. Such a smaller efficiency than that given by the fiducial model comes from more total SMBH mass with the same total AGN energy constraints from quasar luminosity functions. Except for the systematic offset in AGN efficiency and the normalization of SMBH growth histories, the main results in this work are not affected. However, this systematic change in the inferred AGN energy with the normalization of the inferred/assumed local \bhbm{} relation demonstrates 
that assuming a certain fixed SMBH mass normalization could induce inconsistency with other observational datasets. This further justifies our choice to use the \emph{distribution} of $z=0$ \bhbm{} relations among different studies as prior constraints. As is pointed out by \citet{Reines2015}, the stellar mass measurements in \citet{Kormendy2013} could be underestimated, leading to an overestimated \bhbm{} normalization by $\sim 0.33$ dex. The difference between \textsc{Trinity}'s best-fitting \bhbm{} normalization with the \citet{Kormendy2013} value, 0.357 dex, is also in line with this explanation. Given the potential inconsistency issue and bias in stellar mass measurements, we choose to present the results of the ``Kormendy BHBM'' model in this appendix, instead of the main text of this work.

As for the ``BHSM'' model, significantly higher values for \mbh{} appear below $M_\bullet \sim 10^{7} M_\odot$, compared to models that parametrize the \bhbm{} relation. This is due to the ``BHSM'' parametrization's inability to simultaneously reproduce the following with a single power-law: 1) AGN observations constraining the massive end; and 2) The steeper \bhsm{} slope at the low-mass end as in the \bhbm{} parametrizations. We also note that such inter-model differences are more pronounced at $z=8-10$, where no data exist. At these redshifts, our model results are pure extrapolations based on model assumptions and lower-redshift data. At $z=8-10$, the variance in \mbh{}-\mstar{} relations from different models highlights the importance of upcoming high-z observations in constraining early galaxy--SMBH connections.

\begin{figure*}
\subfigure{
\includegraphics[width=0.48\textwidth]{figs/submit_fiducial/BHAR_submit_fiducial.pdf}
}
\subfigure{
\includegraphics[width=0.48\textwidth]{figs/submit_fiducial/BHER_tot_submit_fiducial.pdf}

}
\subfigure{
\includegraphics[width=0.48\textwidth]{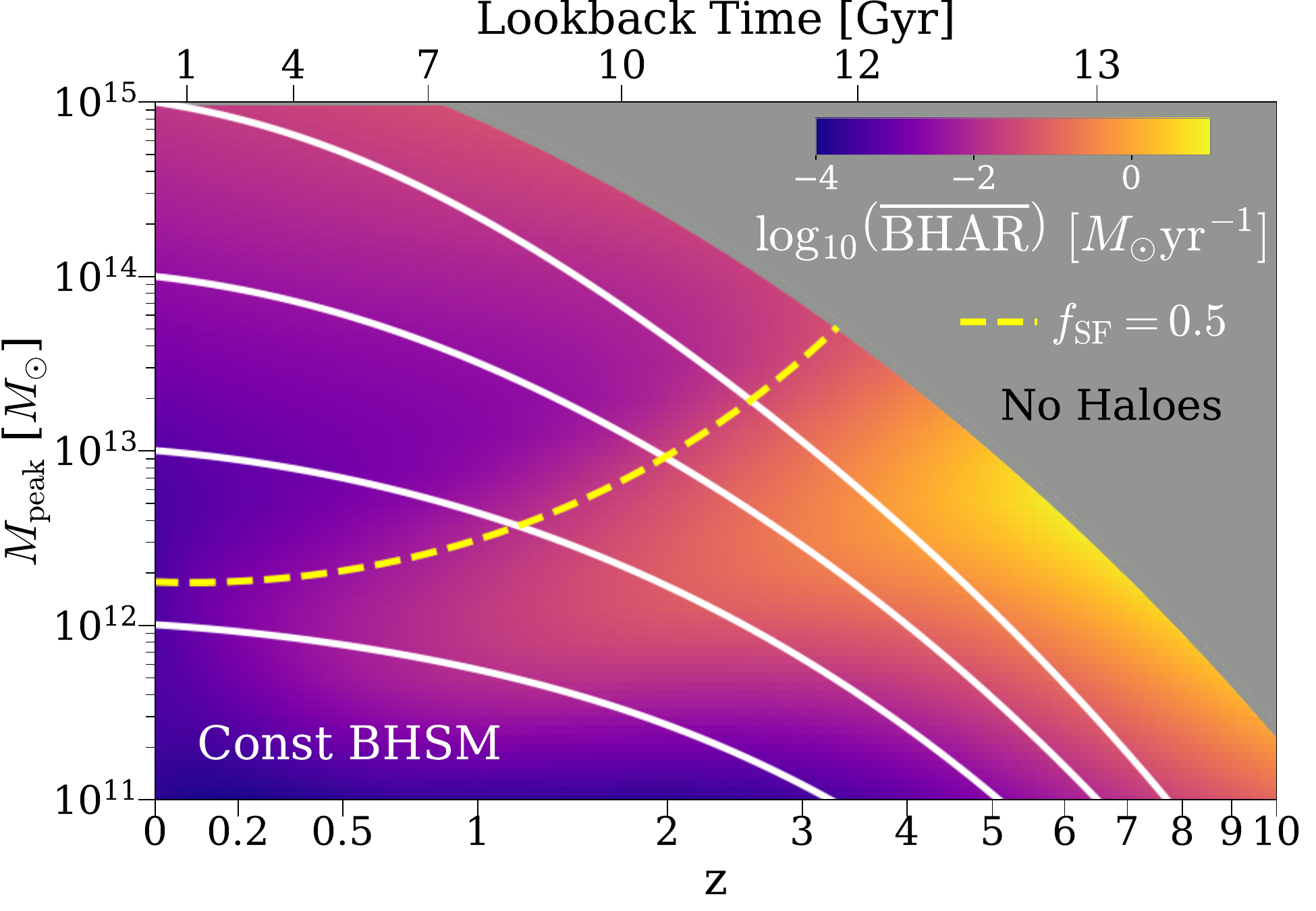}
}
\subfigure{
\includegraphics[width=0.48\textwidth]{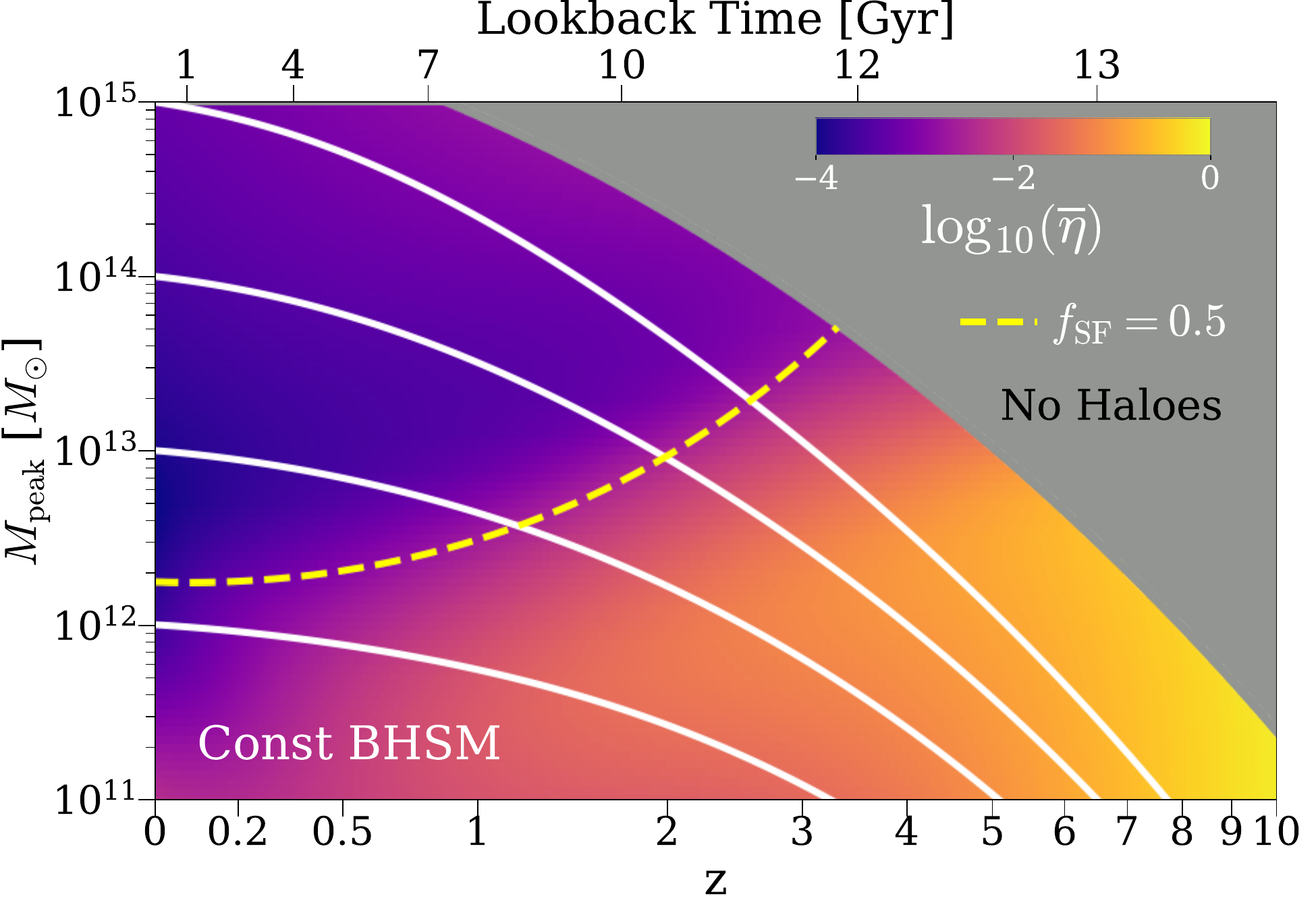}
}
\subfigure{
\includegraphics[width=0.48\textwidth]{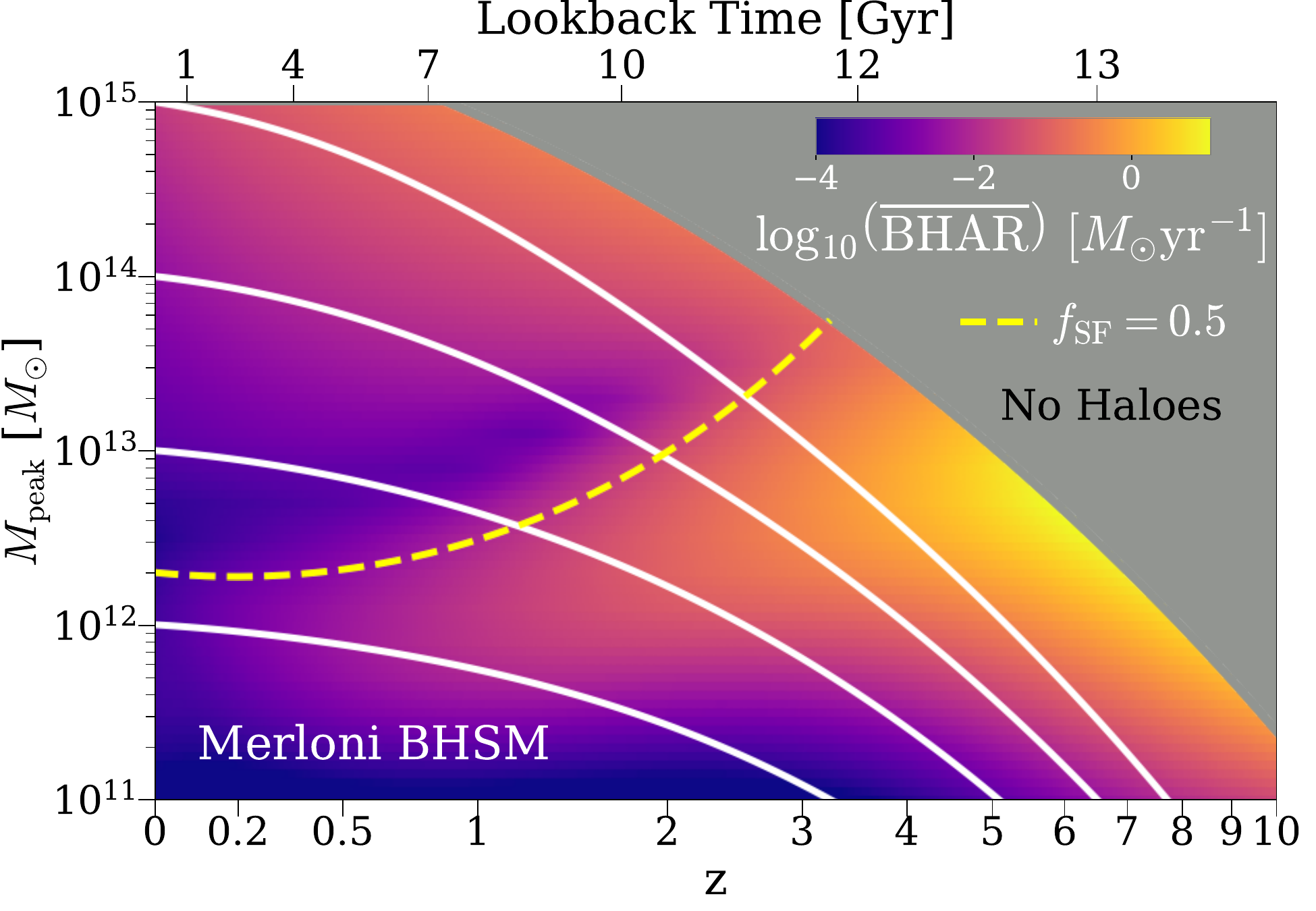}
}
\subfigure{
\includegraphics[width=0.48\textwidth]{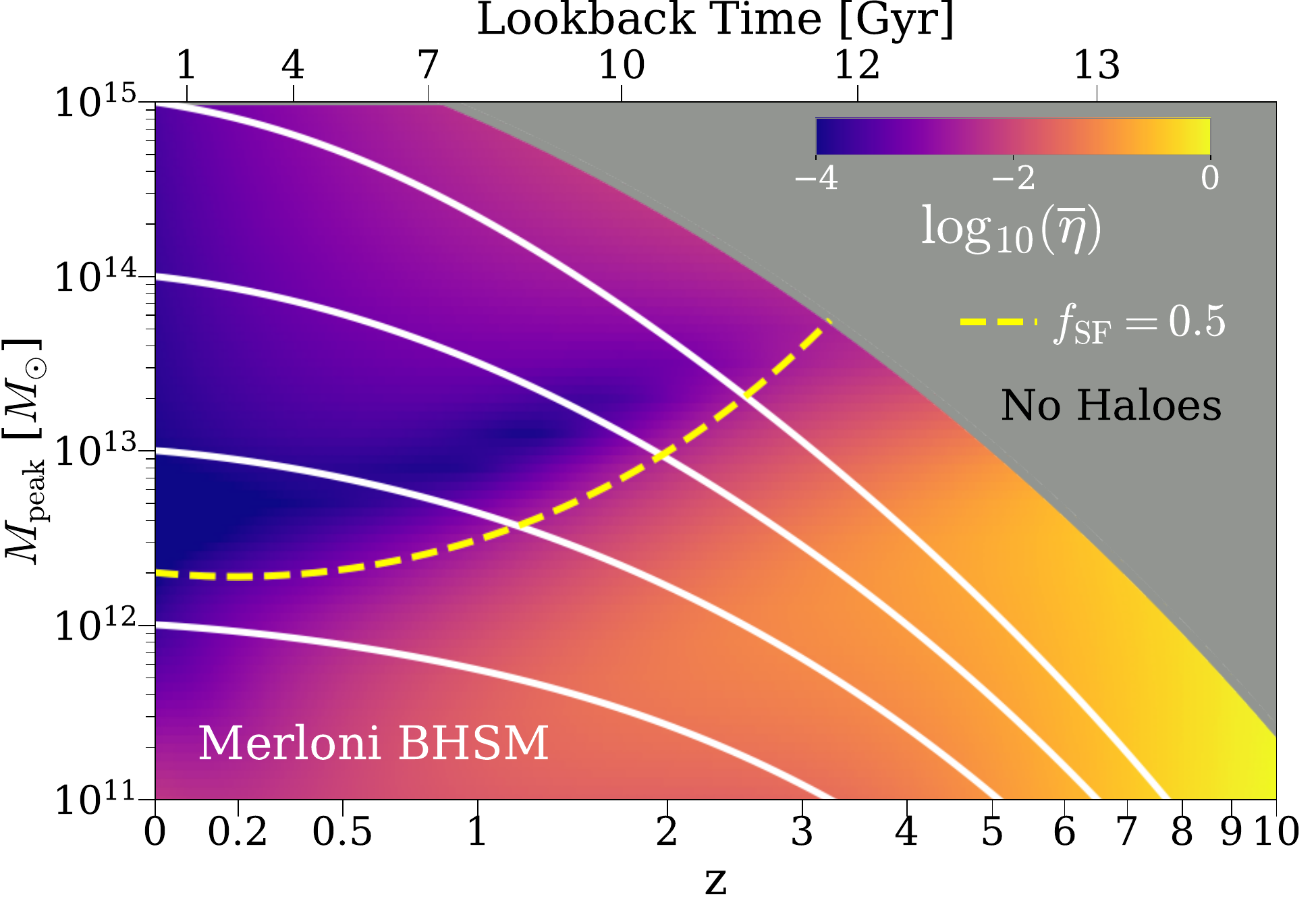}
}

\caption{The average SMBH accretion rates (left column) and average Eddington ratios (right column) as functions of \mpeak{} and $z$, from the fiducial (top panels), the ``Const BHSM'' (middle panels), and the ``Merloni BHSM'' models (bottom panels). See Appendix \ref{a:alt_galaxy_smbh}. All the data used to make this plot can be found \href{https://github.com/HaowenZhang/TRINITY/tree/main/plot_data}{here}.}
\label{f:bhar_bher_alt_bhsm}

\end{figure*}

Although the ``Const BHSM'' and the ``Merloni BHSM'' models have fixed (non-)evolution with redshift, it is still worth checking if they predict qualitatively consistent SMBH accretion rates with the fiducial \textsc{Trinity} model. As shown in Fig.\ \ref{f:bhar_bher_alt_bhsm}, the ``Const BHSM'' and the ``Merloni BHSM'' models both predict average SMBH accretion rates and Eddington ratios as functions of \mpeak{} and $z$. These predictions are qualitatively consistent with the fiducial \textsc{Trinity} model.

Based on these experiments, we therefore argue that our results are relatively independent of the way that the galaxy--SMBH mass connection is parametrized.

\subsubsection{Redshift-independent SMBH mass--bulge mass relations}
\label{a:const_bhbm}

\begin{figure}
\subfigure{
\includegraphics[width=0.48\textwidth]{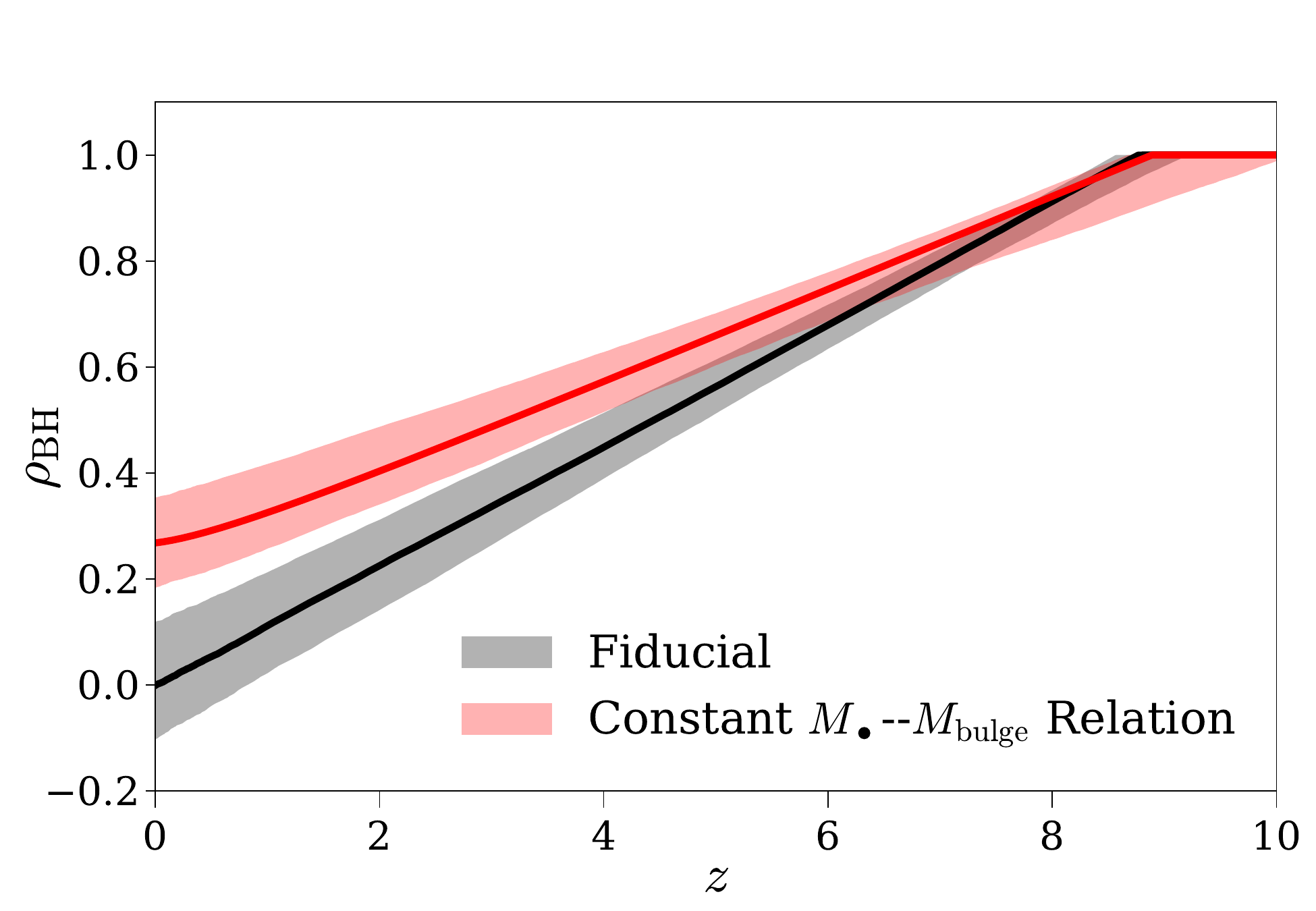}
}
\caption{The correlation coefficient, \rhobh{}, between average SMBH accretion rate and \mbh{} at fixed halo mass from the best-fitting model (black solid line) and the ``constant \bhbm{}'' model. See Appendix \ref{a:const_bhbm}. The shaded regions show the 68\% confidence intervals inferred from the model posterior distribution. The data used to make this plot can be found \href{https://github.com/HaowenZhang/TRINITY/tree/main/plot_data}{here}.}
\label{f:rho_bh_const_bhbm}
\end{figure}

\begin{figure*}
\vspace{-0.5cm}
\subfigure{
\includegraphics[width=0.48\textwidth]{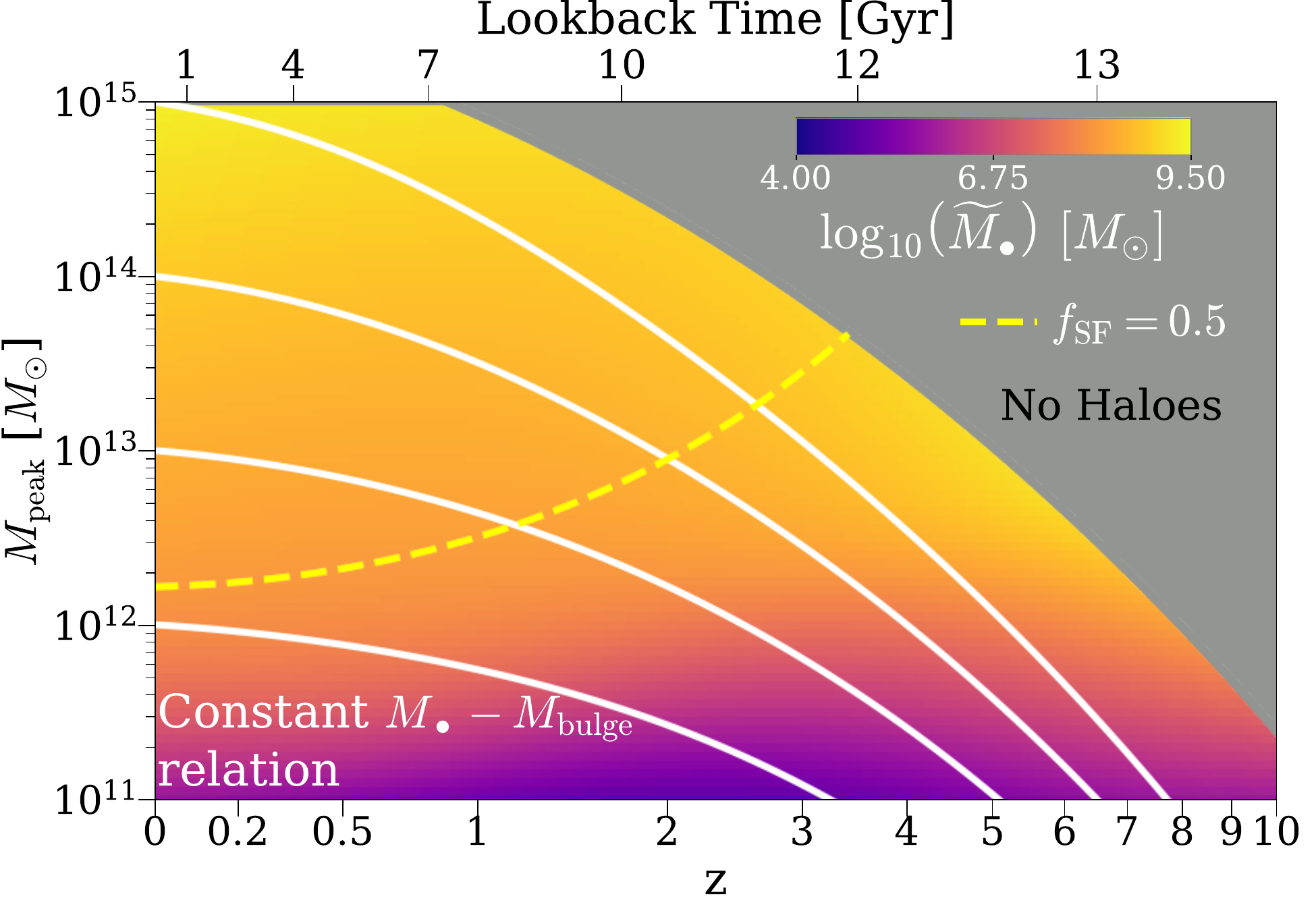}
}
\vspace{-0.5cm}
\subfigure{
\includegraphics[width=0.48\textwidth]{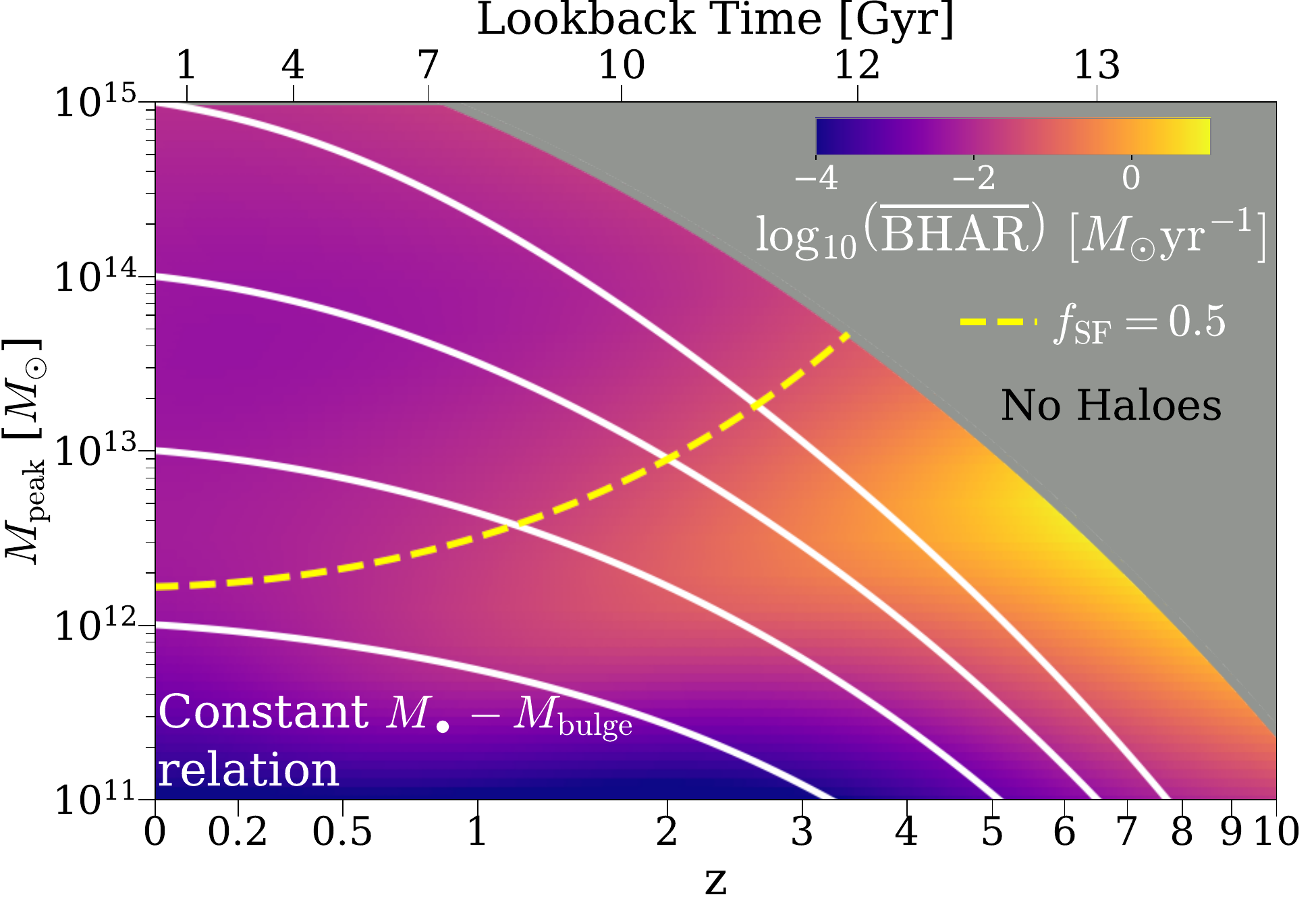}
}
\vspace{-0.5cm}
\subfigure{
\includegraphics[width=0.48\textwidth]{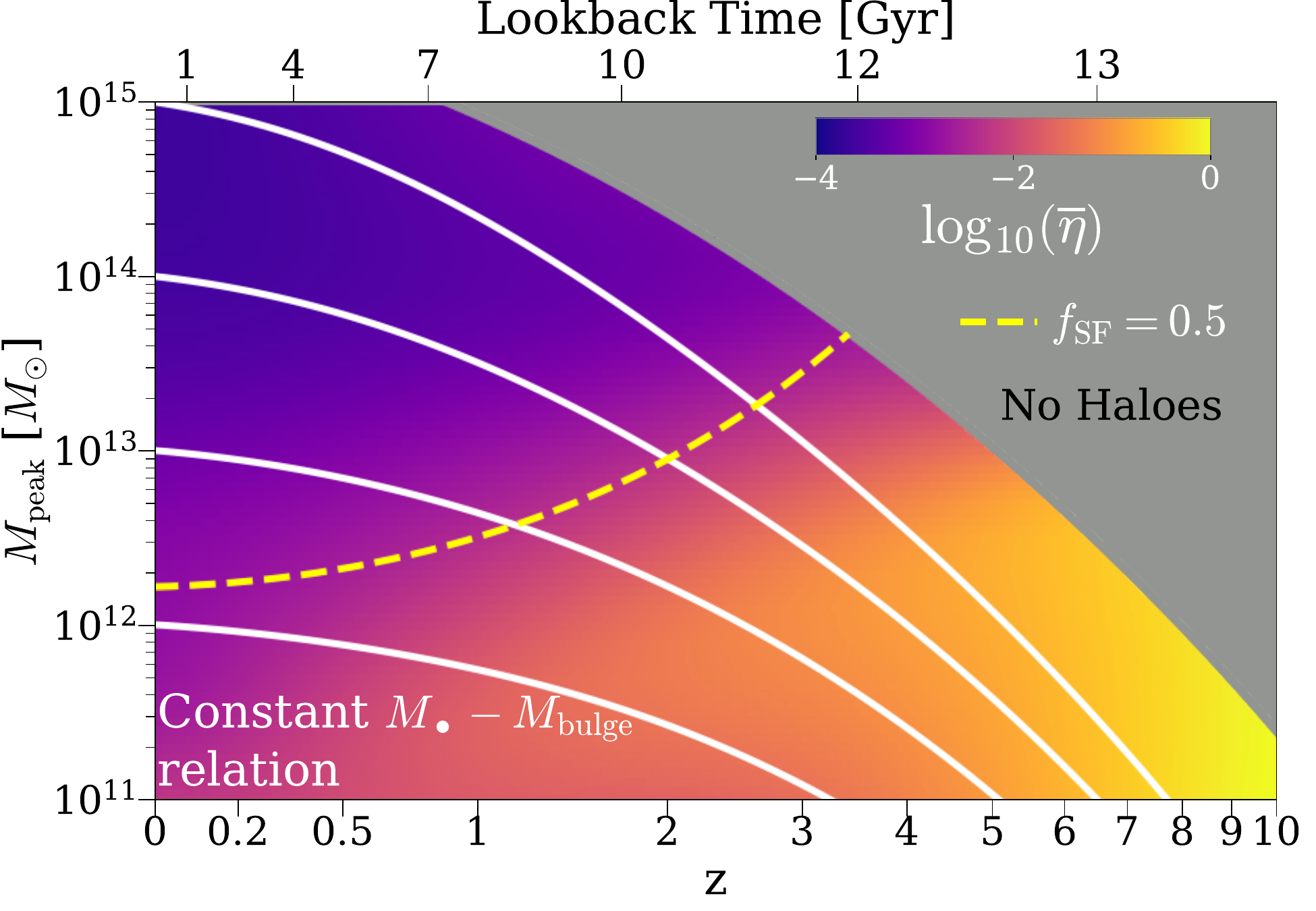}
}
\vspace{-0.5cm}
\subfigure{
\includegraphics[width=0.48\textwidth]{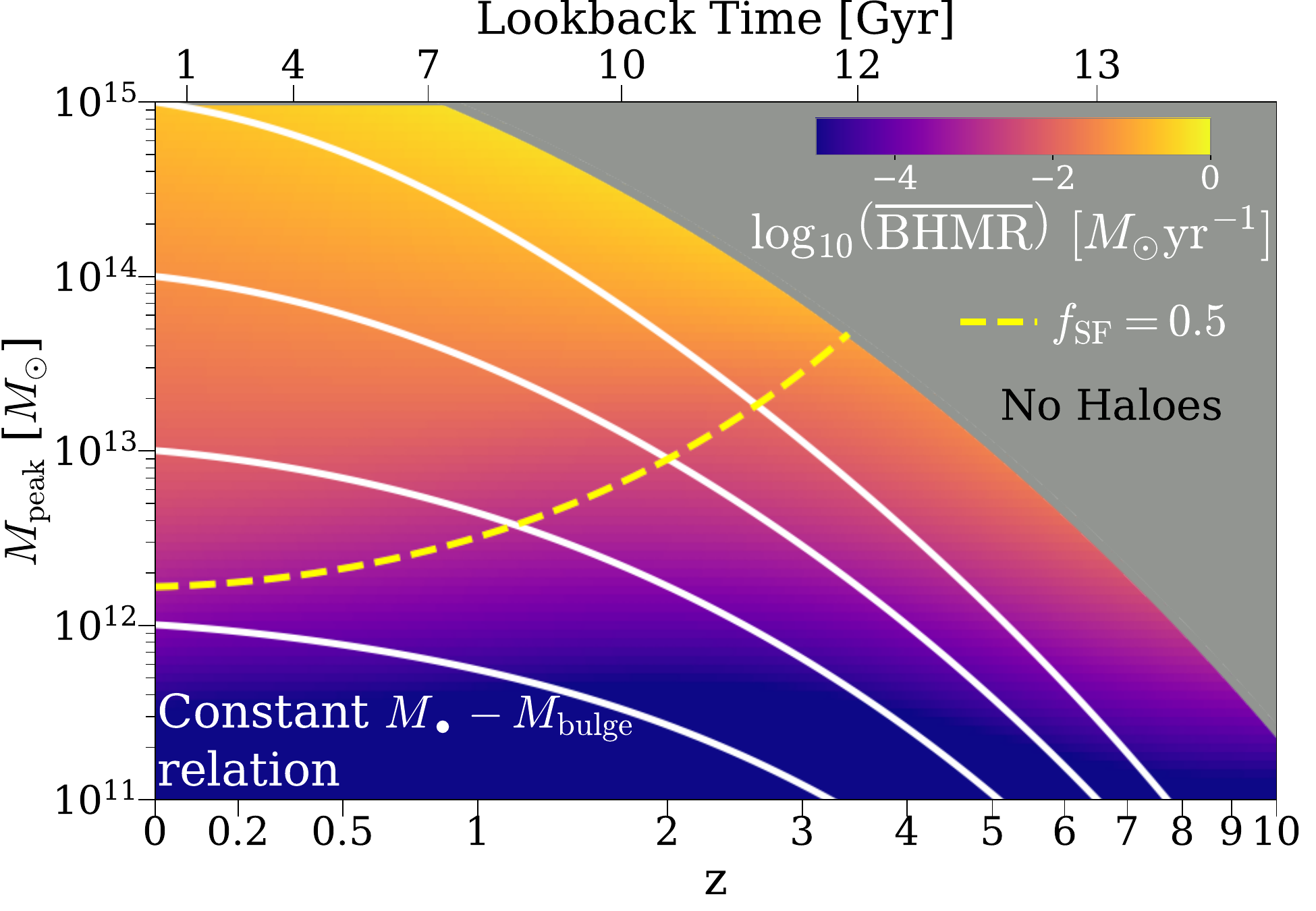}
}
\vspace{0.5cm}
\caption{The average \mbh{}, BHAR, Eddington ratio, and BHMR as functions of \mpeak{} and $z$ from the ``constant \bhbm{}'' model, where the \bhbm{} relation is redshift-independent (see Appendix \ref{a:const_bhbm}). \sfcurve{} \halocurves{} \nohalos{} All the data used to make this plot can be found \href{https://github.com/HaowenZhang/TRINITY/tree/main/plot_data}{here}.}
\label{f:const_bhbm}
\subfigure{
\includegraphics[width=0.48\textwidth]{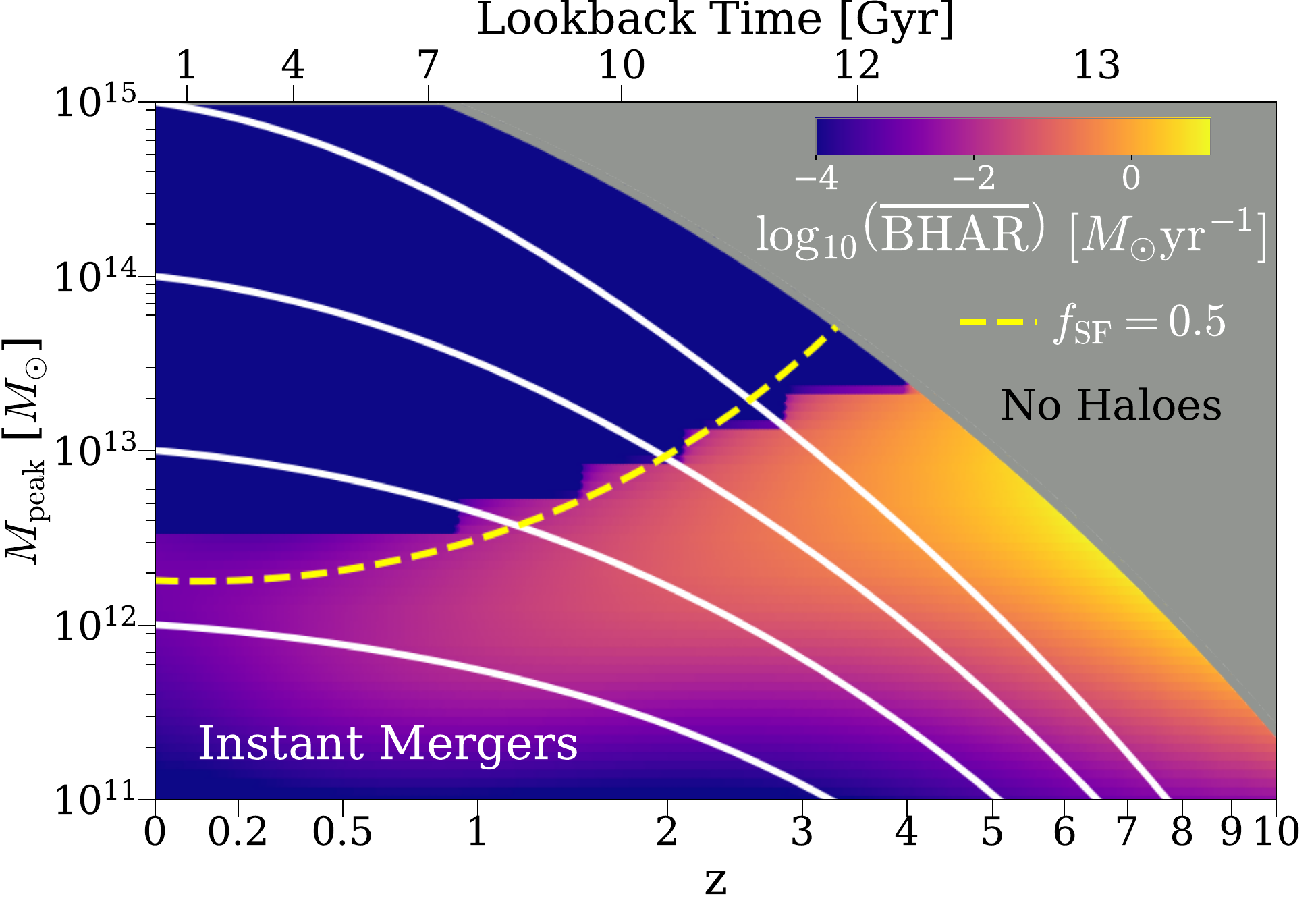}
}
\subfigure{
\includegraphics[width=0.48\textwidth]{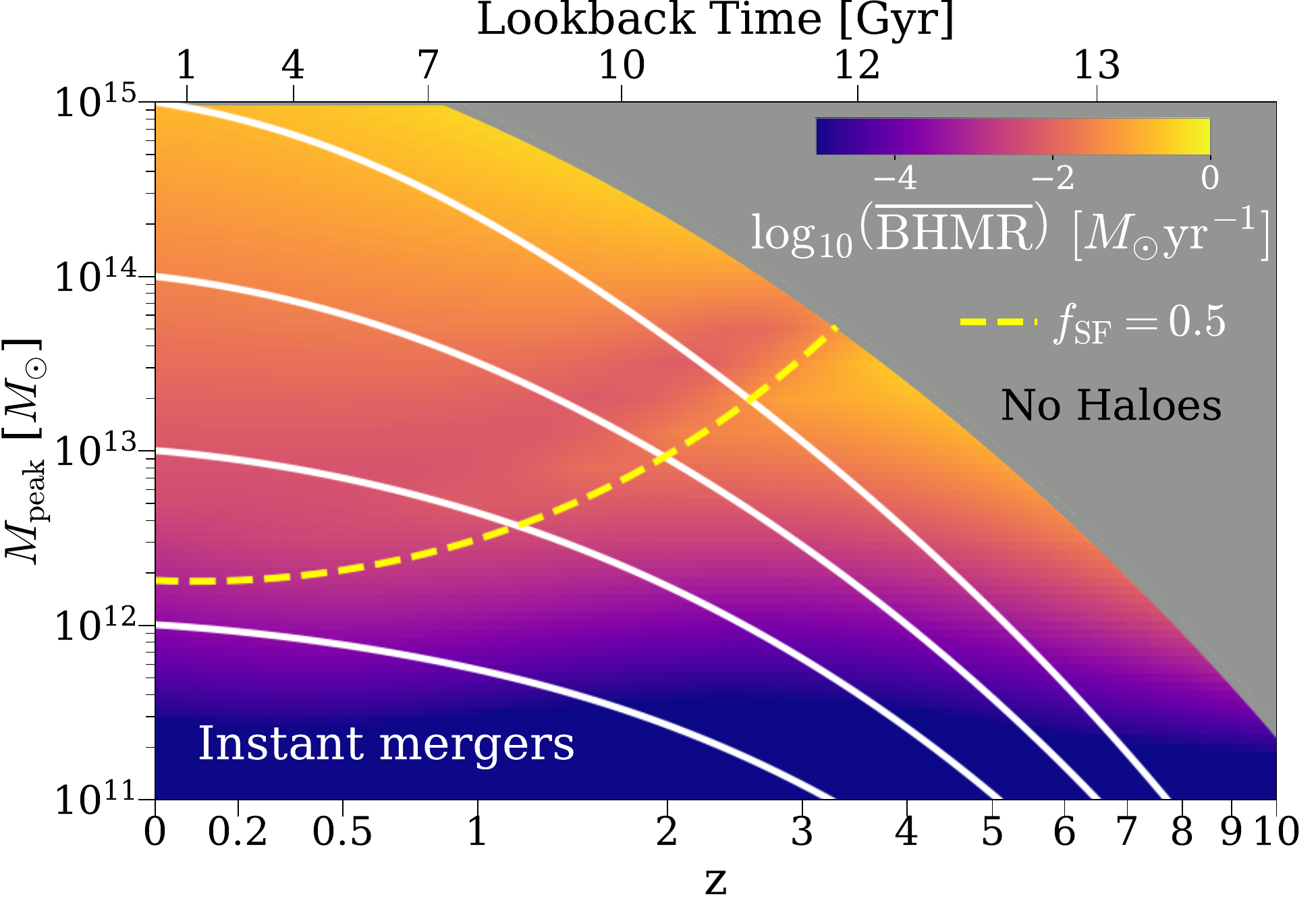}
}

\caption{The average BHAR ($\overline{\mathrm{BHAR}}$, left panel) and  average BHMR ($\overline{\mathrm{BHMR}}$, right panel) as a function of \mpeak{} and $z$ from the ``instant mergers'' model (see Appendix \ref{a:inst_merger}). ``Instant mergers'' means that all the infalling SMBHs in galaxy mergers are consumed immediately by the central SMBHs. \sfcurve{} \halocurves{} \nohalos{} All the data used to make this plot can be found \href{https://github.com/HaowenZhang/TRINITY/tree/main/plot_data}{here}.}
\label{f:bhar_bhmr_inst_merger}
\end{figure*}

\begin{figure}
\subfigure{
\includegraphics[width=0.48\textwidth]{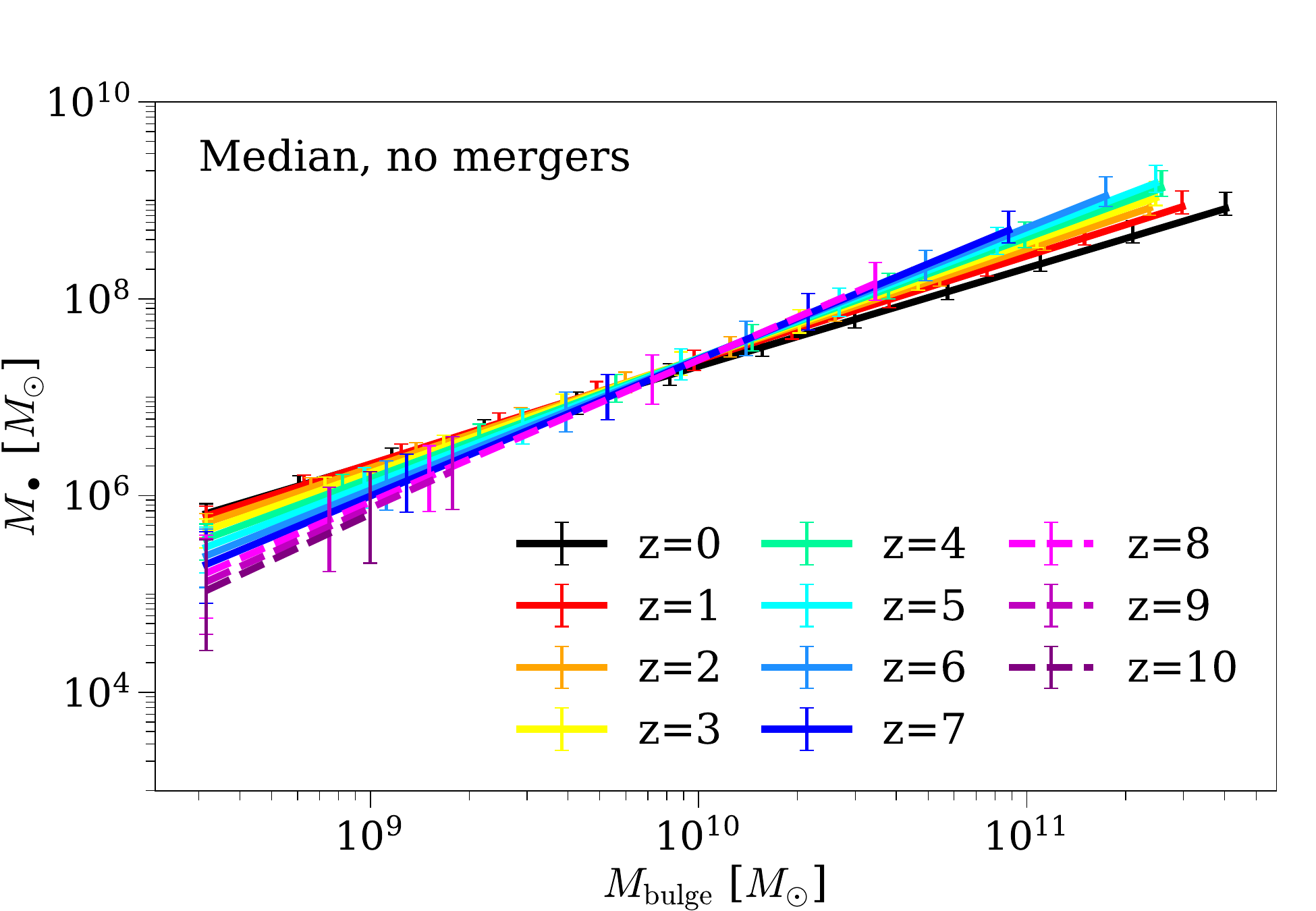}
}
\subfigure{
\includegraphics[width=0.48\textwidth]{figs/submit_fiducial/BHBM_median_submit_fiducial.pdf}
}
\subfigure{
\includegraphics[width=0.48\textwidth]{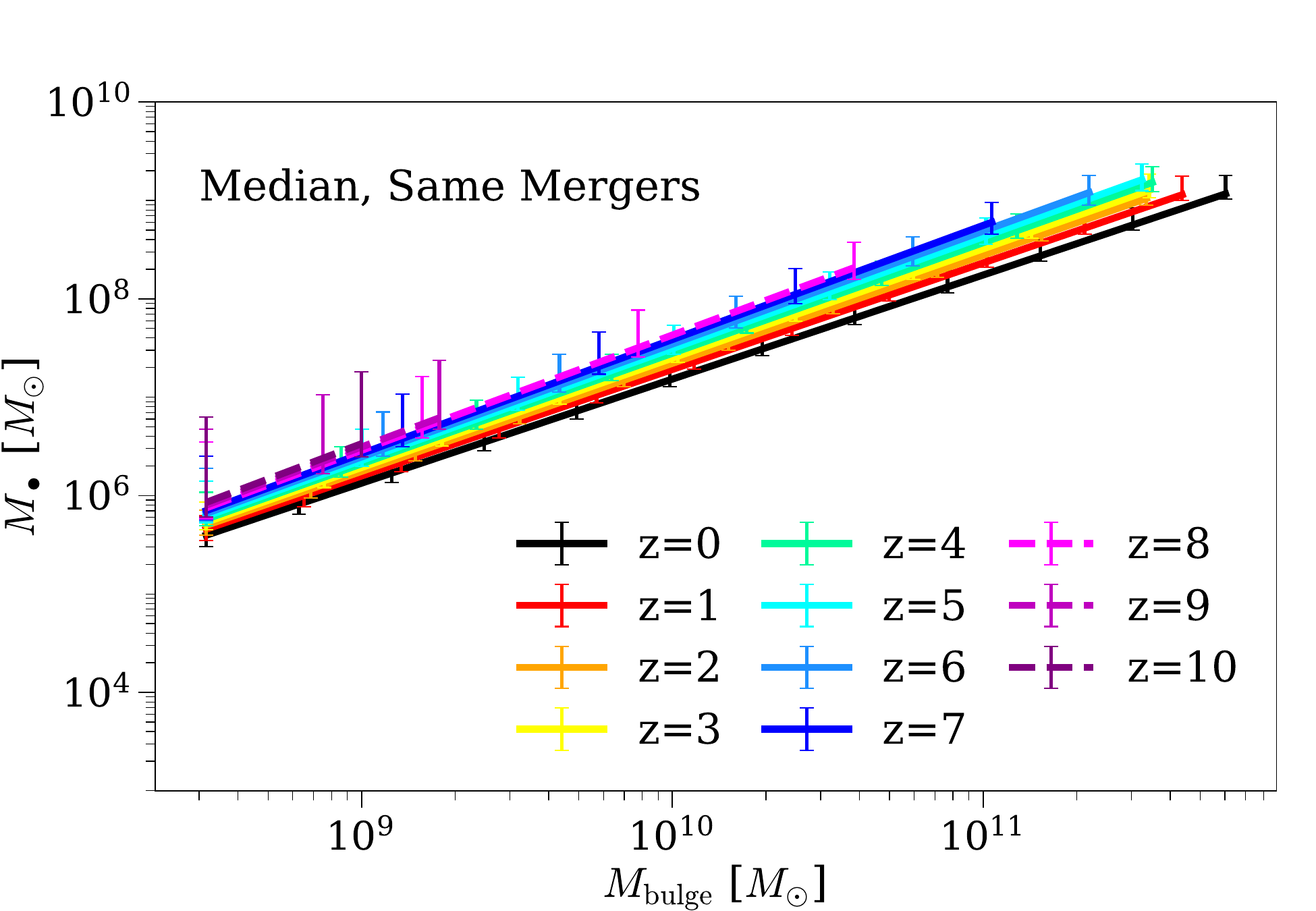}
}
\caption{The median \bhbm{} relations as a function of $z$ from the ``no mergers'' model (top panel, no SMBH mergers take place), the fiducial model (middle panel), and the ``same mergers'' model (bottom panel, the fractional merger contribution to SMBH growth being the same as that for galaxy growth). See Appendix \ref{a:no_or_same_merger}. All the data used to make this plot can be found \href{https://github.com/HaowenZhang/TRINITY/tree/main/plot_data}{here}.}
\label{f:bhbm_no_merger}
\end{figure}

\begin{figure}
\subfigure{
\includegraphics[width=0.48\textwidth]{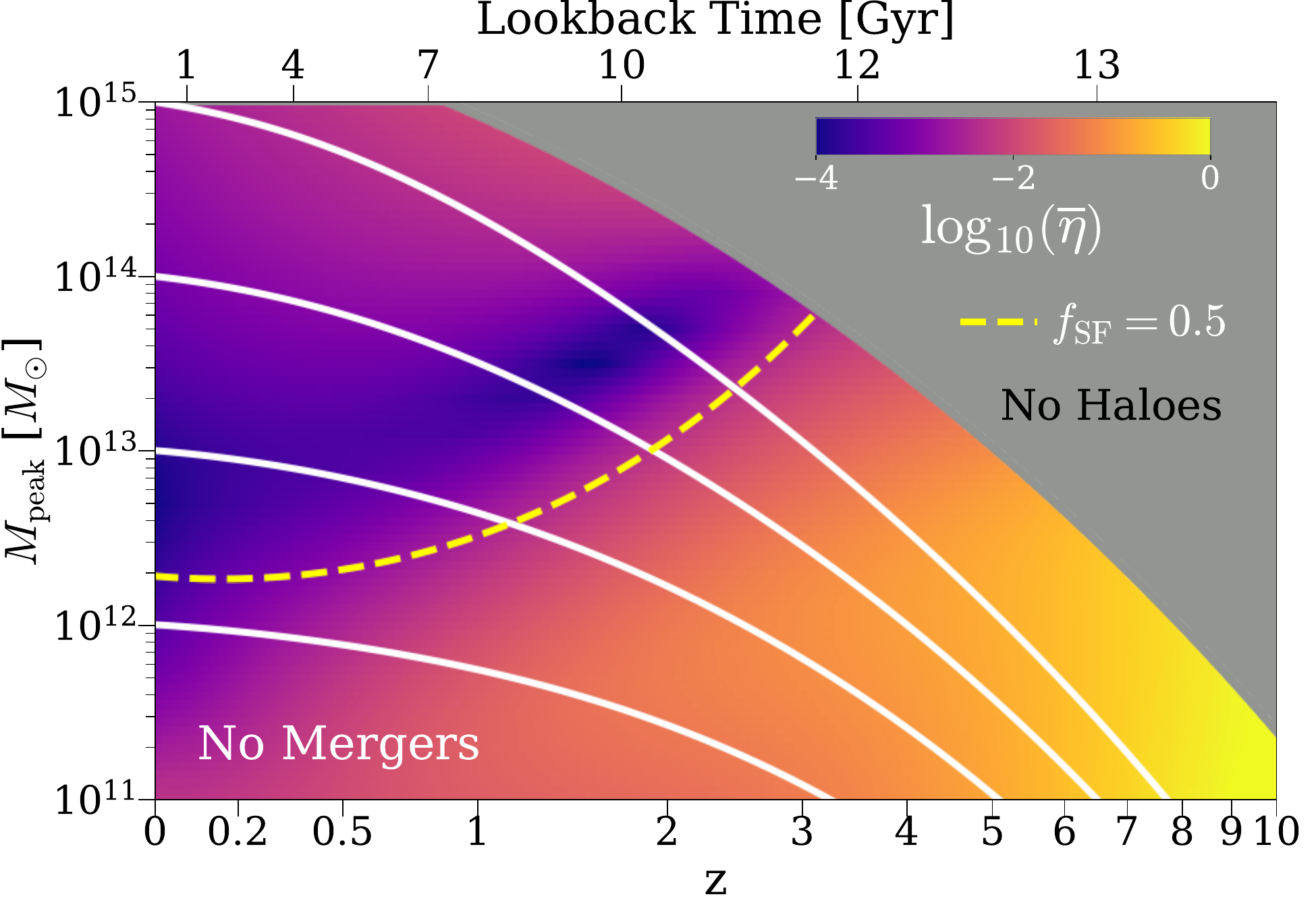}
}
\subfigure{
\includegraphics[width=0.48\textwidth]{figs/submit_fiducial/BHER_tot_submit_fiducial.pdf}
}
\subfigure{
\includegraphics[width=0.48\textwidth]{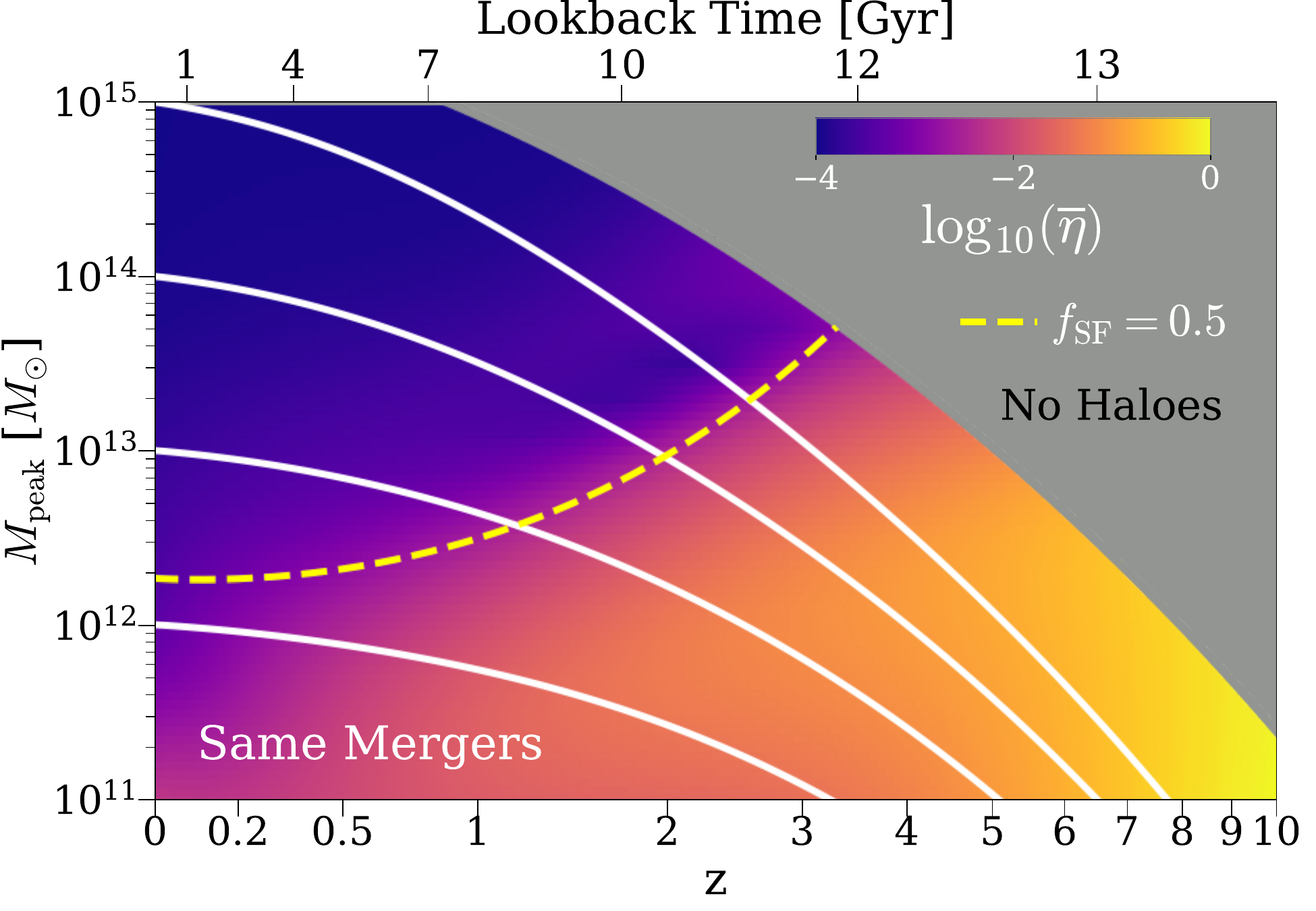}
}
\caption{The average Eddington ratios as functions of \mpeak{} and $z$ from the ``no mergers'' model (top panel), the fiducial model (middle panel), and the ``same mergers'' model (bottom panel). See Appendix \ref{a:no_or_same_merger}. \sfcurve{} \halocurves{} \nohalos{} All the data used to make this plot can be found \href{https://github.com/HaowenZhang/TRINITY/tree/main/plot_data}{here}.}
\label{f:bher_no_merger}
\end{figure}

In the fiducial model, we assume a redshift-dependent \bhbm{} relation. Here, we show the results from the ``constant \bhbm{}'' model, where the redshift dependence is dropped. The best-fitting ``constant \bhbm{}'' model gives $\log_{10} \widetilde{M}_{\bullet}  = 8.378^{+0.161}_{-0.079} + 1.076^{+0.034}_{-0.034} \log_{10} \left(\frac{M_{\mathrm{bulge}}}{10^{11}M_{\odot}}\right)$, which is consistent with the one from the fiducial model: $\log_{10} \widetilde{M}_{\bullet}  = 8.342^{+0.091}_{-0.089} + 1.028^{+0.053}_{-0.035} \log_{10} \left(\frac{M_{\mathrm{bulge}}}{10^{11}M_{\odot}}\right)$ (also see Appendix \ref{a:param_values}). However, these two models differ in the correlation coefficient between SMBH average accretion rate and \mbh{} at fixed host halo mass, \rhobh{}. As shown in Fig.\ \ref{f:rho_bh_const_bhbm}, the ``constant \bhbm{}'' model predicts significantly stronger correlation than the fiducial model. This is because in the fiducial model, the slope of the \bhbm{} relation grows slightly towards higher redshifts, which naturally assigns more accretion to more massive SMBHs. Without this degree of freedom, the ``constant \bhbm{}'' model needs higher \rhobh{} values to reproduce the AGN data from massive galaxies. Fig.\ \ref{f:const_bhbm} shows the average \mbh{}, BHAR, Eddington ratio, and BHMR as functions of \mpeak{} and $z$. The results are qualitatively consistent with the fiducial results. Quantitatively, the ``constant \bhbm{}'' model predicts lower SMBH accretion rates and Eddington ratios at $M_\mathrm{peak} \gtrsim 10^{13} M_\odot$ and $z\lesssim 3$.

\subsection{Different assumptions about galaxy/BH mergers}
\label{a:other_merger_models}

Several previous studies opted to ignore mergers (e.g., \citealt{Marconi2004}), or made simple assumptions by linking SMBH mergers to halo mergers (e.g., \citealt{Shankar2013}). Here, we show the main results from \textsc{Trinity} with alternate assumptions about SMBH mergers.

\subsubsection{Instant SMBH coalescence following halo mergers}
\label{a:inst_merger}

One extreme case is the ``instant mergers'' scenario, i.e.,  there is little delay between halo mergers and the coalescence of SMBHs. In this case, the central SMBH consumes \emph{all} infalling SMBHs, regardless of how much of the infalling stellar mass is merged into the central galaxy vs.\ the intracluster light (ICL) (\S\ref{ss:halo_galaxy_connection}). Fig.\ \ref{f:bhar_bhmr_inst_merger} shows the average BHAR (left panel) and BHMR (right panel) from the ``instant mergers'' model. It is clear that by forcing all the infalling satellite SMBHs to merge with central SMBHs, the vast majority of massive black hole growth at low redshifts must have been due to mergers, leaving little room for accretion. As a result, we see a precipitous drop in BHAR above $M_\mathrm{peak}\sim 10^{13} M_\odot$ below $z\sim 4$. Given that these low BHARs are in conflict with observations like \citet{Hlavacek-Larrondo2015} and \citet{McDonald2021} that show significant massive black hole accretion, we do not show other results from this model.

\subsubsection{No SMBH mergers or identical fractional merger contributions to SMBH and galaxy growth}
\label{a:no_or_same_merger}

In the fiducial model, we assume that the fractional merger contribution to SMBH and galaxy growth are proportional to each other. From the posterior parameter distribution, we found that the merger contribution to SMBH growth is smaller than the contribution to galaxy growth, i.e., $0<f_\mathrm{scale}<1$. Here, we consider two extreme cases.  First, if the delay between galaxy mergers and the ensuing SMBH coalescence is sufficiently long, SMBH mergers would be rare, and the merger contribution to central SMBH growth becomes negligible. In this extreme case, we can assume that no SMBH mergers take place, and all central SMBH growth comes from accretion.  In this ``no mergers'' model, $f_\mathrm{scale}\equiv0$ for all galaxies. The second extreme case we consider is if the fractional merger contributions to SMBH and galaxy growth are identical, i.e., $f_\mathrm{scale}\equiv1$. In the following, we call this scenario the ``same mergers'' model.

Fig.\ \ref{f:bhbm_no_merger} shows the resulting \bhbm{} relations as functions of $z$ from the ``no mergers'' model (top panel), the fiducial model (middle panel), and the ``same mergers'' model (bottom panel). The redshift evolution from all three models is largely consistent at $M_\mathrm{bulge}\gtrsim 10^{10.5} M_\odot$. Below $M_\mathrm{bulge}\sim 10^{10.5} M_\odot$, the ``same mergers'' model predicts quantitatively higher \mbh{} at fixed \mbulge{} (or \mstar{}), and thus less SMBH mass growth. The bigger merger fraction depletes wandering SMBHs in low mass galaxies before the predicted SMBH merger rates are fully accounted for, if the total SMBH growth is kept the same. Therefore, the total SMBH mass growth must be decreased to avoid such depletion.

Fig.\ \ref{f:bher_no_merger} shows the average Eddington ratios as functions of \mpeak{} and $z$ from the ``no mergers'' model (top panel), the fiducial model (middle panel), and the ``same mergers'' model (bottom panel). The main difference between these three models is the average Eddington ratios of halos with $M_\mathrm{peak} \gtrsim 10^{14} M_\odot$ below $z\sim 2$. From the top panel to the bottom panel, \textsc{Trinity} attributes more and more SMBH growth to mergers among these halos, producing lower and lower average Eddington ratios. However, the general ``downsizing'' picture holds qualitatively across all these models.

\section{The systematic effect of varying star formation histories on SMBH growth histories}
\label{a:rho_bhar_sfr}

\begin{figure}
\includegraphics[width=0.48\textwidth]{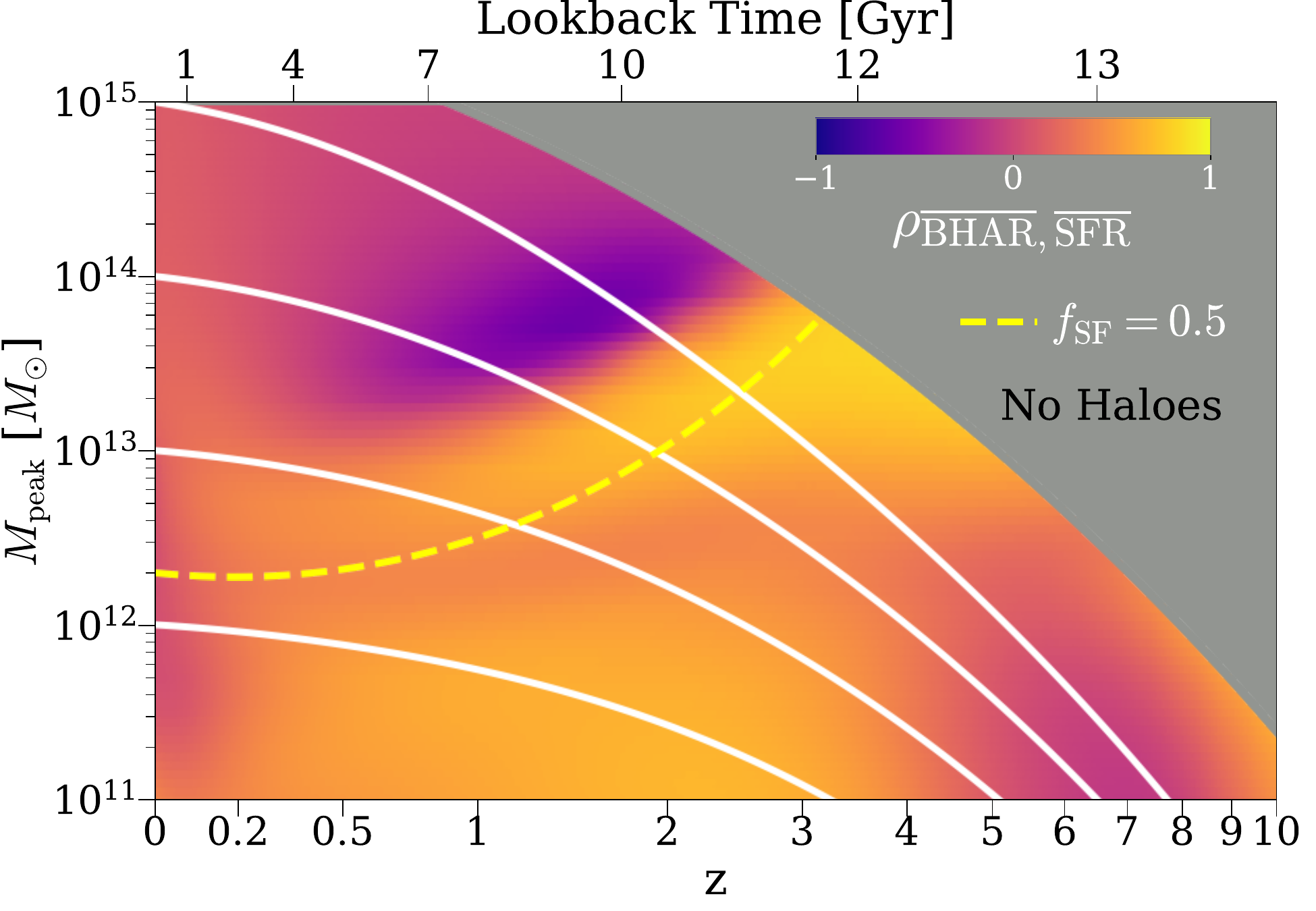}
\caption{The correlation coefficient between average SMBH accretion and average galaxy star formation rate, $\rho_{\overline{\mathrm{BHAR}}, \overline{\mathrm{SFR}}}$, as functions of \mpeak{} and $z$ from the fiducial model. See Appendix \ref{a:rho_bhar_sfr}. \sfcurve{} \halocurves{} \nohalos{} All the data used to make this plot can be found \href{https://github.com/HaowenZhang/TRINITY/tree/main/plot_data}{here}.}
\label{f:rho_bhar_sfr}
\end{figure}

In \textsc{Trinity}, we construct the galaxy--SMBH connection such that $M_\bullet$ is a function of the galaxy stellar mass. Stellar masses are calculated by integrating over galaxies' assembly histories. Consequently, a systematic change in the star formation histories could in principle alter the SMBH growth histories from \textsc{Trinity}. To quantify the sensitivity of SMBH accretion rates to the change in galaxy star formation rates, we: 1) calculate average BHARs and SFRs as functions of \mpeak{} and $z$ for a representative subset of the MCMC chain; and then 2) calculate the correlation coefficient between the log of average BHAR and the log of average SFR, as a function of \mpeak{} and $z$ (Fig.\ \ref{f:rho_bhar_sfr}).

At $0<z\lesssim 3$ and $M_\mathrm{peak} < 13$, there is a  moderate positive correlation between the average BHAR and SFR. This is because in this regime, systematically increasing SFR leads to larger galaxy stellar masses. To reproduce higher QPDF values in more massive stellar mass bins, as suggested by \citet{Aird2018} (\S\ref{ss:agn_observables}), the BHAR needs to increase as well. Over $3<z\lesssim 5$, we technically do not have QPDF constraints for different galaxy mass bins. Therefore, the positive correlation degrades towards higher redshifts. Around $z\sim 6$, the correlation becomes negligible. This is likely because we do not have any observational constraints at such a high redshift, except for the prior against super-Eddington quasars (\S\ref{ss:agn_observables}). With such prior knowledge, \textsc{Trinity} would not be forced to adjust BHAR along with any SFR change in this cosmic era.

It is also worth noting that at $M_\mathrm{peak} \gtrsim 10^{13} M_\odot$, there is a region with apparent negative correlation between the average BHAR and SFR. However, this is also the region where it is hard to robustly constrain SFRs as a function of halo mass. Thus, without better data constraints, we refrain from trying to explain its origin.

\begin{figure}
\includegraphics[width=0.48\textwidth]{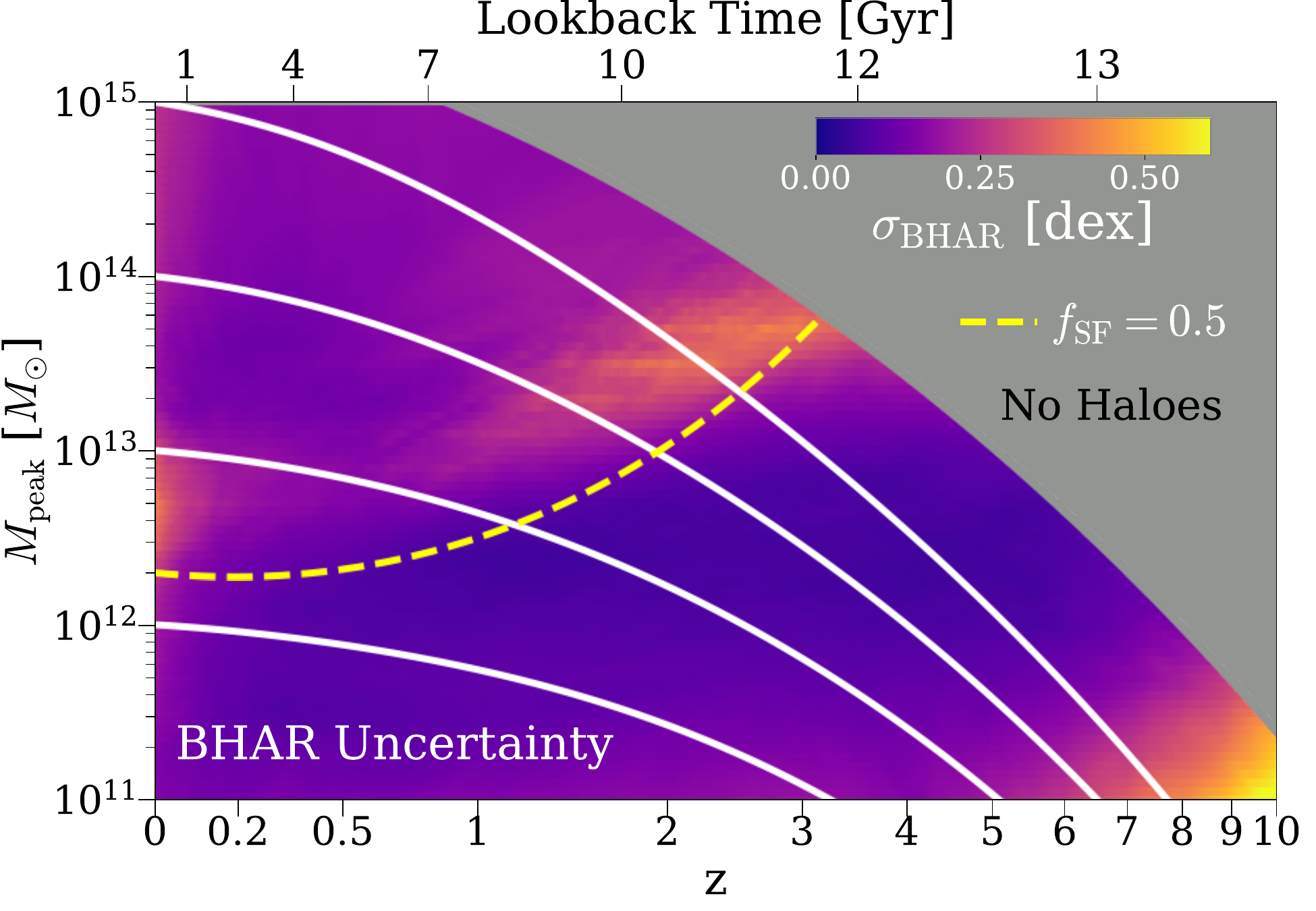}
\caption{The 1$\sigma$ uncertainty (from MCMC) in average SMBH accretion rate, $\sigma_{\mathrm{BHAR}}$ (in dex), as a function of \mpeak{} and $z$ from the fiducial model. See Appendix \ref{a:rho_bhar_sfr}. \sfcurve{} \halocurves{} \nohalos{} All the data used to make this plot can be found \href{https://github.com/HaowenZhang/TRINITY/tree/main/plot_data}{here}.}
\label{f:sigma_bhar}
\end{figure}

Fig.\ \ref{f:sigma_bhar} shows the scatter in average SMBH accretion rate as a function of \mpeak{} and $z$. Below $M_\mathrm{peak}\sim 10^{13} M_\odot$ and $z\sim 6$, the scatter in BHAR remains around 0.1 dex. Above $M_\mathrm{peak}\sim 10^{13} M_\odot$, the \mstar{}--\mpeak{} relation flattens, and thus galaxies with similar \mstar{} can be found in a broader range of halo mass bins. This weakens the QPDFs' ability to constrain BHAR at fixed halo mass, because QPDFs are divided in different \mstar{} bins. Ultimately, the uncertainties in BHAR are higher among more massive haloes. On the other hand, we do not have any constraints for AGNs at $z\gtrsim 6$. Thus, we see a significant increase in $\sigma_{\mathrm{BHAR}}$ with redshift between $6 < z < 10$.

\section{Technical details about the calculation of $\chi^2$}
\label{a:chi2}

Here, we introduce the details of the $\chi^2$ calculation for any given model parameter set. In \textsc{Trinity}, we firstly convert data points and their uncertainties into log units if they are in linear units. For the $i$-th data point with a value of $y_{i\ -e'_{\mathrm{low},i}}^{\ \ +e'_{\mathrm{high},i}}$, we then convolve the error bars with a calculation tolerance of 0.01 dex:

\begin{equation}
    e_{\mathrm{low/high},i} = \sqrt{e_{\mathrm{low/high},i}^{\prime 2} + 0.01^2}\ .\label{e:err_convolve}
\end{equation}
This calculation tolerance is set to prevent the model from overfitting to data points with very small confidence intervals.
For this data point, suppose we have a model prediction, $\hat{y}_i$. If $|\hat{y}_i - y_i| \le \epsilon_\mathrm{fit}\equiv 0.02$, then we assume that the model reproduces the data point sufficiently well, and ignore its contribution to the total $\chi^2$. This error threshold is effectively a tolerance for the deviation of the analytical parametrizations from the actual scaling relations. If $|\hat{y}_i - y_i| > \epsilon_\mathrm{fit}$, we define:

\begin{equation}
    \Delta y_i = 
    \begin{cases}
      \hat{y}_i - y_i - \epsilon_\mathrm{fit}, & \hat{y}_i > y_i\\
      \hat{y}_i - y_i + \epsilon_\mathrm{fit}, & 
      \hat{y}_i < y_i\\
    \end{cases}\ ,
\end{equation}
and the $\chi^2_i$ for this data point is:

\begin{equation}
\label{e:chi2}
    \chi^2_i = 
    \begin{cases}
      \left(\Delta y_i / e_{\mathrm{low},i}\right)^2, & \Delta y_i < -e_{\mathrm{low},i}\\
      \left(\Delta y_i / e_{\mathrm{high},i}\right)^2, & \Delta y_i > e_{\mathrm{high},i}\\
      \left(\Delta y_i / e_{\mathrm{med},i}\right)^2, & \rm otherwise\\
      
    \end{cases}\ ,
\end{equation}
where $e_{\mathrm{med},i}$ is a linear function of $\Delta y_i$:

\begin{equation}
    e_{\mathrm{med},i}(\Delta y_i) = e_{\mathrm{low},i} + \frac{\Delta y_i + e_{\mathrm{low},i}}{e_{\mathrm{high},i} + e_{\mathrm{low},i}} \cdot (e_{\mathrm{high},i} - e_{\mathrm{low},i})\ .
\end{equation}
This definition is adopted to account for asymmetry in error bars, such that $e_{\mathrm{med},i} = e_{\mathrm{low},i}$ when $\Delta y_i = -e_{\mathrm{low},i}$ and $e_{\mathrm{med},i} = e_{\mathrm{high},i}$ when $\Delta y_i = e_{\mathrm{high},i}$. The total $\chi^2$ is a summation of $\chi^2_i$ over all the data points and the priors listed in Table \ref{t:priors}:
\begin{equation}
    \chi^2 = \sum_{i} \chi^2_i\ +\  \mathrm{priors}.
\end{equation}

\section{Best fitting parameter values}
\label{a:param_values}

The resulting best-fitting and 68\% confidence intervals for the posterior distributions follow:
\\

\noindent\textbf{Median Star Formation Rates}: 

Characteristic $v_{\rm Mpeak}$ [km s$^{-1}$] (Eq.\ \ref{e:v_1}):
\begin{flalign*}
\begin{aligned}
    \log_{10}\left(V\right) =&\ 2.289^{+0.017}_{-0.051} + (1.548^{+0.197}_{-0.221})\left(a - 1\right)\\
    &+  (1.218^{+0.147}_{-0.142})\ln\left(1 + z\right) + (-0.087^{+0.021}_{-0.021})z &&
\end{aligned}
\end{flalign*}

Characteristic SFR [$M_{\odot}$ yr$^{-1}$] (Eq.\ \ref{e:epsilon}):
\begin{flalign*}
\begin{aligned}
    \log_{10}\left(\epsilon\right)=&\ 0.556^{+0.045}_{-0.246} + (-0.944^{+1.133}_{-0.481})\left(a - 1\right)\\ &+ (-0.042^{+0.887}_{-0.325})\ln\left(1+z\right) + (0.418^{+0.054}_{-0.132})z &&
\end{aligned}
\end{flalign*}

Faint-end slope of SFR--$v_{\rm Mpeak}$ relation (Eq.\ \ref{e:alpha}):
\begin{flalign*}
\begin{aligned}
    \alpha =&\ -3.907^{+0.148}_{-0.362} + (32.223^{+2.456}_{-1.724})\left(a - 1\right)\\ &+  (20.241^{+1.627}_{-1.117})\ln\left(1 + z\right) +(-2.193^{+0.166}_{-0.175})z &&
\end{aligned}
\end{flalign*}

Massive-end slope of SFR--$v_{\rm Mpeak}$ relation (Eq.\ \ref{e:beta}):
\begin{flalign*}
\begin{aligned}
    \beta =&\ 0.329^{+0.239}_{-0.849} + (2.342^{+1.205}_{-0.953})\left(a - 1\right)\\ &+  (0.492^{+0.190}_{-0.154})z &&
\end{aligned}
\end{flalign*}

\noindent\textbf{Quenched Fractions}:

Characteristic $v_\mathrm{max}$ for quenching [km/s] (Eq.\ \ref{e:m_quenched}):
\begin{flalign*}
\begin{aligned}
    \log_{10} (v_{\rm Q}) =&\ 2.337^{+0.013}_{-0.030} + (0.316^{+0.059}_{-0.143})\left(a - 1\right)\\
    &+ (0.283^{+0.022}_{-0.038})z &&
\end{aligned}
\end{flalign*}

Width in log-$v_\mathrm{max}$ for quenching [dex] (Eq.\ \ref{e:w_quenched}): 
\begin{flalign*}
\begin{aligned}
    w_{\rm Q} =&\ 0.193^{+0.018}_{-0.030} + (0.256^{+0.060}_{-0.126})\left(a - 1\right)\\
    &+ (0.062^{+0.018}_{-0.028})z &&
\end{aligned}
\end{flalign*}

\noindent\textbf{Galaxy Mergers} :

Fraction of merging satellites that are transferred to the central galaxy (Eq.\ \ref{e:mstar_sfr}):
\begin{flalign*}
\begin{aligned}
    \log_{10}\left(f_{\rm merge}\right) =&\ -0.748^{+0.066}_{-0.147}&&
\end{aligned}
\end{flalign*}

\noindent\textbf{The Halo--Galaxy Connection:}

$M_*$ scatter at fixed \mpeak{} [dex]:
\begin{flalign*}
\begin{aligned}
    \sigma_{*} = 0.279^{+0.004}_{-0.028}&&
\end{aligned}
\end{flalign*}

Correlation coefficient between SSFR and $M_*$ at fixed halo mass at $a=0.5$ (i.e., $z=1$) (Eq.\ \ref{e:rho05}):
\begin{flalign*}
\begin{aligned}
    \rho_{0.5} = 0.423^{+0.071}_{-0.100} &&
\end{aligned}
\end{flalign*}

\noindent\textbf{Systematics in Stellar Masses:}

Offset between the true and the measured $M_*$ [dex] (Eq.\ \ref{e:mu_z}):
\begin{flalign*}
\begin{aligned}
    \mu = -0.111^{+0.127}_{-0.023} + (0.159^{+0.053}_{-0.043})\left(a - 1\right) &&
\end{aligned}
\end{flalign*}

Additional systematic offset between the true and the measured SFRs (Eq.\ \ref{e:mu_kappa}):
\begin{flalign*}
\begin{aligned}
    \kappa = 0.259^{+0.035}_{-0.025}&&
\end{aligned}
\end{flalign*}

Scatter between the observed and the true $M_*$ [dex] (Eq.\ \ref{e:sm_scatter}):
\begin{flalign*}
\begin{aligned}
    \sigma = \min \{0.07 + 0.044^{+0.010}_{-0.008}\left(z - 0.1\right), 0.3\} 
\end{aligned}
\end{flalign*}

\noindent\textbf{Galaxy--SMBH Connection:}

Minimum SMBH occupation fraction (Eq.\ \ref{e:focc_min}):
\begin{flalign*}
\begin{aligned}
    \log_{10}(f_{\mathrm{occ,min}}) =&\ -2.640^{+2.285}_{-1.053} + (0.089^{+0.577}_{-1.277})\left(a - 1\right)\\
\end{aligned}
\end{flalign*}
Characteristic halo mass and mass width where the SMBH occupation fraction changes significantly (Eqs.\ \ref{e:mh_c}-\ref{e:wh_c}):
\begin{flalign*}
\begin{aligned}
    \log_{10}(M_\mathrm{h,c}) =&\ 10.804^{+2.792}_{-0.366} + (-14.220^{+6.976}_{-5.409})\left(a - 1\right)\\
    w_\mathrm{h,c} =&\ 3.355^{+0.266}_{-2.276} + (-0.574^{+4.948}_{-0.048})\left(a - 1\right)\\
\end{aligned}
\end{flalign*}

Slope and zero point of the SMBH mass -- bulge mass (\bhbm{}) relation (Eqs.\ \ref{e:beta_bh}-\ref{e:gamma_bh}):
\begin{flalign*}
\begin{aligned}
    \gamma_{\rm BH} =&\ 1.028^{+0.053}_{-0.035} + (0.036^{+0.043}_{-0.125})\left(a - 1\right)\\
    &+ (0.052^{+0.023}_{-0.033})z \\
    \beta_{\rm BH} =&\ 8.343^{+0.091}_{-0.089} + (-0.173^{+0.047}_{-0.012})\left(a - 1\right)\\
    &+ (0.044^{+0.025}_{-0.013})z &&
\end{aligned}
\end{flalign*}

Scatter in the \bhbm{} relation [dex] (Eq.\ \ref{e:scatter_bhhm}):
\begin{flalign*}
\begin{aligned}
    \sigma_{\rm BH} =\ 0.269^{+0.051}_{-0.022}
\end{aligned}
\end{flalign*}

\noindent\textbf{SMBH Mergers:}

The fraction of SMBH growth due to mergers, relative to the fraction of galaxy growth due to mergers (Eq.\ \ref{e:f_merge_bh_z}):
\begin{flalign*}
\begin{aligned}
    \log_{10} \left(f_\mathrm{scale}\right) =&\ -0.192^{+0.127}_{-1.494} + (0.000^{+1.640}_{-0.316})\left(a - 1\right) &&
\end{aligned}
\end{flalign*}

\noindent\textbf{AGN Properties:}

AGN duty cycles (Eqs.\ \ref{e:duty_cycle_mh}-\ref{e:duty_cycle_alpha}):
\begin{flalign*}
\begin{aligned}
    \log_{10}(M_{\mathrm{duty}}) =&\ 11.200^{+0.178}_{-0.003} + 1.269^{+0.049}_{-0.132} \ln \left(1 + z\right)  \\
    \alpha_{\mathrm{duty}} =&\ 4.692^{+0.175}_{-0.531} + (-2.723^{+0.313}_{-0.162}) \ln \left(1 + z\right)  \\
\end{aligned}
\end{flalign*}

Power-law indices of the Eddington ratio distributions (Eqs.\ \ref{e:erdf_c1} and \ref{e:erdf_c2}):
\begin{flalign*}
\begin{aligned}
    c_{1} =&\ 0.527^{-0.023}_{-0.201} + (1.261^{+0.070}_{-0.308})\left(a - 1\right) \\
    c_{2} =&\  2.970^{-0.015}_{-0.339} + (-1.151^{+0.285}_{-0.215})\left(a - 1\right)&&
\end{aligned}
\end{flalign*}

AGN energy efficiencies (Eq.\ \ref{e:eta_calc}):
\begin{flalign*}
\begin{aligned}
    \log_{10}(\epsilon_{\rm tot}) =&\ -1.318^{+0.114}_{-0.010}
\end{aligned}
\end{flalign*}

Correlation coefficient between SMBH accretion rate and mass at fixed halo mass (Eq.\ \ref{e:rho_bh}):
\begin{flalign*}
\begin{aligned}
    \rho_\mathrm{BH} =&\ 0.001^{+0.117}_{-0.105} + (0.071^{+0.025}_{-0.160})\left(a - 1\right)\\ &+  (0.123^{+0.005}_{-0.026})z &&
\end{aligned}
\end{flalign*}

\noindent\textbf{AGN Systematics:}

Offset in the Eddington ratio between \citet{Ueda2014} and \citet{Aird2018} [dex] (Eq.\ \ref{e:xi}):

\begin{flalign*}
\begin{aligned}
    \xi =&\ -0.497^{+0.101}_{-0.058} &&
\end{aligned}
\end{flalign*}

\section{Parameter Correlations}
\label{a:param_corrs}
\begin{figure*}
\includegraphics[width=1.8\columnwidth]{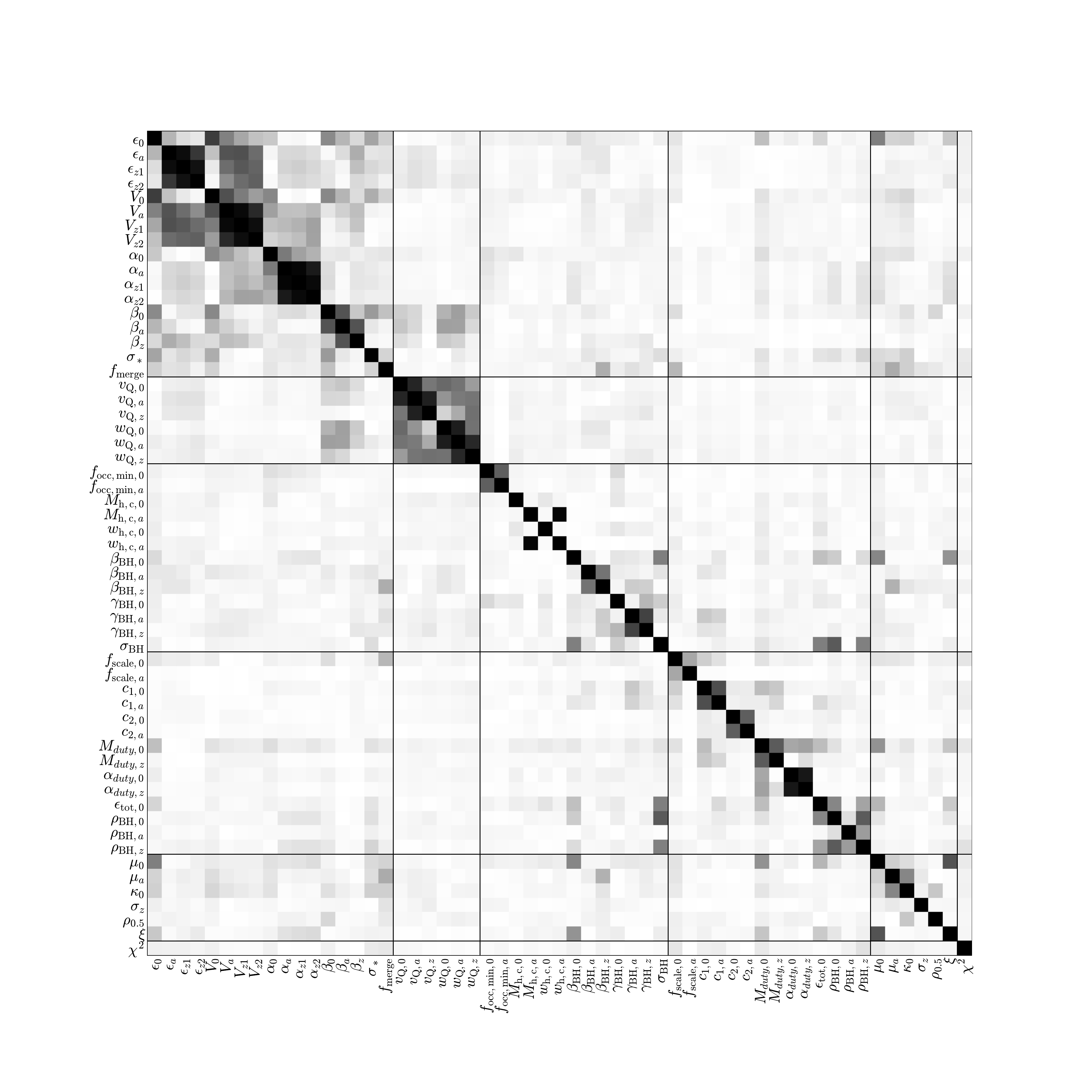}
\vspace{-3ex}
\caption{Rank correlation coefficients in the model posterior distribution. Darker shades indicate higher \emph{absolute values} of correlation coefficients (both positive and negative). See Appendix \ref{a:param_corrs}. All the data used to make this plot can be found \href{https://github.com/HaowenZhang/TRINITY/tree/main/plot_data}{here}.}
\label{f:param_corr}
\end{figure*}

Fig.\ \ref{f:param_corr} shows the rank correlation coefficients between all the model parameters, with darker shades indicating stronger (positive or negative) correlations. It is natural to see correlations between different redshift evolution terms of the same parameter (e.g., $\epsilon_a$ and $\epsilon_{z1}$), as each of them can partially mimic the behavior of others at certain redshift intervals. In other words, different redshift evolution terms are not orthogonal to each other.

%%%%%%%%%%%%%%%%%%%%%%%%%%%%%%%%%%%%%%%%%%%%%%%%%%

% Don't change these lines
\bsp	% typesetting comment
\label{lastpage}

\end{document}